\documentclass[11pt,a4paper]{article}

\usepackage{jheppub}

\usepackage[english]{babel}

\usepackage{subfigure}
\usepackage{bbold}

\allowdisplaybreaks[2]

\newcommand{\rev}[1]{#1}

\newcommand{\nn}{\nonumber}

\newcommand{\half}{{\textstyle\frac{1}{2}}}

\newcommand{\tvec}[1]{\boldsymbol{#1}}

\newcommand{\y}{\tvec{y}}
\newcommand{\ytilde}{\tilde{\tvec{y}}}

\newcommand{\one}{\mathbb{1}}

\renewcommand{\d}{\delta}
\newcommand{\D}{\Delta}

\newcommand{\oone}{\underset{x_1}{\otimes}}

\newcommand{\ul}[1]{\underline{#1}}

\newcommand{\us}{\text{us}}

\newcommand{\tr}{\operatorname{tr}}

\newcommand{\ms}{\mskip 1.5mu}
\newcommand{\bs}{\mskip -1.5mu}

\newcommand{\pr}[2]{{}^{#1}\bs #2}      
\newcommand{\prb}[2]{{}^{#1}\! #2}      
\newcommand{\prn}[2]{{}^{#1} #2}        

\newcommand{\ii}[2]{\varepsilon_{#1}(#2)}
\newcommand{\csgn}[2]{\eta_{#1}(#2)}

\title{Transverse momentum in double parton scattering: \\ factorisation,
  evolution and matching}

\author[a]{Maarten G. A. Buffing,}
\author[a]{Markus Diehl}
\author[b]{and Tomas Kasemets}

\affiliation[a]{Deutsches Elektronen-Synchrotron DESY, Notkestra{\ss}e 85,
  22607 Hamburg, Germany}
\affiliation[b]{Nikhef Theory Group and VU University Amsterdam, De
  Boelelaan 1081, 1081 HV Amsterdam, The Netherlands}

\emailAdd{mbuffing@physics.ucla.edu}
\emailAdd{markus.diehl@desy.de}
\emailAdd{kasemets@uni-mainz.de}

\abstract{We give a description of double parton scattering with measured
  transverse momenta in the final state, extending the formalism for
  factorisation and resummation developed by Collins, Soper and Sterman for
  the production of colourless particles.  After a detailed analysis of
  their colour structure, we derive and solve evolution equations in
  rapidity and renormalisation scale for the relevant soft factors and
  double parton distributions.  We show how in the perturbative regime,
  transverse momentum dependent double parton distributions can be expressed
  in terms of simpler nonperturbative quantities and compute several of the
  corresponding perturbative kernels at one-loop accuracy.  We then show how
  the coherent sum of single and double parton scattering can be simplified
  for perturbatively large transverse momenta, and we discuss to which order
  resummation can be performed with presently available results.  As an
  auxiliary result, we derive a simple form for the square root factor in
  the Collins construction of transverse momentum dependent parton
  distributions.}

\preprint{\vbox{
\hbox{NIKHEF 2016-028, DESY 17-014}}}

\begin{document}

\maketitle


\section{Introduction}
\label{sec:introduction}

Proton-proton collisions at high energies are sensitive to regions of phase
space where partons have small momentum fractions.  This implies high parton
densities and thus increases the importance of events in which two or more
partons in each proton take part in a hard interaction.  The most frequent
and best studied case of such multiple hard interactions is double parton
scattering (DPS).  This mechanism can be especially prominent in cross
sections depending on transverse momenta in the final state.  Its
theoretical description involves double parton distributions (DPDs), which
quantify the joint distribution of two partons inside a proton and contain a
wealth of information on correlations between the proton constituents.

Experimental measurements of double parton scattering contributions to
different final states dates back to experiments at the ISR
\cite{Akesson:1986iv} and SPS~\cite{Alitti:1991rd}.  A wide range of DPS
processes has been investigated at the
Tevatron~\cite{Abe:1993rv,Abe:1997bp,Abe:1997xk,Abazov:2009gc,%
  Abazov:2011rd,Abazov:2014fha,Abazov:2014qba,Abazov:2015fbl,Abazov:2015nnn}
and in run I of the LHC \cite{Aaij:2011yc,Aaij:2012dz,Aaij:2015wpa,%
  Aad:2013bjm,Aad:2014rua,Aad:2014kba,Aaboud:2016dea,Aaboud:2016fzt,%
  Chatrchyan:2013xxa,Khachatryan:2016ydm}.  An overview of most of these
measurements can be found in figure~14 of \cite{Aaboud:2016fzt}.  The
importance of such processes will be even more pronounced at the full LHC
energy (first results from run II are reported in \cite{CMS:2017jwx}) and at
future hadron colliders.

Theoretical and phenomenological analyses of DPS have a long history; an
overview of early work can be found in~\cite{Sjostrand:2004pf}.  Following
the seminal papers \cite{Paver:1982yp} and \cite{Mekhfi:1983az}, significant
effort has been invested in recent years to achieve a systematic theory
description of DPS \cite{Diehl:2011tt,Diehl:2011yj,%
  Manohar:2012jr,Diehl:2015bca}, aiming at the same level of rigour as has
been achieved for the familiar single parton scattering (SPS) mechanism.  A
formalism for combining the single and double parton scattering contributions
to the physical cross section without double counting was recently presented
in \cite{Diehl:2017kgu} -- a different scheme was proposed earlier
\cite{Manohar:2012pe}.  However, there remain several gaps in our
understanding, and the present paper aims at closing some of them.

A crucial aspect for understanding DPS are correlations between the two
partons that are probed in the reaction.  Correlations in spin and in colour
have been classified systematically
\cite{Mekhfi:1985dv,Diehl:2011yj,Manohar:2012jr} and will play an important
role in the present work.  Their size is poorly known, but can be limited by
positivity bounds \cite{Diehl:2013mla,Kasemets:2014yna}, which have similar
theoretical status as positivity constraints on single parton distribution
functions (PDFs).  Quark model calculations \cite{Chang:2012nw,%
  Rinaldi:2013vpa,Rinaldi:2014ddl,Rinaldi:2016jvu,Rinaldi:2016mlk,%
  Broniowski:2013xba,Broniowski:2016trx,Kasemets:2016nio} typically yield
strong correlations in the valence region, but for the region of small
momentum fractions $x$ there is little guidance from models so far.  The
decrease of spin correlations under evolution to higher scales has been
studied in \cite{Diehl:2014vaa}, and their influence on final-state
distributions has been investigated for several processes
\cite{Manohar:2012jr,Kasemets:2012pr,Echevarria:2015ufa}.  The generation of
parton correlations by the splitting of a single parton into two has been
investigated by several groups \cite{Diehl:2011yj,Gaunt:2012dd,%
  Blok:2011bu,Blok:2013bpa,Ryskin:2011kk,Ryskin:2012qx}; a simplified
implementation into the event generator Pythia is described in
\cite{Blok:2015rka,Blok:2015afa}.

At the level of integrated cross sections, double parton scattering is
suppressed by $\Lambda^2/Q^2$ compared to single parton scattering, where $Q$
denotes the scale of the hard scattering and $\Lambda$ the scale of
nonperturbative interactions \cite{Diehl:2011tt,Blok:2010ge}.  However, there
are situations where DPS can nevertheless compete with SPS, for instance when
the latter is suppressed by higher powers in coupling constants.  A prominent
example is same sign $W$ pair production
\cite{Kulesza:1999zh,Gaunt:2010pi,Ceccopieri:2017oqe}, which is an important
search channel for physics beyond the Standard Model.  A generic mechanism
enhancing DPS over SPS in processes involving small parton momentum fractions
$x$ is the fact that with decreasing $x$ the density of two partons increases
roughly like the square of the single parton density.

If the transverse momenta $q_T$ of the particles produced by a hard scattering
are small compared with $Q$, then DPS has the same power behaviour in $1/Q$ as
SPS \cite{Diehl:2011tt,Blok:2010ge}.  If $q_T$ is of order $\Lambda$, then one
needs information about the nonperturbative ``intrinsic'' transverse momentum
of partons.  However, in the region $\Lambda \ll q_T \ll Q$ one can reliably
compute the $q_T$ spectrum in perturbation theory, provided that one resums
the large logarithms of $Q/q_T$ that arise in this regime.  This resummation
is intricately related with evolution in the rapidity of emitted gluons.  For
SPS processes producing colourless particles, like Drell-Yan lepton pairs or a
Higgs boson, a powerful theoretical formalism has long been established
\cite{Collins:1984kg,Catani:2000vq} and been pushed to high perturbative
accuracy, see e.g.~\cite{Catani:2013tia} and references therein.  Formulations
using soft-collinear effective theory (SCET) have been given in
\cite{GarciaEchevarria:2011rb,Becher:2012yn,Neill:2015roa}.
A precise theory for DPS at small measured $q_T$ is of obvious interest.  In
\cite{Blok:2015rka,Blok:2015afa} the DDT formalism \cite{Dokshitzer:1978hw}
has been extended to DPS processes.  The analysis in
\cite{Diehl:2011tt,Diehl:2011yj} was based on transverse-momentum dependent
(TMD) factorisation in the original formulation of Collins and Soper
\cite{Collins:1981uk}, which also underlies the CSS resummation formalism
\cite{Collins:1984kg}.  Since then, an improved version of TMD factorisation
for SPS processes has been formulated by Collins \cite{Collins:2011zzd}; a
brief review is given in \cite{Aybat:2011zv,Collins:2011ca} and the
differences between the old and new versions are described in
\cite{Collins:2017oxh}.  

The aim of the present paper is twofold.  Firstly, we complete the
formulation of DPS in \cite{Diehl:2011tt,Diehl:2011yj} and adapt it to the
new factorisation formalism of \cite{Collins:2011zzd}, providing a
systematic analysis of soft-gluon effects, rapidity evolution and colour
correlations.  Secondly, we show how the theory simplifies for intermediate
transverse momenta $\Lambda \ll q_T \ll Q$, where transverse-momentum
dependent DPDs can be matched on transverse-momentum integrated
distributions, with the transverse-momentum dependence being computed in
perturbation theory.  This significantly increases the predictive power of
the theoretical framework.  Some of our main results have been reported in
\cite{Buffing:2016qql,Buffing:2017sxz}.

TMD factorisation in proton-proton collisions can be established to all orders
in perturbation theory for the production of colourless particles such as a
Higgs boson or electroweak gauge bosons \cite{Collins:2011zzd}.  Because of
serious complications from soft gluon exchange, it is not known if and how
the formalism could be extended to hard-scattering processes with coloured
particles in the final state, such as jets or heavy quarks
\cite{Rogers:2010dm}.  We will therefore limit our discussion of TMD
factorisation in DPS to colourless final states as well.  Important channels
are the production of two electroweak gauge bosons (often called the double
Drell-Yan process), of a Higgs boson and an electroweak gauge boson, or of a
Higgs boson pair.  Instead of a heavy boson, one may also consider a photon
pair of large invariant mass.

This paper is organised as follows.  In section \ref{sec:single_TMD} we
recall some of the concepts and results for TMD factorisation in single
parton scattering.  In section \ref{sec:collinear_DPDs} we discuss
properties of collinear matrix elements and of the soft factor, which are
the ingredients in the definition of transverse-momentum dependent and
transverse-momentum integrated DPDs, which we will call DTMDs and DPDFs,
respectively.  Our definition generalises the combination of collinear and
soft factors in \cite{Collins:2011zzd} to double parton distributions, and
it provides an alternative form of this construction for single parton TMDs.
The colour structure of DPS is significantly more complicated than the one
of SPS, and we show in section \ref{sec:colour} how this structure can be
handled in a general and efficient way.  We find significant simplifications
for transverse-momentum integrated quantities.  In section~\ref{sec:fact} we
present the general factorisation formula for DPS at low $q_T$, its
combination with SPS, as well as the evolution equations for DTMDs and DPDFs
and their general solution.  Section~\ref{sec:short} is devoted to the
region $\Lambda \ll q_T \ll Q$.  We establish the matching of DTMDs onto
different types of transverse-momentum integrated distributions.  The
multi-scale nature of the problem leads to different matching regimes, which
we combine in a consistent way using a subtraction formalism.  In section
\ref{sec:one-loop} we give one-loop expressions for perturbative quantities
that appear in the DPS cross section, extending previous work in the
literature and discussing several technical aspects of the computation. We
summarise our main results in section \ref{sec:conclusions}.  A variety of
technical details and results are given in the appendices.

\section{Reminder: single TMD factorisation}
\label{sec:single_TMD}

To begin with, let us recall a few results from TMD factorisation, as laid
out in \cite{Collins:2011zzd}.  The cross
section depends on TMDs that describe the distribution of partons inside
the proton in both longitudinal and transverse momentum.  Throughout this
work, we consider unpolarised protons.  The factorisation formula for
Drell-Yan production then reads
\begin{align}
  \label{SPS-Xsect}
   \frac{d \sigma}{d x\, d \bar{x}\, d^2 \tvec{q}}
 = \sum_{q} \hat{\sigma}_{q\bar{q}}(Q, \mu)
   \int \frac{d^2 \tvec{z}}{(2\pi)^2}\, e^{-i \tvec{q} \tvec{z}} \,
     W_{q\bar{q}}(x,\bar{x},\tvec{z};\mu)
  + \{ q \leftrightarrow \bar{q} \} \,,
\end{align}
where $\hat{\sigma}_{q\bar{q}}$ is the cross section for $q\bar{q}$
annihilation into an electroweak gauge boson and
\begin{align}
  \label{W-def-single}
W_{q\bar{q}}(x,\bar{x},\tvec{z};\mu) =
  f_q(x,\tvec{z};\mu,\zeta)\,
  f_{\bar{q}}(\bar{x},\tvec{z};\mu,\bar{\zeta})
\end{align}
is the product of a quark and an antiquark TMD.  The invariant mass of the
boson is $Q$ and its transverse momentum is $\tvec{q}$.  The TMDs depend
on longitudinal momentum fractions ($x$~or~$\bar{x}$) that are fixed by
final-state kinematics, and on a transverse distance $\tvec{z}$ that is
Fourier conjugate to the transverse parton momentum (and often denoted by
$\tvec{b}$ in the literature).  They also depend on an ultraviolet
renormalisation scale $\mu$ and on a rapidity parameter ($\zeta$ or
$\bar{\zeta}$) as we will review later.  Notice that the rapidity
parameter dependence cancels in~$W_{q\bar{q}}$.
The parton-level cross section $\hat{\sigma}_{q\bar{q}}$ and the overall
cross section may be taken differential in additional variables if more than
one particle is produced in the hard scattering.  An example is the angular
distribution of the leptons into which the electroweak gauge boson decays in
Drell-Yan production.  In this case one must include the TMDs for transverse
quark and antiquark polarisation in \eqref{W-def-single}.

The dependence of a TMD on the renormalisation scale is given by
\cite{Collins:2011zzd,Aybat:2011zv}
\begin{align}
  \label{RG-single-TMD}
  \frac{\partial}{\partial \log\mu}\, f_a(x,\tvec{z};\mu,\zeta)
  &= \gamma_{F, a}(\mu, \zeta)\, f_a(x,\tvec{z};\mu,\zeta)\, ,
\end{align}
where $a = q, \bar{q}, g$ labels the parton type.  The rapidity dependence
of the anomalous dimension $\gamma_{F,a}$ is given by
\begin{align}
  \label{cusp}
  \frac{\partial}{\partial \log\zeta}\, \gamma_{F,a}(\mu, \zeta)
  &= {}- \frac{1}{2}\ms \gamma_{K,a}(\mu) \,,
\end{align}
where $\gamma_{K,a}$ is called the cusp anomalous dimension.  It depends on
$\mu$ via $\alpha_s(\mu)$, i.e.\ $\gamma_{K,a}(\mu) = \gamma_{K,a}\bigl(
\alpha_s(\mu) \bigr)$.  From~\eqref{cusp} one readily finds
\begin{align}
  \label{cusp-solved}
\gamma_{F,a}(\mu, \zeta) &= \gamma_{a}(\mu)
   - \gamma_{K,a}(\mu)\, \log\frac{\sqrt{\zeta}}{\mu}
\end{align}
with
\begin{align}
\gamma_a(\mu) &= \gamma_{F,a}(\mu, \mu^2) \,.
\end{align}
The evolution of TMDs with the rapidity scale $\zeta$ is governed by the
Collins-Soper equation\footnote{We generally follow the notation of
    \protect\cite{Collins:2011zzd,Collins:2017oxh} in the present paper.  We
    do however not use a tilde to denote quantities in transverse position
    space, thus writing $f$ and $K$ instead of $\tilde{f}$ and
    $\tilde{K}$.}
\begin{align}
  \label{cs-eq}
  \frac{\partial}{\partial\log \zeta}\, f_{a}(x,\tvec{z}; \mu,\zeta)
  &= \frac{1}{2}\ms K_{a}(\tvec{z}; \mu)\,
  f_{a}(x,\tvec{z}; \mu,\zeta) \,,
\end{align}
whose kernel satisfies
\begin{align}
  \label{RG-single-K}
  \frac{\partial}{\partial \log\mu}\,
       {K_{a}(\tvec{z}; \mu)} = {}- \gamma_{K, a}(\mu) \,.
\end{align}
If the transverse boson momentum $\tvec{q}$ is large compared with the
scale $\Lambda$ of non-perturbative interactions, one can use a
short-distance expansion that connects a TMD $f_a(x,\tvec{z})$ at small
$\tvec{z}$ with a conventional collinear PDF $f_b(x)$ according to
\begin{align}
  \label{single-TMD-match}
f_a(x,\tvec{z};\mu,\zeta)
  &= \sum_b {C}_{ab}(x',\tvec{z};\mu,\zeta)
     \underset{x}{\otimes} f_b(x';\mu) \,,
\end{align}
with the convolution product defined by
\begin{align}
	\label{conv-def}
C(x') \underset{x}{\otimes} f(x') &= \int_{x}^1 \frac{dx'}{x'}\, C(x')\,
	f\biggl( \frac{x}{x'} \biggr) \,.
\end{align}
The expansion \eqref{single-TMD-match} has power corrections in the
parameter $\Lambda |\tvec{z}|$.  Combining it with the solution of the
evolution equations in $\mu$ and $\zeta$, one obtains
\begin{align}
  \label{TMD-evo-solved}
f_{a}(x,\tvec{z};\mu,\zeta) &= \exp\, \biggl\{ 
     \int_{\mu_{0}}^{\mu} \frac{d\mu'}{\mu'}\,
          \biggl[ \gamma_{a}(\mu')
            - \gamma_{K,a}(\mu')\, \log\frac{\sqrt{\zeta}}{\mu'} \biggr]
          + K_{a}(\tvec{z};\mu_{0})
	    \log\frac{\sqrt{\zeta}}{\mu_{0}}
   \ms \biggr\}
\nonumber \\[0.3em]
 & \quad \times \sum_{b} 
   {C}_{a b}(x',\tvec{z};\mu_{0}^{},\mu_{0}^2) \underset{x}{\otimes}
             {f_{b}(x';\mu_{0}^{})} \,.
\end{align}
The short-distance coefficient $C$ and the Collins-Soper kernel $K$ should
be evaluated with a scale choice that avoids large logarithms, so that they
can be reliably calculated in fixed-order perturbation theory.  In the
non-perturbative region of $\tvec{z}$ one needs a model ansatz for
$f_a(x,\tvec{z})$. The so-called $b^*$ prescription
\cite{Collins:1981va,Collins:1984kg} can be used to smoothly interpolate
between such an ansatz and the perturbative result \eqref{TMD-evo-solved}.

In the following sections, we will show how these results can be extended
to the case of double parton scattering.  There are several aspects that
make this extension far from trivial.  One is the larger number of
coloured particles involved in the process, which leads to a non-trivial
colour structure of DTMDs and DPDFs.  As a consequence, even DPDFs depend
on the rapidity parameter $\zeta$, unlike PDFs for a single parton
\cite{Diehl:2011yj,Manohar:2012jr}.  A second aspect is that DPS
involves several transverse distances, which makes the analogue of the
short-distance expansion \eqref{single-TMD-match} more complicated.

\section{Defining double parton distributions}
\label{sec:collinear_DPDs}

As reviewed in \cite{Diehl:2015bca}, factorisation of DPS processes
involves separating the leading graphs for the cross sections into
subgraphs that are hard, soft, or collinear to one of the two incoming
protons.  The treatment of the soft subgraph is intimately related with
the rapidity parameter $\zeta$ mentioned in the previous subsection.

There are in fact different alternatives for such a treatment.  The
analysis of the double Drell-Yan process in \cite{Diehl:2011yj} followed
the original procedure for TMD factorisation by Collins and Soper
\cite{Collins:1981uk} and did not work out all relevant aspects of the
problem.  \rev{In the present paper, we perform a systematic analysis of DPS in
the factorisation framework of Collins \cite{Collins:2011zzd}.
We explicitly show how the soft factors relevant for the cross section can
be entirely absorbed into DTMDs or DPDFs, and we derive the resulting
evolution equations in $\zeta$, as well as the ones in $\mu$.  Other schemes
to handle soft factors and rapidity dependence will briefly be discussed in section~\ref{sec:conclusions}.}

The starting point of our discussion is an intermediate expression of the
DPS cross section, given in section~2.1 of \cite{Diehl:2015bca}.  The cross
section for the production of two sets of colourless particles involves a
term
\begin{align}
\label{eq:DPS_colsoft}
H^{}_{1, q\bar{q}}\, H^{}_{2, q\bar{q}}\;
F^T_{\us, \bar{q}\bar{q}}(v_R) \,
  S_{qq}^{-1}(v_L, v_R)\, S_{qq}^{}(v_L, v_R)\, S_{qq}^{-1}(v_L, v_R)\,
F^{}_{\us, qq}(v_L)
\end{align}
for the annihilation of two quarks in one proton with two antiquarks in
the other proton, and corresponding terms for the other parton
combinations.  For definiteness we have only written down the DPDs for
unpolarised partons; polarised terms have the same soft factor.  $H_{i,
q\bar{q}}$ 
denotes the squared hard-scattering amplitudes, with appropriate spin
projections (see section~2.2 in \cite{Diehl:2011yj}) but with the colour
structure removed as specified in \eqref{hard-scatt-sing} below.
$F_{\us}$ denotes unsubtracted collinear matrix elements, and $S$ is a
soft factor.  The inverse of this factor removes contributions of soft
gluons from the unsubtracted collinear matrix elements, so that $S^{-1}\ms
F_{\us}$ receives only contributions from collinear gluons.
For brevity we have omitted momentum fraction and position space arguments
in \eqref{eq:DPS_colsoft}, as well as renormalisation and factorisation
scales and colour indices.  $F_{\us}$ is a row vector in colour space
(with one index for each of the four parton legs), and $S$ is a matrix
with two times four indices.  The spacelike four-vectors $v_L$ and $v_R$
denote the directions of Wilson lines and will be specified later.  As
discussed in \cite{Collins:2011zzd,Diehl:2015bca}, $v_L$ and $v_R$ have to
be chosen such that the effects of so-called Glauber gluon exchange on the
cross section can be subsumed into the soft and collinear factors in
\eqref{eq:DPS_colsoft}.

\rev{The aim of the following sections is to combine soft factors and
collinear matrix elements in such a way that the product \eqref{eq:DPS_colsoft}
takes on a simple form.  After introducing the necessary notation, we derive a
number of symmetry properties of soft factors in section ~\ref{sec:soft-fact}
and then discuss our central hypothesis for their rapidity dependence, given
in \eqref{CS-for-s} and \eqref{S-decomp}.  This leads us to the definition
\eqref{F-sub-def} of DPDs and to their rapidity evolution equation
\eqref{CS-gen} in section~\ref{sec:comb-soft-coll}.  We discuss
renormalisation in section~\ref{sec:ren-DTMDs} and derive the basic evolution
equations \eqref{RG-TMD} and \eqref{Y-dep-gamma-F} relevant for DTMDs.
Applying our construction to TMDs for a single parton in
section~\ref{sec:sub_singleTMD}, we are led to the definition
\eqref{eq:unsub_single} and see that it is equivalent to the definition of
Collins \cite{Collins:2011zzd} by virtue of the relation~\eqref{sqrt-simp}.}

\subsection{Collinear matrix elements}
\label{sec:collinear}

\rev{To begin with, let us} recall the definitions of unsubtracted DTMDs and
DPDFs in terms of proton matrix elements.  These will later be combined with
soft factors in order to define the double parton distributions that appear in
the cross section formula.

For two partons $a_1$ and $a_2$, the unsubtracted DTMDs are defined in
terms of matrix elements as \cite{Diehl:2011yj,Diehl:2011tt}
\begin{align}
\label{eq:dpds}
	F_{\us, a_1a_2}(x_1,x_2,\tvec{z}_1,\tvec{z}_2,\tvec{y}) & = 2p^+
        (x_1\ms p^+)^{-n_1}\, (x_2\ms p^+)^{-n_2} \int
        \frac{dz^-_1}{2\pi}\, \frac{dz^-_2}{2\pi}\, dy^-\,
          e^{i\ms ( x_1^{} z_1^- + x_2^{} z_2^-)\ms p^+}
\nonumber \\
 & \quad \times
    \langle\ms p \ms|\, \mathcal{O}_{a_1}(y,z_1)\, \mathcal{O}_{a_2}(0,z_2)
    \,|\ms p \ms\rangle \,,
\end{align}
where $n_i = 1$ if parton number $i$ is a gluon and $n_i = 0$ otherwise.  We
use light-cone coordinates $w^\pm = (w^0 \pm w^3) /\sqrt{2}$ and the
transverse component $\tvec{w} = (w^1, w^2)$ for any four-vector $w$.  The
definition \eqref{eq:dpds} is natural for a proton moving to the right,
i.e.\ for $p^3>0$.  For a left moving proton, i.e.\ for $p^3 < 0$, one would
interchange the roles of plus and minus coordinates.  It is understood that
$\tvec{p} = \tvec{0}$ in both cases, and that the proton polarisation is
averaged over.  Setting $\tvec{z}_1 = \tvec{z}_2 = \tvec{0}$ in
\eqref{eq:dpds}, one obtains DPDFs, which are relevant for collinear
factorisation.  As we will discuss later, this changes the ultraviolet
behaviour of the operators.

The operators for quarks in a right moving proton read
\begin{align}
\label{eq:quark-ops}
\mathcal{O}_{a}(y,z) &=
  \bar{q}\bigl( y - \half z \bigr)\, W^\dagger \bigl(y-\half z, v_L \bigr) \,
  \Gamma_{a} \, W \bigl(y+\half z, v_L \bigr) \, q\bigl( y + \half z \bigr)
\Big|_{z^+ = y^+_{} = 0}
\end{align}
with spin projections
\begin{align}
  \label{eq:quark-proj}
\Gamma_q & = \half \gamma^+ \,, &
\Gamma_{\Delta q} &= \half \gamma^+\gamma_5 \,, &
\Gamma_{\delta q}^j = \half i \sigma^{j+}_{} \gamma_5  \quad (j=1,2)
\end{align}
onto unpolarised quarks ($q$), longitudinally polarised quarks ($\Delta
q$) and transversely polarised quarks ($\delta q$).  We do not explicitly
display the transverse index $j$ of the operator $\mathcal{O}_{\delta q}$
and of the corresponding DPDs, unless it is needed.  The field with
argument $y + \half z$ in $\mathcal{O}_{q}(y,z)$ is associated with a
quark in the amplitude of a double scattering process and the field with
argument $y - \half z$ with a quark in the complex conjugate amplitude.
The Wilson lines are defined as
\begin{align}
	\label{WL-def}
W(\xi, v) &= \operatorname{P} \exp\biggl[\ms ig\ms t^a
    \int_{-\infty}^0 \!ds\, v A^a(\xi + s v) \ms\biggr] \,,
\end{align}
where $\operatorname{P}$ denotes path ordering, such that fields $t^a
A^a(\xi + s v)$ with smaller $s$ stand further to the left in the expanded
exponential.  Our convention for the strong coupling $g$ is specified in
appendix \ref{sec:Feynman_rules}.  Throughout this work, we only consider
the case $\tvec{v} = \tvec{0}$.  In the matrix element \eqref{eq:dpds} for
a right-moving proton, one takes a direction $v_L$ with $v_L^- \gg -v_L^+
> 0$, and in its analogue for a left-moving proton one has a direction
$v_R$ with $v_R^+ \gg -v_R^- > 0$.  In both cases, the Wilson lines are
past-pointing.
Analogous operators are defined for antiquarks, with some sign changes as
specified in section~2.2 of \cite{Diehl:2011yj}.  For gluons, one has
\begin{align}
\label{eq:gluon-ops}
\mathcal{O}_{a}(y,z) &= \Pi_{a}^{jj'} \,
   G^{+j'}\bigl( y - \half z \bigr)\, W^\dagger\bigl( y-\half z, v_L \bigr)\,
   W\bigl( y+\half z, v_L \bigr)\, G^{+j}\bigl( y + \half z\bigr)
 \Big|_{z^+ = y^+_{} = 0}
\end{align}
with spin projections
\begin{align}
\label{eq:gluon-proj}
\Pi_g^{jj'} &= \delta^{jj'} \,, &
\Pi_{\Delta g}^{jj'} &= i\epsilon^{jj'} \,, &
\bigl[ \Pi_{\delta g}^{kk'} \bigr]{}^{jj'} &= \tau^{jj'\!,kk'}
\end{align}
onto unpolarised gluons ($g$), longitudinally polarised gluons ($\Delta
g$) and linearly polarised gluons ($\delta g$).  The indices $j,j',k,k' =
1,2$ run over transverse components, $\epsilon^{jj'}$ is the antisymmetric
tensor with $\epsilon^{12} = 1$, and $\tau^{jj'\!,kk'}$ is defined as
\begin{align}
  \label{eq:taujjkk}
\tau^{jj'\!,kk'} &=
        \frac{1}{2}\, \Bigl( \delta^{jk}\delta^{j^{\prime}k^{\prime}} +
        \delta^{jk^{\prime}}\delta^{j^{\prime}k} -
        \delta^{jj^{\prime}}\delta^{kk^{\prime}} \Bigr) \,.
\end{align}
The Wilson lines in \eqref{eq:gluon-ops} are in the adjoint representation
rather than in the fundamental one (see section~\ref{sec:soft-fact}).
Making the colour indices of the operators explicit, we have
\begin{align}
  \label{colour-ops}
\mathcal{O}_{q,\, j j'} &=
  \bar{q}_{k'}\, (W^\dagger)_{k' j'} \, \Gamma \, W_{j k}\, q_{k} \,,
&
\mathcal{O}_{g,\, a a'} &=
  \Pi \, G_{b'}\, (W^\dagger)_{b' a'} \, W_{a b}\, G_{b}
\end{align}
for quarks and gluons, respectively.

For the discussion of ultraviolet renormalisation and of the
short-distance expansion, it is useful to introduce operators that
correspond to a definite light-cone momentum fraction $x$ of a parton:
\begin{align}
  \label{x-ops}
O_a(x, \tvec{y}, \tvec{z}) &= 2p^+\ms (x p^+)^{-n} \int \frac{dz^-}{2\pi}\,
  dy^-\; e^{i x z^- p^+}\, \mathcal{O}_a(y,z) \,.
\end{align}
Using translation invariance, one readily finds that unsubtracted single
and double parton TMDs are then given by matrix elements
\begin{align}
  \label{x-matel}
2\pi \delta(p^+ - p'^+)\, 2 p^+ f_{\us, a}(x, \tvec{z})
 &= \langle p'|\, O_a(x, \tvec{0}, \tvec{z}) \,| p \rangle  \,,
\nonumber \\
2\pi \delta(p^+ - p'^+)\, 2 p^+
    F_{\us, a_1 a_2}(x_1,x_2, \tvec{z}_1,\tvec{z}_2, \tvec{y})
 &= \langle p'|\, O_{a_1}(x_1, \tvec{y}, \tvec{z}_1)\,
                  O_{a_2}(x_2, \tvec{0}, \tvec{z}_2)\, | p \rangle \,,
\end{align}
where $\tvec{p}' = \tvec{p} = \tvec{0}$.  The corresponding collinear
distributions are obtained by setting $\tvec{z}$, $\tvec{z}_1$ and
$\tvec{z}_2$ to zero.

We note that in the DPS cross section, there are also distributions
describing the interference between different parton types in the process
amplitude and its complex conjugate, i.e.\ between quarks and antiquarks,
between quarks and gluons, or between quarks of different flavour
\cite{Diehl:2011yj,Manohar:2012jr,Diehl:2013mla,Kasemets:2014yna}.  At low
values of $x_1$ and $x_2$, such interference DPDs are expected to be
negligible, because they have no dynamic cross talk with gluon
distributions, which grow most strongly with decreasing $x$.  Although we
do not consider interference DPDs in this work, they can be treated with
the methods presented in the following.

Let us remark that an analysis of transverse-momentum dependent DPDs in the
small-$x$ limit has recently been given in \cite{Golec-Biernat:2016vbt}.  The
quantities considered in that work are Fourier transformed w.r.t.\ our
variable $\tvec{z}_i$ and -- more importantly -- integrated over $\tvec{y}$.
To make contact with the DPS cross section, one would need to restore the
dependence of these distributions on $\tvec{y}$, or on the Fourier conjugated
momentum.

\subsection{Soft factors}
\label{sec:soft-fact}

\begin{figure}
\begin{center}
\includegraphics[width=0.55\textwidth]{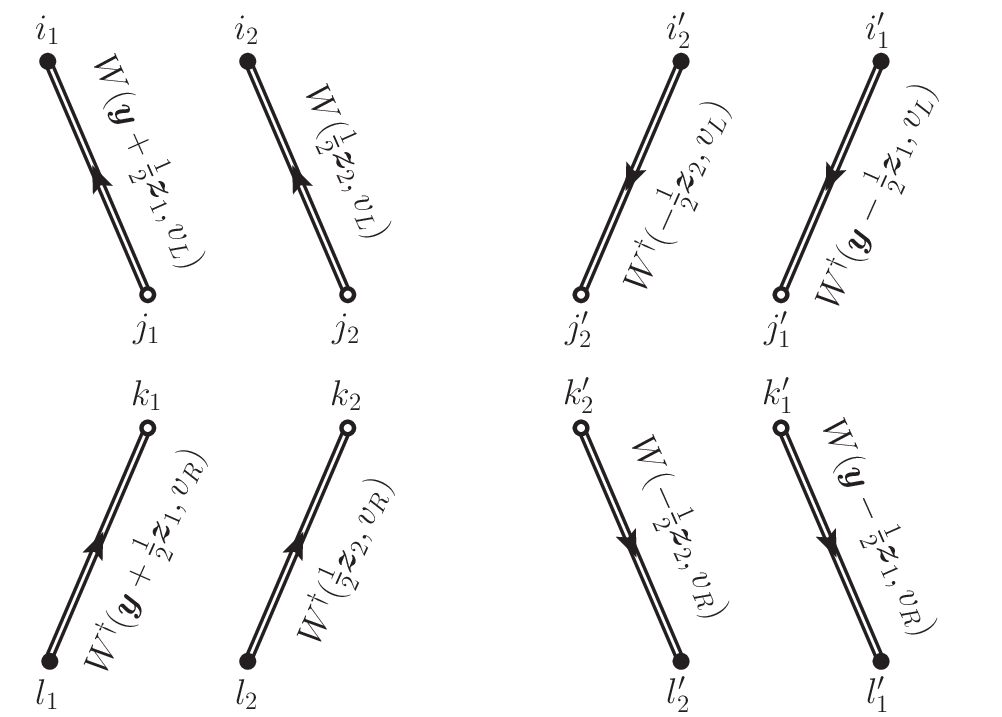}
\caption{\label{fig:soft-fact-indices} Wilson lines and colour indices of
  the extended soft factor defined by \protect\eqref{soft-gen-def} and
  \protect\eqref{WL-ext}.  Wilson lines with subscripts $1$ are
    grouped into the operator $O_{S,q}(\tvec{y},\tvec{z}_1)$ and those
    with subscripts~$2$ into the operator $O_{S,q}(\tvec{0},\tvec{z}_2)$.}
\end{center}
\end{figure}

Before constructing the final DPDs, we must take a closer look at the soft
factor, which is defined as the vacuum expectation value of Wilson lines.
For reasons that will become clear later, we define an ``extended soft
factor'' with open colour indices of all Wilson lines as
\begin{align}
	\label{soft-gen-def}
& \bigl[ S_{qq}(\tvec{z}_1,\tvec{z}_2,\tvec{y}; v_L,v_R) \bigr]^{ i_1^{}
          i_1' i_2^{} i_2', l_1^{} l_1' l_2^{} l_2'}_{ j_1^{} j_1' j_2^{}
          j_2', k_1^{} k_1' k_2^{} k_2'}
\nonumber \\[0.2em]
& \qquad = \bigl\langle\, 0 \,\big|\ms
  \bigl[ O_{S,q}(\tvec{y},\tvec{z}_1; v_L,v_R) \bigr]^{i_1^{} i_1',
         l_1^{} l_1'}_{j_1^{} j_1', k_1^{} k_1'} \,
  \bigr[ O_{S,q}(\tvec{0},\tvec{z}_2; v_L,v_R) \bigr]^{i_2^{} i_2',
    l_2^{} l_2'}_{j_2^{} j_2', k_2^{} k_2'} \,
  \big|\, 0 \,\bigr\rangle \,.
\end{align}
This factor appears in the cross section with two $q\bar{q}$ annihilation
subprocesses, and the subscripts $q$ refer to the right-moving parton in each
subprocess.  The operator
\begin{align}
  \label{WL-ext}
& \bigl[ O_{S,q}(\tvec{y},\tvec{z}; v_L,v_R)
  \bigr]^{i i', l l'}_{j j', k k'}
\nonumber \\
& \qquad
   = W_{ij}^{}(\tvec{y}+\half \tvec{z},v_L)\,
     W_{kl}^\dagger(\tvec{y}+\half \tvec{z},v_R)\,
     W_{l'k'}^{}(\tvec{y}-\half \tvec{z},v_R)\,
     W_{j'i'}^\dagger(\tvec{y}-\half \tvec{z},v_L)
\end{align}
is a product of four Wilson lines, which are 
defined as in \eqref{WL-def}, but with $\xi^+ = \xi^- = 0$, so that their
position arguments are only in the transverse plane and hence written in
boldface.
Primed and unprimed indices $j$ and $k$ in the Wilson lines of
\eqref{WL-ext} correspond to gluon fields $A(\xi)$ at light-cone zero
($\xi^+ = \xi^- = 0$), whereas primed and unprimed indices $i$ and $l$
correspond to gluon fields at light-cone infinity, $A(\xi - \infty v)$.  A
pictorial representation of \eqref{soft-gen-def} is given in
figure~\ref{fig:soft-fact-indices}.

For right-moving antiquarks instead of quarks, one replaces $W^{}_{ij}$ with
$W^*_{ij} = W^\dagger_{ji}$ and vice versa, which corresponds to replacing
$it^a_{ij}$ with $(i t^a_{ij})^* = - i t^a_{ji}$ in the exponential
\eqref{WL-def}.  For gluons, one takes adjoint Wilson lines $W_{bc}$, obtained
by replacing $it^a_{ij}$ in \eqref{WL-def} with $f^{abc}$.  This reflects the
fact that the generators $T^a$ of the colour group in the adjoint
representation are given by $(T^a)_{bc} = -i f^{abc}$.  We therefore have
\begin{align}
  \label{S-star}
S_{qq}^{} &= S_{\bar{q}\bar{q}}^* \,,
&
S_{q\bar{q}}^{} &= S_{\bar{q}q}^* \,,
&
S_{qg}^{} &= S_{\bar{q}g}^* \,,
&
S_{gg}^{} &= S_{gg}^* \,,
\end{align}
where we have omitted colour indices for brevity.
Note that, by construction, the Wilson line operators for all representations
are unitary, i.e.\ $W(\xi,v)\ms W^\dagger(\xi,v) = \one$, where $\one$ is the
unit matrix in the relevant colour space.

In processes producing colourless particles, one needs
\eqref{soft-gen-def} with all index pairs $j$, $k$ contracted, i.e.\
\begin{align}
	\label{soft-def}
& \bigl[ S_{qq}(\tvec{z}_1,\tvec{z}_2,\tvec{y}; v_L, v_R) \bigr]^{ i_1^{}
          i_1' i_2^{} i_2',\, l_1^{} l_1' l_2^{} l_2'}
\nonumber \\[0.2em]
& \qquad = \bigl\langle\, 0 \,\big|\ms
  \bigl[ O_{S,q}(\tvec{y},\tvec{z}_1; v_L,v_R) \bigr]^{i_1^{} i_1',
         l_1^{} l_1'} \,
  \bigr[ O_{S,q}(\tvec{0},\tvec{z}_2; v_L,v_R) \bigr]^{i_2^{} i_2',
         l_2^{} l_2'} \,
  \big|\, 0 \,\bigr\rangle
\end{align}
with
\begin{align}
  \label{WL-ops}
& \bigl[ O_{S,q}(\tvec{y},\tvec{z}; v_L,v_R) \bigr]^{i i', l l'}
\nonumber \\
& \qquad
   = \bigl[ W^{}(\tvec{y}+\half \tvec{z},v_L)\,
     W^\dagger(\tvec{y}+\half \tvec{z},v_R) \bigr]_{il} \,
     \bigl[ W^{}(\tvec{y}-\half \tvec{z},v_R)\,
     W^\dagger(\tvec{y}-\half \tvec{z},v_L) \bigr]_{l'i'} \,.
\end{align}
We simply call this the ``soft factor'' (without the specification
``extended'').  Regarding this as a matrix in the index pairs $(i_1^{}
i_1' i_2^{} i_2')$ and $(l_1^{} l_1' l_2^{} l_2')$, we define the
transposed matrix as
\begin{align}
\bigl[ S_{qq}^T(\tvec{z}_1,\tvec{z}_2,\tvec{y}; v_L, v_R) \bigr]^{ i_1^{}
  i_1' i_2^{} i_2', l_1^{} l_1' l_2^{} l_2'} &= \bigl[
  S_{qq}^{}(\tvec{z}_1,\tvec{z}_2,\tvec{y}; v_L, v_R) \bigr]^{ l_1^{} l_1'
  l_2^{} l_2', i_1^{} i_1' i_2^{} i_2'}
\end{align}
and the Hermitian conjugate as $S_{qq}^{\smash{\dagger}\phantom{T}} =
\bigl[ S_{qq}^{T} \bigr]^*$. Corresponding definitions hold for $S_{gg}$,
$S_{gq}$, $S_{\bar{q}\bar{q}}$ etc.
There are a number of symmetry constraints on the soft factor:
\begin{enumerate}
\item With the restrictions on the directions of Wilson lines in $S(v_L,
  v_R)$ specified below \eqref{WL-def}, one can always perform a
  longitudinal boost such that $v_L = (\alpha, \beta, \tvec{0})$ and $v_R
  = (\beta, \alpha, \tvec{0})$.  A parity transformation interchanges
  plus- and minus components and thus, combined with the boost, exchanges
  $v_L \leftrightarrow v_R$.  Parity and boost invariance therefore give
\begin{align}
	\label{soft-parity}
S_{a_1 a_2}(\tvec{z}_1,\tvec{z}_2,\tvec{y}; v_L, v_R) &= S_{a_1
  a_2}(-\tvec{z}_1,-\tvec{z}_2,-\tvec{y}; v_R, v_L) \nonumber \\ &= S_{a_1
  a_2}(\tvec{z}_1,\tvec{z}_2,\tvec{y}; v_R, v_L) \,,
\end{align}
where $a_1, a_2$ denote the parton types, $q, \bar{q}, g$.  In the final
step we used that the soft factor depends on position arguments only via
scalar products between $\tvec{z}_1, \tvec{z}_2$ and $\tvec{y}$.  This
follows from rotational and parity invariance, combined with the fact that
the Wilson line directions $v_L$ and $v_R$ are purely longitudinal.
\item A combined parity and time reversal transformation reverses the
  vectors $v_L$ and $v_R$, as well as the position arguments. This gives
\begin{align}
	\label{soft-PT}
S_{a_1 a_2}(\tvec{z}_1,\tvec{z}_2,\tvec{y}; v_L, v_R) &= S_{a_1
  a_2}(-\tvec{z}_1,-\tvec{z}_2,-\tvec{y}; -v_L, -v_R) \nonumber \\ &=
S_{a_1 a_2}(\tvec{z}_1,\tvec{z}_2,\tvec{y}; -v_L, -v_R) \,,
\end{align}
where in the last step we used the same symmetry argument as in
\eqref{soft-parity}.
\item Since the soft factor is constructed from products $\bigl[ W^{}(z,
  v_L)\, W^\dagger(z, v_R) \bigr]_{il}$ for quarks, $\bigl[ W^{}(z,
  v_R)\, W^\dagger(z, v_L) \bigr]_{li}$ for antiquarks, and
   $\bigl[ W^{}(z, v_L)\, W^\dagger(z, v_R) \bigr]_{ad}$ for gluons,
  we have
\begin{align}
	\label{soft-symm}
S_{a_1 a_2}(\tvec{z}_1,\tvec{z}_2,\tvec{y}; v_L, v_R) &= S_{a_1
  a_2}^{\dagger}(\tvec{z}_1,\tvec{z}_2,\tvec{y}; v_R, v_L) \nonumber \\ &=
S_{a_1 a_2}^{\dagger}(\tvec{z}_1,\tvec{z}_2,\tvec{y}; v_L, v_R) \,,
\end{align}
where in the second step we have used \eqref{soft-parity}. The soft matrix
is thus Hermitian in the groups of indices for the Wilson lines along the
different directions $v_L$ and $v_R$.
Combining this result with the relations \eqref{S-star}, we find
\begin{align}
  \label{S-transp}
S_{qq}^{} &= S_{\bar{q}\bar{q}}^T \,,
&
S_{q\bar{q}}^{} &= S_{\bar{q}q}^T \,,
&
S_{qg}^{} &= S_{\bar{q}g}^T \,,
&
S_{gg}^{} &= S_{gg}^T \,,
\end{align}
where the matrices on the left and the right hand side have identical
arguments.  We can thus identify the product $F^T_{\us, \bar{q}\bar{q}} \,
S_{qq}^{-1} = ( S_{\bar{q}\bar{q}}^{-1}\, F^{}_{\us, \bar{q}\bar{q}} )^T$ in
the expression \eqref{eq:DPS_colsoft} as the analogue of the product
$S_{qq}^{-1}\ms F^{}_{\us, qq}$.
\end{enumerate}

From now on we restrict ourselves to spacelike Wilson line directions in
$S(v_L, v_R)$ that satisfy
\begin{align}
	\label{v-restrictions}
v_L^- &> 0 \,, & v_R^+ &>0 \,, & Y_L \ll Y_R \,,
\end{align}
where the rapidity $Y_v$ of the spacelike vector $v$ is defined as
\begin{align}
  \label{Y-def}
Y_v = \frac{1}{2} \log \biggl| \frac{v^+}{v^-} \biggr| \,.
\end{align}
Cases other than \eqref{v-restrictions}, such as $v_L^- < 0, \, v_R^+ < 0$,
can be realised using the symmetry relations \eqref{soft-parity} and
\eqref{soft-PT}.  Owing to boost invariance and the fact that the Wilson lines
$W(\xi,v)$ are invariant under rescaling $v\to \lambda v$, the soft factor
depends on $v_L$ and $v_R$ only via the difference of the Wilson line
rapidities.  We can hence write $S(v_L,v_R) = S(Y_R - Y_L)$.

At this point we make an \emph{assumption} on the properties of the soft
factor, which we cannot fully prove (we specify below what can be proven
at present).

For the soft factor $S(Y_R - Y_L)$ with $Y_R - Y_L \gg 1$ (for brevity we
omit the indices $a_1$, $a_2$ and arguments $\tvec{z}_{1}, \tvec{z}_{2},
\tvec{y}$), we assume that there exists a matrix function $s(Y)$
satisfying the following three properties:
\begin{description}
\item[Property 1a:]  The rapidity dependence is given by
\begin{align}
	\label{CS-for-s}
\frac{\partial}{\partial Y}\, s(Y) &= s(Y)\ms K
\end{align}
with $K = K^\dagger$ independent of $Y$.
\item[Property 1b:] $s(Y)$ is nonsingular.  It is sufficient to establish
  this property at an arbitrary value $Y_1$; its validity at any other $Y$
  then follows from the solution of \eqref{CS-for-s}.
\item[Property 1c:]  One has
\begin{align}
	\label{S-decomp}
S(Y) &= s(Y - Y_0)\, s^\dagger(Y_0) & & \text{for $Y \gg 1$ and arbitrary
  $Y_0$} \,.
\end{align}
\end{description}

It is easy to see that $s(Y)$ is not unique, since properties 1a -- 1c
remain true if one redefines $s(Y) \to s(Y)\, U$ and $K
\to U^\dagger K U$ with a $Y$ independent unitary matrix $U$.  Conversely,
if properties 1a -- 1c hold for two matrix functions $s(Y)$ and $s'(Y)$,
one can show that $s'(Y) = s(Y)\, U(Y)$ with a unitary matrix $U$ (which
is not necessarily $Y$ independent).

In appendix~\ref{app:matrix-algebra} we will show that properties 1a -- 1c
are equivalent to the following two properties:
\begin{description}
\item[Property 2a:] $S$ satisfies the Collins-Soper equation
\begin{align}
	\label{CS-for-S}
\frac{\partial}{\partial Y}\, S(Y)
  &= \widehat{K}\ms S(Y) & & \text{for $Y \gg 1$} \,.
\end{align}
$\widehat{K}$ is not necessarily Hermitian, and its relation with $K$ is
specified in~\eqref{K-def}.  Both \eqref{S-decomp} and \eqref{CS-for-S}
are meant to be approximations for $Y\gg 1$.  We can define a matrix
$\widehat{S}(Y)$ that approximates $S(Y)$ for $Y\gg 1$ whilst being an
exact solution of the differential equation \eqref{CS-for-S} at all $Y$.
This equation is then solved by
\begin{align}
	\label{S-hat-solved}
\widehat{S}(Y) &= e^{Y \widehat{K}}\, \widehat{S}(0).
\end{align}
\item[Property 2b:] There is a value $Y_1$ for which $\widehat{S}(Y_1)$ is
  positive definite.
\end{description}
We show in appendix~\ref{app:matrix-algebra} that $\widehat{S}(Y)$ is then
in fact positive definite for all $Y$.  This guarantees that the matrix
$S(Y)$ at large $Y$ has an inverse, which according to
\eqref{eq:DPS_colsoft} is needed in the cross section formula.

Evidence for the properties just discussed comes from perturbation theory,
which can be used to compute the soft factor when the distances
$\tvec{z}_{1}$, $\tvec{z}_{2}$ and $\tvec{y}$ are all small.
\begin{itemize}
\item Properties 2a and 2b are easily checked at one-loop order from our
  explicit calculations in all colour channels, which are reported in
  section~\ref{sec:one-loop}.  In this case one finds $\widehat{S}(0) =
  \one$ and, taking $U = \one$ one has $K = \widehat{K}$ according to
  \eqref{K-def}.
\item Given property 2a, one can motivate property 2b using the
  perturbative expansions $S(Y) = \one + \mathcal{O}(\alpha_s)$ and
  $\widehat{K} = \mathcal{O}(\alpha_s)$.  One may worry that the expansion
  of $S(Y)$ has poor convergence, because higher orders in $\alpha_s$ can
  come with higher powers of the large rapidity $Y$.  However, according
  to \eqref{S-hat-solved} we have $\widehat{S}(0) = \exp(- Y
  \widehat{K})\, S(Y)$ for sufficiently large $Y$.  If the perturbative
  expansions of $S(Y)$ and $\exp(-Y \widehat{K})$ are valid at least in a
  formal sense, we get an expansion $\widehat{S}(0) = \one +
  \mathcal{O}(\alpha_s)$ that is free of any large parameter.  The
  eigenvalues of $\widehat{S}(0)$ then have a perturbative expansion
  around $1$, which supports property 2b with $Y_1=0$, at least for
  sufficiently small $\alpha_s$.
\item $S_{qq}$ and $S_{q\bar{q}}$ have been calculated at two-loop order
  in \cite{Vladimirov:2016qkd}.  The validity of properties 1a and 1c can
  be explicitly verified from the results in sections~4.2 and~4.4 of that
  work.  This requires a translation between the rapidity regulator used
  there and the one used here, which we discuss in
  appendix~\ref{app:delta-reg}.  We note that \cite{Vladimirov:2016qkd}
  defines $S = s^\dagger s$, whilst we choose the order $S = s s^\dagger$
  in \eqref{S-decomp}.

  Property 1b can then be motivated by the perturbative expansion of
  $s(Y)$, arguing along the same lines as in the previous point.

\item \rev{An all-order derivation of the rapidity evolution equation of the
  DPS soft factor has recently been given in \cite{Vladimirov:2017ksc}.  Our
  equation~\eqref{CS-for-S} can be obtained from equation~(5.17) in the arXiv
  version 2 of \cite{Vladimirov:2017ksc} using the relation \eqref{Y-corr}
  between the regulator variables $\delta^+, \delta^-$ and $Y_L, Y_R$.
  However, more work is needed to establish whether the derivation in
  \cite{Vladimirov:2017ksc} holds if one uses either of the associated
  rapidity regulators.}
\end{itemize}
A few more comments are in order.
\begin{itemize}
\item While $S(0) = \one$ by construction (the Wilson line pairs at equal
  positions give unit matrices, $W^\dagger W = \one$), one generally has
  $\widehat{S}(0) \neq \one$.  This implies that the evolution equation
  \eqref{CS-for-S} holds for $Y\gg 1$ but not when $Y$ becomes small.  We
  see this already at one loop:   as follows from sections 3.3.1 and 3.2.2
  of \cite{Diehl:2011yj}, one has $S - \one \propto Y \tanh(Y)$ for the
  one-loop soft factor, whilst \eqref{CS-for-S} truncated to
  $\mathcal{O}(\alpha_s)$ gives $S - \one \propto Y$.
\item For the soft factor needed in collinear factorisation, i.e.\ for
  $\tvec{z}_{1} = \tvec{z}_{2} = \tvec{0}$, we will see in
  section~\ref{sec:soft-simple} that $S(Y)$ is diagonal in the basis of
  colour representations for all $Y$.  As a consequence, $\widehat{K}$ is
  diagonal as well.  In appendix~\ref{app:matrix-algebra} we show that one
  can then also choose $s(Y)$ and $K$ to be diagonal.  For the soft factor
  in collinear factorisation, all matrix multiplications in colour space
  thus become trivial.
\item The matrix $s$ replaces the square roots of soft factors in the
  construction of single parton TMDs by Collins \cite{Collins:2011zzd}, as
  we will show in section~\ref{sec:sub_singleTMD}.
\end{itemize}

\subsection{Definition of DPDs}
\label{sec:comb-soft-coll}

According to \eqref{eq:DPS_colsoft}, the DPS cross section involves a
product of soft matrices, which using \eqref{S-decomp} can be rewritten as
\begin{align}
S^{-1}(v_L,v_R)\, S(v_L,v_R)\, S^{-1}(v_L,v_R)
 &= s^{\dagger -1}(Y_R - Y_C)\, s^{-1}(Y_C - Y_L) \,,
\end{align}
where $Y_C$ is a central rapidity, $Y_L \ll Y_C \ll Y_R$.  Restoring
parton labels, we then define DPDs by
\begin{align}
  \label{F-sub-def}
F^{}_{a_1 a_2}(Y_C) &= \lim_{Y_L\to -\infty}
  s_{a_1 a_2}^{-1}(Y_C - Y_L)\, F^{}_{\text{us}, a_1 a_2}(Y_L) \,,
\nonumber \\
F^{}_{b_1 b_2}(Y_C) &= \lim_{Y_R\to \infty\phantom{-}}
  s_{b_1 b_2}^{-1}(Y_R - Y_C)\, F^{}_{\text{us}, b_1 b_2}(Y_R)
\end{align}
for the distributions in a right-moving and a left-moving proton,
respectively.  An analogous definition in the $\delta$ regulator scheme was
put forward in \cite{Vladimirov:2016qkd}.

From the construction in appendix~\ref{app:matrix-algebra} it follows that
the symmetry relations \eqref{S-star} for $S_{a_1 a_2}$ imply corresponding
relations
\begin{align}
  \label{s-star}
s_{qq}^{} = & s_{\bar{q}\bar{q}}^* \,,
&
s_{q\bar{q}}^{} &= s_{\bar{q}q}^* \,,
&
s_{qg}^{} &= s_{\bar{q}g}^* \,,
&
s_{gg}^{} &= s_{gg}^*
\end{align}
if the matrices $U_{a_1 a_2}$ in \eqref{s-def} are chosen such that they
satisfy the relations \eqref{U-star}.  In the expression
\eqref{eq:DPS_colsoft} for double Drell-Yan production we can thus rewrite
\begin{align}
  \label{CS-Xsect}
   F^T_{\us, \bar{q}\bar{q}}(Y_R)\, s_{qq}^{\dagger -1}(Y_R-Y_C)\,
     s_{qq}^{-1}(Y_C-Y_L)\, F^{}_{\us, qq}(Y_L)
&= F^T_{\bar{q}\bar{q}}(Y_C)\, F^{}_{qq}(Y_C) \,,
\end{align}
where we have omitted the hard-scattering factors $H_{i, q\bar{q}}$ for
brevity.  A corresponding argument leads to the combination
$F^T_{\bar{q}q}(Y_C)\, F^{}_{q\bar{q}}(Y_C)$ in the same process.  The
production of one Higgs and one electroweak gauge boson involves a term
\begin{align}
  \label{CS-Xsect-mix}
   F^T_{\us, \bar{q} g}(Y_R)\, s_{qg}^{\dagger -1}(Y_R-Y_C)\,
     s_{qg}^{-1}(Y_C-Y_L)\, F^{}_{\us, qg}(Y_L)
&= F^T_{\bar{q} g}(Y_C)\, F^{}_{qg}(Y_C) 
\end{align}
with mixed quark-gluon and antiquark-gluon DPDs, and likewise one obtains the
product $F^T_{gg}(Y_C)\, F^{}_{gg}(Y_C)$ in double Higgs boson production.

In \eqref{F-sub-def} we can take the limit of infinite Wilson line
rapidities $Y_L$ and $Y_R$ in the unsubtracted matrix elements.  This
means that the Wilson lines in the parton operators $\mathcal{O}_a$ are
along lightlike paths, which leads to important simplifications as
discussed in chapter~10.11 of \cite{Collins:2011zzd}.  Let us see that
this limit is well behaved in \eqref{F-sub-def}.  The rapidities $Y_L$ and
$Y_R$ in the cross section formula originate from Grammer-Yennie
approximations (see e.g.~\cite{Diehl:2015bca}), and one finds that their
precise values in \eqref{CS-Xsect} and \eqref{CS-Xsect-mix} do not matter
as long as $|Y_L|, |Y_R| \gg 1$.  One can thus take the limit of infinite
$Y_L$ and $Y_R$ in these expressions, and thus also in the individual
factors in \eqref{F-sub-def}.  By contrast, taking infinite Wilson line
rapidities in individual unsubtracted DPDs or in the soft factor would
lead to rapidity divergences.

The dependence of the DPDs on the central rapidity $Y_C$ is easily
obtained from \eqref{CS-for-s}, which gives
\begin{align}
	\label{inverse_s_rapidity}
\frac{\partial}{\partial Y}\, s^{-1}(Y) &= - K\ms s^{-1}(Y)
\end{align}
and thus
\begin{align}
	\label{CS-gen}
	\frac{\partial}{\partial Y_C}\ms F(Y_C) &= - K F(Y_C) 
\end{align}
for DPDs in a right-moving proton, whilst for a left-moving proton one has
the opposite sign on the r.h.s.\ of \eqref{CS-gen}.

The preceding construction can be performed for DTMDs and DPDFs alike.
Important differences between the two types of distributions arise at the
level of ultraviolet renormalisation.  In the next subsection, we discuss
this for the case of DTMDs, postponing the case of DPDFs to
section~\ref{sec:DPDF_evo}.

\subsection{Renormalisation of DTMDs}
\label{sec:ren-DTMDs}

The ultraviolet (UV) renormalisation of TMDs arises from vertex and self
energy corrections associated with composite operators that contain fields
at the same transverse position.  Interactions between fields at different
transverse positions do not give rise to UV divergences: the finite
spacelike distance acts as an effective UV cutoff in the corresponding
graphs.  Operators that require renormalisation are therefore $[ W(\xi)\,
  q(\xi) ]_i$, $[ W(\xi)\, G^{+i}(\xi) ]_a$ and their Hermitian conjugates
in the collinear matrix elements, as well as $[ W(\xi,v_L)\,
  W^\dagger(\xi,v_R) ]_{ij}$, $[ W(\xi,v_L)\, W^\dagger(\xi,v_R) ]_{ab}$
and their Hermitian conjugates in the soft factor.  Due to colour SU(3)
invariance, the corresponding renormalisation factors are all colour
independent.  We could in principle choose a different renormalisation
scale for each of the four parton operators in $F_{\us}$ and for each of
the four corresponding Wilson line products in the soft factor.  We choose
a slightly simpler variant, taking common renormalisation scales $\mu_1$
and $\mu_2$ for the operators associated with partons carrying momentum
fraction $x_1$ and $x_2$, respectively.  Denoting bare quantities with a
subscript $B$, we then have $O_a(x,\tvec{y},\tvec{z}) = Z_{\us, a}\,
O_{B,a}(x,\tvec{y},\tvec{z})$, where $Z_{\us, a}$ is the product of
renormalisation factors for the composite operators at transverse
positions $\tvec{y} - \half \tvec{z}$ and $\tvec{y} + \half \tvec{z}$.
Likewise, for the operator defined in \eqref{WL-ops} we have
$O_{S,a}(\tvec{y},\tvec{z}) = Z_{S,a}\, O_{BS,a}(\tvec{y},\tvec{z})$,
where $Z_{S,a}$ is the product of renormalisation factors for the Wilson
line pairs at equal transverse positions.\footnote{Our convention for
renormalisation factors $Z$ of composite operators corresponds to the one in
\protect\cite{Collins:1984xc,Collins:2011zzd}.  Other authors, such as the
ones of \protect\cite{Peskin:1995ev}, use $Z^{-1}$ instead.}
Both $Z_{\us,a}$ and $Z_{S,a}$ are independent of spin and colour, but
they differ for quarks and gluons (being equal for quarks and antiquarks
due to charge conjugation invariance).  They depend on a scale $\mu$ and
on Wilson line rapidities, and $Z_{\us,a}$ also depends on the
plus-momentum $x p^+$ of the relevant parton.  We thus have
\begin{align}
F_{\us, a_1 a_2}(x_i; \mu_i, Y_L)
  &= Z_{\us, a_1}(\mu_1, Y_L, x_1\ms p^+)\,
     Z_{\us, a_2}(\mu_2, Y_L, x_2\ms p^+)\,
     F_{B, \us, a_1 a_2}(x_i; Y_L) \,,
\nonumber \\
S_{a_1 a_2}(\mu_i, Y) &= Z_{S, a_1}(\mu_1, Y)\, 
                         Z_{S, a_2}(\mu_2, Y)\, S_{B, a_1 a_2}(Y)
\end{align}
with $Y= Y_R-Y_L$. Throughout this work we use the convention that a
function with arguments $x_i$ ($\mu_i$) depends on both $x_1$ and $x_2$
($\mu_1$ and $\mu_2$).  In this and the next subsection, we consider a
right-moving proton for definiteness, and we omit transverse position
arguments for brevity.

We now derive the renormalisation properties of the matrix $s_{a_1 a_2}(Y)$.
Assuming that the $Y$ dependence of $S_B(Y)$ is given by \eqref{CS-for-S} with
a bare kernel $\widehat{K}_{B}$ and defining
\begin{align}
  \label{Lambda-def}
\Lambda_a(Y) &= \frac{\partial}{\partial Y}\, \log Z_{S,a}(Y) \,,
\end{align}
we readily get
\begin{align}
  \label{CS-renorm}
\frac{\partial}{\partial Y}\, S_{a_1 a_2}(Y) &= \widehat{K}_{a_1 a_2}(Y)\,
   S_{a_1 a_2}(Y)
\end{align}
with
\begin{align}
	\label{K-renorm}
\widehat{K}_{a_1 a_2}(Y)
&= \Lambda_{a_1}(Y) \,\one + \Lambda_{a_2}(Y) \,\one
   + \widehat{K}_{B, a_1 a_2} \,,
\end{align}
where we have dropped the dependence on the renormalisation scales for
brevity.  The renormalised soft factor hence satisfies \eqref{CS-for-S} if
$\Lambda_a$ is independent of $Y$.  This is easily shown for the
$\text{MS}$ scheme.  Expanding $Z_S$ in the renormalised coupling, we have
$Z_{S} = 1 + \sum_{n=1}^\infty \alpha_s^n Z_{S}^{(n)}(\epsilon)$ with
coefficients $Z_{S}^{(n)}$ being a finite sum of poles in $\epsilon$ for
each $n$.  The corresponding coefficients in $\Lambda = \sum_{n=1}^\infty
\alpha_s^n \Lambda^{(n)}(\epsilon)$ are hence pure pole terms as well.
Since $S$ is finite for $\epsilon=0$, the same holds for $\widehat{K}$
according to \eqref{CS-renorm}, so that the $\epsilon$ poles of $\Lambda$
must cancel those of $\widehat{K}_{B}$ in \eqref{K-renorm}.  Since the
latter are $Y$ independent, the same holds for $\Lambda$.  In the
$\overline{\text{MS}}$ scheme one can repeat the previous argument after
rescaling $\alpha_s$ in the expansion of $Z_{S}$ by a factor
$S_\epsilon$ (see section~\ref{sec:renorm} for further explanation).

With $\Lambda_a$ being $Y$ independent, the solution of \eqref{Lambda-def}
reads $Z_{S,a}(Y) = Z_{S,a}(0)\, e^{Y \Lambda_a}$.  Using \eqref{s-def} and
\eqref{K-def}, one finds that
\begin{align}
K_{a_1 a_2}(\mu_i) &=
\Lambda_{a_1}(\mu_1) \,\one + \Lambda_{a_2}(\mu_2) \,\one
+ K_{B, a_1 a_2} \,,
\nonumber \\
s_{a_1 a_2}(\mu_i, Y) 
  &= Z_{s, a_1}(\mu_1, Y)\, Z_{s, a_2}(\mu_2, Y)\, s_{B, a_1 a_2}(Y) 
\end{align}
with $Z_{s,a}(Y) = \sqrt{Z_{S,a}(0)}\, e^{Y \Lambda_a}$, where we have
restored the $\mu$ dependence of the factors.  We thus find that the
renormalisation of $s(Y)$ is multiplicative and independent of the colour
channel, just as the one of $S(Y)$.  With bare and renormalised DTMDs
defined as in \eqref{F-sub-def}, we then have
\begin{align}
  \label{sub-F-def}
F_{a_1 a_2}(x_i; \mu_i, Y_C)
  &= Z_{F,a_1}(\mu_1, Y_C, x_1\ms p^+)\,
     Z_{F,a_2}(\mu_2, Y_C, x_2\ms p^+)\, F_{B, a_1 a_2}(x_i; Y_C)
\intertext{with}
  \label{sub-Z-fact}
Z^{}_{F,a}(\mu, Y_C, x p^+) &= \lim_{Y_L\to -\infty}
    Z^{-1}_{s,a}(\mu, Y_C-Y_L)\, Z^{}_{\us,a}(\mu, Y_L , x p^+)
\end{align}
for a right-moving proton, and correspondingly for a left-moving one.  Note
that, as explained in chapter~10.11 of \cite{Collins:2011zzd}, one should
first take the limit of infinite $Y_L$ in the product \eqref{sub-Z-fact} and
in $F_{B}(Y_C)$, and then let $\epsilon$ go to zero on the r.h.s.\ of
\eqref{sub-F-def}.

We can now derive the $\mu$ dependence of the DTMDs and of their
Collins-Soper kernel.  Defining
\begin{align}
\gamma_{F,a}(\mu, Y_C, x p^+)
   &= \frac{\partial}{\partial \log\mu}\, \log Z_{F,a}(\mu, Y_C, x p^+) \,,
&
\gamma_{K,a}(\mu)
   &= - \frac{\partial}{\partial \log\mu}\, \Lambda_{a}(\mu) \,,
\end{align}
we obtain for a right-moving proton
\begin{align}
	\label{RG-TMD}
\frac{\partial}{\partial \log\mu_1}\, F_{a_1 a_2}(x_i; \mu_i, Y_C)
   &= \gamma_{F,a_1}(\mu_1, Y_C, x_1\ms p^+)\,
      F_{a_1 a_2}(x_i; \mu_i, Y_C,) \,,
\nonumber \\
\frac{\partial}{\partial \log\mu_1}\, K_{a_1 a_2}(\mu_i)
   &= {}- \gamma_{K,a_1}(\mu_1)\, \one \,,
\end{align}
and analogous relations for the $\mu_2$ dependence, as well as the
additional condition
\begin{align}
  \label{Y-dep-gamma-F}
\frac{\partial}{\partial Y_C}\, \gamma_{F,a}(\mu, Y_C, x p^+)
     &= \gamma_{K,a}(\mu) \,.
\end{align}
We note that $Z_{F,a}$ and hence $\gamma_{F,a}$ can depend on $Y_C$ and
$x p^+$ only via the boost invariant combination $x p^+ e^{-Y_C}$.  This will
be used to introduce the rapidity parameter $\zeta$ later on (see
\eqref{zeta-def-TMD} and \eqref{zeta-def}).

It is easy to repeat the above arguments with four different renormalisation
scales for the four parton legs of $F$.  One then obtains analogues of
\eqref{RG-TMD} for each scale, with $\gamma_{K,a}$ replaced by
$\gamma_{K,a} /2$ and $\gamma_{F,a}$ by $\gamma_{E,a}$ or $(\gamma_{E,a})^*$,
where $\gamma_{E,a} + (\gamma_{E,a})^* = \gamma_{F,a}$.  The anomalous
dimension $\gamma_{E,a}$ has an imaginary part and is for a parton momentum
leaving $F$, whereas $(\gamma_{E,a})^{*}$ is for a parton momentum entering
$F$.  This can be seen in the study of the quark form factor
\cite{Collins:2017oxh} (see section~10.12.2 of \cite{Collins:2011zzd} for an
explicit calculation at leading order).

\subsection{Definition of single parton TMDs}
\label{sec:sub_singleTMD}

It is instructive to revisit the definition of single parton TMDs in the
framework we have just laid out.  The colour structure is considerably
simplified in this case.  The operator $O_a(x,\tvec{0},\tvec{z})$ in the
matrix element for an unsubtracted TMD in \eqref{x-matel} must carry colour
singlet quantum numbers, which is achieved by contracting the operators in
\eqref{colour-ops} with $\delta_{jj'}$ or $\delta_{aa'}$.
Likewise, the soft factor for single hard scattering involves the colour
singlet projection of the operator $O_{S,a}(\tvec{0},\tvec{z})$ defined in
\eqref{WL-ops}.  One can easily adapt the
arguments in the previous subsections to this case: the matrices $S$,
$\widehat{S}$, $s$, $K$ and $\widehat{K}$ then become single real valued
functions.  One finds $K=\widehat{K}$, whilst the definition \eqref{s-def}
simplifies to
\begin{align}
	\label{s_single_TMD}
s(Y) & = e^{Y K}\, \widehat{S}^{\, 1/2}(0)\,,
\end{align}
which is positive and satisfies \eqref{CS-for-s}.  This gives
\begin{align}
	\label{S-single-decomp}
S(Y) = \widehat{S}(Y) = e^{YK}\, \widehat{S}(0) = s(Y-Y_0)\, s(Y_0)
\end{align}
for $Y\gg1$ and arbitrary $Y_0$.
The ambiguity in the choice of $s$, due to unitary transformations in the
matrix case, is no longer present.  Combining the two previous equations, we
deduce that
\begin{align}
	\label{eq:S-single-simp}
s(Y) = \sqrt{ S(2Y) } \,.
\end{align}
The square-root factor in the construction of single parton TMDs by Collins,
given in equation (13.106) of \cite{Collins:2011zzd}, can be rewritten as
\begin{align}
	\label{sqrt-simp}
\sqrt{\frac{S(Y_R-Y_C)}{S(Y_R-Y_L)\, S(Y_C-Y_L)}}
  &= s^{-1}(Y_C-Y_L)
\end{align}
using \eqref{S-single-decomp} and \eqref{eq:S-single-simp}.  The final TMD is
then given by
\begin{align}
	\label{eq:unsub_single}
f_{a}(Y_C) &= \lim_{Y_L\to -\infty} s_{a}^{-1}(Y_C - Y_L)\, f_{\us,a}(Y_L) \,,
\end{align}
where we have restored the parton label $a$ to denote quarks, antiquarks
and gluons.  This form closely resembles the definition in
\cite{Echevarria:2015byo,GarciaEchevarria:2011rb}; we will expand on this
further in appendix~\ref{app:delta-reg}.  A related discussion has been
given in \cite{Collins:2012uy}, but the separation into different factors
made there differs from ours.

The renormalisation of single parton TMDs is done with the same factors
$Z_{\us,a}$, $Z_{S,a}$ and $Z_{F,a}$ as in section~\ref{sec:ren-DTMDs}, and
hence involves the same anomalous dimensions.  Coming back to our remark after
\eqref{Y-dep-gamma-F} and defining
\begin{align}
  \label{zeta-def-TMD}
  \zeta &= 2\ms (x p^+)^2\, e^{-2 Y_C}
\end{align}
for a right-moving proton, we recover the familiar evolution equations given in
\eqref{RG-single-TMD} to~\eqref{RG-single-K}.

\section{Colour}
\label{sec:colour}

An essential element in the description of DPS is the colour structure, which
is much more involved than in single hard scattering.  To deal efficiently
with matrices in colour space, we introduce projection operators and make use
of the fact that certain colour indices must couple to an overall colour
singlet.  \rev{We derive the combination \eqref{CS-Xsect-final} of colour
projected quantities that enters the DPS cross section in
section~\ref{sec:colour-defs} and establish a number of symmetry relations in
section~\ref{sec:soft-symm}.  In sections~\ref{sec:soft-simple} and
\ref{sec:interlude} we consider the case of collinear factorisation.
Projector identities such as the one in figure~\ref{fig:projector-comm} lead
to dramatic simplifications in this case, with soft matrices that are diagonal
and depend on only three independent functions as specified in
\eqref{coll-soft-sing} to \eqref{soft-octet}.  These simplifications also
allow us to derive the structure \eqref{col-Xsimp2} of the DPS cross section
for produced particles carrying colour.  A reader who is not interested in
technical details may skip the following derivations and will find the
principal results in the equations just mentioned.}

\subsection{Projection operators}
\label{sec:projectors}

For the fundamental representation of the group SU($N$), we introduce the
operators
\begin{align}
	\label{quark-proj}
P_{1}^{ii'\, jj'} &= \frac{1}{N}\ms \delta_{ii'} \delta_{jj'} \,,
\nonumber \\
P_{8}^{ii'\, jj'} &= 2\ms t^a_{ii'} t^a_{jj'} \,,
\end{align}
which project the index pairs $ii'$ and $jj'$ onto a colour singlet or a
colour octet, respectively.\footnote{We use the term ``octet'' to denote
  the adjoint representation, which of course has dimension 8 only for~SU(3).}
In these and all following projection operators, the four indices are
coupled to an overall colour singlet.  We take $T_F=1/2$ throughout this
paper and do not distinguish between upper and lower colour indices.  We
will make repeated use of the colour Fierz identity
\begin{align}
	\label{colour-Fierz}
2\ms t^a_{ii'} t^a_{jj'} &= \delta_{ij'} \delta_{i'j} - \frac{1}{N}\ms
\delta_{ii'} \delta_{jj'} \,.
\end{align}
For the adjoint representation we use
\cite{Bartels:1993ih,Mekhfi:1985dv,Keppeler:2012ih}
\begin{align}
	\label{gluon-proj}
P_{1}^{aa'\, bb'} &= \frac{1}{N^2-1}\ms \delta^{aa'} \delta^{bb'} \,,
\nonumber \\
P_{A}^{aa'\, bb'} &= \frac{1}{N}\ms f^{aa'c} f^{bb'c} \,,
\nonumber \\
P_{S}^{aa'\, bb'} &= \frac{N}{N^2-4}\ms d^{aa'c} d^{bb'c} \,,
\nonumber \\
P_{D}^{aa'\, bb'} &= \frac{1}{2}\ms \bigl( \delta^{ab}
\delta^{a'b'} - \delta^{ab'} \delta^{a'b} \bigr) - P_{A}^{aa'\, bb'} \,,
\nonumber \\
P_{27}^{aa'\, bb'} &= \frac{1}{2}\ms \bigl(
\delta^{ab} \delta^{a'b'} + \delta^{ab'} \delta^{a'b} \bigr)
- P_{S}^{aa'\, bb'} - P_{1}^{aa'\, bb'} \,.
\end{align}
The subscript on $P$ always denotes the colour representation onto which index
pairs $aa'$ and $bb'$ are projected, with $A$ denoting the antisymmetric and
$S$ the symmetric octet.\footnote{For $N>3$ a further representation appears
when decomposing the product of two octets (see table~9.4 in
\protect\cite{Cvitanovic:2008zz}).  This representation is sometimes labelled
as $R=0$, since it has dimension $0$ for $N=3$.  The projector $P_{27}$ in
\protect\eqref{gluon-proj} corresponds to $P_{27} + P_{0}$ in the case of
general $N$ \protect\cite{Keppeler:2012ih}.}
Useful relations for calculating with the $f$ and $d$
tensors are given in \cite{MacFarlane:1968vc}.

The operator $P_D = P_{10} + P_{\;\smash{\overline{10}}}$ projects on the
sum of the decuplet and antidecuplet representations.  For any tensor
$M^{aa'bb'}$ that transforms as an overall colour singlet, the projections
on decuplet and antidecuplet are equal (and can hence be added without loss
of information).  Following appendix~A of \cite{Kasemets:2014yna}, one can
show this by first decomposing $M^{aa'bb'}$ on $P_{R}^{ab\, a'b'}$ with
$R=1,A,S,10,\overline{10},27$ and then projecting on $P_{R}^{aa'\, bb'}$.
This result can be traced back to the fact that $(P_{10})^{\ms aa'\,bb'} -
(P_{\;\smash{\overline{10}}})^{\ms aa'\,bb'}$ is odd but all
$P_{R}^{ab\,a'b'}$ are even under the simultaneous exchange
$(a\leftrightarrow b, a'\leftrightarrow b')$.

We will further need projectors for mixed fundamental and adjoint indices:
\begin{align}
	\label{mixed-proj}
P_1^{ii'\, aa'} &= P_1^{aa'\, ii'} = \frac{1}{\sqrt{N(N^2-1)}}\,
  \delta_{ii'} \delta^{aa'} \,,
\nonumber \\
P_A^{ii'\, aa'} &= P_A^{aa'\, ii'} = \sqrt\frac{2}{N}\,
  t^c_{ii'}\ms f^{aa'c} \,,
\nonumber \\
P_S^{ii'\, aa'} &= P_S^{aa'\, ii'} = \sqrt\frac{2N}{N^2-4}\,
  t^c_{ii'}\ms d^{aa'c} \,.
\end{align}
The normalisation factors are chosen such as to yield the simple symmetry and
projection properties to be discussed next.

To ease our notation, we introduce double indices $\ul{i} = (ii')$, $\ul{a}
= (aa')$.  Here and in the following, $\ul{i}, \ul{j}, \ul{k}$ are in the
fundamental representation, $\ul{a}, \ul{b}, \ul{c}$ are in the adjoint,
whereas $\ul{r}, \ul{s}, \ul{t}, \ul{u}, \ul{v}, \ul{w}$ can belong to
either representation.  Repeated double indices are summed over, so that for
adjoint indices we have
\begin{align}
M_1^{\ul{r} \,\ul{a}}\, M_2^{\ul{a}\, \ul{s}}
   &= M_1^{\ul{r}\, aa'} M_2^{aa'\, \ul{s}} \,.
\end{align}
Some care is required for indices of the fundamental representation.  In
fundamental Wilson lines $W$ and in projection operators $P_R$, the first
index always transforms as a triplet and the second index as an antitriplet.
For such cases, we define matrix multiplication with a transposition
\begin{align}
\label{mat-mul-rule}
M_1^{\ul{r} \,\ul{i}}\, M_2^{\ul{i}\, \ul{s}}
   &= M_1^{\ul{r}\, ii'} M_2^{i'\!i\, \ul{s}} \,.
\end{align}
In soft matrices $S$ and in the matrix elements giving DPDs, the order of
triplet and antitriplet indices depends on the parton channel ($q$ or
$\bar{q}$) and in $S$ also on the direction of the Wilson lines (along $v_L$
or $v_R$).  Contraction of double indices should always be done such that a
triplet index is contracted with an antitriplet one.  This ensures proper
behaviour of the result under gauge transformations.

With these definitions, the projection operators have the symmetry
\begin{align}
  \label{proj-sym}
P_{R}^{\ul{r}\, \ul{s}} = P_{R}^{\ul{s}\, \ul{r}}
\end{align}
and the projection property
\begin{align}
	\label{proj-prop}
P_{R}^{\ul{r}\, \ul{s}}\, P_{R'}^{\ul{s}\, \ul{t}} &= \delta_{R R'}^{}\,
P_{R}^{\ul{r}\, \ul{t}}
\end{align}
for any pair of projectors $P_{R}, P_{R'}$ with a common double index $\ul{s}$
in either the fundamental or the adjoint representation.  In the octet sector,
it is understood that
$\delta_{8 A} = \delta_{A 8} = \delta_{8 S} = \delta_{S 8} = 1$ but
$\delta_{AS} = \delta_{SA} = 0$ in \eqref{proj-prop}.  The normalisation is
given by
\begin{align}
   \label{PR-norm}
P_{R}^{\ul{r}\, \ul{s}}\, P_{R'}^{\ul{r}\, \ul{s}} &= \delta_{R R'}^{}\,
m(R) \,,
\end{align}
where
\begin{align}
  \label{mR-def}
m(R) = P_R^{\ul{r}\, \ul{r}}
\end{align}
is the multiplicity of the representation, i.e.\ $m(1) = 1$, $m(8) = m(A)
= m(S) = 8$, $m(D) = 20$ and $m(27) = 27$ for SU(3). Since the projectors
above form a basis of the space of rank-four tensors $M$ that transform as an
SU(3) singlet, any such tensor $M$ can be decomposed as
\begin{align}
	\label{eq:mat-dec}
M^{\ul{r}\, \ul{s}} &= \sum_{R} \frac{1}{m(R)}\, P_R^{\ul{r}\, \ul{s}}\,
\bigl( P_{R}^{\ul{t}\, \ul{u}} M^{\ul{t}\, \ul{u}}_{\phantom{1}} \bigr)\,,
\end{align}
where the sum is over all representations $R$ given in \eqref{quark-proj},
\eqref{gluon-proj} and \eqref{mixed-proj}, as applicable. The contraction of
two colour singlet tensors is then given by
\begin{align}
	\label{eq:contr_proj}
M_1^{\ul{r}\, \ul{s}}\, M_2^{\ul{r}\, \ul{s}} &= \sum_{R} \frac{1}{m(R)}\,
\bigl( P_{R}^{\ul{r}\, \ul{s}} M_1^{\ul{r}\, \ul{s}} \bigr)\, \bigr(
P_{R}^{\ul{t}\, \ul{u}} M_2^{\ul{t}\, \ul{u}} \bigr)\,.
\end{align}
For representation labels $R$, we will \emph{not} use the summation
convention, i.e.\ summation over $R$ will always be indicated explicitly.

\subsection{Colour structure of the DPS cross section}
\label{sec:colour-defs}

The projection operators just introduced allow us to rewrite the DPS
cross section in a compact way.  We start by defining DPDs that are
projected on a definite colour representation $R$.  Using the general
decomposition in \eqref{eq:mat-dec}, we can write
\begin{align}
\label{dpd-colour-decomp}
F_{a_1 a_2}^{\ul{r}_1 \ul{r}_2} &=
\sum_{R} \frac{1}{\ii{a_1}{R}\, \ii{a_2}{R}}\,
  \frac{1}{\mathcal{N}_{a_1} \mathcal{N}_{a_2}}\,
  \frac{1}{\sqrt{m(R)}}\;
  \pr{R}{F}_{a_1 a_2}\, P_R^{\ul{r}_1 \ul{r}_2}
\end{align}
with
\begin{align}
\label{proj-dpd-def}
\pr{R}{F}_{a_1 a_2} &=
  \ii{a_1}{R}\, \ii{a_2}{R}\;
  \mathcal{N}_{a_1} \mathcal{N}_{a_2}\,
  \frac{1}{\sqrt{m(R)}}\;
  P_{R}^{\ul{s}_1 \ul{s}_2} F_{a_1 a_2
  \phantom{R}}^{\ul{s}_1 \ul{s}_2}  \!\!.
\end{align}
Note that the lower indices $a_1, a_2$ denote the parton species and
polarisation, not colour.  The factors
\begin{align}
  \label{eps-def}
\ii{a}{R} &= \begin{cases}
  \ms i & \text{if $a$ is a gluon and $R=A$} \\
  \ms 1 & \text{otherwise}
\end{cases}
\end{align}
ensure that the collinear distributions $\pr{A}{F_{qg}}(x_i, \tvec{y})$ and
$\pr{A}{F_{\bar{q} g}}(x_i, \tvec{y})$ and their polarised counterparts are
real valued (rather than imaginary), as are the collinear distributions in
all other channels.  This is shown in the next subsection.  The
prefactors $\mathcal{N}_a$ are given by
\begin{align}
\mathcal{N}_{q} = \mathcal{N}_{\bar{q}} &= \sqrt{N} \,, & \mathcal{N}_{g}
&= \sqrt{N^2-1}
\end{align}
and likewise for polarised partons.  They ensure that colour singlet
distributions ${}^{1} F_{a_1 a_2}$ involve a \emph{sum} over the colours of
the two partons.
The definitions here coincide with the ones in
\cite{Diehl:2011yj,Kasemets:2014yna}, with the notational change
\begin{align}
\pr{D}{F}_{gg} \ms\big|_{\text{here}}
&= \pr{10 + \overline{10}}{F}_{gg}
   \ms\big|_{\text{Ref.~\protect\cite{Kasemets:2014yna}}}
 = \frac{1}{\sqrt{2}}\ms \bigl( \pr{10}{F}_{gg}
        + \pr{\overline{10}}{F}_{gg} \bigr)
   \ms\big|_{\text{Ref.~\protect\cite{Diehl:2011yj}}}
\end{align}
and its analogues for polarised gluons.
For the discussion of renormalisation and the short-distance expansion, it is
useful to project the partonic operators \eqref{x-ops} on definite colour
representations as well.  Introducing
\begin{align}
  \label{col-proj-op}
\prn{R}{O}^{\ms\ul{r}}_{a\rule{0pt}{1.15ex}}
 &= \ii{a}{R}\; \mathcal{N}_a\, P_R^{\ul{r} \ul{s}}\,
    O^{\ms\ul{s}}_{a\rule{0pt}{1.15ex}}
\end{align}
and defining unsubtracted DPDs $\pr{R}{F}_{\us, a_1 a_2}$ in analogy to
\eqref{proj-dpd-def}, we obtain
\begin{align}
  \label{col-proj-matel}
2\pi \delta(p^+ - p'^+)\, 2p^+\, \pr{R}{F}_{\us, a_1 a_2}
&=  \frac{1}{\sqrt{m(R)}}\, \big\langle p' \big|\,
       \prn{R}{O}^{\ms\ul{r}}_{a_1}\;
       \prn{R}{O}^{\ms\ul{r}}_{a_2}\, \big| p \big\rangle
\end{align}
from \eqref{x-matel} and the projection property \eqref{proj-prop}.

The soft factor for DPS producing colour singlet particles in the hard
interactions carries two times four indices.  Its projection on different
representations is defined by
\begin{align}
\label{soft-proj}
\prb{RR'}{S_{a_1 a_2}} &=
  \frac{\ii{a_1}{R}\, \ii{a_2}{R}}{\ii{a_1}{R'}\, \ii{a_2}{R'}}
  \frac{1}{\sqrt{m(R)\, m(R')}}\, P_{R}^{\ul{r}_1
  \ul{r}_2}\, S_{a_1 a_2 \phantom{R}}^{\ul{r}_1 \ul{r}_2, \ul{s}_1
  \ul{s}_2}\, P_{R'}^{\ul{s}_1 \ul{s}_2}\,.
\end{align}
$R$ specifies the colour representation for the Wilson lines with positive
rapidity, and $R'$ the one for the Wilson lines with negative rapidity.
Notice that the $\varepsilon$ factors cancel for $R=R'$.
In analogy to \eqref{soft-proj} we define projections $\prb{RR'}{s}$ and
$\prb{RR'}{K}$ of the matrix $s$ and of the Collins-Soper kernel $K$, as
well as projections ${}^{RR'} (S^{-1})$ and ${}^{RR'} (s^{-1})$ of the
inverse matrices $S^{-1}$ and $s^{-1}$ in colour space.
Using the same argument as in \eqref{eq:contr_proj}, one can show that
\begin{align}
   \label{S-inverse-proj}
  \sum_{R'} \pr{RR'}{(S_{a_1 a_2}^{-1})}\; \prb{R'R''}{S_{a_1 a_2}^{}}
= \sum_{R'} \pr{RR'}{(s_{a_1 a_2}^{-1})}\; \prb{R'R''}{s_{a_1 a_2}^{}}
  &= \delta_{RR''}\,.
\end{align}
We see that matrix multiplication in the space of four colour indices
$(\ul{r}_1 \ul{r}_2) = (r_1^{} r_1' r_2^{} r_2')$ turns into matrix
multiplication in the space of colour representations $R$.  The matrix
decomposition \eqref{S-decomp} can thus be rewritten as
\begin{align}
  \label{S-decomp-proj}
\pr{RR''}{S_{a_1 a_2}}(Y)
&= \sum_{R'} \pr{RR'}{s_{a_1 a_2}}(Y-Y_0)\;
   \prb{R'R''}{\bigl( s_{a_1 a_2}^{\dagger}(Y_0) \bigr)} \,,
\end{align}
and the definition \eqref{F-sub-def} of DPDs in a right-moving proton as
\begin{align}
  \label{sub-unsub}
\pr{R}{F_{a_1 a_2}}(Y_C)
&= \lim_{Y_L\to -\infty} \sum_{R'}
   \pr{RR'}{\bigl( s_{a_1 a_2}^{-1}(Y_C-Y_L) \bigr)}\;
     \prb{R'}{F_{\us, a_1 a_2}^{}}(Y_L).
\end{align}

According to \eqref{eq:DPS_colsoft} and \eqref{CS-Xsect}, the cross
section for the production of colour neutral particles involves the
product
\begin{align}
	\label{CS-Xsect-final}
X &= H_{a_1 b_1}\, H_{a_2 b_2}\, F_{b_1 b_2}^{\ul{r}_1 \ul{r}_2}\, F_{a_1
     a_2 \phantom{b}}^{\ul{r}_1 \ul{r}_2} = \frac{H_{a_1
    b_1}}{\mathcal{N}_{a_1}\ms \mathcal{N}_{b_1}}\, \frac{H_{a_2
    b_2}}{\mathcal{N}_{a_2}\ms \mathcal{N}_{b_2}}\,
    \sum_{R} \csgn{a_1 a_2}{R}\;
    \pr{R}{F_{b_1 b_2}} \pr{R}{F_{a_1 a_2}} \,.
\end{align}
Here we have combined the factors $\ii{a_1}{R}\, \ii{b_1}{R}\,
\ii{a_2}{R}\, \ii{b_2}{R}$ into
\begin{align}
  \label{col-sign}
\eta^{}_{a_1 a_2}(R) &=
\frac{1}{\varepsilon^2_{a_1}(R)\, \varepsilon^2_{a_2}(R)} \,,
\end{align}
using that the only channels that produce colourless particles are $q\bar{q}$
and $gg$ annihilation.
Multiplied with a flux factor, the combinations
$H_{ab} /(\mathcal{N}_{a}\ms \mathcal{N}_{b})$ turn into the hard-scattering
cross sections in the final factorisation formula.  The factor
$\mathcal{N}_{a}\ms \mathcal{N}_{b}$ is $N$ for $q\bar{q}$ annihilation and
$N^2-1$ for $gg$ annihilation and ensures that one takes an \emph{average}
over the colour states of the two colliding partons, as is appropriate for a
parton-level cross section.

\subsection{Symmetry properties}
\label{sec:soft-symm}

As already remarked, multiplication of matrices and vectors, $S$ and $F$, in
the space of colour indices $(\ul{r}_1 \ul{r}_2)$ turns into multiplication of
matrices and vectors in the space of representations $R$.  For brevity, we
refer to the former as ``colour space'' and to the latter as ``representation
space''.  Note that our representation space only corresponds to the
part of colour space in which the indices
$(\ul{r}_1 \ul{r}_2)$ are coupled to an overall colour singlet.  We now derive
some important properties of $S$ and of the related matrices $s$ and $K$ in
representation space.

In \eqref{soft-symm} we have seen that the matrix $S_{a_1 a_2}$ is Hermitian
in colour space,
\begin{align}
S_{a_1 a_2}^{\ul{r}_1 \ul{r}_2, \ul{s}_1 \ul{s}_2}
 &= \bigl( S_{a_1 a_2}^{\ul{s}_1 \ul{s}_2, \ul{r}_1 \ul{r}_2} \bigr)^* \,.
\end{align}
Under Hermitian conjugation, an index pair $\ul{i} = (i,i')$ in $S$ changes
its transformation properties: if $i$ transforms as an SU(3) triplet in $S$,
it transforms as an antitriplet in $S^\dagger$ and vice versa.  The colour
projectors satisfy $P_R^{ii'\, jj'} = (P_R^{i'\!i\, j'\!j})^*$,
$P_R^{ii'\, aa'} = (P_R^{i'\!i\, aa'})^*$ and
$P_R^{aa'\, bb'} = (P_R^{aa'\, bb'})^*$, so that the rule for contraction
with the colour projectors stated below \eqref{mat-mul-rule} is satisfied if
$S$ is contracted with $P_R^{}$ and $S^\dagger$ with $P_R^*$.  We thus find
that
\begin{align}
  \label{S-hermit}
\prb{RR'}{S}_{a_1 a_2} &= \left( \pr{R'\!R}{S}_{a_1 a_2} \right)^* \,,
\end{align}
i.e.\ $S$ is also Hermitian in representation space.  In analogy, the fact
that the Collins-Soper kernel $K_{a_1 a_2}$ is Hermitian in colour space
implies that it is Hermitian in representation space as well.

We now turn to the relations between soft factors for quarks and antiquarks.
The relations \eqref{S-star} imply that
\begin{align}
  \label{S-anti}
\prb{RR'}{S}_{a_1 a_2} &=
  \frac{\csgn{a_1 a_2}{R'}}{\csgn{a_1 a_2}{R}}\,
  \left( \prb{RR'}{S}_{\bar{a}_1 \bar{a}_2} \right)^*
= \frac{\csgn{a_1 a_2}{R'}}{\csgn{a_1 a_2}{R}}\;
  \pr{R'\!R}{S}_{\bar{a}_1 \bar{a}_2} \,,
\end{align}
where $\bar{a}$ denotes the antiparton of $a$ (with the convention
$\bar{g} = g$).  The complex conjugation of projection operators implicit in
the first step corresponds to the fact that an index $i$ transforming as a
triplet in $S_{a_1 a_2}$ transforms as an antitriplet in
$S_{\bar{a}_1 \bar{a}_2}$ and vice versa.  Complex conjugation of the
$\varepsilon$ factors in the definition \eqref{soft-proj} gives the factors
$\eta$ in \eqref{S-anti}.  In the second step we have simply used the property
\eqref{S-hermit}.
We now recall that the relations \eqref{S-star} for $S$ translate into
analogous relations \eqref{s-star} for $s$.  By taking the rapidity
derivative, we find corresponding relations for $K$ as well.  In analogy to
\eqref{S-anti} we can thus derive
\begin{align}
  \label{K-anti}
\prb{RR'}{K}_{a_1 a_2} &=
  \frac{\csgn{a_1 a_2}{R'}}{\csgn{a_1 a_2}{R}}\;
  \pr{R'\!R}{K}_{\bar{a}_1 \bar{a}_2} \,.
\end{align}
We will use this in section~\ref{sec:evol-gen} when making rapidity evolution
explicit in the DPS cross section.
A different way to connect  soft factors for quarks and antiquarks is to note
that the operator $O_{S,q}$ in \eqref{WL-ops}  is related with its analogue
$O_{S,\bar{q}}$ for antiquarks by
\begin{align}
\bigl[ O_{S,q}(\tvec{y},\tvec{z}; v_L,v_R) \bigr]^{i i', l l'}
&= \bigl[ O_{S,\bar{q}}(\tvec{y},-\tvec{z}; v_L,v_R) \bigr]^{i'\! i,\, l'\bs l}
   \,.
\end{align}
Multiplying with colour projectors according to the rule stated below
\eqref{mat-mul-rule}, one obtains
\begin{align}
  \label{S-qbar}
\prb{RR'}{S}_{qq}(\tvec{z}_1, \tvec{z}_2, \tvec{y})
&= \prb{RR'}{S}_{q\bar{q}}(\tvec{z}_1, -\tvec{z}_2, \tvec{y})
\nonumber \\
&= \prb{RR'}{S}_{\bar{q}q}(- \tvec{z}_1, \tvec{z}_2, \tvec{y})
 = \prb{RR'}{S}_{\bar{q}\bar{q}}(- \tvec{z}_1, -\tvec{z}_2, \tvec{y}) \,,
\end{align}
where the arguments $v_L,v_R$ are the same in all four expressions.  Relations
analogous to \eqref{S-qbar} hold for $\prb{RR'}{K}_{a_1 a_2}$, provided that
$\pr{RR'}{U}_{a_1 a_2}$ satisfies the requirements \eqref{U-qbar}.

Let us finally investigate the consequences of charge conjugation invariance.
Charge conjugation transforms
$t^a A^{\mu a}(\xi) \to - (t^a)^* A^{\mu a}(\xi)$, which can be derived from
the familiar transformation properties of quark fields and charge conjugation
invariance of the interaction term
$- g\ms \bar{q}\ms t^a\bs A^{\mu a} \gamma_\mu\ms q$ in the QCD Lagrangian.  As
a result, a fundamental Wilson line transforms as
\begin{align}
  \label{WL-C-conj}
W_{ij}^{} \to W_{ij}^{*} \,.
\end{align}
For soft factors $\prb{RR'}{S}$ involving gluons, we rewrite all adjoint
Wilson lines as \cite{Bomhof:2006dp}
\begin{align}
	\label{wils-adj-to-fun}
W^{ab} = 2 \tr ( t^a W t^b W^\dagger ) \,,
\end{align}
where the Wilson lines on the r.h.s.\ are in the fundamental representation,
express $f^{abc}$ and $d^{abc}$ as traces of products of matrices $t^a$,
$t^b$, $t^c$ and then eliminate all these matrices by repeated use of the
colour Fierz identity \eqref{colour-Fierz}.  After these steps, we have for
all parton combinations $a_1$ and $a_2$ a representation of
$\prb{RR'}{S}_{a_1 a_2}$ in terms of traces of fundamental Wilson lines $W$
and $W^\dagger$, where thanks to the $\epsilon$ factors in the definition
\eqref{soft-proj} no explicit factors of $i$ appear.  The transformation
\eqref{WL-C-conj} then readily implies that $\prb{RR'}{S}_{a_1 a_2}$ is real
valued, and with \eqref{S-hermit} we furthermore obtain that it is symmetric in
representation space, i.e.\ we have $\prb{RR'}{S} = \pr{R'\!R}{S}$, where for
brevity we omit the subscripts $a_1 a_2$.

To show that $\prb{RR'}{s}$ and $\prb{RR'}{K}$ are also real valued, we
translate the derivation starting in \eqref{S-hat-def-2} into representation
space, projecting all relevant matrices as specified in \eqref{soft-proj}.
It is then easy to see that $\prb{RR'}{\widehat{S}}$ and
$\prb{RR'}{\widehat{K}}$ are real valued.  To establish that
$\prb{RR'}{(\widehat{S}^{\, 1/2})}$ is real valued, we note that
$\prb{RR'}{\widehat{S}}$ has positive eigenvalues, which is readily seen by
projecting the eigenvalue equation $\widehat{S} v = \lambda v$ from colour
space to representation space and recalling that in colour space
$\widehat{S}$ is a positive matrix.  According to the discussion below
\eqref{S-hat-decomp}, the square root of a positive real matrix is real.
Choosing the matrix $U$ in \eqref{s-def} and \eqref{K-def} such that
$\pr{RR'}{U}$ is real, we find that both $\prb{RR'}{s}$ and $\prb{RR'}{K}$
are real.  Since $\prb{RR'}{K}$ is Hermitian, it is also symmetric:
$\prb{RR'}{K} = \pr{R'\!R}{K}$.

Combining the symmetry relations \eqref{S-anti} and \eqref{S-qbar} with the
fact that $\prb{RR'}{S}$ is symmetric in $R$ and $R'$, we obtain
\begin{align}
  \label{S-even}
\prb{RR'}{S}_{qq}(\tvec{z}_1, \tvec{z}_2, \tvec{y})
  &= \prb{RR'}{S}_{qq}(-\tvec{z}_1, -\tvec{z}_2, \tvec{y}) \,.
\end{align}
An analogous relation holds for $\prb{RR'}{K}_{qq}$.

The quark and gluon operators in \eqref{eq:quark-ops} and
\eqref{eq:gluon-ops} satisfy $\bigl[ \mathcal{O}_a^{rr'} \bigr]^\dagger(y,z)
= \mathcal{O}_a^{r'\!r}(y,-z)$, where the colour index $r$ pertains to the
fields at $y+\half z$ and $r'$ to the fields at $y-\half z$.  For colour
projected unsubtracted DPDFs this yields the relation
\begin{align}
  \label{Fus-real}
\bigl[ \pr{R}{F}_{\us, a_1 a_2}(x_i,\tvec{y}) \bigr]^*
  &= \pr{R}{F}_{\us, a_1 a_2}(x_i,\tvec{y}) \,.
\end{align}
Let us explicitly show this for $\pr{A}{F}_{\us, qg}$.  Using that
$\ii{q}{A}\ms \ii{g}{A} = i$ and
$( P_A^{jj'\, aa'} )^* = - P_A^{j'\!j\; a'\!a}$, we have
\begin{align}
& \bigl[ \pr{A}{F}_{\us, qg}(x_i,\tvec{y}) \bigr]^*
\nonumber \\
&\qquad = c \int\! dz^-_1\, dz^-_2\, dy^-\,  \Bigl[\ms i\ms P_A^{jj'\, aa'}
          e^{i\ms ( x_1^{} z_1^- + x_2^{} z_2^-)\ms p^+}\,
          \langle p|\, \mathcal{O}^{jj'}_{q}(y,z_1)\,
          \mathcal{O}^{aa'}_{g}(0,z_2)\, |p \rangle \Bigr]^*
\nonumber \\
&\qquad = c \int\! dz^-_1\, dz^-_2\, dy^-\, i\ms P_A^{j'\!j\; a'\!a}\,
          e^{- i\ms ( x_1^{} z_1^- + x_2^{} z_2^-)\ms p^+}\,
          \langle p|\, \mathcal{O}^{j'\!j}_{q}(y,-z_1)\,
          \mathcal{O}^{a'\!a}_{g}(0,-z_2)\, |p \rangle
\nonumber \\
&\qquad = \pr{A}{F}_{\us, qg}(x_i,\tvec{y}) \,, \phantom{\int}
\end{align}
where in the last step we used that $\tvec{z}_1 = \tvec{z}_2 = \tvec{0}$.
Here $c = c^*$ is a product of kinematic and numerical factors not essential
for the argument.  We see that the factor $\ii{g}{A} = i$ compensates the
behaviour of $P_A^{jj'\, aa'}$ under complex conjugation.  One easily checks
that
$\bigl[ \ii{a_1}{R}\ms \ii{a_2}{R}\, P_R^{rr'\, ss'} \bigr]^* = \ii{a_1}{R}\ms
\ii{a_2}{R}\, P_R^{r'\bs r\; s'\!s}$ for all projectors, which ensures the
general validity of \eqref{Fus-real}.  Given that $\prb{RR'}{s}$ is real
valued, we finally find that
\begin{align}
\Bigl[\ms \pr{R}{F}_{a_1 a_2}(x_i,\tvec{y}) \ms\Bigr]^*
  &= \pr{R}{F}_{a_1 a_2}(x_i,\tvec{y}) \,.
\end{align}
In all colour channels and for all parton combinations, DPDFs are thus real
valued.

\subsection{Simplification of soft factors in collinear factorisation}
\label{sec:soft-simple}

In the soft factors relevant for collinear factorisation, one should set
$\tvec{z}_1 = \tvec{z}_2 = \tvec{0}$ in the definitions
\eqref{soft-gen-def} and \eqref{soft-def}.  Corresponding Wilson lines in
the scattering amplitude and its complex conjugate are then at the same
transverse position.  This simplifies the colour structure significantly,
as we will now demonstrate.  In the remainder of this subsection, it is
understood that all soft factors are taken at $\tvec{z}_1 = \tvec{z}_2 =
\tvec{0}$.

\begin{figure}
\begin{center}
\includegraphics[width=0.66\textwidth]{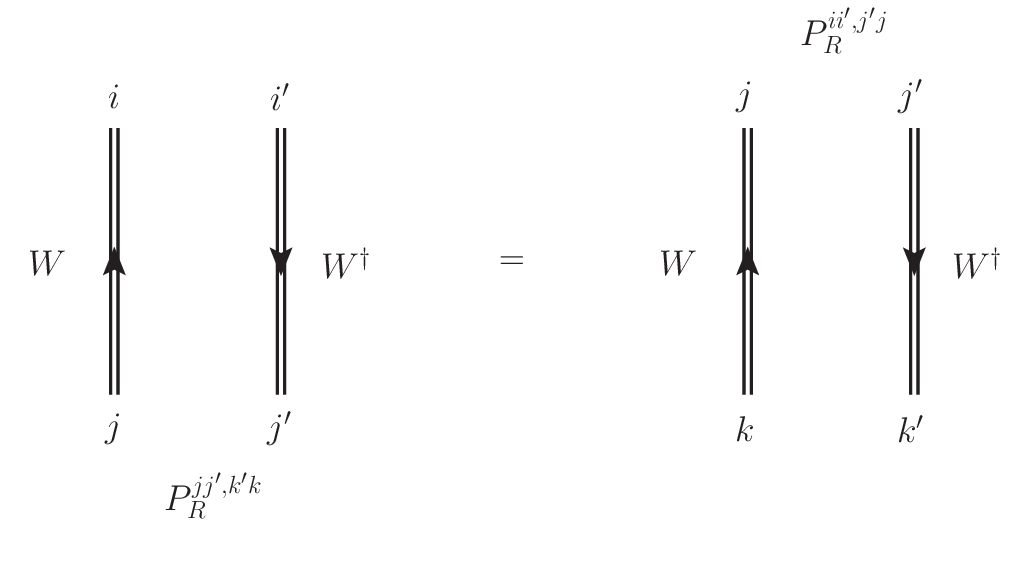}
\caption{\label{fig:projector-comm} Graphical representation of the
  relation \protect\eqref{comm-quark}.  The Wilson lines are understood to
  be interacting, i.e.\ any number of gluons may be attached to them in a
  Feynman graph.}
\end{center}
\end{figure}

The projection operators on Wilson lines obey the relation
\begin{align}
	\label{comm-quark}
W^{}_{ij}\, P_R^{jj',k'k}\, W^\dagger_{j'i'} &= W^{}_{jk}\,
P_R^{ii',j'j}\, W^\dagger_{k'j'}
\end{align}
for $R=1,8$, where it is understood that the Wilson lines $W$ and
$W^\dagger$ are in the same direction and at the same position.  For $R=1$
this is trivial, and for $R=8$ it readily follows from the colour Fierz
identity \eqref{colour-Fierz}.  A graphical representation of
\eqref{comm-quark} is shown in figure~\ref{fig:projector-comm}.
Remarkably, an analogous relation holds also for adjoint Wilson lines
\begin{align}
	\label{comm-glu}
W^{}_{ab}\, P_R^{bb',cc'}\, W^\dagger_{b'a'}
  &= W^{}_{bc}\, P_R^{aa',bb'}\, W^\dagger_{c'b'}
\end{align}
and for the mixed case
\begin{align}
	\label{comm-mixed}
W^{}_{ab}\, P_R^{bb',k'k}\, W^\dagger_{b'a'}
   &= W_{jk}\, P_R^{aa',j'j}\, W^\dagger_{k'j'}
\end{align}
in all relevant representations $R$.  This can be shown by following the steps
in and after \eqref{wils-adj-to-fun}.  For the extended soft factor, the
relations \eqref{comm-quark}, \eqref{comm-glu} and \eqref{comm-mixed} imply
\begin{align}
  \label{comm-soft}
P_R^{\ul{s} \,\ul{t}}\;
   (S_{a_1 a_2})^{\cdots\, \ul{r} \,\cdots}_{\cdots\, \ul{s} \,\cdots}
&= P_R^{\ul{r} \,\ul{s}}\;
   (S_{a_1 a_2})^{\cdots\, \ul{s} \,\cdots}_{\cdots\, \ul{t} \,\cdots} \,,
\end{align}
where we have displayed only one of the four double index pairs
$(\ul{r}_i, \ul{s}_i)$ in
$(S_{a_1 a_2})^{\ul{r}_1 \ul{r}_2, \ul{r}_3 \ul{r}_4}_{\ul{s}_1 \ul{s}_2,
\ul{s}_3 \ul{s}_4}$ for better readability, the other three pairs remaining
untouched.
Instead of making a colour projection for the fields at $\xi^+ = \xi^- = 0$,
one can thus make the same projection for the fields at infinity.  This
yields an important relation between a specific projection of the extended
soft factor in \eqref{soft-gen-def} and the basic one in \eqref{soft-def}.
With \eqref{comm-soft} and the projection property \eqref{proj-prop} we have
\begin{align}
  \label{soft-inner-proj}
	P_{R_1}^{\ul{s}_1 \ul{t}_1}\, P_{R_2}^{\ul{s}_2 \ul{t}_2}\,
        (S_{a_1 a_2})^{\ul{r}_1 \ul{r}_2, \ul{u}_1 \ul{u}_2}_{ \ul{s}_1
          \ul{s}_2, \ul{t}_1 \ul{t}_2} &= P_{R_1}^{\ul{s}_1 \ul{v}_1}\,
        P_{R_1}^{\ul{t}_1 \ul{v}_1}\, P_{R_2}^{\ul{s}_2 \ul{v}_2}\,
        P_{R_2}^{\ul{t}_2 \ul{v}_2}\, (S_{a_1 a_2})^{\ul{r}_1 \ul{r}_2,
          \ul{u}_1 \ul{u}_2}_{ \ul{s}_1 \ul{s}_2, \ul{t}_1 \ul{t}_2}
\nonumber \\
 &= P_{R_1}^{\ul{r}_1 \ul{s}_1}\, P_{R_1}^{\ul{u}_1
          \ul{t}_1} P_{R_2}^{\ul{r}_2 \ul{s}_2}\, P_{R_2}^{\ul{u}_2
          \ul{t}_2} (S_{a_1 a_2})^{\ul{s}_1 \ul{s}_2, \ul{t}_1 \ul{t}_2}_{
          \ul{v}_1 \ul{v}_2, \ul{v}_1 \ul{v}_2} \,,
\end{align}
where $\ul{r}, \ul{s}, \ldots, \ul{v}$ can each be in the fundamental or
adjoint representation.  Projecting this on the remaining open indices, we
obtain
\begin{align}
	\label{soft-factor-comm}
	P_{R}^{\ul{r}_1 \ul{r}_2}\, P_{R'}^{\ul{u}_1 \ul{u}_2}\,
        P_{R_1}^{\ul{s}_1 \ul{t}_1}\, P_{R_2}^{\ul{s}_2 \ul{t}_2}\,
        (S_{a_1 a_2})^{\ul{r}_1 \ul{r}_2, \ul{u}_1 \ul{u}_2}_{ \ul{s}_1
          \ul{s}_2, \ul{t}_1 \ul{t}_2}
 &= \delta^{}_{R R_1}\, \delta^{}_{R
          R_2}\, \delta^{}_{R R'}\, P_{R}^{\ul{s}_1 \ul{s}_2}\,
        P_{R}^{\ul{t}_1 \ul{t}_2}\, (S_{a_1 a_2})^{\ul{s}_1 \ul{s}_2,
          \ul{t}_1 \ul{t}_2}_{ \ul{v}_1 \ul{v}_2, \ul{v}_1 \ul{v}_2}
\nonumber \\
 &= \delta^{}_{R R_1}\, \delta^{}_{R R_2}\,
        \delta^{}_{R R'}\, m(R)\, \pr{RR}{S_{a_1 a_2}} \,,
\end{align}
where in the last step we have used the definition \eqref{soft-proj}.
We furthermore have
\begin{align}
\label{soft-factor-diag}
\prb{RR'}{S_{a_1 a_2}}\,
  \frac{\ii{a_1}{R'}\, \ii{a_2}{R'}}{\ii{a_1}{R}\, \ii{a_2}{R}}\,
  \sqrt{m(R) m(R')}
&= P_{R}^{\ul{r}_1 \ul{r}_2}\,
  P_{R'}^{\ul{t}_1 \ul{t}_2}\, (S_{a_1 a_2})^{\ul{r}_1 \ul{r}_2, \ul{t}_1
  \ul{t}_2}_{ \ul{s}_1 \ul{s}_2, \ul{s}_1 \ul{s}_2}
\nonumber \\
& = P_{R}^{ \ul{r}_2
  \ul{s}_2}\, P_{R'}^{\ul{t}_2 \ul{s}_2}\, (S_{a_1 a_2})^{\ul{r}_1
  \ul{r}_1, \ul{t}_1 \ul{t}_1}_{ \ul{s}_1 \ul{r}_2, \ul{s}_1 \ul{t}_2}
\nonumber \\[0.3em]
& = \delta_{RR'}^{}\, P_{R}^{\ul{r}_2 \ul{t}_2}\, (S_{a_1
  a_2})^{\ul{r}_1 \ul{r}_1, \ul{t}_1 \ul{t}_1}_{ \ul{s}_1 \ul{r}_2,
  \ul{s}_1 \ul{t}_2} \,,
\end{align}
which shows that at $\tvec{z}_1 = \tvec{z}_2 = \tvec{0}$ the soft matrix is
diagonal in the colour representations $R$ and $R'$.  Note that this is not
the case for nonzero $\tvec{z}_1, \tvec{z}_2$, as follows immediately from the
explicit one-loop expressions in section~\ref{sec:CS_kernels}.

The soft factors discussed so far correspond to the same parton species
entering a hard scattering process in the amplitude and its conjugate.  As
already mentioned at the end of section~\ref{sec:collinear}, there is also
interference between $q$ and $\bar{q}$, between $q$ and $g$, and between
$\bar{q}$ and $g$.  The corresponding Wilson line products in the soft
factor are then $W_{ij} W_{i'j'}$, $W_{ij} W_{a'b'}$, etc.  The relevant
projection operators for coupling colour indices in these cases are given
in equations~(9) and (11) of \cite{Mekhfi:1985dv}.  Relations analogous to
\eqref{comm-quark} can be easily derived for these cases, where again the
colour projections for the gluon fields at $\xi^+ = \xi^- = 0$ in the
Wilson lines can be traded for the colour projections of the fields at
infinity.  From this, relations analogous to \eqref{soft-factor-comm} and
\eqref{soft-factor-diag} can be derived for the soft factors associated
with interference between parton species in the hard scattering.

We now relate soft factors in the adjoint representation with those in the
fundamental one. For the mixed quark-gluon soft factor we have
\begin{align}
m(R)\, \prb{RR}{S_{qg}}
 &= P_{R}^{\ul{i}_1 \ul{a}_2}\,
P_{R}^{\ul{k}_1 \ul{c}_2}\, (S_{qg})^{\ul{i}_1 \ul{a}_2, \ul{k}_1
  \ul{c}_2}_{ \ul{j}_1 \ul{b}_2, \ul{j}_1 \ul{b}_2} = P_{R}^{\ul{i}_2
  \ul{b}_2}\, P_{R}^{\ul{k}_2 \ul{b}_2}\, (S_{qq})^{\ul{i}_1 \ul{i}_1,
  \ul{k}_1 \ul{k}_1}_{ \ul{j}_1 \ul{i}_2, \ul{j}_1 \ul{k}_2}
\nonumber \\
 &= P_{R}^{\ul{i}_2 \ul{j}_2}\, P_{R}^{\ul{k}_2 \ul{j}_2}\,
(S_{qq})^{\ul{i}_1 \ul{i}_1, \ul{k}_1 \ul{k}_1}_{ \ul{j}_1 \ul{i}_2,
  \ul{j}_1 \ul{k}_2} = P_{R}^{\ul{i}_1 \ul{i}_2}\, P_{R}^{\ul{k}_1
  \ul{k}_2}\, (S_{qq})^{\ul{i}_1 \ul{i}_2, \ul{k}_1 \ul{k}_2}_{ \ul{j}_1
  \ul{j}_2, \ul{j}_1 \ul{j}_2}
\nonumber \\[0.2em]
 &=  m(R)\, \prb{RR}{S_{qq}}
\end{align}
if $R$ is in the singlet or octet channels, where in the second and the forth
step we have used \eqref{comm-soft} for the indices associated with parton
$a_2$ in $S_{a_1 a_2}$.  Writing
\begin{align}
m(R)\, \prb{RR}{S_{gg}}\
 &= P_{R}^{\ul{a}_1 \ul{a}_2}\,
  P_{R}^{\ul{c}_1 \ul{c}_2}\, (S_{gg})^{\ul{a}_1 \ul{a}_2, \ul{c}_1
  \ul{c}_2}_{ \ul{b}_1 \ul{b}_2, \ul{b}_1 \ul{b}_2}
\nonumber \\
 & = P_{R}^{\ul{i}_1 \ul{a}_1}\, P_{R}^{\ul{i}_1 \ul{a}_2}\, P_{R}^{\ul{k}_1
  \ul{c}_1}\, P_{R}^{\ul{k}_1 \ul{c}_2}\, (S_{gg})^{\ul{a}_1 \ul{a}_2,
  \ul{c}_1 \ul{c}_2}_{ \ul{b}_1 \ul{b}_2, \ul{b}_1 \ul{b}_2}
\end{align}
for $R = 1,A,S$ and using \eqref{comm-soft}
for all index pairs, one can reduce $S_{gg}$ to $S_{qq}$ in the colour
singlet and octet channels as well.  The only independent soft matrix
elements for collinear DPS factorisation (in channels with the same parton
species in the amplitude and its conjugate) are therefore
\begin{align}
  \label{coll-soft-sing}
\pr{11}{S} &= \pr{11}{S_{qq}} = \pr{11}{S_{qg}} = \pr{11}{S_{gq}} =
\pr{11}{S_{gg}} = 1
\intertext{and}
  \label{coll-soft-final}
\pr{88}{S} &= \pr{88}{S_{qq}} = \pr{AA}{S_{qg}} = \pr{SS}{S_{qg}} =
\pr{AA}{S_{gq}} = \pr{SS}{S_{gq}} = \pr{AA}{S_{gg}} = \pr{SS}{S_{gg}} \,,
\nonumber \\ \pr{DD}{S} &= \pr{DD}{S_{gg}} \,, \nonumber \\ \pr{27\,
  27}{S} &= \pr{27\, 27}{S_{gg}} \,,
\end{align}
where one can also replace one or both labels $q$ with $\bar{q}$ in the
soft factors involving quarks.  We recall that each factor in
\eqref{coll-soft-final} depends on $\tvec{y}$, on a rapidity difference, and
on two renormalisation scales $\mu_1$ and $\mu_2$.  In the colour singlet
channel the soft matrix elements are unity because all Wilson lines are
contracted pairwise to $W W^\dagger = W^\dagger W = \one$.

As noted in \cite{Vladimirov:2016qkd}, there is a remarkable identity
\begin{align}
  \label{soft-octet}
\pr{88}{S}(\tvec{y}) &= S_{g}(\tvec{y}) \,,
\end{align}
where $S_g$ is the soft factor for single TMD factorisation with a gluon
initiated hard scattering such as $gg\to H$ \cite{Echevarria:2015uaa}.  Its
definition reads
\begin{align}
	\label{S-g-def}
S_{g}(\tvec{z}) &=
  \bigl\langle 0 \ms\big|\ms P_1^{\ul{a}\, \ul{b}}\,
      \bigl[ O_{S,g}(\tvec{0},\tvec{z}) \bigr]^{\ul{a}, \ul{b}}
  \ms\big|\ms 0 \bigr\rangle \,,
\end{align}
where $O_{S,g}$ is obtained from \eqref{WL-ops} by replacing the fundamental
Wilson lines with adjoint ones.  One can easily prove \eqref{soft-octet} by
applying the colour Fierz identity \eqref{colour-Fierz} to $\pr{88}{S}$ in
the quark representation and to $S_g$ after converting adjoint Wilson lines
into fundamental ones using \eqref{wils-adj-to-fun}.  It is understood that
the two renormalisation scales $\mu_1, \mu_2$ in $\pr{88}{S}(\tvec{y})$ must
be taken equal in \eqref{soft-octet}.

\subsection{Interlude: collinear factorisation for coloured particle production}
\label{sec:interlude}

\begin{figure}
\begin{center}
\includegraphics[width=0.6\textwidth]{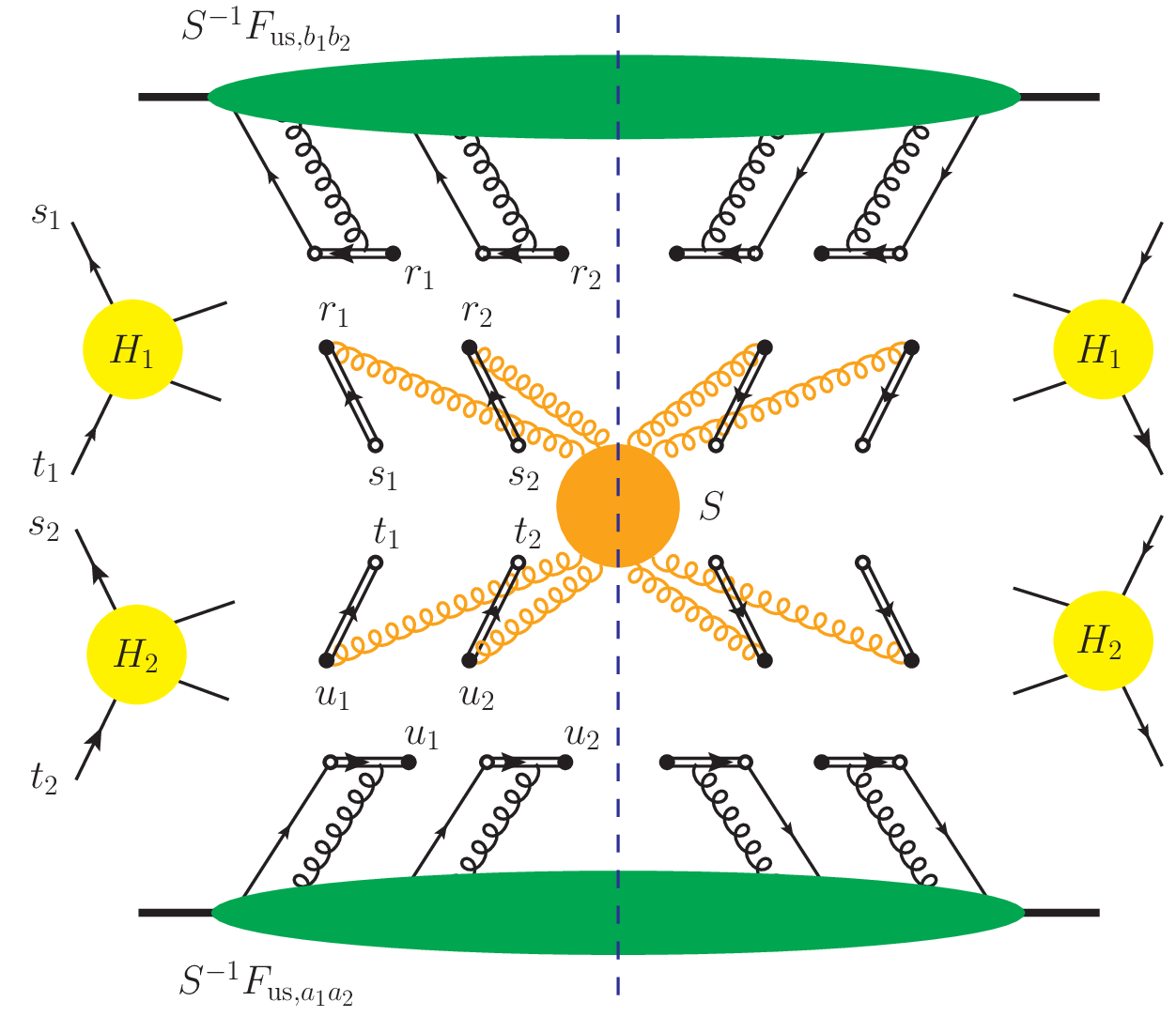}
\caption{\label{fig:jets-fact} Graphical representation of colour index
  contractions in the factorised double dijet cross section.  The indices
  in this figure correspond to the ones given in
  \protect\eqref{col-Xsect-start}.  The corresponding primed indices in
  the complex conjugate part of the graph are not shown.}
\end{center}
\end{figure}

In collinear factorisation, we may consider the production of coloured
particles, a prominent example for DPS being double dijet production.  As
pointed out in section 2.1 of \cite{Diehl:2015bca}, the steps for showing
factorisation in this case lead to the extended soft factor
\eqref{soft-gen-def}.  Instead of the product in \eqref{eq:DPS_colsoft},
the cross section then involves products of the form
\begin{align}
	\label{col-Xsect-start}
X = H_{a_1 b_1}^{\ul{s}_1 \ul{t}_1}\, H_{a_2 b_2}^{\ul{s}_2 \ul{t}_2}\, &
\bigl[ F^T_{\us, b_1 b_2}(Y_R) \, S^{-1}(Y_R-Y_L) \bigr]^{ \ul{r}_1
  \ul{r}_2} \,
\nonumber \\
 & \times \bigl[ S(Y_R-Y_L) \bigr]^{\ul{r}_1
  \ul{r}_2, \ul{u}_1 \ul{u}_2}_{ \ul{s}_1 \ul{s}_2, \ul{t}_1 \ul{t}_2} \,
\bigr[ S^{-1}(Y_R-Y_L)\, F^{}_{\us, a_1 a_2}(Y_L) \bigr]^{ \ul{u}_1
  \ul{u}_2} \,.
\end{align}
A pictorial representation is given in figure \ref{fig:jets-fact}.
Colour indices $s, t$ in the hard scattering factor $H^{s s' t t'}$ are for
the incoming partons in the amplitude, whilst $s', t'$ are for the incoming
partons in the conjugate amplitude.  These indices are contracted with colour
indices of the extended soft factor.  The inverse soft factors $S^{-1}$ in
\eqref{col-Xsect-start} remove soft gluons from the unsubtracted collinear
factors $F_{\us}$ as usual; they are hence the soft factors with left and
right moving Wilson lines colour contracted as in \eqref{soft-def}.  We
have omitted their parton labels, given that after colour projection
they are independent of the parton species as shown in the previous
subsection.  In the extended soft factor, the representation of each Wilson
line must match the parton entering in the corresponding hard-scattering
subprocess.  This includes combinations such as
$W_{ab}^{}(\tvec{\xi},v_L)\, W^\dagger_{kl}(\tvec{\xi},v_R)$, which
corresponds to a right-moving quark scattering on a left-moving gluon.  It is
not difficult to generalise the relation \eqref{soft-factor-comm} between the
extended and usual soft factors to such cases, so that we can omit parton
labels in the extended soft factor as well.  Inserting projection operators as
in \eqref{eq:contr_proj}, we can thus rewrite \eqref{col-Xsect-start} as
\begin{align}
	\label{col-Xsect-int}
X &= \sum_{R, R', R_1, R_2} \frac{P_{R_1}^{\ul{s}_1 \ul{t}_1}\, H_{a_1
    b_1}^{\ul{s}_1 \ul{t}_1}\, P_{R_2}^{\ul{s}_2 \ul{t}_2}\, H_{a_2
    b_2}^{\ul{s}_2 \ul{t}_2}}{m(R)\ms m(R')\ms m(R_1)\ms m(R_2)}\;
    P_{R}^{\ul{r}_1
  \ul{r}_2}\, \bigl[ F^T_{\us, b_1 b_2}(Y_R) \, S^{-1}(Y_R-Y_L) \bigr]^{
  \ul{r}_1 \ul{r}_2}
\nonumber \\[0.3em]
  & \qquad\qquad \times \delta^{}_{R R_1}\, \delta^{}_{R R_2}\,
  \delta^{}_{R R'}\, m(R)\,\pr{RR}{S(Y_R-Y_L)}\;
P_{R'}^{\ul{u}_1 \ul{u}_2}\, \bigr[ S^{-1}(Y_R-Y_L)\, F^{}_{\us, a_1
    a_2}(Y_L) \bigr]^{ \ul{u}_1 \ul{u}_2}
\nonumber \\[0.4em]
&= \sum_{R} \frac{P_{R}^{\ul{s}_1 \ul{t}_1}\, H_{a_1 b_1}^{\ul{s}_1
    \ul{t}_1}\, P_{R}^{\ul{s}_2 \ul{t}_2}\, H_{a_2 b_2}^{\ul{s}_2
    \ul{t}_2}}{m^3(R)}\; P_{R}^{\ul{r}_1 \ul{r}_2}\, F^{\ul{r}_1
  \ul{r}_2}_{\us, b_1 b_2}(Y_R)\; \pr{RR}{S^{-1}(Y_R-Y_L)}\,
P_{R}^{\ul{u}_1 \ul{u}_2}\, F^{\ul{u}_1 \ul{u}_2}_{\us, a_1
  a_2\rule{0pt}{1.3ex}}(Y_L)\,,
\end{align}
where in the last step we used that $\prb{RR'}{S}^{-1}$ is diagonal in
$R$ and $R'$.  Splitting $\pr{RR}{S}$ as in \eqref{S-decomp-proj} and using
that $\pr{RR}{s}$ is real valued, we obtain colour projected DTMDs from
\eqref{sub-unsub} and the analogue of \eqref{proj-dpd-def} for $F_{\us}$.
This yields the main result of this subsection:
\begin{align}
\label{col-Xsimp2}
X &= \sum_{R}
    \frac{1}{\ii{a_1}{R}\, \ii{b_1}{R}\, \ii{a_2}{R}\, \ii{b_2}{R}}\,
     \frac{1}{\mathcal{N}_{a_1}\ms \mathcal{N}_{b_1}\ms
              \mathcal{N}_{a_2}\ms \mathcal{N}_{b_2}}
    \frac{P_{R}^{\ul{s}_1 \ul{t}_1}\, H_{a_1 b_1}^{\ul{s}_1
    \ul{t}_1}\, P_{R}^{\ul{s}_2 \ul{t}_2}\, H_{a_2 b_2}^{\ul{s}_2
    \ul{t}_2}}{m^2(R)}
\nonumber\\[0.3em]
 &\qquad\quad \times \pr{R}{F_{\us, b_1 b_2}(Y_R)}\,
\pr{RR}{s^{-1}(Y_R-Y_C)}\,
\pr{RR}{s^{-1}(Y_C-Y_L)}\, \pr{R}{F_{\us, a_1 a_2}(Y_L)}
\nonumber \\[0.4em]
&= \sum_{R} \pr{R}{H_{a_1 b_1}}\, \pr{R}{H_{a_2 b_2}}\,
            \pr{R}{F_{b_1 b_2}}\, \pr{R}{F_{a_1 a_2}} \,.
\end{align}
The colour structure of the collinear cross section is thus surprisingly
simple: we have the same DPDs in \eqref{col-Xsimp2} as in the formula
\eqref{CS-Xsect-final} for colour singlet production, and the only new
ingredients are the colour projected hard-scattering factors\,\footnote{The
  factors $\pr{R}{H}$ in \protect\eqref{H-col-def} differ from the ones in
  equation (2.11) of \protect\cite{Diehl:2015bca}.  Here we project index
  pairs in the $t$-channel ($s^{}, s'$ etc.) onto a definite colour
  representation $R$, whereas in \protect\cite{Diehl:2015bca} this was done
  for index pairs in the $s$-channel ($s, t$ etc.).}
\begin{align}
  \label{H-col-def}
\pr{R}{H}_{a b} &=
   \frac{1}{\ii{a}{R}\, \ii{b}{R}}\,
   \frac{1}{\mathcal{N}_{a}\, \mathcal{N}_{b}}\,
  \frac{P_{R}^{\ul{s}\, \ul{t}}\, H_{a b}^{\ul{s}\, \ul{t}}}{m(R)} \,.
\end{align}
For the production of a colour singlet system, the hard scattering
subprocess has the structure
\begin{align}
  \label{hard-scatt-sing}
H_{a b}^{\ul{s}\, \ul{t}} &= \delta_{s t}^{}\, \delta_{s' t'}^{}\,
  H_{a b} ^{}\,.
\end{align}
Using \eqref{mR-def} we thus find
\begin{align}
  \label{H-sing-proj}
P_R^{\ul{s}\, \ul{t}}\, H_{a b}^{\ul{s}\, \ul{t}} &= m(R)\, H_{a b} ^{} \,,
\end{align}
so that $\pr{R}{H}_{a b}$ reproduces $H_{ab} /(\mathcal{N}_{a}\ms
\mathcal{N}_{b})$ divided by $\ii{a}{R}\, \ii{b}{R}$, which is absorbed into
the sign factor $\csgn{a_1 a_2}{R}$ in~\eqref{CS-Xsect-final}.

\section{Factorisation at small transverse momenta}
\label{sec:fact}

We are now in a situation to give the TMD factorisation formula for DPS
processes.
Let us consider the production of two particles or groups of particles with
transverse momenta $\tvec{q}_1, \tvec{q}_2$ and invariant masses $Q_1, Q_2$.
Both $Q_1$ and $Q_2$ are required to be large, and we denote their generic
size by $Q$ for the purpose of power counting.  Both transverse momenta, whose
generic size we denote by $q_T$, are required to be small compared with~$Q$.
Up to power corrections in $\Lambda/Q$ and $|\tvec{q}_i|/Q$, the DPS cross
section with measured transverse momenta reads
\begin{align}
	\label{TMD-Xsect}
& \frac{d\sigma_{\text{DPS}}}{dx_1\, dx_2\, d\bar{x}_1\, d\bar{x}_2\,
          d^2\tvec{q}_1\, d^2\tvec{q}_2} = \frac{1}{C}\,
        \sum_{a_1,a_2,b_1,b_2} \!\!\!  \hat{\sigma}_{a_1 b_1}(Q_1,
        \mu_1)\, \hat{\sigma}_{a_2 b_2}(Q_2, \mu_2)\,
\nonumber \\
&       \qquad \times \int \frac{d^2\tvec{z}_1}{(2\pi)^2}\,
        \frac{d^2\tvec{z}_2}{(2\pi)^2}\, d^2\tvec{y}\; e^{-i\tvec{q}_1
          \tvec{z}_1 -i\tvec{q}_2 \tvec{z}_2}\, W_{a_1 a_2 b_1
          b_2}(\bar{x}_i,x_i, \tvec{z}_i,\tvec{y}; \mu_i,\nu) \,,
\end{align}
with
\begin{align}
	\label{W-def}
& W_{a_1 a_2 b_1 b_2}(\bar{x}_i,x_i,\tvec{z}_i,\tvec{y};\mu_i,\nu)
 = \Phi(\nu \tvec{y}_+)\, \Phi(\nu \tvec{y}_-)
\nonumber \\[0.2em]
 &\qquad \times \sum_{R} \csgn{a_1 a_2}{R}\; \pr{R}{ F_{b_1
            b_2}(\bar{x}_i,\tvec{z}_i,\tvec{y};\mu_i,\bar{\zeta})}\,
        \pr{R}{F_{a_1 a_2}(x_i,\tvec{z}_i,\tvec{y};\mu_i,\zeta)}
\end{align}
and
\begin{align}
	\label{y-pm-def}
\tvec{y}_{\pm} = \tvec{y} \pm \half (\tvec{z}_1 - \tvec{z}_2) \,.
\end{align}
We recall that we write arguments $x_i$ if a function depends on both
$x_1, x_2$, and likewise for all other variables.  The combinatorial
factor $C$ is equal to $2$ if the systems produced by the two hard
scatters are identical and equal to $1$ otherwise.  The factor
$\csgn{a_1 a_2}{R}$ was defined in \eqref{col-sign}; it is equal to $-1$
if $R=A$ and exactly one of $a_1$ and $a_2$ is a gluon and equal to $1$
otherwise.  The subprocess cross sections are given by
\begin{align}
\hat\sigma_{a_1 b_1}(Q_1^2, \mu_1^2) &= \frac{1}{2 Q_1^2}\, \frac{H_{a_1
    b_1}(Q_1^2, \mu_1^2)}{ \mathcal{N}_{a_1}\ms \mathcal{N}_{b_1}} \,
\end{align}
and likewise for index $2$.  Here, $1/(2 Q_1^2)$ is the standard flux
factor, and $1/ (\mathcal{N}_{a_1}\ms \mathcal{N}_{b_1})$ implements colour
averaging over the initial state as noted after \eqref{CS-Xsect-final}.
The sum over $a_1,a_2,b_1,b_2$ in \eqref{TMD-Xsect} runs over both parton
species and polarisations, following the notation specified in
section~\ref{sec:collinear}.  Hard-scattering cross sections and DPDs
carry transverse Lorentz indices for transverse quark and linear gluon
polarisation, which must be contracted appropriately.  For later
discussion, it is useful to rewrite the integration measure in
\eqref{TMD-Xsect} as
\begin{align}
  \label{fourier-rewrite}
d^2\tvec{y}\, d^2\tvec{z}_1\, d^2\tvec{z}_2\;
  e^{-i\tvec{q}_1 \tvec{z}_1 -i\tvec{q}_2 \tvec{z}_2}
&=  d^2\tvec{Z}\, d^2\tvec{y}_+\, d^2\tvec{y}_-\;
e^{- i\ms (\tvec{q}_1 + \tvec{q}_2)\ms \tvec{Z}
   - i\ms (\tvec{q}_1 - \tvec{q}_2) (\tvec{y}_+ - \tvec{y}_-) /2}
\end{align}
with
\begin{align}
	\label{z-plus-def}
\tvec{Z} &= \half (\tvec{z}_1 + \tvec{z}_2) \,.
\end{align}

The function $\Phi(\nu \tvec{y}_\pm)$ in \eqref{W-def} regulates the
ultraviolet region.  It is intimately related with the cross talk between
double and single hard scattering and was introduced in the recent work
\cite{Diehl:2017kgu}.  This is discussed in section \ref{sec:TMD-UV-region},
where we show how to combine the contributions from DPS, SPS and their
interference to obtain the physical cross section.  Apart from this function,
the factorisation formula \eqref{TMD-Xsect} is identical to the form that
emerged from the analysis of lowest-order graphs in section~2 of
\cite{Diehl:2011yj}.  This is because our DPD definition in
section~\ref{sec:comb-soft-coll} absorbs all relevant effects of soft gluon
exchange in the cross section, leading to the crucial relations
\eqref{CS-Xsect} and \eqref{CS-Xsect-mix}.

Of course, the scale and rapidity parameter dependence of the DTMDs in
\eqref{TMD-Xsect} does not arise from lowest-order graphs but from the full
analysis in section~\ref{sec:ren-DTMDs}.  \rev{In the next section, we give
the final forms \eqref{RG-TMD-again} and \eqref{CS-TMD} of the corresponding
evolution equations, as well as their explicit solutions \eqref{DTMD-evolved}
and \eqref{W-generic} at the level of DPDs and of the cross section,
respectively.  In section~\ref{sec:DPDF_evo} we derive the evolution equations
\eqref{CS-coll} to \eqref{DGLAP-CS} of DPDFs and give the explicit form
\eqref{coll-zeta-expl} of their rapidity dependence.}

\rev{When combining SPS and DPS in the physical cross section, one must take
care of interference effects and of double counting.  A systematic framework
for this has been proposed in \cite{Diehl:2017kgu}, and in section
\ref{sec:TMD-UV-region} of the present paper we characterise the individual
terms in the master formula \eqref{eq:tot-x-sec} for the cross section with
measured transverse momenta.}

\subsection{DTMD evolution: renormalisation scale and rapidity}
\label{sec:evol-gen}

Let us rewrite the evolution equations for the DTMDs using the projection
operators introduced in section \ref{sec:colour}.  The renormalisation group
equation \eqref{RG-TMD} then reads
\begin{align}
   \label{RG-TMD-again}
\frac{\partial}{\partial \log\mu_1}\,
  \pr{R}{F_{a_1 a_2}(x_i,\tvec{z}_i,\tvec{y};\mu_i,\zeta)}
&= \gamma_{F, a_1}(\mu_1, x_1\zeta/x_2)\, \pr{R}{F_{a_1
      a_2}(x_i,\tvec{z}_i,\tvec{y};\mu_i,\zeta)}
\end{align}
for the scale $\mu_1$, and in analogy for $\mu_2$.  Here we have introduced
the rapidity parameter
\begin{align}
	\label{zeta-def}
\zeta &= 2 x_1 x_2\, (p^+)^2\, e^{-2 Y_C}\,,
\end{align}
where $p$ is the proton momentum.\footnote{A different definition (going back
  to \protect\cite{Ji:2004wu}) was used in \cite{Diehl:2011yj}, namely
  $x_1 x_2 \ms \zeta^2 |_{\text{Ref.~\protect\cite{Diehl:2011yj}}} = \zeta
  |_{\text{here}\,}$.}
This extends the definition \eqref{zeta-def-TMD} for SPS to the case of DPS in
a way that is symmetric in the two momentum fractions $x_1, x_2$.  As
discussed after \eqref{Y-dep-gamma-F}, the correct rapidity argument of
$\gamma_{F,a}$ in \eqref{RG-TMD-again} is
$x_1\zeta/x_2 = 2 (x_1\ms p^+)^2 e^{-2 Y_C}$, since the corresponding UV
divergent subgraphs depend on the momentum of parton $1$ (whose plus-component
is $x_1\ms p^+$), but not on the momentum of parton $2$.  For the DPD of the
left moving proton with momentum $\bar{p}$ we define accordingly
\begin{align}
\bar{\zeta} &= 2 \bar{x}_1 \bar{x}_2\, (\bar{p}^-)^2\, e^{2 Y_C}\,.
\end{align}
Note that
\begin{align}
	\label{zeta-product}
\zeta\, \bar{\zeta} &= x_1 \bar{x}_1 x_2\ms \bar{x}_2 \, (2 p^+
\bar{p}^{\,-})^2 = Q_1^2\, Q_2^2 \,.
\end{align}
Rewriting the rapidity derivative as
$\partial /\partial Y_C = - 2 \ms \partial /\partial \log\zeta = 2
\ms \partial /\partial \log\bar\zeta$, one obtains from \eqref{CS-gen} the
Collins-Soper equation
\begin{align}
	\label{CS-TMD}
\frac{\partial}{\partial\log \zeta} \pr{R}{F_{a_1
    a_2}}(x_i,\tvec{z}_i,\tvec{y}; \mu_i,\zeta) &= \frac{1}{2} \sum_{R'}
\prb{RR'}{K_{a_1 a_2}(\tvec{z}_i,\tvec{y}; \mu_i)}\, \prb{R'}{F_{a_1
    a_2}(x_i,\tvec{z}_i,\tvec{y}; \mu_i,\zeta)}
\end{align}
for colour projected DTMDs in a right moving proton.  The Collins-Soper
equation for a left moving proton has the same form, with $\zeta$ replaced by
$\bar{\zeta}$.  The scale dependence of the Collins-Soper kernel in
\eqref{RG-TMD} now reads
\begin{align}
	\label{CS-TMD-RG}
\frac{\partial}{\partial \log\mu_1}\, \prb{RR'}{K_{a_1
    a_2}(\tvec{z}_i,\tvec{y}; \mu_i)} &= - \gamma_{K, a_1}(\mu_1)\,
\delta_{RR'}
\end{align}
and correspondingly for $\mu_2$.  It is thus only the diagonal elements of
$\prb{RR'}{K}$ that have UV divergences and depend on the renormalisation
scales $\mu_{1}$ and $\mu_{2}$.  Let us recall that the kernel $K_{a_1 a_2}$
and the anomalous dimensions $\gamma_{F,a}$ and $\gamma_{K,a}$ depend on the
colour representation of the parton (quarks or antiquarks vs.\ gluons) but not
on their flavour or polarisation.

It is easy to solve the rapidity and renormalisation scale evolution with
general starting scales $\mu_{01}$, $\mu_{02}$ for $\mu_1$, $\mu_2$ and
$\zeta_0$ for $\zeta$.  Renormalisation scale evolution gives
\begin{align}
	\label{RG-TMD-sol}
& \pr{R}{F_{a_1 a_2}(x_i,\tvec{z}_i,\tvec{y};\mu_i,\zeta)} = \pr{R}{F_{a_1
            a_2}(x_i,\tvec{z}_i,\tvec{y};\mu_{0i},\zeta)} \nonumber
        \\[0.2em] & \qquad \qquad \times \exp\biggl[
          \int_{\mu_{01}}^{\mu_1} \frac{d\mu}{\mu}\, \gamma_{F,a_1}(\mu,
          x_1\zeta/x_2) + \int_{\mu_{02}}^{\mu_2} \frac{d\mu}{\mu}\,
          \gamma_{F,a_2}(\mu, x_2\zeta/x_1) \biggr] \,.
\end{align}
Using the rapidity evolution in \eqref{CS-TMD} we write
\begin{align}
	\label{CS-TMD-sol}
\pr{R}{F}_{a_1 a_2}(x_i,\tvec{z}_i,\tvec{y};\mu_i,\zeta) &= \sum_{R'}
\prb{RR'}{\exp}\biggl[ K_{a_1 a_2}(\tvec{z}_i,\tvec{y};\mu_i) \log
  \frac{\sqrt{\zeta}}{\sqrt{\zeta_0}} \,\biggr] \prb{R'}{F}_{a_1
  a_2}(x_i,\tvec{z}_i,\tvec{y};\mu_i,\zeta_0) \,,
\end{align}
where $\prb{RR'}{\exp}$ is to be understood as a matrix exponential, i.e.\
\begin{align}
\prb{RR'}{\exp}(M) &= \delta_{RR'} + \prb{RR'}{M} +
  \sum_{n=2}^\infty\, \sum_{R_2,\ldots,R_{n}}\,
     \frac{\pr{RR_2}{M} \,\cdots\, \prb{R_{n} R'}{M}}{n!}\,.
\end{align}
We now split
\begin{align}
	\label{split-K-M}
\prb{RR'}{K_{a_1 a_2}(\tvec{z}_i,\tvec{y}; \mu_i)} &= \delta_{RR'}^{}\,
\bigl[ \pr{1}{K}_{a_1}(\tvec{z}_1;\mu_1) +
  \pr{1}{K}_{a_2}(\tvec{z}_2;\mu_2) \bigr] + \prb{RR'}{M_{a_1
    a_2}(\tvec{z}_i,\tvec{y})}\,,
\end{align}
where $\pr{1}{K}_a(\tvec{z};\mu)$ is the kernel in the Collins-Soper equation
\eqref{cs-eq} for a single parton TMD. The colour singlet label $1$ on $K$ is
given for consistency with our later notation in section~\ref{sec:match-evol}.
By virtue of \eqref{RG-single-K} and \eqref{CS-TMD-RG}, the matrix
$\prb{RR'}{M}$ is independent of $\mu_1$ and $\mu_2$.  Combining
\eqref{RG-TMD-sol} and \eqref{CS-TMD-sol} with \eqref{split-K-M}, we obtain
the relation between the DTMD at initial and final scales in the form
\begin{align}
  \label{DTMD-evolved}
  & \pr{R}{F}_{a_1 a_2}(x_i,\tvec{z}_i,\tvec{y};\mu_i,\zeta)
\nonumber \\
  & \quad = \exp\,\biggl\{ \int_{\mu_{01}}^{\mu_1} \frac{d\mu}{\mu}\,
        \biggl[ \gamma_{a_1}(\mu) - \gamma_{K,a_1}(\mu)\,
          \log\frac{\sqrt{x_1\zeta/x_2}}{\mu} \biggr] +
        \pr{1}{K}_{a_1}(\tvec{z}_1;\mu_{01})
    \log\frac{\sqrt{\zeta}}{\sqrt{\zeta_0}}
\nonumber \\
  & \qquad \hspace{1.5em} + \int_{\mu_{02}}^{\mu_2} \frac{d\mu}{\mu}\,
        \biggl[ \gamma_{a_2}(\mu) - \gamma_{K,a_2}(\mu)\,
          \log\frac{\sqrt{x_2\ms \zeta/x_1}}{\mu} \biggr] +
        \pr{1}{K}_{a_2}(\tvec{z}_2;\mu_{02})
    \log\frac{\sqrt{\zeta}}{\sqrt{\zeta_0}} \,\biggr\}
\nonumber \\
  & \qquad \times \sum_{R'} \prb{RR'}{\exp}\,\biggl[ M_{a_1
            a_2}(\tvec{z}_i,\tvec{y})
          \log\frac{\sqrt{\zeta}}{\sqrt{\zeta_0}} \,\biggr]\,
        \prb{R'}{F}_{a_1
          a_2}(x_i,\tvec{z}_i,\tvec{y};\mu_{01},\mu_{02},\zeta_0)\,,
\end{align}
where we have used the explicit form \eqref{cusp-solved} of $\gamma_{F,a}$.
The exponential in the second and third lines is just the generalisation to two
partons of the evolution factor for a single parton TMD, as is readily seen by
solving the system of equations in \eqref{RG-single-TMD} to
\eqref{RG-single-K}.  It resums both double and single logarithms.  The last
line in \eqref{DTMD-evolved} describes the mixing between different
colour representations $R$ under rapidity evolution and involves a single
logarithm.  The double logarithms in the evolution of
$\pr{R}{F}_{a_1 a_2}(x_i,\tvec{z}_i,\tvec{y};\mu_i,\zeta)$ are thus the same
as those for a product of two single TMDs with appropriate arguments.

The symmetry relation \eqref{K-anti} for $\prb{RR'}{K}$ translates into an
analogous relation for $\prb{RR'}{M}$.  Using that $b_i$ is the antiparton of
$a_i$ in the cross section formula \eqref{TMD-Xsect} and that $M_{a_1 a_2}$
does not depend on parton spins, we can thus rewrite
\begin{align}
\label{rewrite-M-exp}
\csgn{a_1 a_2}{R}\; \prb{RR'}{\exp}\,\Biggl[ M_{b_1 b_2}
    \log\frac{\sqrt{\bar{\zeta}\rule{0pt}{2.1ex}}}{\sqrt{\zeta_0}}
    \,\Biggr]\, \prb{R'}{F}_{b_1 b_2}
 &= \csgn{a_1 a_2}{R'}\; \prb{R'}{F}_{b_1 b_2}\ms
    \pr{R'\!R}{\exp}\,\Biggl[ M_{a_1 a_2}
    \log\frac{\sqrt{\bar{\zeta}\rule{0pt}{2.1ex}}}{\sqrt{\zeta_0}} \,\Biggr]
\end{align}
when inserting the evolved form of $\pr{R}{F}_{b_1 b_1}$ into the expression
\eqref{W-def} of $W$.  This allows us to combine the matrix exponentials
associated with the two DPDs.  Further using that one has equal anomalous
dimensions $\gamma_{a_i} = \gamma_{b_i}$ and $\gamma_{K,a_i} =
\gamma_{K,b_i}$ and equal Collins-Soper kernels $\pr{1}{K}_{a_i} =
\pr{1}{K}_{b_i}$, we obtain
\begin{align}
  \label{W-generic}
W_{a_1 a_2 b_1 b_2} &=
  \exp\,\biggl\{ \int_{\mu_{01}}^{\mu_1} \frac{d\mu}{\mu}\,
        \biggl[ \gamma_{a_1}(\mu) - \gamma_{K,a_1}(\mu)\,
          \log\frac{Q_1^2}{\mu^2} \biggr] +
        \pr{1}{K}_{a_1}(\tvec{z}_1;\mu_{01})
    \log\frac{Q_1 Q_2}{\zeta_0}
\nonumber \\
  & \quad \hspace{1.5em} + \int_{\mu_{02}}^{\mu_2} \frac{d\mu}{\mu}\,
        \biggl[ \gamma_{a_2}(\mu) - \gamma_{K,a_2}(\mu)\,
          \log\frac{Q_2^2}{\mu^2} \biggr] +
        \pr{1}{K}_{a_2}(\tvec{z}_2;\mu_{02})
    \log\frac{Q_1 Q_2}{\zeta_0} \,\biggr\}
\nonumber \\[0.5em]
  & \quad \times \Phi(\nu \tvec{y}_+)\, \Phi(\nu \tvec{y}_-)
    \sum_{RR'} \csgn{a_1 a_2}{R}\;
        \pr{R}{F}_{b_1 b_2}(\bar{x}_i,\tvec{z}_i,\tvec{y};
            \mu_{01},\mu_{02},\zeta_0)\,
\nonumber \\
  & \quad \hspace{1.5em} \times
     \prb{RR'}{\exp}\,\biggl[ M_{a_1 a_2}(\tvec{z}_i,\tvec{y})
          \log\frac{Q_1 Q_2}{\zeta_0} \,\biggr]\,
        \prb{R'}{F}_{a_1 a_2}(x_i,\tvec{z}_i,\tvec{y};
            \mu_{01},\mu_{02},\zeta_0)\,.
\end{align}
We have used the relation \eqref{zeta-product} for combining the arguments
of the logarithms and expressing them in terms of the physical scales $Q_1$
and $Q_2$.  The variables $\zeta$ and $\bar{\zeta}$ have thus completely
disappeared from the physical cross section.  This is not the case for the
factorisation scales $\mu_i$ in \eqref{W-generic} and in the hard scattering
cross sections $\hat{\sigma}$.  As is well known, the dependence on these
scales only cancels up to un-calculated higher orders in $\alpha_s$.

\subsection{DPDF evolution}
\label{sec:DPDF_evo}

We now turn to DPDFs, which appear not only in the factorisation formula for
DPS with integrated transverse momenta, but also in the short-distance
expansion of DTMDs, as we will see in section~\ref{sec:small_zi}.  We have
derived in section~\ref{sec:soft-simple} that the soft factor
$\prb{RR'}{S}_{a_1 a_2}$ at $\tvec{z}_1 = \tvec{z}_2 = \tvec{0}$ is diagonal
in $R$ and $R'$ and independent of the parton species.  Following the
construction in appendix~\ref{app:matrix-algebra}, one finds that the matrix
$s$ in \eqref{S-decomp} is also diagonal in $R$ and $R'$ and satisfies the
relation
\begin{align}
  \label{S-double-simp}
\pr{RR}{s}(Y) &= \sqrt{ \pr{RR}{S}(2Y) } \,,
\end{align}
which is the analogue of \eqref{eq:S-single-simp} for SPS.
The definition of DPDFs from unsubtracted collinear matrix elements in
\eqref{F-sub-def} thus turns into
\begin{align}
  \label{F-sub-def-coll}
\pr{R}{F}_{a_1 a_2}(Y_C) &= \lim_{Y_L\rightarrow -\infty}
   \pr{RR}{s}^{-1}(Y_C-Y_L)\, \pr{R}{F}_{\text{us},a_1 a_2}(Y_L)
\end{align}
for a right moving proton, and correspondingly for a left moving one.  The
kernel $K = \widehat{K}$ controlling the $Y$ dependence of $S(Y)$ and
$s(Y)$ is of course also diagonal in $R$ and $R'$.  We denote it by
$\prb{R}{J}$ in order to avoid confusion with the evolution kernel for
TMDs.  In terms of the variable $\zeta$, the rapidity evolution of DPDFs
is then given by
\begin{align}
	\label{CS-coll}
\frac{\partial}{\partial\log \zeta}\, \pr{R}{F}(x_i,\tvec{y}; \mu_i,\zeta)
&= \frac{1}{2}\, \prb{R}{J(\tvec{y}; \mu_i)}\, \pr{R}{F(x_i,\tvec{y};
  \mu_i,\zeta)} \,,
\end{align}
where we have explicitly given all arguments of the functions.  Since in the
colour singlet channel $\pr{11}{S} = 1$, one has $\pr{1}{J} = 0$ for the
corresponding evolution kernel, i.e.\ the collinear colour singlet DPDFs are
independent of the rapidity parameter $\zeta$.

Let us now discuss renormalisation.  Setting
$\tvec{z}_1 = \tvec{z}_2 = \tvec{0}$ in the soft factor $S$ and in the
hadronic matrix elements $F_{\us}$ induces additional UV divergences compared
to the case where these distances are finite.  The graphs requiring
renormalisation are now not only vertex corrections and self energies, but
also involve the exchange of hard gluons between partons or Wilson lines
associated with the scattering amplitude and its complex conjugate.  This is
of course well known from the renormalisation of the twist-two operators that
define ordinary PDFs, which is not purely multiplicative but involves a
convolution in the momentum fraction $x$ and mixing between quark and gluon
operators.  In the renormalisation group equations for PDFs, splitting kernels
$P_{ab}$ take the role of the anomalous dimensions $\gamma_{F,a}$ in the TMD
case.

In the case of DPDFs we must pay particular attention to colour.
Taking the twist-two operators \eqref{x-ops} at $\tvec{z}=\tvec{0}$ and
projecting their colour indices on a definite representation $R$ as in
\eqref{col-proj-op}, we obtain the colour projected matrix element
$\pr{R}{F}_{\us}$ according to \eqref{col-proj-matel}.  Due to gauge
invariance, operators $\prn{R}{O}^{}_a$ belonging to different colour
representations do not mix under renormalisation.
To obtain the soft factor $\pr{RR}{S}$, we evaluate the Wilson line products
\eqref{WL-ops} at $\tvec{z}=\tvec{0}$ and project them on definite colour
channels $R$ as $P_R^{\ul{r}\, \ul{s}}\, O_{S,a}^{\ul{r}, \ul{s}}$.  Again,
operators with different $R$ do not mix under renormalisation.  The
renormalisation of the soft factor remains multiplicative, because it is
identical for quarks and gluons and because it does not depend on $x$.

With this in mind, one can repeat the arguments we
developed in section \ref{sec:ren-DTMDs} for the renormalisation of DTMDs and
of their Collins-Soper kernels.  One then obtains
\begin{align}
  \label{CS-coll-RG}
\frac{\partial}{\partial \log\mu_1}\, \prb{R}{J(\tvec{y}; \mu_i)}
  &= -{}\, \prn{R}{\gamma_J}(\mu_1)
\end{align}
for the Collins-Soper kernel and
\begin{align}
	\label{DGLAP-zeta}
\frac{\partial}{\partial \log\mu_1}\, \pr{R}{F_{a_1
    a_2}(x_i,\tvec{y};\mu_i,\zeta)} &= 2 \sum_{b_1} \pr{R}{P_{a_1
    b_1}(x_1';\mu_1^{},x_1\zeta/x_2)} \underset{x_1}{\otimes}
\pr{R}{F_{b_1 a_2}(x_1',x_2,\tvec{y};\mu_i,\zeta)}
\end{align}
for DPDFs, as well as analogous equations for the $\mu_2$
dependence.\footnote{Regarding factors of 2, our
  convention for splitting kernels follows \protect\cite{Collins:2011zzd}, so
  that $2 P$ takes the place of the anomalous dimension $\gamma_{F,a}$ in TMD
  evolution.  The kernel $\pr{R}{P}$ in \protect\eqref{DGLAP-zeta} should not
  be confused with the colour projector $P_R$.}
Let us emphasise that the evolution equation \eqref{DGLAP-zeta} for DPDFs
depending on $\tvec{y}$ is homogeneous.  As discussed in
\cite{Diehl:2011yj,Diehl:2017kgu}, an additional term for the splitting of
one parton into two arises on the r.h.s.\ if one integrates the distributions
over $\tvec{y}$ or Fourier transforms them w.r.t.\ that variable.  The
resulting inhomogeneous equation has been extensively studied in the
literature \cite{Kirschner:1979im,Shelest:1982dg,Snigirev:2003cq,%
  Gaunt:2009re,Ceccopieri:2010kg}.

By construction, the kernels $\pr{1}{P_{a_1 b_1}}$ for the colour singlet
sector are the ordinary DGLAP kernels for the evolution of PDFs.  They are
hence independent of $\zeta$, which is not the case in the other colour
channels.  Indeed, combining \eqref{CS-coll}, \eqref{CS-coll-RG},
\eqref{DGLAP-zeta} and requiring equality of the derivatives
$\partial/(\partial \log\mu_i)\, \partial/(\partial \log\zeta)\,\pr{R}{F}$ and
$\partial/(\partial \log\zeta)\, \partial/(\partial\log\mu_i)\, \pr{R}{F}$, we
find
\begin{align}
	\label{DGLAP-CS}
\frac{\partial}{\partial \log\zeta}\, \pr{R}{P_{a b}(x;\mu,\zeta)} &= -
\frac{1}{4}\, \delta_{ab}^{}\, \delta(1-x)\, \prn{R}{\gamma_J}(\mu)
\end{align}
and thus
\begin{align}
	\label{DGLAP-zeta-expl}
\pr{R}{P_{a b}(x;\mu,\zeta)} &= - \frac{1}{2}\, \delta_{ab}^{}\,
\delta(1-x)\, \prn{R}{\gamma_J}(\mu) \log\frac{\sqrt{\zeta}}{\mu} +
\pr{R}{P_{a b}(x;\mu,\mu^2)} \,.
\end{align}
These are the analogues of the relations \eqref{cusp} and \eqref{cusp-solved}
for the anomalous dimension $\gamma_{F,a}(\mu,\zeta)$ in TMD evolution.

We can make the $\zeta$ dependence of the DPDFs fully explicit in the form
\begin{align}
  \label{coll-zeta-expl}
\pr{R}{F_{a_1 a_2}(x_i,\tvec{y};\mu_i,\zeta)}
  &= \exp\,\biggl[ -
  \int_{\mu_{0}}^{\mu_1} \frac{d\mu}{\mu}\, \prn{R}{\gamma_J}(\mu)\,
  \log\frac{\sqrt{x_1\zeta/x_2}}{\mu} - \int_{\mu_{0}}^{\mu_2}
    \frac{d\mu}{\mu}\, \prn{R}{\gamma_J}(\mu)\,
    \log\frac{\sqrt{x_2\ms \zeta/x_1}}{\mu}
\nonumber \\
  &\qquad\quad\; +
  \prb{R}{J}(\tvec{y};\mu_0,\mu_0) \log\frac{\sqrt{\zeta}}{\sqrt{\zeta_0}}
  \ms\biggr]\; \pr{R}{\widehat{F}_{a_1 a_2,\,
    \mu_0,\zeta_0}(x_i,\tvec{y};\mu_i)}\, ,
\end{align}
where $\widehat{F}$ is defined by the differential equation
\begin{align}
	\label{DGLAP-red}
& \frac{\partial}{\partial \log\mu_1}\, \pr{R}{\widehat{F}_{a_1 a_2,\,
            \mu_0,\zeta_0}(x_i,\tvec{y};\mu_1,\mu_2)} \nonumber \\ &
        \qquad\qquad = 2 \sum_{b_1} \pr{R}{P_{a_1
            b_1}(x_1';\mu_1^{},\mu_1^2)} \underset{x_1}{\otimes}
        \pr{R}{\widehat{F}_{b_1 a_2,\,
            \mu_0,\zeta_0}(x_1',x_2,\tvec{y};\mu_1,\mu_2)}
\end{align}
and its analogue for $\mu_2$, with the initial condition
\begin{align}
   \label{DGLAP-red-initial}
\pr{R}{\widehat{F}_{a_1 a_2,\, \mu_0,\zeta_0}(x_i,\tvec{y};\mu_0,\mu_0)}
&= \pr{R}{F_{a_1 a_2}(x_i,\tvec{y};\mu_0,\mu_0,\zeta_0)} \,.
\end{align}
Unlike $F$, the distribution $\widehat{F}$ evolves with $\zeta$ independent
kernels.
The correct scale dependence of \eqref{coll-zeta-expl} can be verified by
inserting \eqref{DGLAP-CS} into \eqref{DGLAP-zeta}, and its $\zeta$
dependence can be verified by using \eqref{CS-coll} and
\eqref{CS-coll-RG}.

The integrals over $\mu$ in \eqref{coll-zeta-expl} give rise to double
logarithms, which have been exponentiated to a Sudakov factor in the solution
of the evolution equation.  Since the anomalous dimensions
$\prn{R}{\gamma}_{J}$ are positive for all $R \neq 1$ (see
section~\ref{sec:anom-dims}), the contribution of all colour non-singlet DPDs
to the cross section is Sudakov suppressed for sufficiently large hard scales.
This was recognised long ago in \cite{Mekhfi:1988kj} and confirmed in
\cite{Manohar:2012jr}, where a simple numerical estimate was also given.  In
\cite{Manohar:2012jr}, scale and rapidity evolution was discussed for
collinear DPS factorisation, keeping unsubtracted DPDs and an explicit soft
factor in the cross section.  The individual evolution equations hence differ
between their approach and ours, but must of course give the same effects at
the cross section level.

\subsection{Combining DPS with SPS}
\label{sec:TMD-UV-region}

\begin{figure}
\begin{center}
\subfigure[]{\includegraphics[width=0.32\textwidth]{%
    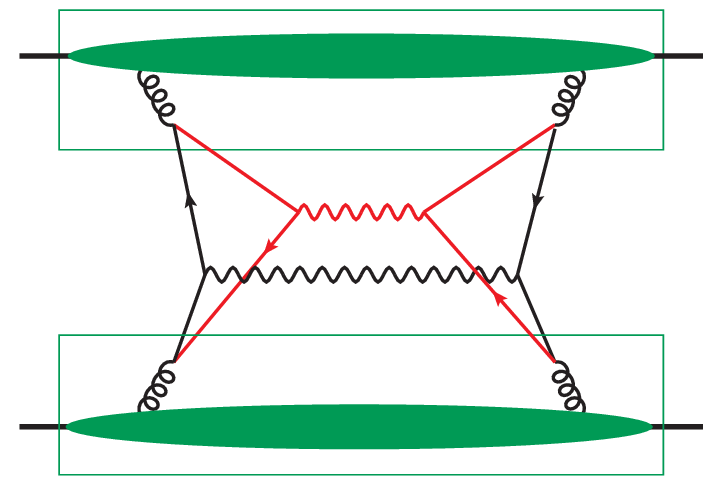}}
\subfigure[]{\includegraphics[width=0.32\textwidth]{%
    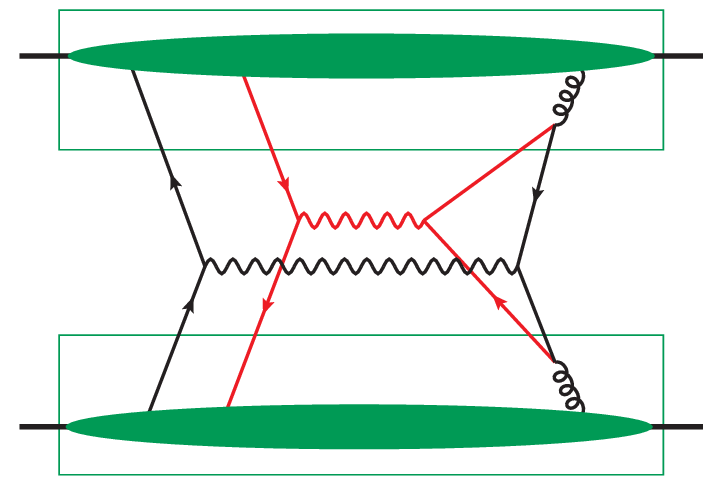}}
\subfigure[]{\includegraphics[width=0.32\textwidth]{%
    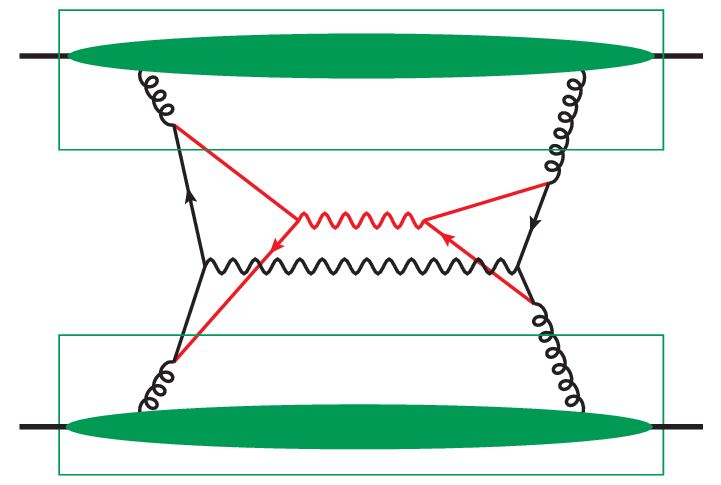}}
\caption{\label{fig:boxed} Graphs that contribute to several terms in the
  cross section~\protect\eqref{eq:tot-x-sec}.  The rectangular boxes
  and oval blobs indicate the relevant hadronic matrix elements, depending
  on whether the $g\to q\bar{q}$ splittings are collinear or hard.}

\vspace{2em}

\subfigure[]{\includegraphics[width=0.32\textwidth]{%
    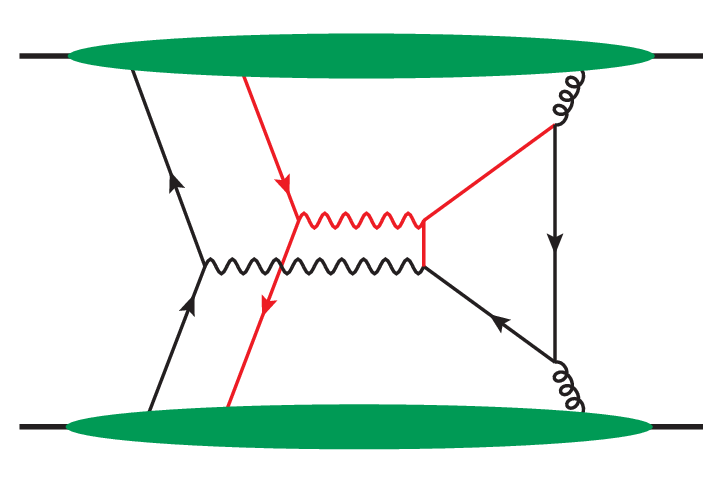}}
\hspace{3em}
\subfigure[]{\includegraphics[width=0.32\textwidth]{%
    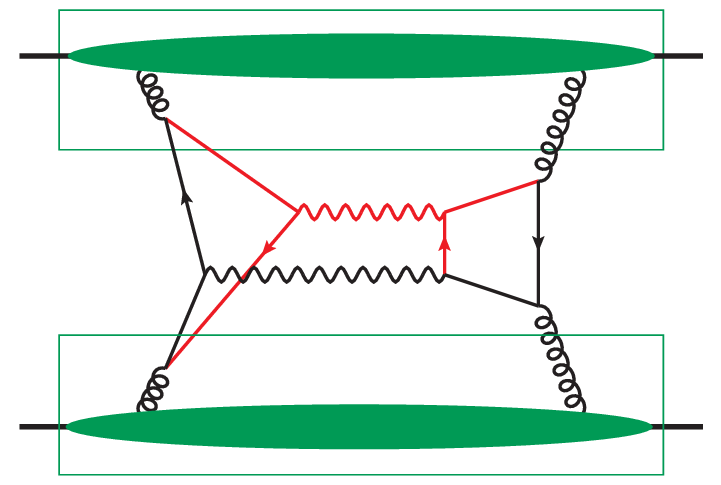}}
\caption{\label{fig:nobox} (a) A graph for the SPS/DPS interference that
  does not have overlap with DPS.  (b) An SPS graph that has no overlap with
  DPS but with the SPS/DPS interference.}
\end{center}
\end{figure}

The DPS cross section given in \eqref{TMD-Xsect} is not a physical observable,
because the same final state can also be produced by SPS.  Dimensional
analysis readily shows that DPS and SPS contribute to the TMD cross section at
the same power in $1/Q$.  The distinction between the two mechanisms is in
fact nontrivial, since there are Feynman graphs that contribute to both of
them.  These are graphs in which one parton splits into two, as shown in
figure~\ref{fig:boxed}a.  The kinematic region where the $q\bar{q}$ pairs are
nearly collinear with their parent gluon is naturally described as DPS, with
the splitting being included in the DPDs.  By contrast, the region where the
quarks and antiquarks have large transverse momenta (and hence large
virtualities) is adequately described as a loop correction to SPS, with gluon
TMDs describing the initial state of the hard scattering.

The interference between SPS and DPS has the same power behaviour in $1/Q$ as
SPS and DPS and hence should also be included in the cross section.  Again,
there is a double counting problem, since graphs like the one in
figure~\ref{fig:boxed}b contribute to either the interference term or to DPS,
depending on whether the splitting takes place in the collinear or in the hard
momentum region.
Note that not all graphs are affected by double counting.  The graphs in
figure~\ref{fig:nobox} for instance have no overlap with DPS, whereas there
is overlap between SPS and the SPS/DPS interference in
figure~\ref{fig:nobox}b.

A systematic formalism for separating the different contributions and for
adding them in the cross section without double counting has been presented in
\cite{Diehl:2017kgu}.  In the following, we briefly recapitulate the
essentials of this scheme for TMD factorisation, referring to that work for
details.  DTMDs and other hadronic matrix elements are used in the transverse
position (rather than transverse momentum) representation.  This is very
convenient for treating Collins-Soper evolution, which is then multiplicative
(rather than involving convolution integrals).  The distinction between
collinear and hard parton splitting then corresponds to the transverse
distance between the partons being large or small, respectively.

\begin{figure}
\begin{center}
\subfigure[]{\includegraphics[width=0.4\textwidth]{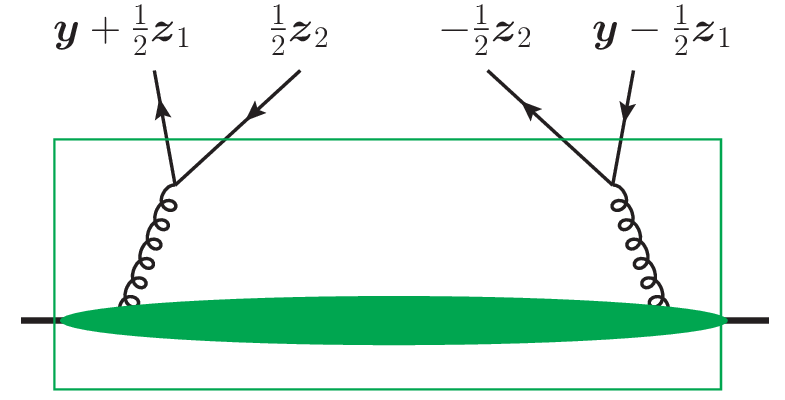}}
\hspace{3em}
\subfigure[]{\includegraphics[width=0.4\textwidth]{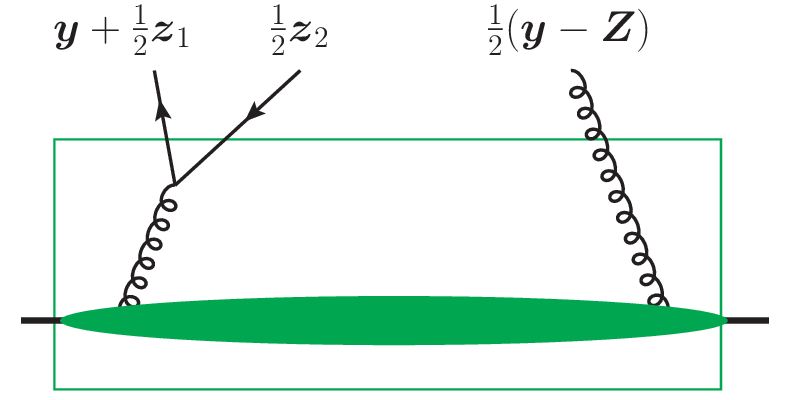}}
\caption{\label{fig:split} Transverse positions of the parton fields in parton
splitting contributions to (a) the quark-antiquark DTMD and (b) the
twist-three TMD $D_{q\bar{q}| g}$.  Due to translation invariance, the latter
depends only on $\tvec{y}_+$ and $\tvec{Z}$.  The positions in (b) correspond
to equation (3.9) in~\protect\cite{Diehl:2017kgu}.}
\end{center}
\end{figure}

As can be seen in figure~\ref{fig:split}a, the relative distance between the
two partons with momentum fractions $x_1$ and $x_2$ in a DTMD is $\tvec{y}_+$
for the partons in the amplitude and $\tvec{y}_-$ for those in the conjugate
amplitude, with the distances defined in \eqref{y-pm-def}.  The contribution
from small distances to the DPS cross section $\sigma_{\text{DPS}}$ is removed
by the factors $\Phi(\nu \tvec{y}_+)$ and $\Phi(\nu \tvec{y}_-)$ in
\eqref{W-def}, which we discuss further at the end of this section.
Corresponding factors are inserted in the interference
$\sigma_{\text{DPS/SPS}} + \sigma_{\text{SPS/DPS}}$ between SPS and DPS.  The
SPS cross section $\sigma_{\text{SPS}}$ is not modified in this scheme and
computed as usual from TMDs and hard scattering cross sections, as described
for instance in \cite{Collins:2011zzd}.  The master formula for the overall
cross section then reads
\begin{align}
  \label{eq:tot-x-sec}
\sigma
  & = \sigma_{\text{DPS}\rule{0pt}{1.3ex}}
    + \left[\ms \sigma_{\text{DPS/SPS}}
    - \sigma_{\text{DPS},\, y_-\to 0 \rule{0pt}{1.3ex}}
    + \sigma_{\text{SPS/DPS}}
    - \sigma_{\text{DPS},\, y_+\to 0 \rule{0pt}{1.3ex}}
    \ms\right]
\nonumber \\
  & \quad + \left[\ms \sigma_{\text{SPS}\rule{0pt}{1.3ex}}
  - \sigma_{\text{DPS/SPS},\, y_+ \rightarrow 0}
  - \sigma_{\text{SPS/DPS},\, y_- \rightarrow 0}
  + \sigma_{\text{DPS},\,y_\pm\rightarrow 0 \rule{0pt}{1.3ex}} \ms\right]\,,
\end{align}
where all terms should be taken differential in $dx_1\, dx_2\, d\bar{x}_1\,
d\bar{x}_2\, d^2\tvec{q}_1\, d^2\tvec{q}_2$.  In this formula, the double
counting problem between different contributions is solved by subtraction
terms, which we will discuss shortly.

Let us start with the interference between DPS and SPS.  As the colour and
polarisation of the partons are not the main focus of the present section, we
will not display these quantum numbers explicitly.  The following equations
hold for specific colour and helicity of each parton on the l.h.s.\ and with
appropriate sums on the r.h.s.  The indices $\alpha_i, \beta_i$ will just
denote the parton species.  Up to kinematic and numeric factors, the
interference cross section is given by
\begin{align}
	\label{eq:DPSSPS}
& \frac{d\sigma_{\text{DPS/SPS}}}{dx_1\, dx_2\, d\bar{x}_1\, d\bar{x}_2\,
  d^2\tvec{q}_1\, d^2\tvec{q}_2}
\,\propto \sum_{\genfrac{}{}{0pt}{1}{\alpha_1,
  \alpha_2,\alpha_0}{\beta_1,\beta_2,\beta_0}} \!\!\!
  H_{\alpha_1 \beta_1}^{}\, H_{\alpha_2 \beta_2}^{}\, H^*_{\alpha_0 \beta_0}
  \int d^2\tvec{Z}\, d^2\tvec{y}_+\;
  e^{- i\ms (\tvec{q}_1 + \tvec{q}_2) \tvec{Z}}
\nonumber \\[0.2em]
&\qquad\qquad \times
  e^{- i\ms (\tvec{q}_1 - \tvec{q}_2) \tvec{y}_+ /2} \;
  \Phi(\nu \tvec{y}_+)\,
  D_{\beta_1 \beta_2| \beta_0}(\bar{x}_i, \tvec{y}_+, \tvec{Z}) \,
  D_{\alpha_1 \alpha_2| \alpha_0}(x_i, \tvec{y}_+, \tvec{Z})
\end{align}
for DPS in the amplitude, and by
\begin{align}
	\label{eq:SPSDPS}
& \frac{d\sigma_{\text{SPS/DPS}}}{dx_1\, dx_2\, d\bar{x}_1\, d\bar{x}_2\,
  d^2\tvec{q}_1\, d^2\tvec{q}_2}
\,\propto \sum_{\genfrac{}{}{0pt}{1}{\alpha_0,
  \alpha_1,\alpha_2}{\beta_0,\beta_1,\beta_2}} \!\!\!
  H^{}_{\alpha_0 \beta_0}\, H^*_{\alpha_1 \beta_1}\, H^*_{\alpha_2 \beta_2}
  \int d^2\tvec{Z}\, d^2\tvec{y}_-\;
  e^{- i\ms (\tvec{q}_1 + \tvec{q}_2) \tvec{Z}}
\nonumber \\[0.2em]
&\qquad\qquad \times
  e^{i\ms (\tvec{q}_1 - \tvec{q}_2) \tvec{y}_- /2} \;
  \Phi(\nu \tvec{y}_-)\,
  D_{\beta_0| \beta_1 \beta_2}(\bar{x}_i, \tvec{y}_-, \tvec{Z}) \,
  D_{\alpha_0| \alpha_1 \alpha_2}(x_i, \tvec{y}_-, \tvec{Z})
\end{align}
for DPS in the complex conjugate amplitude.
Here $D_{\alpha_1 \alpha_2| \alpha_0}$ is a twist-three TMD with two partons
$\alpha_1 \alpha_2$ in the amplitude and one parton $\alpha_0$ in the
conjugate amplitude.  Its definition is similar to the one of DTMDs, with
three instead of four parton operators in the hadronic matrix element.  The
transverse positions of the partons are as indicated in
figure~\ref{fig:split}b.  In $D_{\alpha_0| \alpha_1 \alpha_2}$, the roles of
amplitude and conjugate amplitude are interchanged.  $H_{\alpha_1\beta_1}$ and
$H_{\alpha_2\beta_2}$ are the hard scattering amplitudes for the DPS
processes, whilst $H_{\alpha_0\alpha_0}$ is the amplitude for the hard SPS
interaction.

$\sigma_{\text{DPS/SPS}}$ receives a contribution from the graph in
figure~\ref{fig:boxed}b without the rectangular boxes.  The quark loop
in the conjugate amplitude includes an integration over the full phase
space, including the region where the $g\to q\bar{q}$ splittings become
collinear.  By a Fourier transform, one can rewrite the transverse part of
the loop integration as an integral over $\tvec{y}_-$, after which the
region of collinear splitting corresponds to large $\tvec{y}_-$.  This
contribution is already included in $\sigma_{\text{DPS}}$, so that a
subtraction term is necessary to prevent double counting.  It reads
\begin{align}
	\label{sps-dps-subt}
& \frac{d\sigma_{\text{DPS}, y_-\to 0}}{dx_1\, dx_2\, d\bar{x}_1\,
          d\bar{x}_2\, d^2\tvec{q}_1\, d^2\tvec{q}_2} \,\propto\,
        \sum_{\genfrac{}{}{0pt}{1}{\alpha_1,
            \alpha_2}{\beta_1,\beta_2}}
        H^{}_{\alpha_1 \beta_1}\, H^{}_{\alpha_2 \beta_2}\, H^*_{\alpha_1
  \beta_1}\, H^*_{\alpha_2 \beta_2}\,
\nonumber \\
 &\qquad\qquad
  \times \int d^2\tvec{Z}\, d^2\tvec{y}_+\, d^2\tvec{y}_-\,\,
  e^{- i\ms (\tvec{q}_1 + \tvec{q}_2) \tvec{Z}
     - i\ms (\tvec{q}_1 - \tvec{q}_2) (\tvec{y}_+ -  \tvec{y}_-) /2}\;
   \Phi(\nu \tvec{y}_+)\, \Phi(\nu\tvec{y}_-)\,
   \nonumber \\[0.2em]
&\qquad\qquad\quad \times
     F_{\beta_1 \beta_2,\ms y_- \to 0}(\bar{x}_i, \tvec{y}_\pm, \tvec{Z}) \,
     F_{\alpha_1 \alpha_2,\ms y_- \to 0}(x_i, \tvec{y}_\pm, \tvec{Z}) \,.
\end{align}
Compared with the DPS cross section given by \eqref{TMD-Xsect} and
\eqref{W-def}, we have traded the arguments $\tvec{z}_i$, $\tvec{y}$ of the
DTMDs for $\tvec{y}_\pm$, $\tvec{Z}$ and correspondingly changed the
integration according to \eqref{fourier-rewrite}.  For simplicity we write
$y_-\to 0$ instead of $\tvec{y}_- \to \tvec{0}$ in subscripts.  The DTMDs in
\eqref{sps-dps-subt} correspond to the rectangular boxes in
figure~\ref{fig:boxed}b and are given by
\begin{align}
	\label{TMD-one-split}
F_{\alpha_1 \alpha_2,\, y_-\to 0}(x_i,\tvec{y}_\pm,\tvec{Z})
 &= \frac{\tvec{y}_-^{l'}}{\tvec{y}_-^{2}}\,
\bigl[ U^{l'}_{\alpha_0\to \alpha_1 \alpha_2}(x_i) \bigr]^* \, D_{\alpha_1
  \alpha_2| \alpha_0}(x_i, \tvec{y}_+, \tvec{Z})
+ \mathcal{O}(\alpha_s^{3/2}) \,.
\end{align}
This notation reflects that in the limit $\tvec{y}_- \to \tvec{0}$ the DTMD
is dominated by the contribution where the two partons $\alpha_1$ and
$\alpha_2$ in the conjugate amplitude originate from the splitting of
a single parton $\alpha_0$.  The lowest order perturbative splitting kernel
$U$ includes a factor of $\sqrt{\alpha_s\rule{0pt}{1.5ex}}$.  For given
$\alpha_1$ and $\alpha_2$, the parton species $\alpha_0$ is fixed at this
order, but its helicity must be summed over.
As shown in \cite{Diehl:2017kgu}, the combination
$\sigma_{\text{DPS}\rule{0pt}{1.3ex}} + \sigma_{\text{DPS/SPS}} -
\sigma_{\text{DPS}, y_-\to 0 \rule{0pt}{1.3ex}} $ correctly treats the
contribution of graph~\ref{fig:boxed}b to the cross section, for small and
for large $\tvec{y}_-$, and without double counting.  We will see in more
detail how this happens in section~\ref{sec:reg_sub_simp}.  The other
subtraction term in the first line of \eqref{eq:tot-x-sec} is given by
expressions analogous to \eqref{sps-dps-subt} and \eqref{TMD-one-split},
with $\tvec{y}_-$ replaced by $\tvec{y}_+$ and the roles of amplitude and
conjugate amplitude being interchanged.

The subtraction terms $\sigma_{\text{DPS/SPS},y_+\rightarrow0}$ and
$\sigma_{\text{SPS/DPS},y_-\rightarrow0}$ in the second line of
\eqref{eq:tot-x-sec} are obtained from \eqref{eq:DPSSPS} and
\eqref{eq:SPSDPS} by replacing all distributions $D$ with their perturbative
splitting approximation.  For the case shown in figure~\ref{fig:split}b,
this is given by
\begin{align}
	\label{tw3-split}
D_{\alpha_1\alpha_2|\alpha_0,\, y_+\to 0}(x_i,\tvec{y}_+,\tvec{Z})
 &= \frac{\tvec{y}_+^l}{\tvec{y}_+^2}\;
U_{\alpha_0\rightarrow \alpha_1 \alpha_2}^l(x_i)\,
  f^{}_{\alpha_0}(x_1+x_2,\tvec{Z})
  + \mathcal{O}(\alpha_s^{3/2}) \,,
\end{align}
where on the r.h.s.\ we have a single parton TMD and the same splitting kernel
$U$ as in \eqref{TMD-one-split}.  An analogous expression holds for the limit
$\tvec{y}_- \to \tvec{0}$ of $D_{\alpha_0| \alpha_1 \alpha_2}$.  The
combination
$\sigma_{\text{DPS/SPS}} + \sigma_{\text{SPS}\rule{0pt}{1.3ex}} -
\sigma_{\text{DPS/SPS},y_+\rightarrow 0}$ correctly represents
graph~\ref{fig:boxed}c in the region where the quark loop in the conjugate
amplitude is hard.

Finally, the term $\sigma_{\text{DPS},y_\pm\rightarrow0}$ in
\eqref{eq:tot-x-sec} ensures the correct treatment of the case when both quark
loops in graph~\ref{fig:boxed}a are in the collinear region.  This term is
given by the DPS cross section with each DTMD approximated
for the regime where both parton pairs are produced by perturbative
splitting:
\begin{align}
	\label{F-short-dist_TMD}
F_{\alpha_1 \alpha_2,\, y_\pm \to 0}(x_i,  \tvec{y}_\pm, \tvec{Z})
 &= \frac{\tvec{y}_+^l}{\tvec{y}_+^2}\,
        \frac{\tvec{y}_-^{l'}}{\tvec{y}_-^2}\,
        U^l_{\alpha_0\to \alpha_1 \alpha_2}(x_i) \, \bigl[
        U^{l'}_{\alpha_0\to \alpha_1 \alpha_2}(x_i) \bigr]^* \,
    f^{}_{\alpha_0}(x_1+x_2,\tvec{Z})
    + \mathcal{O}(\alpha_s^2)\,.
\end{align}
This is illustrated in figure~\ref{fig:split}a and can be obtained by
inserting the expression \eqref{tw3-split} into \eqref{TMD-one-split}.
Notice that $\sigma_{\text{DPS},y_\pm\rightarrow0}$ appears with a plus rather
than a minus sign in \eqref{eq:tot-x-sec}.  This is a consequence of the
recursive nature of double counting subtractions, as explained in section~4.2
of \cite{Diehl:2017kgu}.

To make contact with our notation in section~\ref{sec:short-dist-exp}, we
rewrite \eqref{F-short-dist_TMD} in terms of a single splitting kernel
as\footnote{The convention for $T^{jj'}$ here is the same as in
\protect\cite{Diehl:2017kgu} (arXiv version 2) and differs from the one in
\protect\cite{Diehl:2011yj} as specified in
section~\ref{sec:kernels_for_splitting}.}
\begin{align}
  \label{split-TMD}
\pr{R}{F}_{a_1 a_2,\, y_\pm \to 0}(x_i,\tvec{z}_i,\tvec{y})
 &= \frac{\tvec{y}_{+}^{l}\ms
  \tvec{y}_{-}^{l'}}{ \tvec{y}_{+}^{2}\ms \tvec{y}_{-}^{2}}\,
   \sum_{a_0} \frac{\alpha_s}{2\pi^2}\;
   \prn{R}{T}_{a_0\to a_1 a_2}^{ll'}\biggl(
\frac{x_1}{x_1+x_2} \biggr)\, \frac{f_{a_0}(x_1+x_2, \tvec{Z})}{x_1+x_2}
+ \mathcal{O}(\alpha_s^2)\,,
\end{align}
where we have restored the dependence on the colour representation and the
notation with labels $a_i$ that specify parton species and polarisation.
For transverse quark or linear gluon polarisation, the parton distributions
and the kernel $T_{a_0\to a_1 a_2}$ carry additional transverse indices, which
are not displayed here (see section~\ref{sec:kernels_for_splitting}).
The sum over $a_0$ reflects the fact that there are TMDs for these
polarisation states, even if the proton is unpolarised.

We can now discuss the role of the function $\Phi$ introduced earlier.
The short-distance behaviour of the DTMDs and twist-three distributions,
given in \eqref{TMD-one-split} and \eqref{tw3-split}, results in an integral
\begin{align}
	\label{log-int}
\int\limits_{|\tvec{y}_+| \ll 1/\Lambda}
\frac{d^2\tvec{y}_+}{\tvec{y}_+^2}\; e^{- i\ms (\tvec{q}_1 -
  \tvec{q}_2)\ms \tvec{y}_+ /2}\; \Phi(\nu \tvec{y}_+)
\end{align}
and its analogue for $\tvec{y}_-$ in the cross section formulae for DPS and
for the SPS/DPS inter\-ference.  The function $\Phi$ must satisfy
$\Phi(\tvec{u})\to 0$ for $\tvec{u}\to \tvec{0}$ to ensure that these
integrals converge at small distances (rather than having a logarithmic
divergence).  In order to keep the regions of large $\tvec{y}_+$ and
$\tvec{y}_-$ unaffected, one should furthermore have $\Phi(\tvec{u})\to 1$
for $|\tvec{u}| \gg 1$.  A simple choice for $\Phi(\tvec{u})$ is a step
function in $|\tvec{u}|$, which corresponds to a hard cutoff on the
$\tvec{y}_\pm$ integrals.

The integral \eqref{log-int} goes like $\log(\nu/q_T)$ and thus depends on the
artificial parameter $\nu$ that controls which distances $\tvec{y}_\pm$ are
included in what we \emph{define} to be DPS rather than SPS.  The $\nu$
dependence of the different terms in the cross section~\eqref{eq:tot-x-sec}
cancels to the perturbative order of the calculation, in close analogy with
the familiar case of renormalisation and factorisation scale dependence.  As
discussed in \cite{Diehl:2017kgu}, a suitable choice for $\nu$ is the lowest
hard scale, $\text{min}(Q_1,Q_2)$.  With this choice, $\sigma_{\text{DPS}}$
contains a squared logarithm $\log^2(Q/q_T)$, where we recall that $Q$
denotes the generic size of $Q_1$ and $Q_2$.  After the subtractions in the
first line of \eqref{eq:tot-x-sec}, the SPS/DPS interference contains only a
single $\log(Q/q_T)$ and the subtracted SPS term in the second line has
no such logarithm at the corresponding order in $\alpha_s$.  This has an
important consequence.  If one is satisfied with leading logarithmic accuracy,
then the terms in square brackets in \eqref{eq:tot-x-sec} can be neglected at
the order in $\alpha_s$ where they have overlap with DPS, because at that
order it is $\sigma_{\text{DPS}}$ that has the highest power of
$\log(Q/q_T)$.

We should note that the structure of the higher-order terms in
\eqref{TMD-one-split}, \eqref{tw3-split} and \eqref{F-short-dist_TMD} is
currently unknown.  In particular, it cannot be excluded that singularities
will appear at points other than $\tvec{y}_+ = \tvec{0}$ and
$\tvec{y}_- = \tvec{0}$.  The treatment of ultraviolet divergences and of the
associated double counting issues may thus be more involved beyond the leading
order discussed here.

To conclude this section, we briefly discuss the scale evolution of the
twist-three TMDs~$D$.  As already noted, they are defined in terms of hadronic
matrix elements, with three parton fields and the corresponding Wilson lines
instead of the four fields we have in the unsubtracted DTMDs in
\eqref{eq:dpds}.  The three parton fields are at different transverse
positions (if all three positions are equal, one has collinear twist-three
distributions).  It is easy to see that the analysis of factorisation for the
SPS/DPS interference will yield a soft factor, with three products of Wilson
lines associated with the three parton fields in the hadronic matrix elements
(rather than four products of Wilson lines as in
figure~\ref{fig:soft-fact-indices}).  One can readily adapt the discussion in
sections~\ref{sec:soft-fact} to \ref{sec:ren-DTMDs} to this case and absorb
the soft factor into the twist-three distributions.  The colour algebra is
much simpler than in the DPS case, with a single colour representation in the
$q\bar{q}g$ channel, and two for the channel with three gluons (one
constructed with $f^{abc}$ and the other with $d^{abc}$).  The rapidity
dependence of $D_{\alpha_1 \alpha_2 |\alpha_0}$ is finally given by a
Collins-Soper equation as in \eqref{CS-TMD} with a kernel
$K_{\alpha_1 \alpha_2 |\alpha_0}$.  If one takes different renormalisation
scales $\mu_1, \mu_2$ and $\mu_0$ for the three parton legs, then the
dependence of $D$ and $K$ on these scales is given by the analogues of
\eqref{RG-TMD-again} and \eqref{CS-TMD-RG}, with $\gamma_{K,a}$ replaced by
$\gamma_{K,a} /2$ and $\gamma_{F,a}$ by $\gamma_{E,a}$ or $(\gamma_{E,a})^*$
as discussed at the end of section~\ref{sec:ren-DTMDs}.
The rapidity parameter $\zeta$ is defined by \eqref{zeta-def} and must be
properly rescaled in the argument of $\gamma_{F,a}$, namely by
$x_1 \zeta/x_2$, $x_2\ms \zeta/x_1$ and $(x_1+x_2)^2 \ms \zeta /(x_1 x_2)$ for
the partons with momentum fractions $x_1$, $x_2$ and $x_1+x_2$, respectively.

\section{Matching for small but perturbative transverse momenta}
\label{sec:short}

In the multi-scale regime $\Lambda \ll q_T \ll Q$, where the transverse
momenta $|\tvec{q}_i|$ are small compared with $Q_i$ but large compared with
the scale $\Lambda$ of nonperturbative physics, the DPS cross section, as well
as its combination with SPS can be considerably simplified.  The reason is
that the additional scale $q_T$ allows for more perturbative calculations,
which greatly enhances the predictive power of the theory.

\rev{We start our analysis by discussing the different transverse distance
regions in the DPS factorisation formula and find that two distinct regions of
$\tvec{y}$ are relevant.  The corresponding short-distance expansions of DTMDs
are derived in sections~\ref{sec:small_zi} and \ref{sec:small-y},
respectively.  In section~\ref{sec:combine-y} we discuss the master
formula~\eqref{Xsect-DPS} for combining the two types of expansion in the
cross section, where care must be taken to avoid double counting.  After this,
we revisit the combination between DPS and SPS just discussed in
section~\ref{sec:TMD-UV-region}.  In section~\ref{sec:resummation_orders} we
go through the perturbative ingredients that are necessary for a given
resummation order as specified in table~\ref{tab:orders}; their availability
determines the perturbative accuracy of cross section computations.}

\subsection{Regions of transverse momenta and distances}
\label{sec:regions}

In the DPS cross section \eqref{TMD-Xsect}, contributions from distances
$|\tvec{z}_i| \gg 1 /|\tvec{q}_i|$ are suppressed by oscillations of the
Fourier exponent $e^{-i\tvec{q}_1 \tvec{z}_1 - i\tvec{q}_2 \tvec{z}_2}$.  If
$\tvec{q}_1$ and $\tvec{q}_2$ are sufficiently large, this keeps
$\tvec{z}_1$ and $\tvec{z}_2$ in the region where perturbation theory has
predictive power for the dependence on these variables.

Care is required if $\tvec{q}_1$ and $\tvec{q}_2$ are perturbatively large but
$|\tvec{q}_1 + \tvec{q}_2|$ is of order $\Lambda$.  According to
\eqref{fourier-rewrite} the sum $|\tvec{z}_1 + \tvec{z}_2|$ can then reach
large values of order $1/\Lambda$, as long as $|\tvec{z}_1 - \tvec{z}_2|$
remains small.  The individual oscillations of $e^{-i\tvec{q}_1 \tvec{z}_1}$
and $e^{- i\tvec{q}_2 \tvec{z}_2}$ cancel in that case.  Likewise,
$|\tvec{z}_1 - \tvec{z}_2|$ can become as large as $1/\Lambda$ if $\tvec{q}_1$
and $\tvec{q}_2$ are perturbatively large but $|\tvec{q}_1 - \tvec{q}_2|$ is
of order $\Lambda$.
The perturbative splitting form \eqref{split-TMD} of DTMDs has a factorised
dependence on the variables $\tvec{z}_1 + \tvec{z}_2$ and
$\tvec{z}_1 - \tvec{z}_2$, and the short-distance expansion
\eqref{split-TMD-coll} we will derive from it requires both distances to be
small.
We therefore do not consider the particular phase space regions just
mentioned and require $|\tvec{q}_1|$ and $|\tvec{q}_2|$ as well as
$|\tvec{q}_1 + \tvec{q}_2|$ and $|\tvec{q}_1 - \tvec{q}_2|$ to be much
larger than $\Lambda$.  The dominant contribution to the $\tvec{z}_1$ and
$\tvec{z}_2$ integrals in the cross section formula \eqref{TMD-Xsect} is
then in the perturbative region.
Even with this requirement, one could still consider a region where
$|\tvec{q}_1 + \tvec{q}_2|$ is much smaller than $|\tvec{q}_1|$ and
$|\tvec{q}_2|$.  This multi-scale regime has been discussed in section~5.2.1
of \cite{Diehl:2011yj} and in \cite{Blok:2011bu,Blok:2013bpa}.  We shall not
investigate it in the present work.

Other observables in which sensitivity to large $\tvec{z}_1$ and
$\tvec{z}_2$ can be avoided are obtained by integrating the cross section
over one or both of $\tvec{q}_1$ and $\tvec{q}_2$ up to $q_{\text{max},1}$ and
$q_{\text{max},2}$ respectively.  In
\begin{align}
  \int d^2\tvec{q}\; \theta\bigl( q_{\text{max}}^2 - \tvec{q}^2
  \bigr) \int \frac{d^2\tvec{z}}{(2\pi)^2}\; e^{-i \tvec{q} \tvec{z}}\,
  W(\tvec{z}, \ldots) &= q_{\text{max}} \int \frac{d^2\tvec{z}}{2\pi
    |\tvec{z}|}\, J_1\bigl( q_{\text{max}} |\tvec{z}| \bigr)\, W(\tvec{z},
  \ldots)
\end{align}
the region of $|\tvec{z}| \gg 1/q_{\text{max}}$ is damped by oscillations,
as it is by the exponential factors in the differential cross section.  To
avoid that the product of the two Bessel functions has a
non-oscillating component for $|\tvec{z}_1| \approx |\tvec{z}_2|$, one
must take $q_{\text{max},1} \neq q_{\text{max},2}$.

We henceforth assume that both $|\tvec{z}_1|$ and $|\tvec{z}_2|$ are small,
namely of order $1/q_T$.  Since $\tvec{y}$ is integrated over without any
Fourier exponent in \eqref{TMD-Xsect} we must still consider the two cases
$|\tvec{y}| \sim 1/\Lambda$ and $|\tvec{y}| \sim 1/q_T$, which are referred
to as large $\tvec{y}$ and small $\tvec{y}$ in the following.  For even
smaller $|\tvec{y}| \ll 1/q_T$, the approximations for DPS are not valid and
the subtraction formalism sketched in section~\ref{sec:TMD-UV-region} comes
into play.  The short-distance expansion of DPDs in the two DPS regions just
mentioned is quite different and will now be discussed in turn.

\subsection{The large-\texorpdfstring{$\tvec{y}$}{y} region}
\label{sec:small_zi}

For small $\tvec{z}_1$, $\tvec{z}_2$ and large $\tvec{y}$, the DTMD
$F(x_i, \tvec{z}_i,\tvec{y})$ can be computed in terms of DPDFs
$F(x_i,\tvec{y})$, in close analogy to the case of single parton
distributions.  This is commonly referred to as ``matching''.  For DTMDs there
are complications from colour, but our results from
section~\ref{sec:soft-simple} lead to considerable
simplifications. \rev{Generalising the relations \eqref{single-TMD-match} and
\eqref{TMD-evo-solved} between TMDs and PDFs, we can derive the general
matching equation \eqref{full-match} and the explicit solution
\eqref{small-z-evolved} of the evolution equations for DTMDs.  This solution
is promoted to the cross section level in \eqref{W-large-y}.  We emphasise
that the simple structure of \eqref{small-z-evolved} and \eqref{W-large-y} is
due to the additive form \eqref{CS-gen-match} of the Collins-Soper matrix
kernel in the short-distance limit considered here.}

\subsubsection{Short-distance expansion}

We now derive the approximation of DTMDs in the region $|\tvec{z}_1|,
|\tvec{z}_2| \ll 1/\Lambda$, with $\tvec{y}$ kept fixed at a value such
that $|\tvec{z}_1|, |\tvec{z}_2| \ll |\tvec{y}|$.  Corrections to this
approximation are suppressed by powers of $|\tvec{z}_i| \Lambda$ and of
$|\tvec{z}_i| /|\tvec{y}|$.  To make the analogy between DTMDs and single
parton TMDs transparent, we formulate the matching in terms of operator
product expansions around $\tvec{z}_i = \tvec{0}$.

\begin{figure}
\begin{center}
\includegraphics[width=0.3\textwidth]{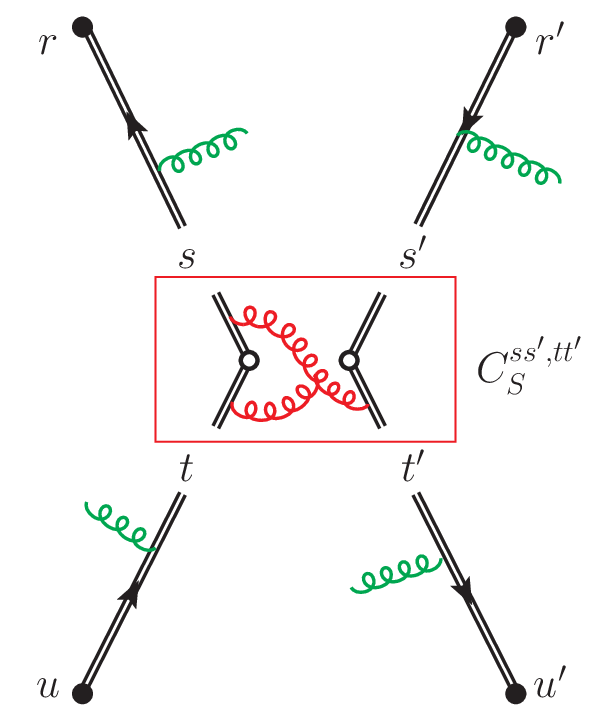}
\caption{\label{fig:soft-match} Interactions of Wilson lines in the operator
  $O_{S,a}(\tvec{y},\tvec{z})$ in the limit of small $\tvec{z}$.
  Short-distance interactions are located in the box at the centre, whereas
  gluons outside the box indicate possible interactions with other Wilson
  lines at distances much larger than $\tvec{z}$.  In the soft factor $S_{a_1
    a_2}(\tvec{z}_i, \tvec{y})$ these distances are of order $\tvec{y}$ for
  small $\tvec{z}_1$ and $\tvec{z}_2$.}
\end{center}
\end{figure}

Let us start with the expansion of the soft factor.  We need the
small-$\tvec{z}$ limit of the operators $O_{S,a}(\tvec{y},\tvec{z})$ defined
in \eqref{WL-ops}, which contain two Wilson line pairs separated by a
transverse distance $\tvec{z}$.  The colour structure of the short-distance
expansion can be deduced from the corresponding Feynman graphs.  To minimise
the number of eikonal propagators that carry large momentum, interactions at
the hard scale $1/|\tvec{z}|$ must take place closest to the points where the
left- and right-moving Wilson lines meet, as shown in
figure~\ref{fig:soft-match}.  In terms of the operators in \eqref{WL-ext} and
\eqref{WL-ops}, we thus find the structure
\begin{align}
  \label{soft-op-start}
\bigl[ {O}_{S,a}(\tvec{y},\tvec{z})
       \bigr]^{\ul{r}, \ul{u}}
  &= C_{S,a}^{\ul{s}\, \ul{t}}(\tvec{z})\; \bigl[ O_{S,a}(\tvec{y})
      \bigr]^{\ul{r}, \ul{u}}_{\ul{s}, \ul{t}}
\end{align}
with a short-distance coefficient $C_{S,a}$ that can be computed in
perturbation theory.  Here and in the following, the absence of the
argument $\tvec{z}$ in operators or functions means that one has set
$\tvec{z} = \tvec{0}$.  The analogue of this convention will be used for
arguments $\tvec{z}_i$.  Introducing colour projected coefficients
$\prn{R}{C}_{S,a}^{} = P_R^{\ul{s}\, \ul{t}}\, C_{S, a}^{\ul{s}\, \ul{t}}
\big/ m(R)$ and using \eqref{eq:mat-dec}, we obtain
\begin{align}
  \label{soft-op-match}
\bigl[ O_{S,a}(\tvec{y},\tvec{z})
       \bigr]^{\ul{r}, \ul{u}}
 &= \sum_{R} \prn{R}{C}_{S,a}^{}(\tvec{z})\, P_{R}^{\ul{s}\, \ul{t}}\,
    \bigl[ O_{S,a}(\tvec{y})
      \bigr]^{\ul{r}, \ul{u}}_{\ul{s}, \ul{t}}
\nonumber \\
 &= \sum_{R} \prn{R}{C}_{S,a}^{}(\tvec{z})\,
      P_{R}^{\ul{r}\, \ul{s}}\, P_{R}^{\ul{u}\, \ul{t}}\,
      \bigl[ O_{S,a}(\tvec{y})
      \bigr]^{\ul{s}, \ul{t}}_{\ul{v}, \ul{v}} \,,
\end{align}
where the second step is completely analogous to the derivation of the
relation \eqref{soft-inner-proj} between the extended soft
factor and the usual one.  Taking the vacuum expectation value of
$O_{S,a_1}(\tvec{y},\tvec{z}_1)\, O_{S,a_2}(\tvec{0},\tvec{z}_2)$ with
appropriate colour projections, we obtain
\begin{align}
\label{soft-fact-match}
\prb{RR'}{S_{a_1 a_2}(\tvec{z}_i, \tvec{y})} &=
\prn{R}{C_{S, a_1}(\tvec{z}_1)}\, \prn{R}{C_{S, a_2}(\tvec{z}_2)}\,
\pr{RR}{S(\tvec{y})}\, \delta_{RR'}^{} \,.
\end{align}
For computing the short-distance coefficients, it will be convenient to
contract \eqref{soft-op-match} with $P_{R'}^{\ul{r}\, \ul{u}}$, which
gives
\begin{align}
  \label{soft-op-proj}
P_{R}^{\ul{r}\, \ul{u}}
\bigl[ O_{S,a}(\tvec{y},\tvec{z}) \bigr]^{\ul{r}, \ul{u}}
 &= \prn{R}{C}_{S,a}^{}(\tvec{z})\, P_{R}^{\ul{r}\, \ul{u}}\,
      \bigl[ O_{S,a}(\tvec{y}) \bigr]^{\ul{r}, \ul{u}} \,.
\end{align}

For approximating the unsubtracted collinear matrix element $F_{\us}$, we need
the small-$\tvec{z}$ limit of the parton operators in \eqref{x-ops}.  We use
their projections \eqref{col-proj-op} on definite colour representations,
which cannot mix with each other due to gauge invariance.  The short-distance
expansion involves a convolution integral over momentum fractions and
reads\footnote{The lower limit of the convolution in
\protect\eqref{x-op-match} is $x'=0$.  When matrix elements between hadrons
with plus momentum $p^+$ are taken, the lower limit becomes $x'=x$, as in
\protect\eqref{conv-def}.}
\begin{align}
  \label{x-op-match}
\prn{R}{O}^{\ms\ul{r}}_a(x,\tvec{y},\tvec{z})
  &= \sum_{b} \prn{R}{C}_{\us,ab}(x',\tvec{z})^{}\,
     \underset{x}{\otimes} \prn{R}{O}^{\ms\ul{r}}_b(x',\tvec{y}) \,.
\end{align}
If $a$ or $b$ refer to transverse quark or linear gluon polarisation, then
the associated operators carry additional transverse indices, and the same
holds of course for $C_{\us, ab}$.  Taking matrix elements as in
\eqref{col-proj-matel}, we obtain an approximation
\begin{align}
  \label{F-us-match}
\pr{R}{F_{\us, a_1 a_2}(x_i, \tvec{z}_i, \tvec{y})} &= \sum_{b_1,
b_2} \prn{R}{C_{\us, a_1 b_1}(x_1', \tvec{z}_1)} \underset{x_1}{\otimes}
\prn{R}{C_{\us, a_2 b_2}(x_2', \tvec{z}_2)} \underset{x_2}{\otimes}
\pr{R}{F_{\us, b_1 b_2}(x_i', \tvec{y})}
\end{align}
of unsubtracted DTMDs in terms of unsubtracted DPDFs and perturbative
coefficients $C_{\us, ab}(x,\tvec{z})$.

We now combine the soft factor with the unsubtracted DTMD.  Using the relation
\eqref{S-double-simp} between $s$ and $S$, we obtain
\begin{align}
  \label{eq:s_smallz}
\pr{RR'}{s}_{a_1 a_2}(\tvec{z}_i,\tvec{y}; \mu_i, Y)
 & = \sqrt{\prn{R}{C}_{S, a_1}(\tvec{z}_1; \mu_1, 2Y)}\,
     \sqrt{\prn{R}{C}_{S, a_2}(\tvec{z}_2; \mu_2, 2Y)}\;
     \pr{RR}{s}(\tvec{y}; \mu_i,Y)\, \delta_{RR'}^{}
\end{align}
from \eqref{soft-fact-match}, where we have restored the dependence on
renormalisation scales and rapidities.  Combining this with \eqref{F-us-match}
and using the definition \eqref{sub-unsub} of DPDs, we get
\begin{align}
	\label{combine-match}
\pr{R}{F_{a_1 a_2}(x_i, \tvec{z}_i, \tvec{y};\mu_i,Y_C)}
 & = \sum_{b_1, b_2}
    \prn{R}{C_{a_1 b_1}(x_1\!\bs\smash{'},
      \tvec{z}_1;\mu_1,Y_C)} \underset{x_1}{\otimes}
    \prn{R}{C_{a_2 b_2}(x_2\!\bs\smash{'},
      \tvec{z}_2;\mu_2,Y_C)}
\nonumber \\
 & \qquad \underset{x_2}{\otimes} \lim_{Y_L \to -\infty}
    \pr{R}{F_{\us, b_1 b_2}(x_i\!\bs\smash{'}, \tvec{y};\mu_i,Y_L)}
     \big/ \pr{RR}{s}(\tvec{y}; \mu_i,Y_C - Y_L)
\end{align}
for a DPD in a right moving proton, where the short-distance coefficients
are defined as
\begin{align}
	\label{full-C-def}
\prn{R}{C_{a b}(x, \tvec{z};\mu,Y_C)} &= \lim_{Y_L \to -\infty}
   \frac{\prn{R}{C_{\text{us}, a b}(x, \tvec{z}; \mu,Y_L)}}{
   \sqrt{\prn{R}{C}_{S, a}(\tvec{z}; \mu,2Y_C - 2Y_L)\rule{0pt}{1.9ex}}} \,.
\end{align}
Switching variables from $Y_C$ to $\zeta$ and again using the definition
\eqref{sub-unsub} of DPDs, we obtain the final form of the matching equation:
\begin{align}
	\label{full-match}
\pr{R}{F_{a_1 a_2}(x_i, \tvec{z}_i, \tvec{y};\mu_i,\zeta)}
 &= \sum_{b_1, b_2}
    \prn{R}{C_{a_1 b_1}(x_1\!\bs\smash{'},
      \tvec{z}_1;\mu_1, x_1\zeta/x_2)}
\nonumber \\
 & \qquad\quad \underset{x_1}{\otimes}
    \prn{R}{C_{a_2 b_2}(x_2\!\bs\smash{'},
      \tvec{z}_2;\mu_2, x_2\zeta/x_1)} \underset{x_2}{\otimes}
    \pr{R}{F_{b_1 b_2}(x_i\!\bs\smash{'}, \tvec{y}; \mu_i,\zeta)} \,.
\end{align}
The rescaling factors $x_1/x_2$ or $x_2/x_1$ of $\zeta$ in the
short-distance coefficients arise for the same reason as discussed after
\eqref{zeta-def}.  Note that they involve the parton momentum fractions
$x_1$ or $x_2$ and not the integration variables $x_1'$ or $x_2'$ of the
convolution \eqref{conv-def}.

Let us emphasise that \eqref{full-match} involves mixing between quark and
gluon distributions.  Therefore, the combination of $\pr{RR}{s(\tvec{y})}$
and $\pr{R}{F_{\us, b_1 b_2}}(x_i,\tvec{y})$ into a DPDF works only
because $\pr{RR}{S(\tvec{y})}$ is the same for quarks and gluons (for
representations $R$ accessible to quarks).

Let us finally set $R=1$ and show that in the colour singlet channel, the
coefficient $\prn{1}{C}_{ab}(x,\tvec{z})$ in \eqref{full-match} is identical
to the coefficient $C_{ab}(x,\tvec{z})$ in the small-$\tvec{z}$ expansion
\eqref{single-TMD-match} of a single parton TMD.  The corresponding identity
for the coefficient $\prn{1}{C}_{\us, ab}(x,\tvec{z})$ readily follows from
taking the matrix element of the operator relation \eqref{x-op-match}
between two proton states, which is related to the unsubtracted TMD
$f_{\us,a}$ via the first relation in \eqref{x-matel}.  For
$\prn{1}{C}_{S,a}(\tvec{z})$ the identity is obtained by taking the vacuum
expectation value of \eqref{soft-op-proj}, which gives the soft factor
$S_a(\tvec{z})$ relevant for constructing the TMD $f_a(x,\tvec{z})$ via
\eqref{eq:S-single-simp} and \eqref{eq:unsub_single}.  The short-distance
coefficient for $f_a(x,\tvec{z})$ is then obtained as in~\eqref{full-C-def}.

Notice that the vacuum expectation value of the colour projected operator on
the r.h.s.\ of \eqref{soft-op-proj} gives unity for $R=1$, because the
Wilson lines along both $v_L$ and $v_R$ appear in the combination $W
W^\dagger = \one$.  This reflects the fact that no soft factor appears in
the definition of ordinary PDFs.  We thus find
\begin{align}
  \label{soft-singlet}
S_a(\tvec{z};\mu,Y) = \prn{1}{C}_{S,a}(\tvec{z};\mu,Y)
\end{align}
for small $\tvec{z}$, i.e.\ the soft factor for single TMD factorisation
is purely perturbative at small distances.

\subsubsection{Evolution equations and their solution}
\label{sec:match-evol}

Let us now establish the consequences of the short-distance expansion
\eqref{full-match} for scale and rapidity evolution.  From
\eqref{eq:s_smallz} we deduce that
\begin{align}
& \frac{\partial}{\partial Y}
       \log\sqrt{ \prn{R}{C}_{S,a_1}(\tvec{z}_1;\mu_1, 2Y) }
+ \frac{\partial}{\partial Y}
       \log\sqrt{ \prn{R}{C}_{S,a_2}(\tvec{z}_2;\mu_2, 2Y) }
\nonumber \\
& \quad
 = \frac{\partial}{\partial Y}
       \log \pr{RR}{s}_{a_1 a_2}(\tvec{z}_i,\tvec{y};\mu_i, Y)
 - \frac{\partial}{\partial Y}
       \log \pr{RR}{s}(\tvec{y};\mu_i, Y)
\nonumber \\
& \quad
 = \pr{RR}{K}_{a_1 a_2}(\tvec{z}_i,\tvec{y}; \mu_i)
   - \prb{R}{J}(\tvec{y}; \mu_i) \phantom{\frac{\partial}{\partial}}
\end{align}
with the kernels $\pr{RR}{K}_{a_1 a_2}$ and $\prb{R}{J}$ from
\eqref{CS-for-s} and \eqref{CS-coll}.  The rapidity derivative of
$\prn{R}{C}_{S,a}(\tvec{z};\mu, 2Y)$ must thus be independent of $Y$.
Defining
\begin{align}
  \label{C-soft-CS}
\pr{R}{K}_a(\tvec{z};\mu) &=
  \frac{\partial}{\partial Y} \log \prn{R}{C}_{S,a}(\tvec{z};\mu,Y) \,,
\end{align}
we obtain
\begin{align}
	\label{CS-coeff}
\frac{\partial}{\partial \log\zeta}\,
   \prn{R}{C_{ab}(x, \tvec{z}; \mu,\zeta)}
 &= \frac{1}{2}\, \pr{R}{K_a}(\tvec{z}; \mu)\,
    \pr{R}{C_{a b}(x, \tvec{z}; \mu,\zeta)}
\end{align}
for the coefficients in \eqref{full-C-def}.  Inserting this into
\eqref{full-match}, we find
\begin{align}
	\label{CS-gen-match}
\prb{RR'}{K_{a_1 a_2}(\tvec{z}_i,\tvec{y}; \mu_i)}
&= \delta_{RR'}^{}\, \bigl[ \pr{R}{K_{a_1}(\tvec{z}_1;\mu_1)} +
   \pr{R}{K_{a_2}(\tvec{z}_2;\mu_2)} + \prb{R}{J(\tvec{y}; \mu_i)} \bigr]\,.
\end{align}
This represents a significant simplification of the Collins-Soper equation
for DTMDs in the small $\tvec{z}_i$ limit, both in the colour structure and
in the separation of the three distance variables.

Taking the derivative $\partial/\partial \log\mu_1$ of \eqref{CS-gen-match}
and using the renormalisation group equations \eqref{CS-TMD-RG} and
\eqref{CS-coll-RG} for $\prb{RR'}{K}_{a_1 a_2}$ and $\prb{R}{J}$, we deduce
that
\begin{align}
	\label{AD-sum}
\gamma_{K,a}(\mu) &= \prn{R}{\gamma_{K,a}}(\mu) + \prn{R}{\gamma_J}(\mu)
\end{align}
for all $R$, where we have introduced the anomalous dimension of the kernel
$\pr{R}{K}_a$,
\begin{align}
\label{gamma-K-def}
 \frac{\partial}{\partial \log\mu}\, \pr{R}{K_a}(\tvec{z}; \mu)
&= {}- \prn{R}{\gamma_{K,a}}(\mu)\,.
\end{align}
In the colour singlet sector, we have $\pr{1}{J} = 0$ and hence
$\pr{1\,}{\gamma_J} = 0$, which implies $\pr{1\,}{\gamma_{K,a}} =
\gamma_{K,a}$.  Indeed, $\pr{1}{K_a(\tvec{z};\mu)}$ is identical to the
ordinary Collins-Soper kernel $K_a(\tvec{z};\mu)$ in \eqref{cs-eq}, which is
readily shown by taking the rapidity derivative of \eqref{soft-singlet}.

Taking the $\mu_1$ derivative of the short-distance expansion
\eqref{full-match} and using the evolution equations \eqref{RG-TMD-again}
and \eqref{DGLAP-zeta} for DTMDs and DPDFs, we obtain
\begin{align}
  \label{DGLAP-C}
  & \frac{\partial}{\partial \log\mu}
\prn{R}{C_{ac}(x,\tvec{z};\mu,\zeta)} \nonumber \\ & \qquad = \sum_{b}
\prn{R}{C_{ab}(x',\tvec{z};\mu,\zeta)} \underset{x}{\otimes} \Bigl[
\delta_{bc}\, \delta(1-x')\, \gamma_{F,c}(\mu,\zeta) - 2\,
\pr{R}{P_{bc}(x';\mu,\zeta)} \Bigr] \,.
\end{align}
Having computed the short-distance kernel $\prn{R}{C_{ab}}$ at a certain
order in $\alpha_s$, we can use this relation to reconstruct the evolution
kernel $\pr{R}{P_{ab}}$ at the same order.  We will do this in
section~\ref{sec:oneloop-nonsing}.

With the short-distance limit \eqref{CS-gen-match} of the Collins-Soper
kernel, the general solution \eqref{DTMD-evolved} of the evolution
equations for DTMDs takes the form
\begin{align}
  \label{small-z-start}
 & \pr{R}{F_{a_1 a_2}(x_i,\tvec{z}_i,\tvec{y};\mu_i,\zeta)}
\nonumber \\
 & \quad = \exp\,\biggl\{ \int_{\mu_{01}}^{\mu_1}
\frac{d\mu}{\mu}\, \biggl[ \gamma_{a_1}(\mu) - \gamma_{K,a_1}(\mu)
\log\frac{\sqrt{x_1\zeta/x_2}}{\mu} \biggr]
\nonumber \\
 & \hspace{3.55em} + \int_{\mu_{02}}^{\mu_2} \frac{d\mu}{\mu}\,
\biggl[ \gamma_{a_2}(\mu) - \gamma_{K,a_2}(\mu)
\log\frac{\sqrt{x_2\zeta/x_1}}{\mu} \biggr]
\nonumber \\
 & \hspace{3.55em} + \Bigl[ \pr{R}{K}_{a_1}(\tvec{z}_1;\mu_{01}) +
\pr{R}{K}_{a_2}(\tvec{z}_2;\mu_{02}) + \prb{R}{J}(\tvec{y};\mu_{0i}) \Bigr]
\log\frac{\sqrt{\zeta}}{\sqrt{\zeta_0}} \biggr\} \,
 \pr{R}{F_{a_1 a_2}(x_i,\tvec{z}_i,\tvec{y};\mu_{0i},\zeta_0)}  \,,
\end{align}
where different colour channels no longer mix with each other.  Using
\eqref{CS-coeff}, we can rewrite the rapidity dependence of the
short-distance kernels as
\begin{align}
\label{CS-evolve-C}
 \prn{R}{C}_{a_1 b_1}(x_1',\tvec{z}_1;\mu_{01},x_1\zeta_{0}/x_2)
&= \exp\,\biggl[ \pr{R}{K}_{a_1}(\tvec{z}_1;\mu_{01})
   \log\frac{\sqrt{x_1\zeta_0/x_2}}{\mu_{01}} \biggr]\,
   \prn{R}{C}_{a_1 b_1}(x_1',\tvec{z}_1^{};\mu_{01}^{},\mu_{01}^2)
\end{align}
and likewise for the index $2$, which reduces the number of independent
scales in $C_{ab}$.  Using the short-distance matching \eqref{full-match}
for $\pr{R}{F_{a_1 a_2}(x_i,\tvec{z}_i,\tvec{y};\mu_{0i},\zeta_0)}$ in
\eqref{small-z-start}, we obtain our master formula for DTMDs at small
$\tvec{z}_i$ and large $\tvec{y}$:
\begin{align}
	\label{small-z-evolved}
 & \pr{R}{F_{a_1 a_2}(x_i,\tvec{z}_i,\tvec{y};\mu_i,\zeta)}
\nonumber \\[0.2em]
 &\quad = \exp\, \biggl\{ \int_{\mu_{01}}^{\mu_1} \frac{d\mu}{\mu}\, \biggl[
\gamma_{a_1}(\mu) - \gamma_{K,a_1}(\mu)
\log\frac{\sqrt{x_1\zeta/x_2}}{\mu} \biggr] +
\pr{R}{K}_{a_1}(\tvec{z}_1;\mu_{01}) \log\frac{\sqrt{x_1\zeta/x_2}}{\mu_{01}}
\nonumber \\
 & \qquad\quad\;\; + \int_{\mu_{02}}^{\mu_2} \frac{d\mu}{\mu}\,
\biggl[ \gamma_{a_2}(\mu) - \gamma_{K,a_2}(\mu)
\log\frac{\sqrt{x_2\ms\zeta/x_1}}{\mu} \biggr] +
\pr{R}{K}_{a_2}(\tvec{z}_2;\mu_{02}) \log\frac{\sqrt{x_2\ms\zeta/x_1}}{\mu_{02}}
\nonumber \\
 & \qquad\quad\;\; + \prb{R}{J}(\tvec{y};\mu_{0i})
\log\frac{\sqrt{\zeta}}{\sqrt{\zeta_0}} \biggr\}
\nonumber \\
 & \quad\; \times \sum_{b_1, b_2}
   \prn{R}{C}_{a_1 b_1}(x_1',\tvec{z}_1^{};\mu_{01}^{},\mu_{01}^2)
\underset{x_1}{\otimes} \prn{R}{C}_{a_2
b_2}(x_2',\tvec{z}_2^{};\mu_{02}^{},\mu_{02}^2) \underset{x_2}{\otimes}
\pr{R}{F_{b_1 b_2}(x_i',\tvec{y};\mu_{0i}^{},\zeta_0^{})} .
\end{align}
In the colour singlet channel, the result \eqref{small-z-evolved} is a
simple copy of its analogue \eqref{TMD-evo-solved} for a single-parton
TMD, with separate short-distance coefficients $C_{ab}$ and Sudakov
exponentials for each parton.  For colour non-singlet channels, the
Collins-Soper kernels $\pr{R}{K}_{a}(\tvec{z})$ acquire a colour dependence,
and there is an additional term $\prb{R}{J}(\tvec{y})$ in the exponent.  The
latter is also present in collinear DPS factorisation: in fact one simply
has
\begin{align}
	\label{eq:colzeta}
\exp\,\biggl[ \prb{R}{J}(\tvec{y};\mu_{0i})\,
\log\frac{\sqrt{\zeta}}{\sqrt{\zeta_0}} \,\biggr]\,
   \pr{R}{F_{b_1 b_2}(x_i',\tvec{y};\mu_{0i},\zeta_0)}
 &= \pr{R}{F_{b_1 b_2}(x_i',\tvec{y};\mu_{0i},\zeta)}
\end{align}
according to \eqref{CS-coll}.  In this sense, the value of $\zeta_0$ in
\eqref{small-z-evolved} is irrelevant (contrary to the choice of $\mu_{0i}$,
which appears in quantities that are computed using fixed-order perturbation
theory).  The choice of $\zeta_0$ does matter when this is the scale at
which one formulates an ansatz or model for the DPDFs (which is of course
inevitable for concrete calculations).

After inserting \eqref{small-z-evolved} and its counterpart for the left
moving proton into the definition~\eqref{W-def} of $W$, we can combine the
logarithms of $\zeta$ and $\bar{\zeta}$ in the exponentials, as we did in
the generic expression \eqref{W-generic}.  We thus obtain
\begin{align}
	\label{W-large-y}
& W_{\text{large $y$}} = \sum_{R} \csgn{a_1 a_2}{R}
\nonumber \\[0.2em]
 & \qquad \times
   \exp\, \biggl\{ \int_{\mu_{01}}^{\mu_1} \frac{d\mu}{\mu}\,
\biggl[ \gamma_{a_1}(\mu) - \gamma_{K,a_1}(\mu)\,
\log\frac{{Q_1^2}}{\mu^2} \biggr] + \pr{R}{K}_{a_1}(\tvec{z}_1;\mu_{01})
\log\frac{{Q_1^2}}{\mu_{01}^2}
\nonumber \\[0.2em]
 & \hspace{4.45em} +
\int_{\mu_{02}}^{\mu_2} \frac{d\mu}{\mu}\,
\biggl[ \gamma_{a_2}(\mu) - \gamma_{K,a_2}(\mu)\,
\log\frac{{Q_2^2}}{\mu^2} \biggr] +
\pr{R}{K}_{a_2}(\tvec{z}_2;\mu_{02})\, \log\frac{{Q_2^2}}{\mu_{02}^2} \biggr\}
\nonumber \\[0.5em]
 & \qquad \times \sum_{c_1, c_2, d_1, d_2}
   \prn{R}{C}_{b_1 d_1}(\bar{x}_1',\tvec{z}_1^{};\mu_{01}^{},\mu_{01}^2)
   \underset{\bar{x}_1}{\otimes}
   \prn{R}{C}_{b_2 d_2}(\bar{x}_2',\tvec{z}_2^{};\mu_{02}^{},\mu_{02}^2) \,
\nonumber \\[0.2em]
 & \hspace{5.1em} \underset{\bar{x}_2}{\otimes}
   \prn{R}{C}_{a_1 c_1}(x_1',\tvec{z}_1;\mu_{01}^{},\mu_{01}^2)
   \underset{x_1}{\otimes}
   \prn{R}{C}_{a_2 c_2}(x_2',\tvec{z}_2;\mu_{02}^{},\mu_{02}^2)
\nonumber \\[0.2em]
 & \qquad\quad \underset{x_2}{\otimes}
   \bigl[ \Phi(\nu \tvec{y}) \bigr]^2\, \exp\,\biggl[
   \prb{R}{J}(\tvec{y};\mu_{0i})\ms \log\frac{Q_1 \ms Q_2}{\zeta_0}
   \,\biggr]\, \pr{R}{F_{d_1 d_2}(\bar{x}_i',\tvec{y};\mu_{0i}^{},\zeta_0^{})}\,
   \pr{R}{F_{c_1 c_2}(x_i',\tvec{y};\mu_{0i}^{},\zeta_0^{})}\,,
\end{align}
which is the main result of this section.  The product of regulator
functions $\Phi(\nu \tvec{y}_+)\ms \Phi(\nu \tvec{y}_-)$ has been
approximated as appropriate for $|\tvec{z}_1|, |\tvec{z}_2| \ll |\tvec{y}|$.
Note that the dependence on $\tvec{y}$, $\tvec{z}_1$ and $\tvec{z}_2$ is
completely factorised in \eqref{W-large-y}.  The $\tvec{y}$ integral can
hence be performed separately.  For unpolarised or longitudinally polarised
partons, the short-distance coefficients $C$ are independent of the
direction of $\tvec{z}_1, \tvec{z}_2$ by rotation invariance. The angular
part of the $\tvec{z}_i$ integrations can then readily be performed and
turns the Fourier exponentials $e^{-i \tvec{q}{}_i\tvec{z}_i}$ into Bessel
functions.  The situation for transverse quark or linear gluon polarisation,
where $C$ (as well as $F$ and $\hat{\sigma}$) carries transverse indices,
can be discussed along similar lines.

Each of the two partonic cross sections $\hat{\sigma}_{i}$ in
\eqref{TMD-Xsect} and the four short-distance kernels $\prn{R}{C}$ in
\eqref{W-large-y} has an $\alpha_s$ expansion starting at order
$\alpha_s^0$.  It is natural (although not mandatory) to truncate their
product to the highest order in $\alpha_s$ at which the individual factors
are computed.

To evaluate the kernels $\pr{R}{K}$ and $\prn{R}{C}$ in \eqref{W-large-y} in
fixed-order perturbation theory, one should choose the scales $\mu_{0i}$
such that no large logarithms appear at higher orders.  A standard choice
for single hard scattering is to take $\mu_{0i}^2 = b_0^2
\ms/\tvec{z}_i^2$, which makes the one-loop expression of
$\pr{R}{K}(\tvec{z}_i; \mu_{0i})$ vanish.  Here
\begin{align}
  \label{b0-def}
b_0 &= 2 e^{-\gamma_E}
\end{align}
with $\gamma_E$ being the Euler constant.  While the cross section
\eqref{TMD-Xsect} is dominated by $|\tvec{z}_i| \sim 1/ |\tvec{q}_i|$ due
to the Fourier exponentials $e^{-i \tvec{q}{}_i\tvec{z}_i}$, one still must
integrate over the full range of these distances.  To avoid evaluating the
DPDs at unreasonably small factorisation scales $\mu_{0i}$, one may modify
the above scale choice and take instead $\mu_{0i} = \mu_{z_i}$, where for
any transverse distance vector $\tvec{b}$ we define
\begin{align}
	\label{z-star-def}
\mu_{b}^2 &= \frac{b_0^2}{\tvec{b}^{*2}} \,,
&
\text{with ~$\tvec{b}^*(\tvec{b}) \xrightarrow[\tvec{b}\to \tvec{0}]{}
\tvec{b}$~ and ~$|\ms\tvec{b}^*(\tvec{b})| \xrightarrow[|\tvec{b}|\to \infty]{}
b_{\text{max}}$} \,,
\end{align}
\rev{where $b_{\text{max}}$ is chosen such that $\tvec{b}^*$ remains in the
region where one trusts perturbative theory even when $\tvec{b}$ becomes
large.  A possible choice for $\tvec{b}^*$ is}
\begin{align*}
\tvec{b}^* &= \frac{\tvec{b}}{\sqrt{1 + \tvec{b}^2/b_{\text{max}}^2}} \,,
\end{align*}
\rev{which was proposed long ago \cite{Collins:1981va,Collins:1984kg} and is
extensively used in TMD phenomenology, but other functional forms have also
been explored \cite{Bacchetta:2015ora}.  We note that it is not mandatory to
take factorisation scales proportional to inverse distances, referring to
\cite{Becher:2011xn,Becher:2012yn} and
\cite{DAlesio:2014mrz,Echevarria:2015uaa} for examples of scale setting in
momentum space and to section 7.1.2 of \cite{Ebert:2016gcn} for a brief
synopsis.  Most results in our paper do not depend on choosing scales as in
\eqref{z-star-def}.  An exception is the discussion of scale setting in
section~\ref{sec:combine-y} and appendix~\ref{app:diff_scales}, which needs to
be adapted if a different choice is made.}

In the colour nonsinglet sector, the DPDs $\pr{R}{F}$ in \eqref{W-large-y}
still involve the different scales $\mu_{0i}$ and $\zeta_0$.  To disentangle
this dependence further, one may use \eqref{coll-zeta-expl} and
\eqref{eq:colzeta} to rewrite $\pr{R}{F}$ in terms of the distribution
$\pr{R}{\widehat{F}}$, whose DGLAP evolution does not involve a separate
rapidity scale, and which is therefore better suited to make the separation
of scales explicit.  At the cross section level, one can then replace the
last line of \eqref{W-large-y} by
\begin{align}
  \label{eq:coll_rap_sep}
 &\bigl[ \Phi(\nu \tvec{y}) \bigr]^2\,
\exp\,\biggl[ - \int_{\mu_0}^{\mu_{01}} \frac{d\mu}{\mu}\;
\prn{R}{\gamma}_{J}(\mu)\, \log\frac{{Q_1^2}}{\mu^2}
 - \int_{\mu_0}^{\mu_{02}}
\frac{d\mu}{\mu}\; \prn{R}{\gamma}_{J}(\mu)\, \log\frac{{Q_2^2}}{\mu^2}
\nonumber \\
 & \qquad + \prb{R}{J}(\tvec{y};\mu_{0},\mu_0)\ms \log\frac{Q_1 \ms
Q_2}{\zeta_0} \,\biggr]\, \pr{R}{\widehat{F}}_{d_1
d_2,\,\mu_0,\zeta_0}(\bar{x}_i',\tvec{y};\mu_{0i}^{})\,
 \pr{R}{\widehat{F}}_{c_1
c_2,\,\mu_0,\zeta_0}(x_i',\tvec{y};\mu_{0i}^{}) \,,
\end{align}
where we have explicitly indicated that the two renormalisation scales in
$\prb{R}{J}$ are taken equal.  It is natural to take the starting scale of
evolution as $\mu_0 = \sqrt{\zeta_0}$.  A particular choice is $\mu_0 =
\mu_y$, which for large $\tvec{y}$ saturates at a hadronic scale $\mu_0 =
b_0/b_{\text{max}}$.  This choice ensures that the initial conditions for
DPD evolution do not involve widely different scales as $\tvec{y}$ becomes
small.

The factor $\prb{R}{J}(\tvec{y};\mu_{0i})$ in \eqref{W-large-y} is a
nonperturbative function in the large-$\tvec{y}$ region, but we can ensure
that it has the correct perturbative small-$\tvec{y}$ behaviour by copying
\rev{part of the so-called $b^*$ trick} formulated for ordinary TMDs in
\cite{Collins:1981va,Collins:1984kg}.  We thus write
\begin{align}
	\label{eq:ystar_trick}
\prb{R}{J}(\tvec{y};\mu_{01},\mu_{02}) &=
-\, \pr{R}{g}_J(\tvec{y}) + \prb{R}{J}(\tvec{y}^*\bs ;\mu_y,\mu_y) -
\int_{\mu_y}^{\mu_{01}} \frac{d\mu}{\mu}\; \prn{R}{\gamma_J}(\mu) -
\int_{\mu_y}^{\mu_{02}} \frac{d\mu}{\mu}\; \prn{R}{\gamma_J}(\mu) \,,
\end{align}
where the nonperturbative information is contained in
\begin{align}
  \pr{R}{g}_J(\tvec{y}) &=
    \prb{R}{J}(\tvec{y}^*\bs ;\mu_{i}) - \prb{R}{J}(\tvec{y};\mu_{i}) \,.
\end{align}
Because of \eqref{CS-coll-RG}, the dependence on the renormalisation scales
drops out in $\pr{R}{g}_J$, which satisfies $\pr{R}{g}_J(\tvec{0}) = 0$ by
construction.  The term $\prb{R}{J}(\tvec{y}^*\! ;\mu_y,\mu_y) $ in
\eqref{eq:ystar_trick} can be evaluated in fixed-order perturbation theory;
according to the results in section~\ref{sec:CS_kernels} it is zero up to
terms of $\mathcal{O}(\alpha_s^2)$.  Of course, the form
\eqref{eq:ystar_trick} with $\mu_{0i}$ replaced by $\mu_0$ can be used
in~\eqref{eq:coll_rap_sep}.

In the colour octet sector, the relation \eqref{soft-octet} implies that
\begin{align}
  \label{J-octet}
\pr{8}{J}(\tvec{y};\mu,\mu) &= K_g(\tvec{y};\mu) \,,
\end{align}
where $K_g$ is the Collins Soper kernel for single gluon TMDs.  While little
is known about this kernel for nonperturbative distances $\tvec{y}$, one may
construct a model by connecting it with its counterpart $K_q$ in the
quark-antiquark channel, for which there is a considerable body of
phenomenology from single Drell-Yan production and from semi-inclusive DIS,
see e.g.~\cite{Aidala:2014hva,Collins:2014jpa} and references therein.  A
simple possibility would be to assume Casimir scaling, $K_{g} /C_A = K_{q}
/C_F$. This scaling holds perturbatively up to $\mathcal{O}(\alpha_s^3)$, as
can for instance be seen in appendix~D of
\cite{Echevarria:2016scs}.\footnote{Just recently it has been found
  \protect\cite{Moch:2017uml} that Casimir scaling for the cusp anomalous
  dimension, $\gamma_{K,g} /C_A = \gamma_{K,q} /C_F$, is broken at
  $\mathcal{O}(\alpha_s^4)$.  Equation \eqref{RG-single-K} implies the
  breaking of Casimir scaling for $K_a$ at the same order.  For a related
  calculation see \protect\cite{Grozin:2017css}.}

\subsubsection{Extrapolation to large \texorpdfstring{$\tvec{z}_1$}{z1} and
  \texorpdfstring{$\tvec{z}_2$}{z2}}
\label{sec:large-z}

If $\tvec{q}_1$ and $\tvec{q}_2$ are not sufficiently large to ensure
dominance of small $\tvec{z}_i$ in the TMD cross section, then a more
realistic description of the integrand for large $\tvec{z}_i$ is necessary.
\rev{Of course, a corresponding statement holds already for single hard
scattering.  A widely used procedure in that case is the $b^*$ trick of
\cite{Collins:1981va,Collins:1984kg}.  Let us briefly show how it can be
adapted to DPS.}  We recall the solution \eqref{CS-TMD-sol} of the rapidity
evolution equation,
\begin{align*}
\pr{R}{F}_{a_1 a_2}(x_i,\tvec{z}_i,\tvec{y};\mu_i,\zeta) &=
\sum_{R'} \prb{RR'}{\exp}\biggl[ K_{a_1 a_2}(\tvec{z}_i,\tvec{y};\mu_i) \log
\frac{\sqrt{\zeta}}{\sqrt{\zeta_0}} \,\biggr] \prb{R'}{F}_{a_1
a_2}(x_i,\tvec{z}_i,\tvec{y};\mu_i,\zeta_0)
\end{align*}
and introduce
\begin{align}
\label{g-function-defs}
\pr{R}{g}^{}_{F, a_1 a_2}(x_i,\tvec{z}_i,\tvec{y};\zeta) &=
\log\, \frac{\pr{R}{F}_{a_1 a_2}(x_i,\tvec{z}^*_i,\tvec{y};
  \mu_i,\zeta)}{\pr{R}{F}_{a_1 a_2}(x_i,\tvec{z}_i,\tvec{y};
  \mu_i,\zeta)} \,,
\nonumber \\[0.4em]
\prb{RR'}{g}_{K, a_1 a_2}^{}(\tvec{z}_i, \tvec{y}) &=
  \prb{RR'}{K}_{a_1 a_2}(\tvec{z}^*_i,\tvec{y};\mu_i)
- \prb{RR'}{K}_{a_1 a_2}(\tvec{z}_i,\tvec{y};\mu_i)
\end{align}
with $\tvec{z}^*$ defined by \eqref{z-star-def}.  Both $g_{F}$ and $g_K$ are
independent of $\mu_i$ according to \eqref{RG-TMD-again} and
\eqref{CS-TMD-RG}, and both functions depend on the parameter
$b_{\text{max}}$ via $\tvec{z}^*_i$. By construction, both functions vanish
at the point $\tvec{z}_1 = \tvec{z}_2 = \tvec{0}$.  We can then write
\begin{align}
	\label{b-star-DPDs} \pr{R}{F}_{a_1
a_2}(x_i,\tvec{z}_i,\tvec{y};\mu_i,\zeta) &= \sum_{R'} \prb{RR'}{\exp}\biggl[
{}- g_{K, a_1 a_2}(\tvec{z}_i,\tvec{y})
  \log \frac{\sqrt{\zeta}}{\sqrt{\zeta_0}}\,
+ K_{a_1 a_2}(\tvec{z}_i^*,\tvec{y};\mu_i) \log
\frac{\sqrt{\zeta}}{\sqrt{\zeta_0}} \,\biggr]
\nonumber \\[0.2em]
 &\quad \times \exp\bigl[ {}- \prb{R'}{g}^{}_{F, a_1
a_2}(x_i,\tvec{z}_i,\tvec{y};\zeta_0) \bigr]\, \prb{R'}{F}_{a_1
a_2}(x_i,\tvec{z}^*_i,\tvec{y};\mu_i,\zeta_0)\,.
\end{align}
In this expression, $\prb{RR'}{K}_{a_1 a_2}$ and $\prb{R'}{F}_{a_1 a_2}$ can
be evaluated using the short-distance expansion discussed in the previous
subsection, provided that $|\tvec{z}_1^*|, |\tvec{z}_2^*| \ll |\tvec{y}|$.  As
a consequence, $\prb{RR'}{K}_{a_1 a_2}$ is diagonal in $R$ and $R'$. By
contrast, the functions $\prb{RR'}{g}_{K, a_1 a_2}$ and
$\pr{R}{g}_{F, a_1 a_2}$ are entirely nonperturbative and need to be modelled.
\rev{This is a daunting task, because they depend on several transverse
variables, there are many functions in the different colour channels and there
is less guidance from data.  Using the $b^*$ trick for computing DPS processes
with $\tvec{q}_1$ and $\tvec{q}_2$ in the nonperturbative region would
therefore require additional theory input, or strong simplifying assumptions.}

\rev{The $b^*$ trick is not the only way to handle the region of large
transverse distances, and for TMD factorisation in SPS a variety of other
methods have been employed
\cite{Laenen:2000de,Kulesza:2002rh,Bozzi:2005wk,Qiu:2000hf,
Becher:2011xn,Becher:2012yn,%
DAlesio:2014mrz,Echevarria:2015uaa,Scimemi:2016ffw,Scimemi:2017etj}.  It would
be interesting to study if and how they could be adapted to DPS.  This is
however beyond the scope of the present work, where we focus on transverse
momenta in the perturbative region.}

\subsection{The small-\texorpdfstring{$\tvec{y}$}{y} region}
\label{sec:small-y}

We now consider the region where $|\tvec{y}|$ is of the same order as
$|\tvec{z}_1|$ and $|\tvec{z}_2|$, with all distances being small compared
with $1/\Lambda$.  Again, we will derive double parton analogues of the
relations \eqref{single-TMD-match} and \eqref{TMD-evo-solved} between TMDs and
PDFs.  \rev{They are given in \eqref{split-TMD-coll}, \eqref{intr-TMD-tw4} and
\eqref{small-yz-evolved} and involve collinear PDFs and twist-four
distributions as nonperturbative input, which also appear in the expansion
\eqref{coll-DPD-exp} of DPDFs at small $\tvec{y}$.  At the cross section
level, we then obtain the expression \eqref{W-small-y} in which all large
logarithms are explicitly resummed.  An important finding is that the
different combinations of collinear PDFs and twist-four distributions have
different power behaviour in the cross section, specified in
\eqref{power-small-y} and compared with the contribution from large $\tvec{y}$
in \eqref{power-large-y}.}

\subsubsection{Short-distance expansion}
\label{sec:short-dist-exp}

\begin{figure}
\begin{center}
\includegraphics[width=0.64\textwidth]{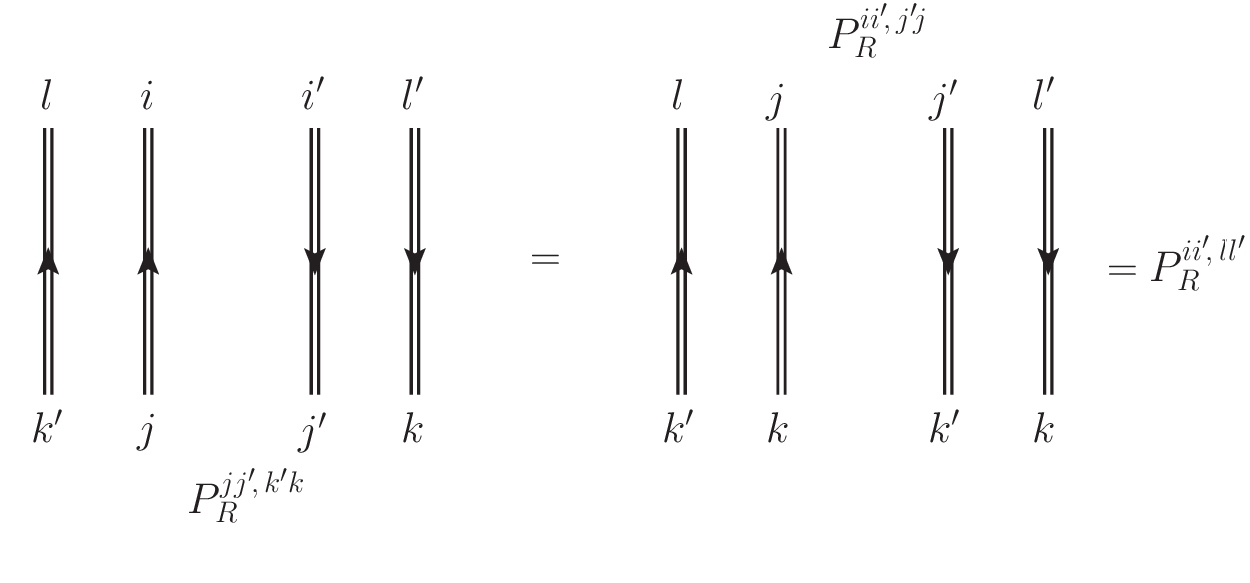}
\caption{\label{fig:soft-match-2} Simplification in the extended soft factor
  $S_{qq}$ with all Wilson lines at the same transverse position, using the
  identity in figure~\protect\ref{fig:projector-comm} and the unitarity of
  Wilson lines.  Analogous relations hold for indices in the adjoint colour
  representation.}
\end{center}
\end{figure}

Let us first establish that the soft factor, and hence the Collins-Soper
kernel $\prb{RR'}{K}_{a_1 a_2}$, can be fully computed in perturbation theory
when all transverse distances become small.  Repeating the arguments that lead
to \eqref{soft-op-start}, we obtain
\begin{align}
  \label{soft-match-colour-2}
  (S_{a_1 a_2})^{\ul{r}_1 \ul{r}_2, \ul{u}_1
\ul{u}_2}(\tvec{z}_i, \tvec{y}) &= C_{S, a_1 a_2 \phantom{|}}^{\ul{s}_1
\ul{s}_2, \ul{t}_1 \ul{t}_2}(\tvec{z}_i, \tvec{y})\, (S_{a_1 a_2})^{\ul{r}_1
\ul{r}_2, \ul{u}_1 \ul{u}_2}_{ \ul{s}_1 \ul{s}_2, \ul{t}_1
\ul{t}_2}(\tvec{0})\,,
\end{align}
which involves the extended soft factor \eqref{soft-gen-def} with all distance
arguments set to zero and a short-distance coefficient $C_{S, a_1 a_2}$
describing interactions between all Wilson lines.  Inserting colour projectors
as in \eqref{eq:contr_proj}, we get
\begin{align}
 (S_{a_1 a_2})^{\ul{r}_1 \ul{r}_2, \ul{u}_1 \ul{u}_2}(\tvec{z}_i,
\tvec{y}) &= \sum_{RR'} P_{R}^{\ul{v}_1 \ul{v}_2}\, P_{R'}^{\ul{w}_1
\ul{w}_2}\, C_{S, a_1 a_2 \phantom{|}}^{\ul{v}_1 \ul{v}_2, \ul{w}_1
\ul{w}_2}(\tvec{z}_i, \tvec{y})\, \frac{P_{R}^{\ul{s}_1 \ul{s}_2}\,
P_{R'}^{\ul{t}_1 \ul{t}_2}\, (S_{a_1 a_2})^{\ul{r}_1 \ul{r}_2, \ul{u}_1
\ul{u}_2}_{ \ul{s}_1 \ul{s}_2, \ul{t}_1 \ul{t}_2}(\tvec{0})}{ m(R)\, m(R')}
\nonumber \\ &= \sum_{RR'} P_{R}^{\ul{v}_1 \ul{v}_2}\, P_{R'}^{\ul{w}_1
\ul{w}_2}\, C_{S, a_1 a_2 \phantom{|}}^{\ul{v}_1 \ul{v}_2, \ul{w}_1
\ul{w}_2}(\tvec{z}_i, \tvec{y})\, \frac{P_{R}^{\ul{r}_1 \ul{s}_1}\,
P_{R'}^{\ul{u}_1 \ul{t}_1}\, (S_{a_1 a_2})^{\ul{s}_1 \ul{r}_2, \ul{t}_1
\ul{u}_2}_{ \ul{s}_2 \ul{s}_2, \ul{t}_2 \ul{t}_2}(\tvec{0})}{ m(R)\, m(R')}\,,
\end{align}
where in the second step the identity \eqref{comm-soft} has been used for the
indices associated with parton $a_1$ in $S_{a_1 a_2}$. The sum over $\ul{s}_2$
on the r.h.s.\ now ties together pairs of conjugate Wilson lines along the
same direction and at the same transverse position, which results in unit
matrices according to $W(\tvec{\xi}, v) W^\dagger(\tvec{\xi}, v) =
\one$. For $S_{qq}$ this
is represented pictorially in figure~\ref{fig:soft-match-2}. The same holds
for the sum over $\ul{t}_2$. The extended soft factor thus collapses to a
product of unit matrices, and we have
\begin{align}
 (S_{a_1 a_2})^{\ul{r}_1 \ul{r}_2, \ul{u}_1 \ul{u}_2}(\tvec{z}_i,
\tvec{y}) &= \sum_{RR'} P_{R}^{\ul{v}_1 \ul{v}_2}\, P_{R'}^{\ul{w}_1
\ul{w}_2}\, C_{S, a_1 a_2 \phantom{|}}^{\ul{v}_1 \ul{v}_2, \ul{w}_1
\ul{w}_2}(\tvec{z}_i, \tvec{y}) \frac{P_{R}^{\ul{r}_1 \ul{r}_2}\,
P_{R'}^{\ul{u}_1 \ul{u}_2}}{ m(R)\, m(R')} \,.
\end{align}
Projecting this on definite representations for $(\ul{r}_1 \ul{r}_2)$ and
for $(\ul{u}_1 \ul{u}_2)$, we simply obtain
\begin{align}
  \label{soft-fact-small-y}
  \prb{RR'}{S_{a_1 a_2}}(\tvec{z}_i, \tvec{y}; \mu_i,Y)
&= \prb{RR'}{C_{S, a_1 a_2}}(\tvec{z}_i, \tvec{y}; \mu_i,Y)\,,
\end{align}
where the projection $\pr{RR'}{C}$ is defined in analogy to \eqref{soft-proj}
and where we have restored all arguments.  In analogy to the case
\eqref{soft-singlet} for SPS, we thus find that when all transverse distances
become small, the soft factor for DPS is equal to its matching coefficient and
thus can be entirely computed in perturbation theory. The same then holds for
the Collins-Soper kernel $\prb{RR'}{K}$ and for the matrix $\prb{RR'}{s}$,
which can be constructed from $\prb{RR'}{S}$ as shown in
appendix~\ref{app:matrix-algebra}.  An explicit check of this result is the
fact that the one-loop expression for $\prb{RR'}{K}$ in
section~\ref{sec:CS_kernels} is free of infrared divergences, as is the
two-loop soft factor for DPS computed in \cite{Vladimirov:2016qkd}.

We now turn to the short-distance limit of the collinear matrix elements,
which can be discussed in close analogy to the small-$\tvec{y}$ limit of
collinear DPDs in section~3.3 of \cite{Diehl:2017kgu}.  In the limit of
small $\tvec{z}_i$ and $\tvec{y}$, the unsubtracted distributions
$\prb{R'}{F}_{\us}$ are given by short-distance coefficients times
proton matrix elements of operators with all fields at transverse
position~$\tvec{0}$.  The latter are given by collinear parton distributions
of different twist.  Multiplying with the perturbative expression of
$\prb{RR'}{s}$, one obtains the expansion of $\pr{R}{F}$.  Adapting the
notation in \cite{Diehl:2017kgu}, we write\footnote{The distributions on
  the r.h.s.\ of \protect\eqref{DPD-split-intr} are respectively denoted
  by $F_{\text{spl,pt}}$, $F_{\text{int,pt}}$ and $F_{\text{tw3,pt}}$ in
  \protect\cite{Diehl:2017kgu}.  For brevity we omit the specification
  ``pt'' here.}
\begin{align}
  \label{DPD-split-intr}
  \pr{R}{F} &= \pr{R}{F}_{\text{spl}} + \pr{R}{F}_{\text{tw3}} +
      \pr{R}{F}_{\text{int}} \,,
\end{align}
where the short-distance expansion of the terms on the r.h.s.\ involves
proton matrix elements of operators with twist two, twist three and twist
four, respectively.  Following the discussion in section~2.1 of
\cite{Diehl:2017kgu}, we neglect the contribution with collinear
twist-three distributions, which for unpolarised protons are chiral odd.
They have no dynamic cross talk with gluon distributions and are therefore
expected to be small compared with twist-two and twist-four distributions
at small $x$.

The splitting contribution $F_{\text{spl}}$ describes the case where a
single parton splits into partons $a_1$ and $a_2$, as shown in
figure~\ref{fig:split}a at lowest order in $\alpha_s$.  It can be obtained
from~\eqref{split-TMD} by taking the limit of small $\tvec{Z}$.  For the
$\mathcal{O}(\alpha_s)$ term, this simply corresponds to replacing
the TMD $f_{a_0}(x_1+x_2, \tvec{Z})$ by the PDF $f_{a_0}(x_1+x_2)$.  Since
the proton is unpolarised, the parton $a_0$ in the PDF is unpolarised as
well.  We thus obtain
\begin{align}
  \label{split-TMD-coll}
\pr{R}{F}_{a_1 a_2,\, \text{spl}}(x_i,\tvec{z}_i,\tvec{y}; \mu,\mu,\zeta)
&= \frac{\tvec{y}_{+}^{l}\ms \tvec{y}_{-}^{l'}}{ \tvec{y}_{+}^{2}\ms
  \tvec{y}_{-}^{2}}\, \frac{\alpha_s(\mu)}{2\pi^2}\;
\prn{R}{T}_{a_0\to a_1 a_2}^{ll'}
\biggl( \frac{x_1}{x_1+x_2} \biggr)\,
\frac{f_{a_0}(x_1+x_2; \mu)}{x_1+x_2} + \mathcal{O}(\alpha_s^2) \,,
\end{align}
where we have restored the dependence on the UV renormalisation scale,
taken equal for the partons with momentum fraction $x_1$ and $x_2$.  A
$\zeta$ dependence appears only at order $\alpha_s^2$. To avoid
large logarithms of the higher-order terms in the region of small
$|\tvec{z}_1| \sim |\tvec{z}_2| \sim |\tvec{y}|$, any choice where $1/\mu$
and $1/\sqrt{\zeta}$ are of the order of these distances will do.  A
particular choice is given in \eqref{small-y-scales} below.

The term $F_{\text{int}}$ in \eqref{DPD-split-intr} is referred to as the
``intrinsic'' contribution to the DPD and may be thought of as describing
parton pairs $a_1$, $a_2$ in the ``intrinsic'' proton wave function.  It
is the only contribution starting at order $\alpha_s^0$ and reads
\begin{align}
  \label{intr-TMD-tw4}
  \pr{R}{F}_{a_1 a_2,\, \text{int}}(x_i,\tvec{z}_i,\tvec{y};
  \mu,\mu,\zeta) &= \prn{R}{G}_{a_1 a_2}(x_1,x_2,x_2,x_1;\mu)
  + \mathcal{O}(\alpha_s) \,,
\end{align}
where $\prn{R}{G}$ denotes a collinear twist-four distribution.  The
lowest-order term is simply obtained by setting
$\tvec{z}_i=\tvec{y} = \tvec{0}$ in the matrix element \eqref{eq:dpds}
(this must be done before renormalisation and hence in $D=4-2\epsilon$
dimensions).  The $\mathcal{O}(\alpha_s)$ term involves short-distance
interactions and is briefly discussed in section~3.3 of
\cite{Diehl:2017kgu}.

Inserting \eqref{DPD-split-intr} with \eqref{split-TMD-coll} and
\eqref{intr-TMD-tw4} into the cross section formula \eqref{TMD-Xsect}, we
can derive the power behaviour of the contribution from the region
$|\tvec{z}_1| \sim |\tvec{z}_2| \sim |\tvec{y}| \sim 1/q_T$, using that
the twist-four distribution scales like $G \sim \Lambda^2$.  For the
scaled DPS cross section we obtain
\begin{align}
  \label{power-small-y} Q^4\,
\frac{d\sigma_{\text{DPS}}}{d^2\tvec{q}_1\, d^2\tvec{q}_2}
  \bigg|_{\text{small $y$}} & \sim
\begin{cases}
  \alpha_s^2 /q_T^2 &
  \text{from $F_{\text{spl}} \times F_{\text{spl}}$ ~(1v1)} \\
  \alpha_s \ms \Lambda^2/q_T^4 &
  \text{from $F_{\text{spl}} \times F_{\text{int}}$ ~(2v1)} \\
  \Lambda^4/q_T^6 &
  \text{from $F_{\text{int}} \times F_{\text{int}}$ ~(2v2)} \\
\end{cases}
\intertext{where we have also indicated the lowest order
in $\alpha_s$ for each contribution. For the contribution from large
$|\tvec{y}| \sim 1/\Lambda$ discussed in section~\ref{sec:small_zi}, we
have}
  \label{power-large-y}
Q^4\, \frac{d\sigma_{\text{DPS}}}{d^2\tvec{q}_1\, d^2\tvec{q}_2}
  \bigg|_{\text{large $y$}} & \sim \Lambda^2/q_T^4 \,,
\end{align}
as follows from the small-distance expansion \eqref{full-match} and the power
behaviour $F \sim \Lambda^2$ of a DPDF at large $\tvec{y}$.  We note that
\eqref{power-small-y} and \eqref{power-large-y} do not hold in the kinematic
region
$|\tvec{q}_1 + \tvec{q}_2| \sim \Lambda$, which we have excluded from our
considerations, and which has a different power counting as seen from table~1
in \cite{Diehl:2011yj}.  Notice also that some parton combinations, such as
$(a_1 a_2) = (d \bar{u})$, cannot be obtained by parton splitting at
$\mathcal{O}(\alpha_s)$.  In this case, $F_{\text{spl}}$ starts
at~$\mathcal{O}(\alpha_s^2)$ and the powers of $\alpha_s$ in
\eqref{power-small-y} must be adjusted accordingly.

\begin{figure}
\begin{center}
\subfigure[]{\includegraphics[width=0.32\textwidth]{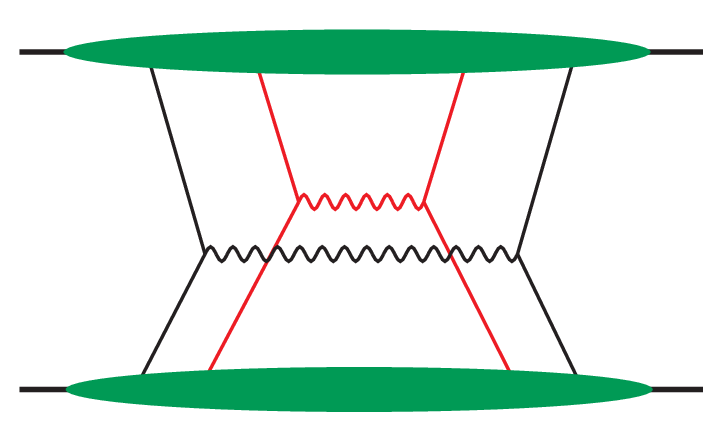}}
\hspace{3em}
\subfigure[]{\includegraphics[width=0.32\textwidth]{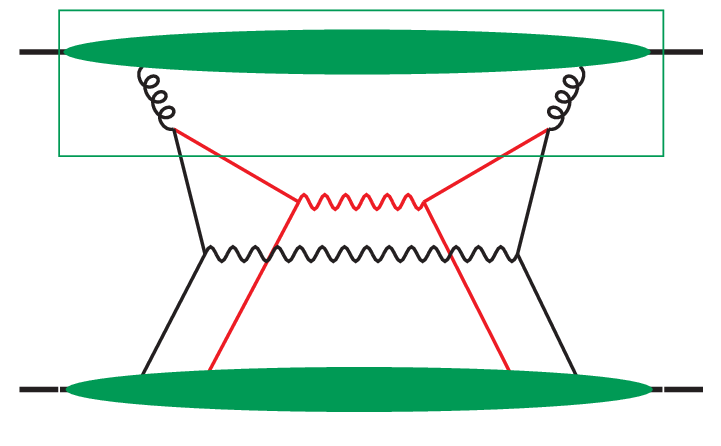}}
\caption{\label{fig:2vn} Counterparts to the 1v1 graph in
  figure~\protect\ref{fig:boxed}a.  In the 2v2 graph (a) there is no
  perturbative splitting, and in the 2v1 graph (b) one has a perturbative
  splitting in one of the two protons.  The box encloses the DTMD of the
  upper proton and indicates that the splittings take place at transverse
  distances of order $1/q_T$.}
\end{center}
\end{figure}

The three contributions in \eqref{power-small-y} are referred to as 1v1, 2v1
and 2v2 terms as indicated.  The 1v1 (read ``one versus one'') term is
associated with graphs as in figure~\ref{fig:boxed}a, where one parton in
each proton initiates a sequence of short-distance interactions (a splitting
at scale $q_T$ followed by hard scattering at scale $Q$).  The 2v2 term
corresponds to graphs where two ``intrinsic'' partons in each proton
directly enter the hard subprocesses, as shown in figure~\ref{fig:2vn}a.  In
the 2v1 term one starts with two ``intrinsic'' partons in one proton,
whereas in the other proton one starts with one parton that splits into two
at scale $q_T$.  This is shown in figure~\ref{fig:2vn}b.  It is natural to
associate the term in \eqref{power-large-y} with 2v2 as well, since at large
$\tvec{y}$ one cannot identify any perturbative splitting contribution.

Among the terms in \eqref{power-small-y} and \eqref{power-large-y}, only the
term with $F_{\text{spl}} \times F_{\text{spl}}$ is leading as far as powers
of $\Lambda^2/q_T^2$ are concerned.  However, the other terms involve fewer
powers in $\alpha_s$, and they rise more strongly as the momentum fractions
$x$ become small, because one roughly expects the intrinsic part of the DPDF
to grow like the square of a single PDF in that limit.  From this perspective,
one may still discard the term $F_{\text{int}} \times F_{\text{int}}$, which
is power suppressed compared with the large-$\tvec{y}$ contribution but has
the same power in $\alpha_s$ and a similar small $x$ behaviour.  However, the
inclusion of this term in the cross section will in general not require much
additional effort, and it renders the combination of the contributions from
small and large $\tvec{y}$ more straightforward (see
section~\ref{sec:combine-y}).

The expansion of the DPDFs $F(x_i,\tvec{y})$ for small $\tvec{y}$ proceeds
in exactly the same manner.  It can be obtained from
\eqref{split-TMD-coll} and \eqref{intr-TMD-tw4} by setting
$\tvec{z}_1 = \tvec{z}_2 = \tvec{0}$ before performing UV renormalisation
and setting $D=4$.  This gives
\begin{align}
  \label{coll-DPD-exp}
\pr{R}{F}_{a_1 a_2,\, \text{spl}}(x_i,\tvec{y};
\mu,\mu,\zeta) &=
\frac{\tvec{y}^{l}\ms \tvec{y}^{l'}}{\tvec{y}^{4}}\,
\frac{\alpha_s(\mu)}{2\pi^2}\; \prn{R}{T}_{a_0\to a_1 a_2}^{ll'}
\biggl( \frac{x_1}{x_1+x_2} \biggr)\, \frac{f_{a_0}(x_1+x_2; \mu)}{x_1+x_2} +
\mathcal{O}(\alpha_s^2) \,,
\nonumber \\[0.2em]
\pr{R}{F}_{a_1 a_2,\, \text{int}}(x_i,\tvec{y}; \mu,\mu,\zeta) &=
\prn{R}{G}_{a_1 a_2}(x_1,x_2,x_2,x_1;\mu) + \mathcal{O}(\alpha_s) \,,
\end{align}
where for $F_{\text{int}}$ the difference between \eqref{intr-TMD-tw4} and
\eqref{coll-DPD-exp} only appears at $\mathcal{O}(\alpha_s)$.  We note that
only the symmetric part of $T_{a_0\to a_1 a_2}^{ll'}$ survives the
contraction with $\tvec{y}^{l}\ms \tvec{y}^{l'}$.  As a consequence,
combinations of $a_1, a_2$ with exactly one longitudinal polarisation give
zero, since one then has $T^{ll'} \propto \epsilon^{ll'}$ (see
section~\ref{sec:kernels_for_splitting}).

Little is known in phenomenology about twist-four distributions.  The
computation using models of light-cone wave functions in \cite{Braun:2011aw}
can give guidance at large but not at small $x$. If one has a model for
${F}_{\text{int}}(x_i,\tvec{y})$ and is content with leading-order accuracy
in $\alpha_s$, one may combine \eqref{intr-TMD-tw4} and \eqref{coll-DPD-exp}
to approximate
\begin{align}
 \pr{R}{F}_{\text{int},\, a_1 a_2}(x_i,\tvec{z}_i,\tvec{y};
\mu,\mu,\zeta) &= \pr{R}{F}_{\text{int},\, a_1 a_2}(x_i,\tvec{y};
\mu,\mu,\zeta) + \mathcal{O}(\alpha_s)
\end{align}
and use the intrinsic part of the DPDF on the r.h.s.\ in the small-$\tvec{y}$
contribution to the cross section, without further approximation. The
scales $\mu$ and $\zeta$ should of course be chosen appropriately.

\subsubsection{Evolved DPDs and cross section}
\label{sec:short-dist-Xsect}

Inserting the decomposition \eqref{DPD-split-intr} of the DTMDs into
\eqref{DTMD-evolved} and neglecting twist-three distributions, we
obtain
\begin{align}
  \label{small-yz-evolved}
  & \pr{R}{F}_{a_1
    a_2}(x_i,\tvec{z}_i,\tvec{y};\mu_i,\zeta)
\nonumber \\
  & \quad = \exp\,\biggl\{ \int_{\mu_{0}}^{\mu_1} \frac{d\mu}{\mu}\, \biggl[
\gamma_{a_1}(\mu) - \gamma_{K,a_1}(\mu)\,
\log\frac{\sqrt{x_1\zeta/x_2}}{\mu} \biggr] +
\pr{1}{K}_{a_1}(\tvec{z}_1;\mu_{0})\, \log\frac{\sqrt{\zeta}}{\sqrt{\zeta_0}}
\nonumber \\
& \qquad \hspace{1.5em} + \int_{\mu_{0}}^{\mu_2}
\frac{d\mu}{\mu}\, \biggl[ \gamma_{a_2}(\mu) - \gamma_{K,a_2}(\mu)\,
\log\frac{\sqrt{x_2\ms\zeta/x_1}}{\mu} \biggr] +
\pr{1}{K}_{a_2}(\tvec{z}_2;\mu_{0})\, \log\frac{\sqrt{\zeta}}{\sqrt{\zeta_0}}
\,\biggr\}
\nonumber \\
& \qquad \times \sum_{R'} \prb{RR'}{\exp}\,\biggl[ M_{a_1
a_2}(\tvec{z}_i,\tvec{y})\ms \log\frac{\sqrt{\zeta}}{\sqrt{\zeta_0}} \,\biggr]\,
\nonumber\\
&\qquad\quad \times \Bigl[ \prb{R'}{F}_{a_1 a_2,\, \text{spl}}({x}_i,
\tvec{z}_i,\tvec{y}; \mu_0,\mu_0,\zeta_0)
+ \prb{R'}{F}_{a_1 a_2,\, \text{int}}({x}_i, \tvec{z}_i,\tvec{y};
\mu_0,\mu_0,\zeta_0) \Bigr]
\end{align}
as our master formula for DTMDs at small $\tvec{z}_i$ and small $\tvec{y}$.
For the scales in the DTMDs we take
\begin{align}
  \label{small-y-scales}
\mu_0^{} &= \mu_{Z}^{} \,,
&
\zeta_0^{} &= \mu_Z^2
\end{align}
with $\mu_Z$ defined as in \eqref{z-star-def}.  We can then use the
short-distance approximations \eqref{split-TMD-coll} and
\eqref{intr-TMD-tw4} for $F_{\text{spl}}$ and $F_{\text{int}}$.  The
choice \eqref{small-y-scales} will turn out to be well suited for
combining the different contributions to the overall cross section.
In contrast to \eqref{small-z-evolved}, we have taken the two
renormalisation scales $\mu_{01}$ and $\mu_{02}$ equal in
\eqref{small-yz-evolved}, given that we do not have an optimised
short-distance expansion of ${F}$ for different scales.  One can however
optimise the resummation of logarithms in the exponent by writing
\begin{align}
\pr{1}{K}_{a_1}(\tvec{z}_1;\mu_{0}) &=
\pr{1}{K}_{a_1}(\tvec{z}_1;\mu_{01}) + \int_{\mu_{0}}^{\mu_{01}}
\frac{d\mu}{\mu}\, \gamma_{K,a_1}(\mu)
\end{align}
and likewise for $\pr{1}{K}_{a_2}$.  A natural choice for the additional
scales is $\mu_{0i} = \mu_{z_i}$, as already suggested for
\eqref{small-z-evolved}.
Inserting the evolved DTMDs into \eqref{W-generic}, we have
\begin{align}
	\label{W-small-y}
W_{\text{small $y$}} &= \exp\, \biggl\{
\int_{\mu_{0}}^{\mu_1} \frac{d\mu}{\mu}\, \biggl[ \gamma_{a_1}(\mu) -
\gamma_{K,a_1}(\mu)\, \log\frac{{Q_1^2}}{\mu^2} \biggr] +
\pr{1}{K}_{a_1}(\tvec{z}_1,\mu_{0})\, \log\frac{Q_1\ms Q_2}{\zeta_{0}}
\nonumber \\ & \qquad\;\; + \int_{\mu_{0}}^{\mu_2}
\frac{d\mu}{\mu}\, \biggl[ \gamma_{a_2}(\mu) - \gamma_{K,a_2}(\mu)\,
\log\frac{{Q_2^2}}{\mu^2} \biggr] + \pr{1}{K}_{a_2}(\tvec{z}_2,\mu_{0})\,
\log\frac{Q_1 \ms Q_2}{\zeta_{0}} \biggr\}
\nonumber \\[0.3em]
 & \quad \times \sum_{RR'}
  \Bigl[ \pr{R}{F}_{b_1 b_2,\, \text{spl}}(\bar{x}_i,
\tvec{z}_i,\tvec{y}; \mu_0,\mu_0,\zeta_0)
+ \pr{R}{F}_{b_1 b_2,\, \text{int}}(\bar{x}_i, \tvec{z}_i,\tvec{y};
\mu_0,\mu_0,\zeta_0) \Bigr]
\nonumber \\
& \qquad \times \Phi(\nu\tvec{y}_+)\ms \Phi(\nu\tvec{y}_-) \,
  \csgn{a_1 a_2}{R}\;
\prb{RR'}{\exp}\,\biggl[ M_{a_1 a_2}(\tvec{z}_i,\tvec{y})
\log\frac{Q_1\ms Q_2}{\zeta_{0}} \,\biggr]\, \nonumber \\[0.3em]
& \qquad \times \Bigl[ \prb{R'}{F}_{a_1 a_2,\, \text{spl}}({x}_i,
\tvec{z}_i,\tvec{y}; \mu_0,\mu_0,\zeta_0)
+ \prb{R'}{F}_{a_1 a_2,\, \text{int}}({x}_i,
  \tvec{z}_i,\tvec{y}; \mu_0,\mu_0,\zeta_0) \Bigr]
\end{align}
at the cross section level.

At present, one can only use the leading order (LO) expressions
\eqref{split-TMD-coll} and \eqref{intr-TMD-tw4} of the DTMD expansions,
because the next-to-leading order (NLO) terms have not been computed for
either case.  The leading $\mathcal{O}(\alpha_s)$ term of $F_{\text{spl}}$
depends on $\tvec{y}_\pm$ but not on $\tvec{Z}$, and the leading
$\mathcal{O}(\alpha_s^0)$ expression for $F_{\text{int}}$ is independent
of any transverse distance.  A $\tvec{Z}$ dependence via $\mu_0$ and
$\zeta_0$ cancels in $W_{\text{small $y$}}$, up to orders in $\alpha_s$
beyond the accuracy of the calculation.  However, $W_{\text{small $y$}}$
acquires a dependence on $\tvec{Z} $ through the $\tvec{z}_1$ and
$\tvec{z}_2$ dependence of the Collins-Soper kernels
$\pr{1}{K}_{a_1}(\tvec{z}_1;\mu_{0})$,
$\pr{1}{K}_{a_2}(\tvec{z}_2;\mu_{0})$ and
$M_{a_1 a_2}(\tvec{z}_i,\tvec{y})$.  Since these kernels start at
$\mathcal{O}(\alpha_s)$, the corresponding $\tvec{Z}$ dependence comes
with a large logarithm $\log(Q_1 Q_2/\zeta_0)$ for each power of $\alpha_s$
and is thus enhanced compared to the $\alpha_s$ corrections in
$F_{\text{spl}}$ and $F_{\text{int}}$, so that it is consistent to keep
the former while neglecting the latter.  In this way, rapidity evolution
provides a nontrivial $\tvec{Z} $ dependence and thus a nontrivial
dependence on the conjugate momentum $\tvec{q}_1 + \tvec{q}_2$ in the
cross section, even if one uses the short-distance expansions
\eqref{split-TMD-coll} and \eqref{intr-TMD-tw4} at leading order.  In
physical terms, this is because the exponential in
\eqref{W-small-y} resums graphs for the soft factor that have a
real gluon emission enhanced by large rapidity logarithms, even though the
graphs giving the leading-order expansions of $F_{\text{spl}}$ and
$F_{\text{int}}$ have no real gluon emission into the final state.

\subsection{Combining large and small \texorpdfstring{$\tvec{y}$}{y}}
\label{sec:combine-y}

As we have seen in the previous subsections, the different regions of
$\tvec{z}_1, \tvec{z}_2$ and $\tvec{y}$ contributing to the cross section
involve different approximations, whose results are given in \eqref{W-large-y}
and \eqref{W-small-y}.  An important point is that the approximated
expressions give leading-power contributions not only in the regions for which
they are designed, but also outside these regions.  With the short-distance
behaviour $F(x_i,\tvec{y}) \sim 1/\tvec{y}^2$ following from
\eqref{coll-DPD-exp} we see for instance that the integral of
$W_{\text{large $y$}}$ over $\tvec{y}$ extends down to the smallest
values allowed by the cutoff function $\Phi(\nu \tvec{y})$, where the
approximations used to obtain the expression are clearly invalid.

To deal with this problem we adapt the subtraction formalism discussed in
chapter~10 of \cite{Collins:2011zzd}, which we briefly sketch now.  The
formalism is formulated for a given Feynman graph $\Gamma$ in momentum space.
It starts with the smallest regions of loop momenta and works its way towards
increasingly larger regions.  In this context, a region $r'$ is called smaller
than $r$ (i.e.\ $r'<r$) if hard momenta in $r$ are collinear or soft in $r'$,
or if collinear momenta in $r'$ are soft in $r$.  A more general definition of
the relation $r'<r$ is discussed in chapters~5.4.1 and 10.1.3 of
\cite{Collins:2011zzd}.  For each region $r$ that gives a leading contribution
to the cross section, a set of approximations $T_r$ is applied to the graph.
In each approximated term $T_r\ms \Gamma $ one still integrates over
\emph{all} loop momenta, thus avoiding momentum cutoffs.  This leads to double
counting of contributions from the overlap between different regions, which
are recursively removed by subtraction terms.  The full graph is then
approximated by
\begin{align}
  \label{Collins-subtr}
  \Gamma &\approx \sum_{r} C_r\ms \Gamma & \text{with}
&&
  C_r\ms \Gamma &= T_r\ms \Gamma - \sum_{r'<r} T_r\ms C_{r'}\ms \Gamma\,.
\end{align}
$C_r\ms \Gamma$ provides a valid approximation of $\Gamma$ in the region $r$
and in all smaller regions $r'<r$, thanks to the subtraction terms
$T_r\ms C_{r'}\ms \Gamma$, which are obtained by applying the approximations
for both $r$ and the smaller regions.  In the case where one has only two
regions $r'<r$, the subtraction term is simply $T_{r}\ms T_{r'}\ms \Gamma$.  In
\cite{Collins:2011zzd} it was shown that if the approximation $T_r$ is good up
to power corrections in $\Lambda/Q$ in its design region $r$, then
$\sum_r C_r\ms \Gamma$ is good with the same accuracy for the full graph.

In \cite{Diehl:2017kgu}, this formalism was applied to the double counting
problem between single and double parton scattering, leading to the
formulation we already presented in section~\ref{sec:TMD-UV-region}.  At
variance to the original formulation in \cite{Collins:2011zzd}, momentum
regions in \eqref{Collins-subtr} were replaced by regions of the transverse
distances $\tvec{y}_+$ and $\tvec{y}_-$, starting from the region of large
distances and going towards small ones in the recursive construction of
subtraction terms.

We now use the same procedure to remove the double counting between the
regions of ``large'' and ``small'' $\tvec{y}$ as characterised in
table~\ref{tab:y-regions}.  The treatment of the regions where
$|\tvec{y}_+|$ or $|\tvec{y}_-|$ are of order $1/Q$ is postponed to the next
subsection.  It is always understood that $|\tvec{z}_1|$ and $|\tvec{z}_2|$
are in the perturbative region, of order $1/q_T$, as discussed in
section~\ref{sec:regions}.

\begin{table}
\begin{center} \renewcommand{\arraystretch}{1.2}
\begin{tabular}{ccc} \hline
  region & power counting & approximations \\ \hline
DPS, large $\tvec{y}$ & $|\tvec{y}|, |\tvec{y}_+|, |\tvec{y}_-| \sim
  1/\Lambda$ & $|\tvec{z}_i| \ll |\tvec{y}|, 1/\Lambda$ \\
DPS, small $\tvec{y}$ & $|\tvec{y}|, |\tvec{y}_+|, |\tvec{y}_-| \sim
  1/q_T$ & $|\tvec{z}_i|,|\tvec{y}| \ll 1/\Lambda$ \\
SPS & $|\tvec{y}_+|, |\tvec{y}_-| \sim 1/Q$ &
  $|\tvec{y}_+|, |\tvec{y}_-| \ll |\tvec{z}_i| \ll 1/\Lambda$ \\ \hline
\end{tabular}
\caption{\label{tab:y-regions} Regions of $\tvec{y}$ discussed in the
text.  For power counting, $|\tvec{z}_i| \sim 1/q_T$ is always assumed.}
\end{center}
\end{table}

According to the general construction, the subtraction term for large
$\tvec{y}$ in the small-$\tvec{y}$ term is obtained by applying the
approximations in the first and second row of table~\ref{tab:y-regions},
i.e.\ by using both $|\tvec{z}_i| \ll |\tvec{y}|$ and $|\tvec{y}| \ll
1/\Lambda$.  For a given Feynman graph (and thus for any finite sum over
graphs) the order in which these approximations are made does not matter.
This suggests that we can start from either $W_{\text{large $y$}}$ or
$W_{\text{small $y$}}$ and then apply the approximation for the other
region.  At this point, we must however note that the expressions of $W$ are
not for fixed-order graphs but contain logarithms resummed to all orders in
the strong coupling.  This is reflected in the scale dependence of
$\alpha_s$ and of the nonperturbative functions.  We must therefore make
sure that the choice of scales in the different terms is compatible with the
way in which the subtraction formalism works.  In $W_{\text{small $y$}}$,
this concerns the scales $\mu_0$ and $\zeta_0$ at which one evaluates the
short-distance expansions \eqref{split-TMD-coll} and \eqref{intr-TMD-tw4} of
DTMDs at fixed order in $\alpha_s$.

We now define the double counting subtraction term as
\begin{align}
  \label{TMD-subtr-term}
W_{\text{sub}} &= W_{\text{small $y$}} \,
\bigl|_{\text{approx.~for}~|\tvec{z}_i| \ll |\tvec{y}|}
\end{align}
with the small-$\tvec{y}$ expression for $W$ given in \eqref{W-small-y}.  We
should take the limit $|\tvec{z}_1|, |\tvec{z}_2| \ll |\tvec{y}|$ in all
parts of that expression.  In particular, this means replacing
$\tvec{y}_{\pm}$ with $\tvec{y}$ in $\Phi$ and in the leading-order term of
the short-distance expansion \eqref{split-TMD-coll} of $F_{\text{spl}}$, and
replacing $\prb{RR'}{M}_{a_1 a_2}$ with $\delta_{RR'}\, \bigl[ \pr{R}K_{a_1}
  + \pr{R}{K}_{a_2} - \pr{1}K_{a_1} - \pr{1}{K}_{a_2} + \prb{R}{J} \bigr]$
according to \eqref{split-K-M} and \eqref{CS-gen-match}.

We note at this point that the integral of \eqref{W-small-y} over $\tvec{y}$
is actually divergent at large~$\tvec{y}$.  Up to logarithmic corrections,
the 2v2 term $F_{\text{int}} \times F_{\text{int}}$ is constant and the 2v1
term $F_{\text{spl}} \times F_{\text{int}}$ goes like $1/\tvec{y}^2$ in that
limit, and for $R=R'=1$ there is no large-$\tvec{y}$ suppression from the
exponential since $\pr{1}{J} = 0$. The construction \eqref{TMD-subtr-term}
ensures that the region $|\tvec{y}| \gg 1/q_T$ cancels in $W_{\text{small
    $\tvec{y}$}} - W_{\text{sub}}$.  When evaluating the cross section
numerically, one may want to use an upper cutoff of order $1/\Lambda$ on the
integral over $|\tvec{y|}$, thus avoiding to evaluate terms in a region
where they cancel anyway.

For the subtraction formalism to work, we need that
\begin{align}
  \label{TMD-subtr-cond}
W_{\text{sub}} &\approx W_{\text{large
$\tvec{y}$}} & \text{for \, $|\tvec{y}| \sim |\tvec{z}_1|, |\tvec{z}_2|$}
\end{align}
up to terms in $\alpha_s$ that are beyond the accuracy of the calculation.  In
the specified region of $\tvec{y}$, we can use the short-distance limit of
${J}(\tvec{y})$ and of the DPDFs in \eqref{W-large-y}, the latter being
obtained by adding $F_{\text{spl}}$ and $F_{\text{int}}$ from
\eqref{coll-DPD-exp}.  If one retains only
the leading-order terms for $F_{\text{spl}} + F_{\text{int}}$ on both sides of
\eqref{TMD-subtr-cond}, one finds manifest equality of the two expressions
up to differences due to scale dependence, which we will discuss shortly.

The full DPS cross section is then obtained from
\begin{align}
  \label{Xsect-DPS}
W_{\text{DPS}}(\nu) &= W_{\text{large $y$}}(\nu') -
  W_{\text{sub}}(\nu') + W_{\text{small $y$}}(\nu) \,,
\end{align}
where we make explicit the choice of cutoff parameters for the $\tvec{y}$
integration, taking $\nu' \sim q_T$ and $\nu \sim Q$. In the region
$|\tvec{y}| \gg |\tvec{z}_1|, |\tvec{z}_2|$ the last two terms cancel by
virtue of \eqref{TMD-subtr-term}, and one is left with the first term, which
is designed to give a correct approximation of the cross section there. For
$|\tvec{y}| \sim |\tvec{z}_1|, |\tvec{z}_2|$, the first and second terms
cancel according to \eqref{TMD-subtr-cond}, and the third term gives a correct
approximation of the cross section.  In this way, $W_{\text{DPS}}$ leads to a
correct approximation of the cross section for $|\tvec{y}|$ of order $1/q_T$
and larger.

We can take a more restrictive cutoff parameter $\nu' \sim q_T$ in the first
two terms, because according to \eqref{TMD-subtr-cond} the sum of these terms
is already suppressed for $|\tvec{y}| \sim 1/q_T$, and the dependence on the
exact value of $\nu'$ is thus beyond the accuracy of the calculation. Taking
$\nu'$ lower than $Q$ is also more economical for numerical calculations since
it excludes a $\tvec{y}$ region that is not needed. By contrast, $\nu$ in the
last term should be of order $Q$ (see the next subsection).

Returning to the issue of scale dependence, we should first specify what is
to be used for the nonperturbative quantities in the expression
\eqref{W-large-y} of $W_{\text{large $y$}}$.  For the Collins-Soper kernel
we take the form in \eqref{eq:ystar_trick} with $J(\tvec{y}^*;\mu_y,\mu_y)$
and $\gamma_J$ evaluated at fixed perturbative order; this ensures the
correct small-$\tvec{y}$ limit.  Using the DGLAP equations, the DPDFs are to
be evolved from $\mu_{0i}$ to $\mu_0 = \sqrt{\zeta_0} = \mu_y$, and it is at
that scale that one makes an ansatz which at small $\tvec{y}$ tends to the
fixed-order form specified by \eqref{coll-DPD-exp}.\footnote{An explicit
  example for such an ansatz in the colour singlet sector can be found in
  section~9.2 of \cite{Diehl:2017kgu}.}
As usual, this scale choice ensures that higher-order corrections are not
accompanied by large logarithms.

We now recall that in $W_{\text{small $y$}}$, and hence by construction
also in $W_{\text{sub}}$, we take $\mu_0 = \sqrt{\zeta_0} = \mu_Z$ for the
scales at which we perform the short-distance expansion of the DTMDs.  The
difference between the two sides in \eqref{TMD-subtr-cond} is thus due to
different scales of the short-distance matching and to the different scales at
which the DPDs and the Collins-Soper kernels are evaluated in
\eqref{W-large-y} and \eqref{W-small-y}.  However, all these scales, $\mu_y$,
$\mu_{z_i}$ and $\mu_Z$ are of the same order in the region
$|\tvec{y}| \sim |\tvec{z}_i|$ where we need \eqref{TMD-subtr-cond}.  Using
the same methods as in section~6.2 of \cite{Diehl:2017kgu}, one can show that
the approximation \eqref{TMD-subtr-cond} holds up to higher-order terms in
$\alpha_s$.  These terms are {not} accompanied by large logarithms and beyond
the accuracy of the calculation.

\subsection{Combining DPS with SPS at short distances}
\label{sec:reg_sub_simp}

In section~\ref{sec:TMD-UV-region} we recalled how to combine DPS, SPS
and their interference in TMD factorisation at generic values of $q_T$.  We
now derive the form of the double counting subtraction terms for the case
where $q_T \gg \Lambda$.  We restrict ourselves to the lowest order of
$\alpha_s$ in the perturbative short-distance coefficients.  An extension to
higher orders is beyond the scope of the present work.

Let us start by discussing the relevant transverse distance scales and the
approximations they imply.  We recall that the DPS cross section for $q_T\gg
\Lambda$ is dominated by distances $|\tvec{z}_1|, |\tvec{z}_2| \ll
1/\Lambda$ thanks to the Fourier exponentials in \eqref{TMD-Xsect}.  In full
analogy, the Fourier exponentials in~\eqref{eq:DPSSPS} and
~\eqref{eq:SPSDPS} ensure that $\sigma_{\text{DPS/SPS}}$ is dominated by
$|\tvec{Z}|, |\tvec{y}_+| \ll 1/\Lambda$ and $\sigma_{\text{SPS/DPS}}$ by
$|\tvec{Z}|, |\tvec{y}_-| \ll 1/\Lambda$.  Likewise, $\sigma_{\text{SPS}}$
is dominated by $|\tvec{Z}| \ll 1/\Lambda$, where in the following we use
the variable $\tvec{Z}$ instead of $\tvec{z}$ in the SPS formula
\eqref{SPS-Xsect}.
According to the formalism recalled in the previous subsection, the
subtraction terms for the overlap between double and single parton
scattering are obtained by applying the approximations appropriate for the
SPS region to the DPS term in the cross section.  A corresponding
prescription holds for the SPS/DPS interference.  The limits $\tvec{y}_+ \to
\tvec{0}$, $\tvec{y}_- \to \tvec{0}$ or $\tvec{y}_\pm \to \tvec{0}$ in the
subtraction terms hence imply that these distances are taken to be much
smaller than $1/q_T$, because $q_T$ is set to zero when one computes the
hard scattering process for SPS.
We thus obtain the approximations given in table~\ref{tab:y-regions-UV}.
The constraint on $\tvec{y}_+$ in the third row results from $\tvec{y}_+ =
\tvec{y}_- + (\tvec{z}_1 - \tvec{z}_2)$ and the constraints on $\tvec{y}_-$,
$\tvec{z}_1$ and $\tvec{z}_2$.  We thus find that all terms in the overall
cross section \eqref{eq:tot-x-sec} are dominated by distances in the
perturbative region, except for $\sigma_{\text{DPS}}$.

\begin{table}
\begin{center} \renewcommand{\arraystretch}{1.2}
\begin{tabular}{ccc} \hline
partial cross section & approximations \\ \hline
$\sigma_{\text{DPS/SPS}}$ & $|\tvec{y}_+| \ll 1/\Lambda$ \\
$\sigma_{\text{DPS/SPS},\, y_+ \to 0}$ & $|\tvec{y}_+| \ll 1/q_T$ \\
$\sigma_{\text{DPS},\, y_- \to 0}$ & $|\tvec{y}_-| \ll 1/q_T$, $|\tvec{y}_+| \ll
                                     1/\Lambda$ \\
$\sigma_{\text{DPS},\, y_\pm \to 0}$ & $|\tvec{y}_-|, |\tvec{y}_+| \ll 1/q_T$
\\ \hline
\end{tabular}
\caption{\label{tab:y-regions-UV} Conditions on $\tvec{y}_\pm$ for different
  terms in the overall cross section \protect\eqref{eq:tot-x-sec} if $q_T
  \gg \Lambda$.  Analogous relations hold for $\sigma_{\text{SPS/DPS}}$ and
  the associated subtraction terms.  In all cases one has $|\tvec{Z}| \ll
  1/\Lambda$.}
\end{center}
\end{table}

In all terms that are dominated by short distances, we can use the
perturbative splitting expressions \eqref{tw3-split} and
\eqref{F-short-dist_TMD} of the twist-three TMDs and the DTMDs.  Notice that
the LO terms in these expressions have a factorised dependence on
$\tvec{y}_+$, $\tvec{y}_-$ and $\tvec{Z}$, so that they do not depend on the
relative size of these distances (this will no longer hold at NLO).
Moreover, we can take the short-distance limit of the TMDs
$f_a(x,\tvec{Z};\mu,\zeta)$.  At LO accuracy we can simply replace them with
collinear PDFs $f_a(x; \mu)$, provided that $\mu^2 \sim \zeta \sim
1/\tvec{Z}^2$.  Collinear PDFs are then the only nonperturbative input
necessary to compute the cross section, apart of course from the DPDs and
Collins-Soper kernels $J$ in $\sigma_{\text{DPS}}$.  In a complete
treatment, one would also have collinear twist-three distributions, but we
neglect these as explained in section~\ref{sec:TMD-UV-region}.

From the dependence on $\tvec{y}_\pm$ of the splitting expressions
\eqref{tw3-split} and \eqref{F-short-dist_TMD} and from the cross section
formulae~\eqref{SPS-Xsect}, \eqref{eq:DPSSPS} and \eqref{eq:SPSDPS} we deduce
that the SPS cross section, the SPS/DPS interference and all associated
subtraction terms have the same power behaviour as the 1v1 term of the DPS
cross section in~\eqref{power-small-y}, namely
$Q^4\, d\sigma/ (d^2\tvec{q}_1\, d^2\tvec{q}_2) \sim 1/q_T^2$.

We must now take a closer look at the scale dependence of the distributions.
Let us denote by $\mu_h$ the hard scale chosen in the computation of the SPS
amplitude, and by $\mu_1$ and $\mu_2$ the scales for the two DPS amplitudes.
To simplify the presentation, we first assume that all hard scales are
equal, $\mu = \mu_h = \mu_1 = \mu_2$.  The adaptation to the case of
different hard scales is discussed in appendix~\ref{app:diff_scales}.
As we have explicitly shown for single and double parton TMDs, the evolution
in $\mu$ and $\zeta$ of these distributions is multiplicative:
\begin{align}
  \label{TMD-evol}
  f_{a}(x,\tvec{z}; \mu,\zeta) &= E_{2; a}(\tvec{z};
\mu,\zeta; \mu_0,\zeta_0)\, f_{a}(x,\tvec{z}; \mu_0,\zeta_0) \,, \nonumber \\
F_{a_1 a_2}(x_i,\tvec{z}_i,\tvec{y}; \mu,\zeta)
  &= E_{4; a_1 a_2}(x_i,\tvec{z}_i,\tvec{y}; \mu,\zeta;
\mu_0,\zeta_0)\, F_{a_1 a_2}(x_i,\tvec{z}_i,\tvec{y}; \mu_0,\zeta_0) \,,
\end{align}
where it is understood that $F$ is a vector and the evolution factor $E_4$ a
matrix in the space of colour representations $R$.  From the discussion at the
end of section~\ref{sec:TMD-UV-region} if follows that for the twist-three
TMDs, we have
\begin{align}
  \label{TMD-tw3-evol}
D_{\alpha_1 \alpha_2|
  \alpha_0}(x_i,\tvec{y}_+,\tvec{Z}; \mu,\zeta)
 &= E_{3; \alpha_1 \alpha_2|
   \alpha_0}(x_i,\tvec{y}_+,\tvec{Z}; \mu,\zeta; \mu_0,\zeta_0)\,
 D_{\alpha_1 \alpha_2|
   \alpha_0}(x_i,\tvec{y}_+,\tvec{Z}; \mu_0,\zeta_0)
 \rule[-1.8ex]{0pt}{0pt}
\end{align}
and an analogous equation for $D_{\alpha_0 | \alpha_1 \alpha_2}$.

Let us now specify the evolution and scale dependence of the distributions
appearing in $\sigma_{\text{SPS}}$, the interference terms and
$\sigma_{\text{DPS}}$ in the region where all transverse distances are small
compared with $1/\Lambda$.  In a schematic notation, we have
\begin{align}
  \label{eq:x-secs_onescale}
\sigma_{\text{SPS}}: && f(x_1+x_2,\tvec{Z}; \mu) &\sim E_2(\mu;\mu_Z)\,
f(\mu_Z) \,,
\nonumber \\
\sigma_{\text{DPS/SPS}}: &&
D(x_i,\tvec{y}_+,\tvec{Z}; \mu) &\sim E_3(\mu;\mu_Z)\, U(\mu_Z)\,
f(\mu_Z) \,,
\nonumber \\
\sigma_{\text{SPS/DPS}}: &&
D(x_i,\tvec{y}_-,\tvec{Z}; \mu) &\sim E_3(\mu;\mu_Z)\, U^*(\mu_Z)\,
f(\mu_Z) \,,
\nonumber \\
\sigma_{\text{DPS}}: &&
F(x_i,\tvec{z}_i,\tvec{y}; \mu)
   &\sim E_4(\mu;\mu_Z)\, U(\mu_Z)\, U^*(\mu_Z)\, f(\mu_Z)
    + F_{\text{int}}(\mu) \,,
\end{align}
where $f(\mu_Z)$ on the r.h.s.\ is a collinear PDF.  We have specified only
the splitting part $F_{\text{spl}}$ of $F$, because the detailed form of
$F_{\text{int}}$ will not be needed.  For brevity we have omitted parton
labels, as well as momentum fraction and transverse position arguments on
the r.h.s.  The factors $\tvec{y}_+^{l} /\tvec{y}_+^2$ and $\tvec{y}_-^{l'}
/\tvec{y}_-^2$ which accompany the splitting kernels $U(\mu_Z)$ and
$U^*(\mu_Z)$ have been omitted as well.
For brevity, we display neither the rapidity parameter $\zeta$ in the
evolved distributions nor the starting value $\zeta_0$.  For the latter, we
follow \eqref{small-y-scales} and take $\zeta_0 = \mu_Z^2$.  Since $\zeta$
is defined by \eqref{zeta-def}, it must be rescaled by $(x_1+x_2)^2 \ms
\zeta/(x_1 x_2)$ when used in $f(x_1+x_2,\tvec{Z})$ and in $E_2$.  The same
forms as in \eqref{eq:x-secs_onescale} are to be taken for the distributions
in the left-moving proton, with $x_i$ replaced by $\bar{x}_i$ and $\zeta$ by
$\bar{\zeta}$.

The subtraction terms should be such that, up to higher orders in $\alpha_s$
and up to power corrections, the overall cross section \eqref{eq:tot-x-sec}
is given by
\begin{align}
 d\sigma_{\text{DPS}} & \text{ \,for\, }
   |\tvec{y}_+| \sim |\tvec{y}_-| \gg 1/\nu \,,
\nonumber \\
 d\sigma_{\text{SPS}} & \text{ \,for\, }
   |\tvec{y}_+| \sim |\tvec{y}_-| \sim 1/\nu \,,
\nonumber \\
 d\sigma_{\text{DPS/SPS}} & \text{ \,for\, }
   |\tvec{y}_+| \gg |\tvec{y}_-| \sim 1/\nu \,,
\nonumber \\
 d\sigma_{\text{SPS/DPS}} & \text{ \,for\, }
   |\tvec{y}_-| \gg |\tvec{y}_+| \sim 1/\nu \,,
\end{align}
where it is understood that $\nu \sim Q$.  In the subtraction terms,
renormalisation and rapidity scales should be chosen such that the
cancellations required to avoid double counting are not spoiled by large
logarithms at higher orders.

To achieve this, we take so-called profile scales, which vary smoothly
between $\mu_Z$ and $\mu$ as a function of the appropriate transverse
distance.  We therefore introduce a profile function $p$ that satisfies
\begin{align}
  \label{profile-scales-gen}
  p({u};\mu_a,\mu_b) &\approx \mu_a & & \text{for ${u} \sim 1$,}
\nonumber \\
  p({u};\mu_a,\mu_b) &\approx \mu_b & & \text{for ${u} \gg 1$;}
\end{align}
a concrete example is given below.  For the subtraction terms between DPS
and the SPS/DPS interference, we now take
\begin{align}
  \label{interf-subtr-terms}
\sigma_{\text{DPS/SPS}, y_+\to 0} &: D \sim E_3(\mu;\hat{\mu})\,
  U(\hat{\mu})\, E_2(\hat{\mu};\mu_Z)\, f(\mu_Z) & &
\mbox{with }
  \hat{\mu} = p\bigl( \nu\ms |\tvec{y}_+|; \mu_Z, \mu \bigr) \,,
\nonumber \\
\sigma_{\text{SPS/DPS}, y_-\to 0} &: D \sim E_3(\mu;\hat{\mu})\,
  U^*(\hat{\mu})\, E_2(\hat{\mu};\mu_Z)\, f(\mu_Z) & &
\mbox{with }
  \hat{\mu} = p\bigl( \nu\ms |\tvec{y}_-|; \mu_Z, \mu \bigr) \,.
\end{align}
This corresponds to a two-step matching: at the scale $\mu_Z$ a twist-two
TMD is matched onto a PDF.  The twist-two TMD is then evolved up to the
scale $\hat{\mu}$, where a twist-three TMD is matched onto the twist-two TMD
using the lowest order term in \eqref{tw3-split}.  The twist-three TMD is
then evolved up to the final scale $\mu$.  It is understood that at the
intermediate scale $\hat{\mu}$ the rapidity parameter is taken equal to
\begin{align}
  \label{zeta-hat}
\widehat{\zeta} &= p(u; \mu_Z^2, \zeta) \,,
\end{align}
where the first argument $u$ is the same as the first argument in
$\hat{\mu}$.
We thus find that
\begin{align}
  \label{interf-limits}
  \sigma_{\text{DPS/SPS}, y_+\to 0} \approx
\begin{cases}
  \sigma_{\text{DPS/SPS}} & \mbox{for $|\tvec{y}_+| \sim 1/\nu$} \\
  \sigma_{\text{SPS}}     & \mbox{for $|\tvec{y}_+| \gg 1/\nu$}
\end{cases}
\end{align}
where we have used that for distances $|\tvec{y}_+| \gg 1/Q$ the hard
scattering amplitude for SPS can be written as
\begin{align}
  \label{amp-matching}
  H_{\alpha_0 \beta_0} \big|_{y_+ \gg 1/Q}
&=  \frac{\tvec{y}_+^k}{\tvec{y}_+^2}\, \frac{\tvec{y}_+^l}{\tvec{y}_+^2}\;
    U^k_{\alpha_0\to \alpha_1 \alpha_2} \,
    U^l_{\beta_0\to \beta_1 \beta_2} \,
    H_{\alpha_1 \beta_1}^{}\, H_{\alpha_2 \beta_2}^{}
\end{align}
at leading order in $\alpha_s$.  To make this relation explicit, one needs to
Fourier transform the appropriate transverse momentum in the SPS graph to
position space.  For diagrams that have kinematic overlap between SPS and the
SPS/DPS interference, but not with DPS, such as the one in
figure~\ref{fig:nobox}b, the behaviour in \eqref{interf-limits} and in its
analogue for $\sigma_{\text{SPS/DPS}, y_-\to 0}$ solve the double counting
problem completely.  If there is also overlap with the DPS region, as in the
three graphs of figure~\ref{fig:boxed}, then we need the DPS subtraction
terms, where we choose scales as follows:
\begin{align}
	\label{1v1-subtr-terms}
\sigma_{\text{DPS}, y_-\to 0} & : F \sim
E_4(\mu;\hat{\mu})\, U^*(\hat{\mu})\, E_3(\hat{\mu};\mu_Z)\, U(\mu_Z)\,
f(\mu_Z) \,, \nonumber \\ \sigma_{\text{DPS}, y_+\to 0} & : F \sim
E_4(\mu;\hat{\mu})\, U(\hat{\mu})\, E_3(\hat{\mu};\mu_Z)\, U^*(\mu_Z)\,
f(\mu_Z) \,, \nonumber \\ \sigma_{\text{DPS}, y_\pm\to 0} & : F \sim
E_4(\mu;\hat{\mu})\, U(\hat{\mu})\ U^*(\hat{\mu})\, E_2(\hat{\mu};\mu_Z)\,
f(\mu_Z)
\intertext{with}
\hat{\mu} &=
   p\bigl( \nu \min\{ |\tvec{y}_+|, |\tvec{y}_-| \}; \mu_Z, \mu \bigr) \,.
\end{align}
In all three terms we have a two-step matching, with a twist-three TMD
at the intermediate stage in the first two cases and with a twist-two TMD in
the third case.  If both $|\tvec{y}_+|$ and $|\tvec{y}_-|$ are large
compared with $1/\nu$, we thus have
\begin{align}
  \label{DPS-limits}
   \sigma_{\text{DPS}, y_-\to 0}   &\approx \sigma_{\text{DPS/SPS}} \,,
 & \sigma_{\text{DPS}, y_+\to 0}   &\approx \sigma_{\text{SPS/DPS}} \,,
 & \sigma_{\text{DPS}, y_\pm\to 0} &\approx \sigma_{\text{SPS}}
\end{align}
so that the sum in \eqref{eq:tot-x-sec} leaves only $\sigma_{\text{DPS}}$,
as it should.
It remains to discuss the regions where one or both of $|\tvec{y}_+|$ and
$|\tvec{y}_-|$ is of order $1/\nu$.  In each of these regions,
$\sigma_{\text{DPS}}$ can be obtained from $W_{\text{small $y$}}$, because
$|\tvec{y}| \ll 1/\Lambda$.  Furthermore, the integral over $W_{\text{small
    $y$}}$ in these regions is dominated by the 1v1 term, with the 2v1 and
2v2 terms being power suppressed by at least a relative factor $\Lambda^2
/(\nu q_T)$.  This follows from the dependence of $F_{\text{spl}}$ and
$F_{\text{int}}$ on $\tvec{y}_+$ and $\tvec{y}_-$.  Finally, one finds that
all three terms in \eqref{1v1-subtr-terms} are approximately equal to
the 1v1 term in $\sigma_{\text{DPS}}$, which is thus cancelled in the
combination \eqref{eq:tot-x-sec}.  This leaves us with
$\sigma_{\text{SPS}}$, $\sigma_{\text{DPS/SPS}}$ or
$\sigma_{\text{SPS/DPS}}$ as appropriate.  Overall we thus have achieved a
correct description of the cross section in all relevant regions of
$\tvec{y}_+$ and $\tvec{y}_-$.

Profile scales have been used in other contexts, see
e.g.~\cite{Ligeti:2008ac,Abbate:2010xh}.  For our purpose here, a suitable
profile $p$ for $\hat{\mu}$ is given in equation (6.33) of
\cite{Diehl:2017kgu}.  An alternative choice is
\begin{align}
p(u ;\mu_a,\mu_b) =
\begin{cases}
  \mu_a & \mbox{ \,for\, } u \le u_a \\
  \bigl[ 1 - \frac{1}{2} (1- \cos u' \pi)\bigr] \ms \mu_a
           + \frac{1}{2} (1- \cos u' \pi)\, \mu_b &
  \mbox{ \,for\, } u_a < u < u_b \\
  \mu_b & \mbox{ \,for\, } u \ge u_b \\
\end{cases}
\end{align}
with
\begin{align}
  u' = \frac{u-u_a}{u_b - u_a}\,.
\end{align}
The transition points for the $u$ dependence should be such that $u_a \sim
1$ and $u_b \gg 1$.  A natural choice is $u_a = {\nu}/{\mu}$ and $u_b =
{\nu}/{\mu_Z}$.

\subsection{Perturbative accuracy}
\label{sec:resummation_orders}

Equations \eqref{TMD-evo-solved}, \eqref{small-z-evolved} and
\eqref{small-yz-evolved} express TMDs in terms of collinear distributions
and perturbative quantities, and they permit the resummation of large
logarithms in the cross section.  In the following, we give a brief overview
of the order in $\alpha_s$ at which the different quantities are available.
We also take a closer look at the type of logarithms that are resummed at
all orders.  To this end, we count all large scales $\mu_i, \zeta, \nu$ as
order $Q$.  Logarithms of $|\tvec{z}_i|$ turn into logarithms of
$|\tvec{q}_i|$ after Fourier transformation, and we count $\mu_{i0} \sim 1
/|\tvec{z}_i| \sim q_T$.

An important point in this context is that a variety of schemes for TMD
factorisation and for the associated resummation are used in the literature.
Some care is needed when converting perturbative expressions from one scheme
to another.  A systematic discussion of this issue is given in
\cite{Collins:2017oxh}, along with a comprehensive list of higher-order
results for the quark channel in the scheme we use here (referred to as CSS2
in that paper).

\subsubsection*{SPS}

The SPS cross section can be calculated within the single TMD formalism
outlined in section~\ref{sec:single_TMD}.  To compute the spectrum in
$\tvec{q}_1$ and $\tvec{q}_2$, one needs the relevant parton-level cross
sections, e.g.\ for gauge boson pair production, as a function of the
relative transverse momentum $\tvec{q}_1 - \tvec{q}_2$.  Logarithms of the
net transverse momentum $\tvec{q}_1 + \tvec{q}_2$ are resummed by the
exponential in \eqref{TMD-evo-solved}.

Perturbative ingredients in the cross section are the anomalous dimensions
$\gamma_{K}$, $\gamma_{a}$, the Collins-Soper kernel $\pr{1}{K}$, the
short-distance (or matching) coefficients $\prn{1}{C}$, and finally the
coefficient $C_H^2$ of the hard-scattering cross section, defined as
$\hat{\sigma}$ divided by its tree-level value.\footnote{For definiteness we
  include the colour-singlet labels on $K$ and $C$ here, although this is the
  only colour channel that appears in SPS.}
In table~\ref{tab:orders} we specify different levels of accuracy,
following the scheme of \cite{Abbate:2010xh,Berger:2010xi}.
After expanding $\alpha_s(\mu')$ in $\gamma_{a}(\mu')$ and $\gamma_K(\mu')$
around $\alpha_s(\mu)$, the integral over~$\mu'$ in \eqref{TMD-evo-solved}
gives powers of $\log(\mu/\mu_0)$.  We thus find that $\gamma_K$ goes with
double logarithms $\alpha_s(Q) \log^2(Q/q_T)$, whilst $\gamma_a$ and
$\pr{1}{K}$ go with single logarithms $\alpha_s(Q) \log(Q/q_T)$.  One
therefore requires one order higher for $\gamma_K$ than for $\gamma_a$ and
$\pr{1}{K}$ (see table~\ref{tab:orders}).  Likewise, one finds that one
needs the same perturbative order for the $\beta$ function as for
$\gamma_K$.  At accuracy N${}^{k}$LO one then has control over all terms
from $\alpha_s^{n} \log^{n+1}$ to $\alpha_s^{n+k} \log^{n+1}$ in the
exponent.  $\gamma_K$, $\gamma_{a}$ and $\pr{1}{K}$ all start at
$\mathcal{O}(\alpha_s)$ and are known up to $\mathcal{O}(\alpha_s^3)$, both
in the quark and in the gluon channel.  A compilation of these results can
for instance be found in appendix~D of \cite{Echevarria:2016scs}.  The
notation in that work is related to the one we are using by
\begin{align}
  \label{notation-dict}
\gamma_{K,a} &= 2\ms \Gamma_{\text{cusp}}^a \,,
&
\gamma_{a} &= {}- \gamma_V^a \,,
&
\gamma_{F,a}(\mu,\zeta) &= \gamma^a(\mu,\zeta)
\end{align}
and
\begin{align}
\pr{1}{K}_{a}(\tvec{z},\mu) &= - 2 \mathcal{D}^a(\mu,\tvec{b}_T^{})
  && \text{with } \tvec{z}=\tvec{b}_T \,,
\end{align}
where $a=q,g$ and it is understood that the strong coupling is taken at scale
$\mu$.  Results for the quark channel are also given in
\cite{Collins:2017oxh}, where the same notation as here is used (except that
the Collins-Soper kernel is denoted by $\tilde{K}$ instead of $K$).

\begin{table}
  \center
  \setlength\tabcolsep{1em}
  \renewcommand{\arraystretch}{1.1}
\begin{tabular}{ccccccc} \hline
  & $\gamma_K$ & $\gamma_{a}$ & $\pr{R}{K}$, $\prb{RR'}{M}$
  & $\prn{R}{C}$, $\pr{RR'}{C}_{\text{int}}$
  & $C_H^2\rule[-1.1ex]{0pt}{1ex}$ & $\pr{R}{P}$ \\ \hline
  LL  & $\alpha_s$ & $-$ & $-$ & $\alpha_s^0$ & $\alpha_s^0$ & $\alpha_s$ \\
  NLL & $\alpha_s^2$ & $\alpha_s$ &
        $\alpha_s$ & $\alpha_s^0$ & $\alpha_s^0$ & $\alpha_s$ \\
  NLL$'$ & $\alpha_s^2$ & $\alpha_s$ &
           $\alpha_s$ & $\alpha_s$ & $\alpha_s$ & $\alpha_s^2$ \\
  NNLL & $\alpha_s^3$ & $\alpha_s^2$ &
         $\alpha_s^2$ & $\alpha_s$ & $\alpha_s$ & $\alpha_s^2$ \\
  NNLL$'$ & $\alpha_s^3$ & $\alpha_s^2$ &
            $\alpha_s^2$ & $\alpha_s^2$ & $\alpha_s^2$ & $\alpha_s^3$ \\
  \hline
\end{tabular}
\caption{\label{tab:orders} Different levels of accuracy and the
  associated orders of $\alpha_s$ for the perturbative ingredients in the SPS
  cross section.  The order required for the $\beta$ function is the same as
  for $\gamma_K$.}
\end{table}

At lowest order one has $\prn{1}{C}_{ab} = \delta_{ab}\, \delta(1-x)$ and
$C_H^2 = 1$.  These coefficients are not associated with the resummation of
large logarithms, and conventionally one requires them at one order in
$\alpha_s$ lower than $\gamma_a$.  This corresponds to counting
$\log(Q/q_T)$ as order $1/\alpha_s$.  Taking them at the same order as
$\gamma_a$ is indicated by a prime in the table.  For unpolarised quarks or
gluons, $\prn{1}{C}$ has been computed up to $\mathcal{O}(\alpha_s^2)$ in
\cite{Echevarria:2016scs} and \cite{Gehrmann:2014yya}.  With unpolarised
protons this is sufficient for $q\bar{q}$ annihilation processes, whereas
for $gg$ initiated processes one also requires the coefficient
$\prn{1}{C}_{\delta g\ms g}$ for linear gluon polarisation in the TMDs.
This is well known from the process $gg\to H$ (see
e.g.~\cite{Echevarria:2015uaa} and references therein).

To the best of our knowledge, the hard-scattering coefficient $C_H^2$ has
not been given in the literature for final states relevant in our context,
such as $W^+ W^-$, $Z H$, $HH$, etc.  It should however be possible to
extract it from the virtual corrections to the relevant amplitudes, which
are known up to $\mathcal{O}(\alpha_s^2)$ in some channels (for gauge boson
pair production see e.g.\ \cite{Gehrmann:2015ora,Caola:2015ila} and
references therein).  Without this additional effort, the SPS contribution
to the cross section can currently be evaluated at NLL accuracy.

There is one more type of large logarithms hidden in \eqref{TMD-evo-solved},
namely the logarithms of $q_T/\Lambda$ that are resummed by DGLAP evolution
of the PDFs on the r.h.s., from a typical starting scale of order $\Lambda$
up to $\mu_0 \sim 1/|\tvec{z}|$.  A customary choice is to take the DGLAP
kernels $\prn{1}{P}$ at one power in $\alpha_s$ higher than the matching
coefficients $\prn{1}{C}$, as shown in the table.  This amounts to using LO
evolution together with LO matching coefficients, etc.  The perturbative
accuracy of $\prn{1}{C}$ is then retained even if $\log(q_T/\Lambda)$ is as
large as $1/\alpha_s$.  For unpolarised PDFs, the DGLAP kernels are known up
to $\mathcal{O}(\alpha_s^3)$ \cite{Moch:2004pa,Vogt:2004mw}.

\subsubsection*{Interference between SPS and DPS}

The perturbative ingredients required for computing the SPS/DPS interference
term, as laid out in section~\ref{sec:reg_sub_simp}, are the Collins-Soper
kernels $K_{\alpha_1 \alpha_2| \alpha_0}$ for twist-three TMDs, the splitting
kernels $U$ appearing in their short-distance approximation, and the
interference between SPS and DPS hard-scattering amplitudes for the process
under investigation.  None of these have been calculated so far.  It is easy
to compute the $\mathcal{O}(\alpha_s)$ expressions of
$K_{\alpha_1 \alpha_2| \alpha_0}$ and $U$ by adapting the corresponding
calculations for DPDs in \cite{Diehl:2011yj}.  Combining the known
hard-scattering amplitudes for the relevant SPS and DPS processes is a
straightforward exercise (but can be tedious, especially for heavy gauge
bosons due to their three helicities).

\subsubsection*{DPS: small-\texorpdfstring{$\tvec{y}$}{y} contribution}
\label{sec:orders-small-y}

Let us now discuss the perturbative ingredients in the small-$y$ expression
\eqref{W-small-y} for DPS.  $\gamma_K$, $\gamma_a$ and $\pr{1}{K}$ are the
same as in the SPS formula, and in addition one needs the matrix
$\prb{RR'}{M}$, which starts at $\mathcal{O}(\alpha_s)$ and goes with single
logarithms $\log(Q/q_T)$.  The expression for the matching
$\pr{R}{F}_{\text{spl}}$ on PDFs involves a coefficient
$\prn{R}{C}_{\text{spl}}$, and the matching of $\pr{R}{F}_{\text{int}}$ on
twist-four distributions involves a coefficient $\pr{RR'}{C}_{\text{int}}$.
The lowest order of ${C}_{\text{spl}}$ and ${C}_{\text{int}}$ is
$\mathcal{O}(\alpha_s)$ and $\mathcal{O}(\alpha_s^0)$, respectively --- their
explicit definitions are not needed here.
The hard-scattering subprocesses for DPS are simpler than for SPS with the
same final state.  For Drell-Yan and for Higgs boson production the
hard-scattering coefficients $C_H^2$ are known up to $\mathcal{O}(\alpha_s^3)$.
They can for instance be found in \cite{Collins:2017oxh} for Drell-Yan
production and in \cite{Ebert:2017uel} for both channels.

The evolution kernel $\prb{RR'}{M}$ is known at $\mathcal{O}(\alpha_s^2)$ in
the two-quark channel \cite{Vladimirov:2016qkd}; its $\mathcal{O}(\alpha_s)$
expressions for all channels are given in section~\ref{sec:CS_kernels}.  The
matching coefficients $\prn{R}{C}_{\text{spl}}$ have been calculated at
$\mathcal{O}(\alpha_s)$ in \cite{Diehl:2011yj}, and the related kernels
$\prn{R}{T}$ are compiled in section~\ref{sec:kernels_for_splitting} here.
For $\pr{RR'}{C}_{\text{int}}$, only the trivial order $\alpha_s^0$ is
currently known, which corresponds to matching $F_{\text{int}} = G$.  The
kernels for the DGLAP evolution of the distributions $G$ are known at
$\mathcal{O}(\alpha_s)$ and given in \cite{Bukhvostov:1985rn}; for a more
general discussion of twist-four evolution we refer to \cite{Braun:2009vc}.
At a given level of accuracy, their order should be the same as the one of
$\pr{R}{P}$ in table~\ref{tab:orders}.

When denoting the level of perturbative accuracy, one might think of
combining equal orders in $\alpha_s$ for $C_{\text{spl}}$ and
$C_{\text{int}}$.  On the other hand, $C_{\text{spl}}$ starts one order
  higher in $\alpha_s$ than $C_{\text{int}}$, and moreover the different
combinations of $F_{\text{spl}}$ and $F_{\text{int}}$ in the cross section
have different power behaviour in $\Lambda^2 /q_T^2$ according to
\eqref{power-small-y}.  We thus find it more natural to take one power in
$\alpha_s$ more for $C_{\text{spl}}$ than for $C_{\text{int}}$.  With this
naming convention, the small-$y$ expression of DPS can currently be
evaluated to NLL accuracy.  Missing ingredients for achieving NNLL in pure
quark channels or NLL$'$ in channels with gluons are the matching
coefficients $C_{\text{spl}}$ and $C_{\text{int}}$ for DTMDs, as well as the
two-loop DGLAP kernels for twist-four evolution.

\subsubsection*{DPS: large-\texorpdfstring{$\tvec{y}$}{y} contribution}

We now turn our attention to the large-$y$ expression \eqref{W-large-y} of
DPS.  The quantities $\gamma_K$, $\gamma_a$ and $C_H^2$ are the same as in
the small-$y$ expressions of DPS.  The resummation of single logarithms
$\log(Q/q_T)$ requires the Collins-Soper kernels $\pr{R}{K}$ in the different
colour channels, which can be obtained from the perturbative expression of
$\prb{RR'}{M}(\tvec{z}_i, \tvec{y})$ by taking the limit
$|\tvec{z}_i| \ll |\tvec{y}|$ and using that in this limit one has the
structure \eqref{CS-gen-match}.  In the two-quark channel one can thus obtain
the $\mathcal{O}(\alpha_s^2)$ expression for $\pr{R}{K}$ from the results of
\cite{Vladimirov:2016qkd}; in all other channels one has the one-loop
expressions, which we list in section~\ref{sec:CS_kernels}.  For the matching
coefficients $\prn{R}{C}$ we will derive $\mathcal{O}(\alpha_s)$ results in
all colour and polarisation channels in the next section; most of them have
been given in the literature before.  One can thus evaluate gauge boson pair
production with NNLL accuracy, whereas for DPS processes with one or two Higgs
bosons in the final state, NLL$'$ accuracy is currently achievable.

Let us finally take a closer look at the evolution of the DPDFs, which
generates logarithms involving a hadronic starting scale $\Lambda$.  The
situation is now more involved than the one discussed for SPS above.  In
analogy to usual DGLAP evolution, there are single logarithms
$\log(q_T/\Lambda)$, whose resummation requires the $\zeta$ independent part
of the evolution kernels $\pr{R}{P}$.  We will give $\mathcal{O}(\alpha_s)$
expressions for these kernels in all polarisation channels in
section~\ref{sec:oneloop-nonsing}.  The $\zeta$ dependence of the DPDFs,
which has no counterpart for ordinary parton distributions, is made explicit
in \eqref{eq:coll_rap_sep}.  Expanding $\alpha_s(\mu)$ in the anomalous
  dimension $\prn{R}{\gamma}_J(\mu)$ around a fixed scale, we can perform
the integral and obtain double logarithms $\log^2 (Q/\Lambda) - \log^2
(Q/q_T) = \log(\Lambda q_T /Q^2)\, \log(\Lambda /q_T)$ in the exponential.
If both $\log(Q/q_T)$ and $\log(q_T/\Lambda)$ are counted as order
$1/\alpha_s$, one therefore needs $\prn{R}{\gamma}_J$ at the same order as
the cusp anomalous dimension $\gamma_K$.  For smaller values of
$\log(q_T/\Lambda)$ one may instead be content with $\prn{R}{\gamma}_J$ at
the same order as $\gamma_a$ and $\pr{R}{K}$.  From the two-loop results in
\cite{Vladimirov:2016qkd} one can extract the $\mathcal{O}(\alpha_s^2)$ term
of $\pr{8}{\gamma}_J$, whereas for higher colour representations we
currently only know the $\mathcal{O}(\alpha_s)$ results given in
section~\ref{sec:CS_kernels}.  We note that in \eqref{eq:coll_rap_sep} there
are also single logarithms $\log(Q/q_T)$ multiplied with
$\prb{R}{J}(\tvec{y})$, which at large $\tvec{y}$ is beyond the control of
perturbation theory.

\section{One-loop results}
\label{sec:one-loop}

In this section, we present the perturbative ingredients necessary for
evaluating DPS at NLL accuracy (see table~\ref{tab:orders}).  \rev{Some of the
expressions are quoted from the literature to make the presentation
self-contained, whereas for evolution kernels and matching coefficients we
verify previous and derive new results.}

\rev{In section~\ref{sec:TMD_hard-scattering} we discuss the hard-scattering
cross sections for Drell-Yan and for Higgs production via gluon fusion, and in
section~\ref{sec:CS_kernels} we give the one-loop expressions of the
Collins-Soper kernels $\prb{RR'}{K}_{a_1 a_2}$ for all parton and colour
channels.  Section~\ref{sec:Coefficient_functions} is devoted to the
$\mathcal{O}(\alpha_s)$ matching coefficients $\prn{R}{C}_{ab}$ of DTMDs in
the large-$\tvec{y}$ region.  After explaining a number of computational
details, we present results for all polarisation and colour combinations in
sections~\ref{sec:oneloop-sing} and \ref{sec:oneloop-nonsing}.  As a corollary
we obtain the one-loop DGLAP kernels \eqref{DGLAP-R-LO} for colour non-singlet
DPDFs.  In section~\ref{sec:kernels_for_splitting} we list the
$\mathcal{O}(\alpha_s)$ splitting kernels $\prn{R}{T}_{a_0\to a_1 a_2}$ of
DTMDs in the small-$\tvec{y}$ region, adapting the results of
\cite{Diehl:2011yj} to the notation used here.}

\subsection{TMD hard-scattering cross sections}
\label{sec:TMD_hard-scattering}

We define the coefficients $C_H^2$ of the hard-scattering cross section as
\begin{align}
  \label{hard-coeff}
  \hat{\sigma}(Q^2,\mu^2) = \hat{\sigma}_{0}(Q^2)\,
  C_{H}^2(Q^2,\mu^2)
\end{align}
for a given partonic channel, where $\hat{\sigma}_0$ is the cross section at
lowest order in $\alpha_s$.  The expressions for $\hat{\sigma}_0$ can be found
in many places, for instance in \cite{Kasemets:2012pr} for the production of
an electroweak gauge boson $V$ and its subsequent decay into a lepton pair,
and in \cite{Echevarria:2015uaa} for Higgs production via $gg$ fusion in
the limit of large top mass.  The hard-scattering coefficients read
\begin{align}
  \label{hard-coeffs}
C_{H,\, q\bar{q}\to V}^2(Q^2,\mu^2) & = 1 + \frac{\alpha_s(\mu)}{2\pi}\,
  C_F\, \biggl[
   - \ln^2\frac{Q^2}{\mu^2} + 3 \ln\frac{Q^2}{\mu^2} - 8 + \frac{7\pi^2}{6}
  \ms\biggr]\, ,
\nonumber\\
C_{H,\, gg\to H}^2(Q^2,\mu^2) & = \biggl(
   1 + \frac{\alpha_s(\mu)}{2\pi}\, C_A\,
   \biggr[ - \ln^2\frac{Q^2}{\mu^2} + \frac{7\pi^2}{6} \ms\biggr] \biggr)\,
   \biggl( 1 + \frac{\alpha_s(\mu)}{2\pi}\, \bigl[5 C_A - 3 C_F\bigr]
   \biggr)
\end{align}
and are known up to $\mathcal{O}(\alpha_s^3)$, see for instance appendix~A of
\cite{Ebert:2017uel}.  The colour factors are $C_F = (N^2-1) /(2N)$ and $C_A
= N$ as usual.  The first factor in $C_{H,\, gg\to H}^2$ is for two-gluon
fusion via the local current $F^{\mu\nu\ms a} F_{\mu\nu}^a$ and the second
factor for the coupling of this current to the Higgs boson via a top quark
loop.  Throughout this section we drop the explicit indication of
higher-order corrections in equations, omitting a term $\!{}+
\mathcal{O}(\alpha_s^2)$ in the first line of \eqref{hard-coeffs} and so
forth.

An important property of the coefficients $C_H^2$ for Drell-Yan and Higgs
production is their independence of the incoming parton polarisations.  This
holds at all orders in $\alpha_s$.  For $gg\to H$ it follows from the fact
that angular momentum conservation only allows for two non-vanishing helicity
amplitudes, which are related by parity invariance.  Therefore, all
$\alpha_s$ corrections can be expressed by a single hard-scattering
coefficient.  The argument holds for the production of any scalar or
pseudoscalar boson via $gg$ fusion.

For $q\bar{q}$ annihilation into a boson via the vector current, one
additionally needs the conservation of quark chirality, which together with
angular momentum and parity conservation leaves a single independent helicity
amplitude.  To appeal to quark chirality conservation in a calculation using
dimensional regularisation, one can adapt the argument given in
\cite{Collins:1999un}.  As long as one works in $D = 4-2\epsilon$ dimensions,
one can avoid specifying the polarisation of the incoming partons and instead
compute the hard scattering as a matrix in Dirac space.  Chirality
conservation then technically means that the hard-scattering amplitude
contains an odd number of $\gamma_\mu$ matrices at any perturbative order.
The same holds for the terms that must be subtracted to remove ultraviolet and
infrared divergences.  After these subtractions have been done, one can take
$D=4$ and use the relation between quark helicity and chirality.

The extension of this argument to $q\bar{q}$ annihilation via the axial vector
current is simple in $D=4$ dimensions: anticommuting the $\gamma_5$ matrix
from the current, one finds $A_\mu = V_\mu \gamma_5$ for the Dirac matrix of
the hard-scattering amplitudes of the two currents (with their divergences
subtracted).  Their independent helicity amplitudes are hence the same (or
opposite in sign).  To complete the argument, one needs to discuss the
implementation of the electroweak axial current in dimensional regularisation,
which we shall not do here.

\subsection{Collins-Soper kernels and anomalous dimensions}
\label{sec:CS_kernels}

In the limit of small $\tvec{z}_1$, $\tvec{z}_1$ and $\tvec{y}$, the soft
matrix for DPS can be directly computed in perturbation theory, as we
derived in \eqref{soft-fact-small-y}.  Setting $U = \one$ in \eqref{K-def},
we simply have $K = \widehat{K} = d S/ dY$ at $\mathcal{O}(\alpha_s)$.  The
one-loop calculation of $S$ in the $qq$ channel is presented in detail in
section~3.3.2 of \cite{Diehl:2011yj}.  We can rewrite the results there
  in such a way that we have a separate dependence on the two scales $\mu_1$
  and $\mu_2$.  Splitting off a diagonal part according to
\eqref{split-K-M}, we get
\begin{align}
\label{K-qq-LO}
K_{qq} &= C_F\ms \bigl[ K(\tvec{z}_1;\mu_1) + K(\tvec{z}_2;\mu_2) \bigr]
	\begin{pmatrix} 1~ & 0 \\ 0~ & 1 \end{pmatrix} + M_{qq}
\end{align}
with
\begin{align}
\label{M-qq-LO}
M_{qq} &= \begin{pmatrix}
  0 & \frac{\sqrt{N^2-1}}{2N}\, K_d \\[0.3em]
  \frac{\sqrt{N^2-1}}{2N}\, K_d ~
  & {}- \frac{N}{2}\ms K_y - \frac{1}{N}\ms K_d\,
\end{pmatrix}\, ,
\end{align}
where the matrices are in colour representation space, with $R = (1, 8)$. We
recall that $M_{qq}$ is renormalisation scale independent.  The scale
independent combinations of kernels read
\begin{align}
\label{Ky-Kd-def}
K_y(\tvec{z}_i,\tvec{y}) &= K(\tvec{z}_1;\mu) + K(\tvec{z}_2;\mu) -
K\Bigl( \tvec{y} + \half (\tvec{z}_1+\tvec{z}_2); \mu \Bigr) - K\Bigl(
\tvec{y} - \half (\tvec{z}_1+\tvec{z}_2); \mu \Bigr) \, , \nonumber
\\ K_d(\tvec{z}_i,\tvec{y}) &= K\Bigl( \tvec{y} + \half
(\tvec{z}_1+\tvec{z}_2); \mu \Bigr) + K\Bigl( \tvec{y} - \half
(\tvec{z}_1+\tvec{z}_2); \mu \Bigr) \nonumber \\ &\quad - K\Bigl( \tvec{y}
+ \half (\tvec{z}_1-\tvec{z}_2); \mu \Bigr) - K\Bigl( \tvec{y} - \half
(\tvec{z}_1-\tvec{z}_2); \mu \Bigr) \, ,
\end{align}
where
\begin{align}
	\label{eq:K-def_in_z_mu}
K(\tvec{z};\mu) &= - \frac{\alpha_s(\mu)}{\pi}\,
\log\frac{\tvec{z}^2\mu^2}{b_0^2}
\end{align}
is the one-loop Collins-Soper kernel for single parton TMDs, with the colour
factor removed.  We recall that $b_0 = 2 e^{-\gamma_E}$.
We note that the two-loop result in section~4 of \cite{Vladimirov:2016qkd}
has the form given by \eqref{K-qq-LO}, \eqref{M-qq-LO} and \eqref{Ky-Kd-def}
if one replaces $C_F\ms K(\tvec{z};\mu)$ with the two-loop Collins-Soper kernel
for a single quark TMD.

In the solution \eqref{DTMD-evolved} of the evolution equations, one needs
the matrix exponential of $M_{qq}$ times $L = \log\sqrt{\zeta /\zeta_0}$,
which reads (see section 3.4.2 of \cite{Diehl:2011yj})
\begin{align}
\label{eq:solexpML}
\exp[\ms L M_{qq} \ms] &= \frac{1}{d_+ - d_-}
\begin{pmatrix} d_+ e^{L d_-} - d_-
  e^{L d_+} ~ & \frac{\sqrt{N^2-1}}{2N}\, K_d\, \bigl( e^{L d_+} - e^{L
    d_-} \big) \, \\[0.3em]
  \frac{\sqrt{N^2-1}}{2N}\, K_d\, \bigl( e^{L d_+} - e^{L
    d_-} \big) ~ & d_+ e^{L d_+} - d_- e^{L d_-}
\end{pmatrix}
\end{align}
with
\begin{align}
d_{\pm} &= \frac{1}{2N} \Biggl[ - \frac{N^2}{2} K_y - K_d \pm \sqrt{ \biggl(
    \frac{N^2}{2} K_y + K_d \biggr)^2 + (N^2-1) K_d^2 } \;\Biggr] \, .
\end{align}
An interesting situation arises if $|K_d|\ll \sqrt{N^2-1}\; |K_y|$, which
holds for instance in the large-$N$ limit.  As shown in section 3.4.2 of
\cite{Diehl:2011yj}, one then finds that $\pr{8}{F}(\zeta)$ is suppressed
compared with $\pr{1}{F}(\zeta)$ for $\zeta\gg \zeta_0$.  Note that the above
condition always holds in the limit $\tvec{z}_1 = \tvec{z}_{2} = \tvec{0}$,
where $K_d$ tends to zero.

It is easy to extend the above results to other parton channels, because the
one-loop graphs remain the same up to colour factors.  For the quark-gluon
channels, we obtain
\begin{align}
\label{K-qg-LO}
K_{qg}(\tvec{z}_i,\y;\mu_i) &= \bigl[\ms C_F\ms K(\tvec{z}_1;\mu_1)
  + C_A\ms K(\tvec{z}_2;\mu_2) \ms\bigr] \one + M_{qg}
\end{align}
with
\begin{align}
  \label{M-qg-LO}
M_{qg} & = \begin{pmatrix}
  0 & \frac{1}{\sqrt{2}}\ms K_d & 0
\\[0.4em]
  \frac{1}{\sqrt{2}}\ms K_d ~ & - \frac{N}{2}\ms (K_y + \half K_d)
  & \frac{\sqrt{N^2-4}}{4}\ms K_d
\\[0.4em]
  0 & \frac{\sqrt{N^2-4}}{4}\ms K_d
  & - \frac{N}{2}\ms (K_y + \half K_d) \,
\end{pmatrix}\, .
\end{align}
Here the space of colour representations is $R = (1,A,S)$, corresponding to
the projectors in \eqref{mixed-proj}, and $\one$ is the unit matrix in that
space.
For the two-gluon channel we get the kernel
\begin{align}
  \label{K-gg-LO}
K_{gg}(\tvec{z}_i,\y;\mu_i) &= C_A\ms \bigl[ K(\tvec{z}_1;\mu_1) +
  K(\tvec{z}_2;\mu_2) \bigr] \one + M_{gg}
\end{align}
with
\begin{align}
  \label{M-gg-LO}
  M_{gg} & = \begin{pmatrix}
    0 & \frac{N}{\sqrt{N^2 -1}}\ms K_d & 0 & 0 & 0
  \\[0.3em]
    \frac{N}{\sqrt{N^2-1}}\ms K_d ~~ & - \frac{N}{2}\ms (K_y + \half K_d) &
    \frac{N}{4}\ms K_d & 0 & \sqrt{\frac{3}{8}}\ms K_d
  \\[0.3em]
    0 & \frac{N}{4}\ms K_d &
    - \frac{N}{2}\ms (K_y + \half K_d) & - \frac{3}{\sqrt{10}}\ms K_d & 0
  \\[0.3em]
    0 & 0 & - \frac{3}{\sqrt{10}}\ms K_d
    & - 3 (K_y + \half K_d) & - \sqrt{\frac{3}{5}}\ms K_d
  \\[0.3em]
    0 & \sqrt{\frac{3}{8}}\ms K_d & 0
    & - \sqrt{\frac{3}{5}}\ms K_d & - 4 (K_y + \half K_d) \,
\end{pmatrix}
\end{align}
in the space of colour representations $R = (1,A,S,D,27)$, corresponding to
the projectors in \eqref{gluon-proj}.  In the rows and columns for $D$ and
$27$, we have given the numerical coefficients for $N=3$.
The Collins-Soper kernels in channels with antiquarks can be obtained from the
above expressions by using the relation \eqref{K-anti} and the analogue of
\eqref{S-qbar} for ${K}_{a_1 a_2}$.

The matrices $M_{qg}$ and $M_{gg}$ are non-singular for generic values of the
functions $K_y$ and $K_d$.  The eigenvalues of $M_{qg}$ can be calculated
analytically, but the expressions are rather lengthy and we refrain from
giving them here.  For $N=3$, the eigenvalues of $M_{gg}$ read
\begin{align}
- 3 \bigl( K_y + \half K_d \bigr) \,,
&& - \tfrac{3}{2} K_y \,,
&& - \tfrac{3}{2} ( K_y + K_d ) \,,
&& - (2 K_y + K_d) \pm \sqrt{  (2 K_y + K_d)^2 + 3 K_d^2 } \,.
\end{align}

\subsubsection{Limit of small \texorpdfstring{$\tvec{z}_1$}{z1} and
  \texorpdfstring{$\tvec{z}_2$}{z2}}
\label{sec:anom-dims}

We now consider the limit where $|\tvec{z}_1|, |\tvec{z}_2| \ll |\tvec{y}|$.
Expanding \eqref{K-qq-LO}, \eqref{K-qg-LO} and \eqref{K-gg-LO} in powers of
$|\tvec{z}_1| /|\tvec{y}|$ and $|\tvec{z}_2| /|\tvec{y}|$ and comparing the
leading terms with the structure in \eqref{CS-gen-match}, where the full
kernel $\prb{RR'}{K_{a_1 a_2}(\tvec{z}_i,\tvec{y}; \mu_i)}$ is split into
three separate contributions, we can deduce the LO expressions of the kernels
$\pr{R}{K}_{a}$ and $\prn{R}{J}$.  We obtain
\begin{align}
\label{RK-LO}
\pr{1}{K}_q(\tvec{z};\mu) &= C_F\ms K(\tvec{z};\mu) \,, &
\pr{8}{K}_q(\tvec{z};\mu) &= - \frac{1}{2N}\ms K(\tvec{z};\mu) \,,
\nonumber\\[0.2em]
\pr{1}{K}_g(\tvec{z};\mu) &= C_A\ms K(\tvec{z};\mu) \,, &
\pr{A}{K}_g(\tvec{z};\mu) &= \pr{S}{K}_g(\tvec{z};\mu) = \frac{N}{2}\ms
K(\tvec{z};\mu)
\intertext{and}
\label{RJ-LO}
\pr{1}{J} &= 0 \,, & \pr{8}{J}(\tvec{y};\mu_i) &= \frac{N}{2}\bigl[
  K(\tvec{y};\mu_1) + K(\tvec{y};\mu_2) \bigr]
\end{align}
with $K(\tvec{z};\mu)$ defined in \eqref{eq:K-def_in_z_mu}.  For the higher
gluon representations we find
\begin{align}
\pr{D}{K}_g(\tvec{z};\mu) &= 0 \,, & \pr{27}{K}_g(\tvec{z};\mu) &= -
K(\tvec{z};\mu)
\intertext{and}
\label{J-LO-higher}
\pr{D}{J}(\tvec{z};\mu_i) &= 3 \bigl[ K(\tvec{y};\mu_1) + K(\tvec{y};\mu_2)
  \bigr]\,, &
\pr{27}{J}(\tvec{z};\mu_i) &= 4 \bigl[ K(\tvec{y};\mu_1) +
  K(\tvec{y};\mu_2) \bigr]\,,
\end{align}
where again we have set $N=3$.  The expressions for $J$ are of course only
valid at perturbatively small $\tvec{y}$.

Using the explicit form of $K(\tvec{z};\mu)$, we obtain anomalous dimensions
\begin{align}
 \pr{1}{\gamma}_{K,q} &= 2 C_F \,\frac{\alpha_s}{\pi} \, ,
&
 \pr{8}{\gamma}_{K,q} &= - \frac{1}{N}\, \frac{\alpha_s}{\pi}\, ,
\nonumber \\[0.3em]
 \pr{1}{\gamma}_J &= 0 \,,
&
 \pr{8}{\gamma}_J &= N \,\frac{\alpha_s}{\pi}\,.
\end{align}
The expressions for other colour representations are readily obtained from
\eqref{RK-LO} and \eqref{RJ-LO}.  We note that $\pr{8\,}{\gamma}_{K,q}$ and
$\pr{27\,}{\gamma}_{K,g}$ are negative.  Furthermore, we find that
$\pr{R\,}{\gamma}_J > 0$ for all $R \neq 1$.

For completeness we also give the anomalous dimensions $\gamma_{a}$, which
can be taken from the literature for single TMDs, e.g.\ from
\cite{Echevarria:2016scs} (see our equation~\eqref{notation-dict} for
notation).  The one-loop expressions are
\begin{align}
\label{eq:anomdim_Fqg}
\gamma_{q}
  &= \frac{3}{2} \ms  C_F \ms \frac{\alpha_s}{\pi}\, ,
&
\gamma_{g}
  &= \frac{\beta_0}{2}\, \frac{\alpha_s}{\pi}
\end{align}
with $\beta_0 = \frac{11}{3}\ms C_A - \frac{2}{3}\ms n_F$.
Finally, one has
\begin{align}
\gamma_{K,q} &= 2 C_F\ms \frac{\alpha_s}{\pi} + \frac{C_F}{2}\ms
\biggl( \frac{67 - 3 \pi^2}{9}\ms C_A - \frac{10}{9}\ms n_F \biggr) \biggl(
\frac{\alpha_s}{\pi} \biggr)^2\, , \nonumber \\
\gamma_{K,g} &=
\frac{C_A}{C_F}\, \gamma_{K,q}
\end{align}
for the cusp anomalous dimension at two-loop accuracy.  Using the all-order
relation
\begin{align}
\pr{8}{\gamma}_J &= \frac{1}{2}\, \gamma_{K,g} \,,
\end{align}
which follows from \eqref{J-octet}, we readily get the two-loop expression of
$\pr{8}{\gamma_J}$ as well.

\subsection{Matching coefficients for DTMDs}
\label{sec:Coefficient_functions}

We now turn to the short-distance coefficients for matching DTMDs on DPDFs in
the large-$\tvec{y}$ regime.  In the colour singlet channel, they are equal to
the coefficients for matching TMDs on PDFs, which for most polarisation
combinations have been computed by several
groups~\cite{Collins:2011zzd,Aybat:2011zv,%
Bacchetta:2013pqa,Echevarria:2015uaa,Gutierrez-Reyes:2017glx}.  In addition to
extending these results to colour non-singlet channels, we have computed the
colour singlet coefficients for all polarisation channels.  \rev{In the
following two subsections, we describe in detail our method, which allows us
to compute the one-loop matching coefficients from real emission graphs alone
and to handle gluon polarisation without using the Levi-Civita tensor in
$D= 4-2\epsilon$ dimensions.  Subtleties of renormalisation in both the quark
and gluon sector are discussed in section~\ref{sec:renorm}.  We give our final
results for the colour singlet coefficients in section~\ref{sec:oneloop-sing}
and explain their extension to other colour channels in
section~\ref{sec:oneloop-nonsing}.}

\subsubsection{General procedure}
\label{sec:gen-proc}

To compute the matching coefficients, we take matrix elements of the operator
relations in \eqref{soft-op-proj} and \eqref{x-op-match}, using either the
vacuum or a single parton as external states.  We define
\begin{align}
  \label{gen-matel}
2\pi \delta(p^+ - p'^+)\, 2p^+ \,
\pr{R}{\mathcal{M}}_{ab}(x,\tvec{z})
& = \frac{1}{\mathcal{N}_b}\, \frac{1}{m(R)}\,
   \bigl\langle b,p',r' \big|\ms \prn{R}{O}^{\ms\ul{r}}_a(x,\tvec{y},\tvec{z})
   \ms\big| b,p,r \bigr\rangle
\nonumber \\[0.2em]
&= \ii{a}{R}\; \frac{\mathcal{N}_a}{\mathcal{N}_b}\,
   \frac{P_R^{\ul{r}\, \ul{s}}}{m(R)}\,
   \bigl\langle b,p',r' \big|\ms O^{\ms\ul{s}}_a(x,\tvec{y},\tvec{z})
   \ms\big| b,p,r \bigr\rangle \,,
   \hspace{4em}
\nonumber \\[0.3em]
\pr{R}{\mathcal{M}}_{S,a}(\tvec{z})
&= \frac{P_R^{\ul{r}\, \ul{s}}}{m(R)}\, \bigl\langle 0 \big|\ms
     O_{S,a}^{\ms \ul{r}, \ul{s}}(\tvec{y},\tvec{z})
     \ms\big|\ms 0 \bigr\rangle \,,
\end{align}
where $|\ms b,p,r \rangle$ is a parton state $b$ with momentum $p$ and colour
index $r$, and the colour projected operator $\prn{R}{O}^{\ms\ul{r}}_a$ was
introduced in \eqref{col-proj-op}.  We recall that $\ul{r} = (r,r')$ and that
repeated colour indices are summed over.  We gloss over the issue of parton
polarisation for the time being.
In the colour singlet sector, one has $\ii{a}{1} /m(1) = 1$ and the
contraction with $P_1^{\ul{r}\, \ul{s}}\, \mathcal{N}_a / \mathcal{N}_b$ in
$\pr{1}{\mathcal{M}}_{ab}$ implies a sum over the colour indices of the
operator and an average over the colour of the parton state.
Thus, $\pr{1}{\mathcal{M}}_{ab}(x,\tvec{z})$ is the
unsubtracted TMD for parton $a$ in target $b$ according to~\eqref{x-matel}.
Likewise, $\pr{1}{\mathcal{M}}_{S,a}(\tvec{z})$ is the soft factor for single
TMDs.  In analogy to \eqref{gen-matel} we define matrix elements
$\pr{R}{\mathcal{M}}_{ab}(x)$ and $\pr{R}{\mathcal{M}}_{S,a}$ without argument
$\tvec{z}$ from the operators with $\tvec{z}=\tvec{0}$; thus
$\pr{1}{\mathcal{M}}_{ab}(x)$ is the PDF for parton $a$ in target $b$.  It is
understood that all operators are renormalised, and we recall that the
operators at $\tvec{z}=\tvec{0}$ require additional renormalisation compared
with those at finite $\tvec{z}$.

According to the operator matching equations \eqref{x-op-match} and
\eqref{soft-op-proj}, we thus have
\begin{align}
  \label{match-matel}
\pr{R}{\mathcal{M}}^{}_{ac}(x,\tvec{z})
&= \sum_{b} \prn{R}{C}^{}_{\us,ab}(x',\tvec{z})\,
   \underset{x}{\otimes} \pr{R}{\mathcal{M}}^{}_{bc}(x') \,,
\nonumber \\
\pr{R}{\mathcal{M}}^{}_{S,a}(\tvec{z})
&= \prn{R}{C}_{S,a}^{}(\tvec{z})\, \pr{R}{\mathcal{M}}^{}_{S,a} \,.
\end{align}
We now expand the quantities in these relations up to
$\mathcal{O}(\alpha_s)$, denoting the zeroth and first orders with
superscripts $(0)$ and $(1)$, respectively.  At tree level, the matrix
elements $\langle b,p',r' \ms|\ms O^{\ms\ul{s}}_a \ms| b,p,r \rangle$ and
$\langle 0 |\ms O_{S,a}^{\ms \ul{r}, \ul{s}} \ms| 0 \rangle$ have the colour
structure $\delta_{rs}\, \delta_{r's'}$, which gives
\begin{align}
\pr{R}{\mathcal{M}}^{(0)}_{ab}(x,\tvec{z})
&= \pr{R}{\mathcal{M}}^{(0)}_{ab}(x) = \delta_{ab}\, \delta(1-x) \,,
&
\pr{R}{\mathcal{M}}^{(0)}_{S,a}(\tvec{z})
&= \pr{R}{\mathcal{M}}^{(0)}_{S,a} = 1
\end{align}
for the matrix elements on both sides of \eqref{match-matel}, which results in
zeroth-order matching coefficients
\begin{align}
\prn{R}{C}_{\us,ab}^{(0)}(x,\tvec{z}) &= \delta_{ab}\, \delta(1-x) \,,
&
\prn{R}{C}_{S,a}^{(0)}(\tvec{z}) &= 1 \,.
\end{align}
Inserting this into the $\mathcal{O}(\alpha_s)$ terms of \eqref{match-matel},
we obtain
\begin{align}
\prn{R}{C}_{\us,ab}^{(1)}(x,\tvec{z})
&= \pr{R}{\mathcal{M}}^{(1)}_{ab}(x,\tvec{z})
   - \pr{R}{\mathcal{M}}^{(1)}_{ab}(x) \,,
\nonumber \\
\prn{R}{C}_{S,a}^{(1)}(\tvec{z})
&= \pr{R}{\mathcal{M}}^{(1)}_{S,a}(\tvec{z})
   - \pr{R}{\mathcal{M}}^{(1)}_{S,a} \,.
\end{align}
The one-loop matching coefficients are thus given as the difference of
$\mathcal{O}(\alpha_s)$ matrix elements at finite and at zero $\tvec{z}$.
Virtual graphs cancel in this difference, because they do not depend on
$\tvec{z}$, so we can limit our calculation to real graphs.

To obtain the matching coefficient $\prn{R}{C}_{a b}$ of the TMD, we need the
square root of $\prn{R}{C}_{S,a}$ according to \eqref{full-C-def}.  At
$\mathcal{O}(\alpha_s)$, we have
\begin{align}
\sqrt{ \prn{R}{C}_{S,a}^{(1)}(\tvec{z}; 2Y) } &=
1 + \frac{1}{2}\, \prn{R}{C}_{S,a}^{(1)}(\tvec{z}; 2Y)
= 1 + \prn{R}{C}_{S,a}^{(1)}(\tvec{z}; Y) \,.
\end{align}
In the second step, we have used that $\prn{R}{C}_{S,a}^{(1)}(\tvec{z}; Y)$ is
linear in $Y$, which readily follows from its evolution equation
\eqref{C-soft-CS} and the fact that $\pr{R}{K}_a = \mathcal{O}(\alpha_s)$.
Inserting this into \eqref{full-C-def}, we thus obtain at
$\mathcal{O}(\alpha_s)$
\begin{align}
	\label{full-C-NLO2}
& \prn{R}{C_{a b}(x, \tvec{z}; \zeta)} = \delta_{ab}\, \delta(1-x)\,
\nonumber \\
&\qquad + \lim_{Y_L\to -\infty} \Bigl[\ms
          \prn{R}{C^{(1)}_{\text{us}, a b}(x,\tvec{z}; Y_L)}
          - \delta_{ab}^{}\, \delta(1-x)\, \prn{R}{C^{(1)}_{S,
              a}(\tvec{z}; Y_C-Y_L)} \ms\Bigr] \,,
\end{align}
where in the present section we use the definition \eqref{zeta-def-TMD} of
$\zeta$, which refers to a single parton with plus-momentum $x p^+$.  We have
omitted the argument $\mu$ in all functions for brevity.

\subsubsection{Calculation of gluon-gluon matching coefficients}
\label{sec:gg-coeffs}

\begin{figure}
\centering
\subfigure[]{\includegraphics[height=0.19\textwidth]{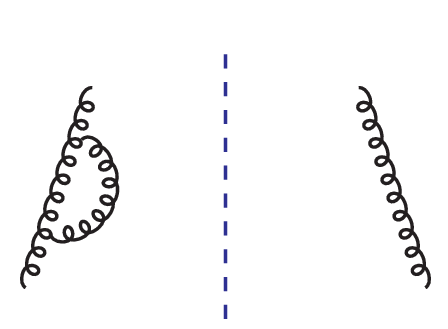}}
\hspace{2.2em}
\subfigure[]{\includegraphics[height=0.19\textwidth]{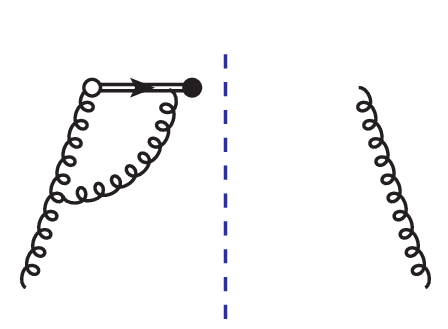}}
\hspace{2.2em}
\subfigure[]{\includegraphics[height=0.19\textwidth]{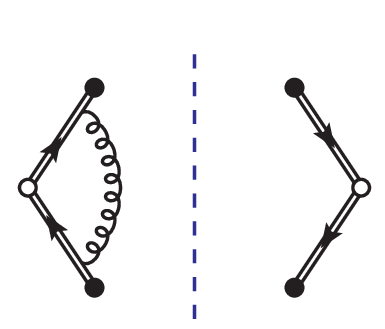}}
\caption{\label{fig:Cgg-virt} Virtual graphs contributing to the matrix
elements $\mathcal{M}^{(1)\, jj'\!, ii}_{gg}$ (a and b) and
$\mathcal{M}_{S,g}^{(1)}$ (c).  Not shown are complex conjugate graphs and
the analogue of (a) with a quark instead of a gluon loop.
All eikonal lines are in the adjoint representation.}

\vspace{2em}

\subfigure[]{\includegraphics[height=0.19\textwidth]{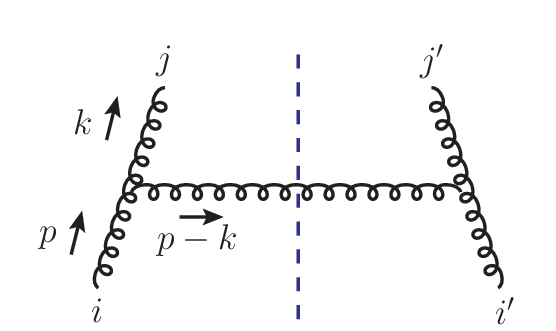}}
\hspace{1.9em}
\subfigure[]{\includegraphics[height=0.19\textwidth]{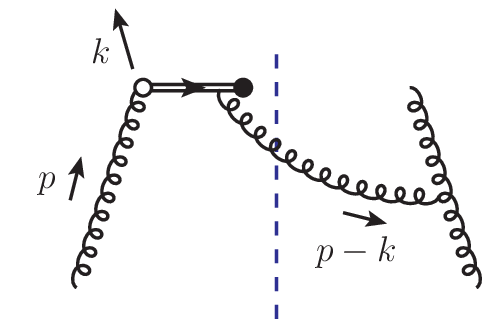}}
\hspace{1.9em}
\subfigure[]{\includegraphics[height=0.19\textwidth]{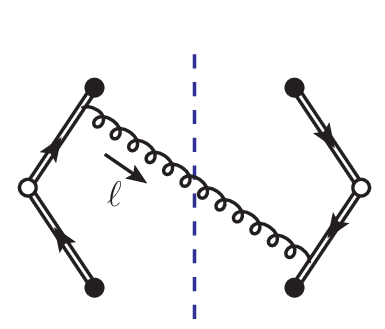}}
\caption{\label{fig:Cgg-real} Real graphs contributing to the matrix elements
$\mathcal{M}^{(1)\, jj'\!, ii}_{gg}$ (a and b) and $\mathcal{M}_{S,g}^{(1)}$
(c).  Not shown are the complex conjugates of graphs (b) and (c).}
\end{figure}

We now describe in detail our procedure to compute the one-loop matching
coefficients in the pure gluon channel.  Throughout this subsection, we
consider the colour singlet sector, $R=1$, and drop the corresponding index.
In turn, we now make gluon polarisation explicit.  We take gluon operators
$O_g^{jj'}(x,\tvec{y},\tvec{z})$ with open polarisation indices, obtained from
\eqref{x-ops} and \eqref{eq:gluon-ops} by dropping the projection operator
$\Pi^{jj'}_a$ in the latter equation.  We also take open polarisation indices
$i$, $i'$ for the gluons in the corresponding matrix element
\eqref{gen-matel}, which thus carries four Lorentz indices,
$\mathcal{M}^{jj'\!, ii'}_{gg}$.  All of them are restricted to be in the
$D-2$ transverse dimensions.  The subscripts $g$ refer of course only to the
parton type here, and not to its polarisation.  For the gluon matrix elements,
we thus have
\begin{align}
  \label{match-gg-matel}
\mathcal{M}^{jj'\!, ll'}_{gg}(x,\tvec{z})
&= C^{jj'\!, ii'}_{\us, gg}(x',\tvec{z})\,
   \underset{x}{\otimes} \mathcal{M}^{ii'\!, ll'}_{gg}(x')
   + \text{\{quark-gluon mixing terms\}}
\end{align}
and
\begin{align}
  \label{gluon-lo-tens}
\mathcal{M}^{(0)\, jj'\!, ii}_{gg}(x,\tvec{z})
&= \mathcal{M}^{(0)\, jj'\!, ii'}_{gg}(x)
 = C^{(0)\, jj'\!, ii}_{gg}(x,\tvec{z})
 = \delta^{ij}\, \delta^{i'j'} \delta(1-x) \,,
\end{align}
which is combined with the matching of the soft factor into
\begin{align}
  \label{match-gg-combine}
C^{(1)\, jj'\!, ii}_{gg}(x,\tvec{z}; \zeta)
 &= \lim_{Y_L\to -\infty} \Bigl[\ms
       C^{(1)\, jj'\!, ii}_{\text{us}, gg}(x,\tvec{z}; Y_L)
       - \delta^{ij}\, \delta^{i'j'} \delta(1-x)\, C^{(1)}_{S,
              a}(\tvec{z}; Y_C-Y_L) \ms\Bigr] \,.
\end{align}
The virtual graphs contributing to the one-loop matrix elements are shown in
figure~\ref{fig:Cgg-virt}.  They are independent of $\tvec{z}$ (the Fourier
conjugated distance to the transverse momentum carried by partons or eikonal
lines) and hence cancel in the matching coefficient.  Therefore we only
compute the real graphs in figure~\ref{fig:Cgg-real}.  We omit so-called
Wilson line self interactions, where a gluon is exchanged between eikonal
lines along the same direction $v$, referring to chapter~13.7 of
\cite{Collins:2011zzd} for further discussion.

We work in Feynman gauge and compute cut graphs (where the emitted gluon is
explicitly put on shell), using the Feynman rules given in
appendix~\ref{sec:Feynman_rules}.  For graph~\ref{fig:Cgg-real}b, we then
obtain
\begin{align}
& C_{\text{us},gg}^{(1)\,jj^\prime,\,ii^\prime}(x,\tvec{z}; Y_L\to -\infty)
  \ms\big|_{\text{fig.~\protect\ref{fig:Cgg-real}b}}
= \frac{1}{x p^{+}} \int\!
  \frac{d^{4-2\epsilon} k}{(2\pi)^{4-2\epsilon}} \,
  \bigl(e^{i\tvec{k}\tvec{z}} - 1\bigr) \;
  2\pi\delta\big( (p-k)^2 \big)\, \delta(k^+ - x p^+)\,
\nonumber \\
& \qquad \times
\frac{\delta_{aa^\prime} \delta_{d e^\prime}}{N^2-1} \;
\bigl( g\mu^\epsilon n^\nu f^{dbc} \bigr) \;
\bigl( - g_{\nu\nu^\prime}\ms \delta_{b b^\prime} \bigr) \;
\frac{i\delta_{c^\prime d^\prime}\ms g_{\rho^\prime \tau^\prime}}{k^2-i0} \,
\frac{i \delta_{ce}}{(p-k)\ms n-i0}
\nonumber \\
& \qquad \times
  \Bigl( g\mu^{\epsilon}f^{a^{\prime}b^{\prime}c^{\prime}} \,
  \bigl[ (p-2k)^{i^{\prime}}g^{\nu^{\prime}\rho^{\prime}} +
    (k+p)^{\nu^{\prime}}g^{i^{\prime}\rho^{\prime}} +
    (k-2p)^{\rho^{\prime}}g^{i^{\prime}\nu^{\prime}} \bigr] \Bigr)
\nonumber \\[0.4em]
& \qquad \times
 \Bigl( -i \bigl[ k n \, g^{ij}-p^{j} n^{i} \bigr]\, \delta_{ae} \Bigr) \,
 \Bigl( i \ms \bigl[k n \, g^{\tau^\prime j^\prime}
   -k^{j^\prime}n^{\tau^\prime} \bigr]\, \delta_{d^\prime e^{\prime}} \Bigr)
  + C_{\us, \epsilon} \ms\big|_{\text{fig.~\protect\ref{fig:Cgg-real}b}} \,,
\label{e:coef_g/g_gen_pol_diag_b_calc_A}
\end{align}
where $n$ is the four-vector with $n^- = 1$, $n^+ = 0$ and $\tvec{n} =
\tvec{0}$.  The assignment of polarisation indices is shown in
figure~\ref{fig:Cgg-real}a.  All primed indices refer to the right of the
final-state cut, and $a, b, \ldots, e$ are colour octet indices.  The factor
$e^{i\tvec{k}\tvec{z}} - 1$ represents the difference between the matrix
element for nonzero and zero $\tvec{z}$.  In accordance with our choice of
light-cone coordinates, the components $p^-$ and $\tvec{p}$ of the target
momentum are zero.  The factor $(x p^+)^{-1}$ comes from the definition
\eqref{x-ops} of the operator $O_g$.  The ultraviolet counterterm $C_{\us,
  \epsilon}$ will be discussed below.

In \eqref{e:coef_g/g_gen_pol_diag_b_calc_A}, we have taken the limit $Y_L
\to -\infty$ by setting $v_L = n$ in the eikonal propagator.  (Only then can
we use the Feynman rule given in figure~\ref{fig:Feynman_rules_gluon}).
This gives a factor $p^+ - k^+ = p^+ (1-x)$ in the denominator.  The
resulting rapidity divergences in the convolution of $C_{\us,gg}$ with a PDF
or a DPDF are cancelled by corresponding divergences in the soft factor, as
we shall see shortly.  In this sense, one should understand the limit $Y_L
\to -\infty$ in $C_{\us, gg}$ as being taken in the combination
\eqref{match-gg-combine}.

Adding the complex conjugate of \eqref{e:coef_g/g_gen_pol_diag_b_calc_A} and
the expression for graph~\ref{fig:Cgg-real}a, we get
\begin{align}
& C_{\us, gg}^{(1)\, jj'\!, ii'}(x,\tvec{z}; Y_L\to -\infty)
 = \frac{\alpha_s\ms C_A}{\pi^2}\, (2\pi \mu)^{2\epsilon}
\int \frac{d^{2-2\epsilon}\tvec{k}}{\tvec{k}^{2}} \,
\bigl( e^{i\tvec{k} \tvec{z}} - 1 \bigr)
\nonumber \\
& \qquad \times \Bigg\{\bigg[\frac{x}{1-x} +
\frac{(1-x)(1+x^2)}{x}\bigg]\,
\frac{1}{2(1-\epsilon)}\, \delta^{ii^{\prime}}\delta^{jj^{\prime}}
\nonumber \\
& \qquad\quad
+ \biggl[ \frac{x}{1-x} + \frac{2(1-x)}{1-\epsilon} \bigg]\,
\frac{1}{2}\ms \Big(\delta^{ij}\delta^{i^{\prime}j^{\prime}}
- \delta^{ij^{\prime}}\delta^{i^{\prime}j}\Big)
\nonumber \\
& \qquad\quad
+ \frac{x}{1-x}\; \frac{1}{2}\ms
\biggl(\delta^{ij}\delta^{i^{\prime}j^{\prime}} +
\delta^{ij^{\prime}}\delta^{i^{\prime}j} -
\frac{1}{1-\epsilon}\,
\delta^{ii^{\prime}}\delta^{jj^{\prime}} \biggr)
\nonumber \\
& \qquad\quad
+ x(1-x)\, \frac{k^{ii^{\prime}}}{\tvec{k}^2}\, \delta^{jj^{\prime}}
+ \frac{1-x}{x}\, \delta^{ii^{\prime}}\, \frac{k^{jj^{\prime}}}{\tvec{k}^2}
\nonumber \\
& \qquad\quad
+ (1-x)\, \biggl( \delta^{ij}\frac{k^{i^\prime j^\prime}}{\tvec{k}^2}
+ \delta^{i^\prime j^\prime}\, \frac{k^{ij}}{\tvec{k}^2}
- \delta^{ij^\prime}\, \frac{k^{i^\prime j}}{\tvec{k}^2}
- \delta^{i^\prime j}\, \frac{k^{ij^\prime}}{\tvec{k}^2}\biggr) \Bigg\}
+ C_{\us, \epsilon} \,,
\label{Cus-kspace}
\end{align}
where we define the symmetric and traceless tensor
\begin{align}
  \label{tens-def}
  v^{ij} &= \tvec{v}^{i} \tvec{v}^{j}
  - \frac{\tvec{v}^2}{2(1-\epsilon)}\; \delta^{ij}
\end{align}
for a given vector $\tvec{v}$ in $2(1-\epsilon)$ transverse dimensions.
Using the integral relations in appendix~\ref{sec:Useful_relations}, we
perform the transverse integration and obtain
\begin{align}
& C_{\us, gg}^{(1)\,jj'\!, ii'}(x,\tvec{z};-\infty)
= - \frac{\alpha_s\ms C_A}{\pi}\,
  (2\pi \mu)^{2\epsilon} \; \Gamma(1-\epsilon)\,
  \biggl( \frac{\tvec{z}^2}{4\pi} \biggr)^{\epsilon}
\nonumber\\
& \qquad \times
\Bigg\{ \, \frac{1}{\epsilon}\,
\biggl[ \frac{x}{1-x} + \frac{(1-x)(1+x^2)}{x} \biggr]\,
\frac{1}{2(1-\epsilon)}\, \delta^{ii^{\prime}}\delta^{jj^{\prime}}
\nonumber \\
& \qquad\quad
+ \frac{1}{\epsilon}\, \biggl[ \frac{x}{1-x} +
\frac{2(1-x)}{1-\epsilon} \biggr]\,
\frac{1}{2}\, \Big( \delta^{ij}\delta^{i^{\prime}j^{\prime}}
-\delta^{ij^{\prime}}\delta^{i^{\prime}j} \Big)
\nonumber \\
& \qquad\quad
+ \frac{1}{\epsilon}\, \frac{x}{1-x}\;
  \frac{1}{2}\ms \biggl( \delta^{ij}\delta^{i^{\prime}j^{\prime}}
+ \delta^{ij^{\prime}}\delta^{i^{\prime}j} -
\frac{1}{1-\epsilon}\, \delta^{ii^{\prime}}\delta^{jj^{\prime}} \biggr)
\nonumber \\
& \qquad\quad
+ x(1-x)\, \frac{z^{ii^{\prime}}}{\tvec{z}^2}\, \delta^{jj^{\prime}} +
\frac{1-x}{x}\, \delta^{ii^{\prime}}\, \frac{z^{jj^{\prime}}}{\tvec{z}^2}
\nonumber \\
& \qquad\quad
+ (1-x)\, \biggl( \delta^{ij}\, \frac{z^{i^\prime
      j^\prime}}{\tvec{z}^2} + \delta^{i^\prime
    j^\prime}\, \frac{z^{ij}}{\tvec{z}^2}
  - \delta^{ij^\prime}\, \frac{z^{i^\prime
      j}}{\tvec{z}^2} - \delta^{i^\prime
    j}\, \frac{z^{ij^\prime}}{\tvec{z}^2} \biggr) \Bigg\}
+ C_{\us, \epsilon}\, .
\label{Cus-zspace}
\end{align}
In \eqref{Cus-kspace} and \eqref{Cus-zspace} we have organised the tensor
structure into terms that are diagonal or antisymmetric or symmetric and
traceless in the index pair $jj'$, for reasons that will become clear in
section~\ref{sec:renorm}.  We note that the tensor in parentheses in the
last line of \eqref{Cus-zspace} is nonzero in a generic number $2 -
2\epsilon$ of transverse dimensions but vanishes for $\epsilon = 0$.  This
is easily seen by contracting this tensor with itself, which gives an
expression proportional to~$\epsilon$.

We see that \eqref{Cus-zspace} has poles in $1/\epsilon$.  They correspond
to ultraviolet divergences of the matrix element at $\tvec{z} = \tvec{0}$
and are removed by the counterterm $C_{\us, \epsilon}$.  If one splits the
$\tvec{k}$ integration in \eqref{Cus-kspace} into the parts going with
$e^{i\tvec{k}\tvec{z}}$ and $-1$, then the $1/\epsilon$ poles arise as
infrared divergences of the term with $e^{i\tvec{k}\tvec{z}}$, which is
ultraviolet finite.  The term with $-1$ involves scaleless integrals, in
which the ultraviolet and infrared divergences add up to zero in dimensional
regularisation.  In the sum of terms, the infrared divergences cancel and the
ultraviolet ones remain.  The $\overline{\text{MS}}$ counterterm reads
\begin{align}
C_{\us, \epsilon} =
\frac{\alpha_s\ms C_{A}}{\pi}\, \frac{S_{\epsilon}}{\epsilon}\;
& \Bigg\{ \bigg[ \frac{x}{1-x} + \frac{(1-x)(1+x^2)}{x} \bigg]\,
  \frac{1}{2(1-\epsilon)}\, \delta^{ii^{\prime}}\delta^{jj^{\prime}}
\nonumber \\
& + \bigg[\frac{x}{1-x} + 2(1-x) \bigg]\,
  \frac{1}{2}\, \Big(\delta^{ij}\delta^{i^\prime j^\prime} -
  \delta^{ij^\prime}\delta^{i^\prime j}\Big)
\nonumber \\
& + \frac{x}{1-x}\; \frac{1}{2}\,
  \biggl( \delta^{ij}\delta^{i^{\prime}j^{\prime}} +
\delta^{ij^{\prime}}\delta^{i^{\prime}j} - \frac{1}{1-\epsilon}\,
\delta^{ii^{\prime}}\delta^{jj^{\prime}} \biggr) \Bigg\} \,,
\label{Cus-counter}
\end{align}
where
\begin{align}
  \label{S-eps-def}
S_{\epsilon} = \bigl( 4\pi e^{-\gamma_E} \bigr)^{\epsilon} \,.
\end{align}
The origin of the factor $1/(1-\epsilon)$ multiplying
$\delta^{ii^{\prime}} \delta^{jj^{\prime}}$ is explained in
section~\ref{sec:renorm}; note that the naive implementation of
$\overline{\text{MS}}$ counterterms as ``the residue of $1/\epsilon$ times
$S_\epsilon$'' would lead to an incorrect result.

We now turn to the matching coefficient of the soft factor, which comes from
the graph in figure~\ref{fig:Cgg-real}c and its complex conjugate.  Taking
again the limit $Y_L \to -\infty$, we obtain
\begin{align}
& C_{S,g}^{(1)}(\tvec{z}; Y_C + \infty)
 = \int\frac{d^{4-2\epsilon}\ell}{(2\pi)^{4-2\epsilon}}\;
\bigl(e^{-i\tvec{\ell} \tvec{z}} - 1\bigr)\;
2\pi\delta\bigl(\ell^2\bigr)\, \theta\bigl(\ell^{+}\bigr)\;
\frac{\delta_{aa^\prime} \delta_{cc^\prime}}{N^2-1}
\nonumber \\[0.2em]
& \qquad \times \frac{i}{\ell\ms v_C + i0}\;
\bigl(-g \mu^{\epsilon} v_C^{\nu^\prime}
    f^{a^{\prime} b^{\prime} c^{\prime}} \bigr)\;
\bigl(-g_{\nu\nu^\prime}\ms \delta_{bb^{\prime}}\bigr)\;
\bigl(g \mu^{\epsilon} n^\nu f^{c b a}\bigr)\;
\frac{i}{\ell\ms n - i0} + \text{c.c.}
+ C_{S, \epsilon}
\nonumber \\[0.3em]
& \quad =
\frac{\alpha_s\ms C_A}{4 \pi^2}\, (2\pi \mu)^{2\epsilon}
\int d^{2-2\epsilon}\tvec{\ell}\;
\bigl(e^{-i\tvec{\ell} \tvec{z}} - 1\bigr)\,
\Bigg[\int_0^{\infty} \frac{d\ell^{+}}{\ell^{+}}\,
  \frac{2 v_C^{+}}{2 (\ell^{+})^2\, v_C^{-} + \tvec{\ell}^2\, v_C^{+} -i0}
\nonumber \\
& \hspace{18em}
- \frac{2}{\tvec{\ell}^2} \int_0^{x p^{+}} \frac{d\ell^{+}}{\ell^{+}} +
  \frac{2}{\tvec{\ell}^2} \int_0^{x p^{+}} \frac{d\ell^{+}}{\ell^{+}}
\ms\Bigg]+ \text{c.c.} + C_{S,\epsilon}
\nonumber \\[0.3em]
& \quad =
- \frac{\alpha_s\ms C_A}{2\pi^2}\, (2\pi \mu)^{2\epsilon}
\int \frac{d^{2-2\epsilon}\tvec{\ell}}{\tvec{\ell}^2}\;
\bigl(e^{-i\tvec{\ell} \tvec{z}} - 1\bigr) \,
\Bigg[ \log \frac{2 (x p^{+})^2\, e^{-2 Y_C}}{\tvec{\ell}^2} -
  2 \int_0^{x p^{+}} \frac{d\ell^{+}}{\ell^{+}} \ms\Bigg] +
  C_{S,\epsilon}\, ,
\label{e:F_g/g_diagram_S1_1}
\end{align}
where $C_{S,\epsilon}$ is the ultraviolet counterterm and ``c.c.'' denotes
the complex conjugate term (which results in an extra factor of $2$ in the
last line).  The eikonal propagator depending on~$n$ gives rise to the
divergent integral over $1/\ell^+$ in the third line.  To combine this
efficiently with the divergence from $C_{\us, gg}$, we have subtracted and
added a divergent integral in the fourth line.  The first two terms in
square brackets then combine to a finite integral, which gives the logarithm
in the last line.    Having rewritten $v_C^- / v_C^+ = - e^{-2Y_C}$, we
recognise the variable $\zeta$ from \eqref{zeta-def-TMD} appearing in
this logarithm.

In the remaining divergent integral, we now change variables as $\ell^+ =
p^+ (1-x')$, motivated by the corresponding form of the gluon plus-momentum
$p^+ - k^+ = p^+ (1-x)$ crossing the cut in figure~\ref{fig:Cgg-real}b.
Performing the Fourier transform using \eqref{e:FT_power} and
\eqref{e:FT_power_and_log}, we then obtain
\begin{align}
C_{S,g}^{(1)}(\tvec{z}; Y_C + \infty)
 &= - \frac{\alpha_s\ms C_{A}}{2\pi}\, (4\pi)^\epsilon\,
\biggl(\frac{\mu^2 \tvec{z}^2}{4}\biggr)^{\epsilon}\;
\frac{\Gamma(1-\epsilon)}{\epsilon}
\nonumber \\
& \quad \times
\Bigg[ \log \frac{4}{\zeta \tvec{z}^2}
   - \gamma_E + \psi(-\epsilon)
   + 2\int_{1-x}^{1} \frac{dx^\prime}{1-x^{\prime}}
\Bigg] + C_{S,\epsilon} \, ,
\label{e:S_g/g^[1]_c}
\end{align}
where $\psi(z) = \frac{d}{dz} \log\Gamma(z)$ is the digamma function.  Its
expansion around $\epsilon=0$ gives $\psi(-\epsilon) = 1/\epsilon +
\ldots\,$, so that overall we obtain a double pole in $\epsilon$.  We recall
that for a one-loop quantity that has the form
\begin{align}
\frac{R_{-2}}{\epsilon^2} + \frac{R_{-1}}{\epsilon} +
\mathcal{O}(\epsilon^0)\, , \label{e:expl_A}
\end{align}
the $\overline{\text{MS}}$-counterterm is given by
\begin{align}
  \label{double-ct}
- S_{\epsilon} \left(\frac{R_{-2}}{\epsilon^2} + \frac{R_{-1} -
    S_{1}\ms R_{-2}}{\epsilon}\right)
\end{align}
times the tree-level term $C_{S,g}^{(0)} = 1$, where the coefficient $S_1$
comes from the Taylor expansion $S_{\epsilon} = (4\pi
\exp(-\gamma_E))^{\epsilon} = 1 + S_{1}\ms \epsilon +
\mathcal{O}(\epsilon^2)$.  The additional contribution $S_{\epsilon}\ms
S_{1}\ms R_{-2} /\epsilon$ compensates the $1/\epsilon$ term that arises
from Taylor expanding $S_{\epsilon}\ms R_{-2} /\epsilon^2$.  Implementing
this prescription, we obtain the final form for the soft matching
coefficient:
\begin{align}
C_{S,g}^{(1)}(\tvec{z}; Y_C + \infty) & = -
\frac{\alpha_s\ms C_A}{2\pi}\, \bigg[ -\frac{1}{2} L^2 + L\,
  \log \frac{\mu^2}{\zeta} - \frac{\pi^2}{12} + 2L
  \int_{1-x}^{1}\frac{dx^\prime}{1-x^{\prime}} \bigg]
\label{e:S1_g/g_A}
\end{align}
with
\begin{align}
L & = \log \frac{\mu^2 \tvec{z}^2}{b_0^2} \,,
\label{e:S1_g/g_logarithm}
\end{align}
where $b_0$ is given in \eqref{b0-def}.  We note in passing that $C_{S,q}$ is
given by the same expression, with the colour factor $C_A$ replaced by $C_F$.
Defining
\begin{align}
  \label{double-log-def}
S_L &= -\frac{1}{2} L^2 +L\,\log \frac{\mu^2}{\zeta} - \frac{\pi^2}{12}
\end{align}
and combining $C_{\us,gg}$ with $C_{S,g}$ according to
\eqref{match-gg-combine}, we obtain the full matching coefficient in tensor
form:
\begin{align}
& C_{gg}^{(1)\,jj'\!, ii'}(x,\tvec{z}; \mu,\zeta)
  \nonumber \\
& \quad\! =
\frac{\alpha_s\ms C_A}{2\pi} \, \biggl\{
  - 2L\, \bigg[\ms \frac{x}{(1-x)_+} + \frac{(1-x)(1+x^2)}{x} \bigg]
  + S_L\, \delta(1-x) \ms\biggr\} \,
  \frac{1}{2}\, \delta^{ii^\prime} \delta^{jj^\prime}
\nonumber \\[0.2em]
& \qquad\! + \frac{\alpha_s\ms C_A}{2\pi} \, \biggl\{
  - 2L\, \bigg[\ms \frac{x}{(1-x)_+} + 2 (1-x) \bigg]
  - 4 (1-x) + S_L\, \delta(1-x) \ms\biggr\} \,
  \frac{1}{2}\, \Big(\delta^{ij}\delta^{i^\prime j^\prime} -
    \delta^{ij^\prime}\delta^{i^\prime j}\Big)
\nonumber \\[0.2em]
& \qquad\! + \frac{\alpha_s\ms C_A}{2\pi} \, \biggl\{
  - 2L\; \frac{x}{(1-x)_+} + S_L\, \delta(1-x) \ms\biggr\} \,
  \frac{1}{2}\, \Bigl( \delta^{ij}\delta^{i^\prime
  j^\prime} + \delta^{ij^\prime}\delta^{i^\prime j} -
  \delta^{ii^\prime} \delta^{jj^\prime} \Bigr)
\nonumber \\[0.2em]
& \qquad\! - \frac{\alpha_s\ms C_A}{2\pi}\;
  2 x(1-x)\, \frac{z^{ii^\prime}}{\tvec{z}^2}\, \delta^{jj^\prime}
  - \frac{\alpha_s\ms C_A}{2 \pi}\; \frac{2(1-x)}{x}\,\delta^{ii^\prime}\,
  \frac{z^{jj^\prime}}{\tvec{z}^2}\,.
\label{Cgg-final}
\end{align}
The rapidity divergences in the $C_{\us,gg}$ and $C_{S,g}$ have been
combined into the plus-distribution
\begin{align}
\frac{1}{(1-x)_{+}} & = \frac{1}{1-x} -
\delta(1-x)\int_{0}^{1}\frac{dx^\prime}{1-x^{\prime}} \,,
   \label{e:plusdistribution}
\end{align}
which gives a finite result in convolution integrals.  The lower integration
limit of the $x'$ integral has changed from $1-x$ in \eqref{e:S1_g/g_A} to
$0$ in \eqref{e:plusdistribution} because $C_{S,g}$ is multiplied with
$\delta(1-x)$ in the combination formula \eqref{match-gg-combine}.

Note that the method we have used avoids introducing the antisymmetric
tensor $\epsilon_{\lambda\mu\nu\rho}$ in $D=4-2\epsilon$ dimensions.  With
the final result \eqref{Cgg-final} in the $D=4$ physical dimensions, it is
straightforward to perform a projection onto unpolarised or longitudinally
polarised gluons.  On the operator side, one uses the spin projectors
$\delta^{jj'}$ or $i \epsilon^{jj'}$ from \eqref{eq:gluon-proj}, and for the
gluon states in the matrix element, one contracts with $\delta^{ii'}/2$ or
$-i \epsilon^{ii'}/2$, which gives the average or half the difference of the
two helicity states, respectively.\footnote{The sign difference for
  longitudinal polarisation reflects that one has polarisation vectors
  $\varepsilon$ for incoming gluons and $\varepsilon^{*}$ for outgoing
  ones.}
For linear gluon polarisation, we see that $\tau^{jj'\!,kk'}$ in
\eqref{eq:gluon-proj} projects on the symmetric and traceless part in the
indices $jj'$, with the result depending on two transverse indices $kk'$.
Matching coefficients involving linear gluon polarisation thus carry
indices, as already remarked earlier.

We will give the results of projecting \eqref{Cgg-final} onto definite gluon
polarisations after discussing in more detail the renormalisation
counterterms we have used in our calculations.

\subsubsection{Subtleties of renormalisation}
\label{sec:renorm}

As explained earlier, the renormalisation of TMDs involves only virtual
graphs, which drop out in the calculation of the one-loop matching
coefficients.  It thus remains to discuss the renormalisation of the matrix
elements $\mathcal{M}^{jj'\!, ll'}_{ab}(x)$ and $\mathcal{M}_{S,a}^{}$,
which are respectively related with PDFs and with the soft factor at
$\tvec{z}=\tvec{0}$.

\paragraph{Gluon polarisation.}

We begin with renormalisation of gluon PDFs, which gives the counterterm
\eqref{Cus-counter} for the real graphs in figure~\ref{fig:Cgg-real}.  We
recall that the renormalisation of a scalar PDF operator (e.g.\ the one for
unpolarised gluons) reads
\begin{align}
  \label{op-ren}
{O}(x) = Z(x') \underset{x}{\otimes} {O}_{B}(x')
\end{align}
and gives
\begin{align}
  \label{matel-ren}
\mathcal{M}^{(1)}(x)
 &= Z^{(0)}(x') \underset{x}{\otimes} \mathcal{M}_B^{(1)}(x')
  + Z^{(1)}(x') \underset{x}{\otimes} \mathcal{M}_B^{(0)}(x') \,,
\end{align}
where $Z^{(0)}(x) = \delta(1-x)$ and the first-order term $Z^{(1)}(x)$ is
given by the $\overline{\text{MS}}$ prescription.  Bare quantities are
denoted with a subscript $B$, and we ignore mixing with quark operators for
the time being.  For the gluon operators $O^{jj'}(x)$ with open indices,
which we are now interested in, the renormalisation factor $Z$ in
\eqref{op-ren} turns into a tensor with four indices.  As explained in
section 6.5 of \cite{Collins:1984xc}, renormalisation is significantly
simplified by first splitting the tensor operator into its antisymmetric,
traceless symmetric diagonal terms, which in $2 - 2\epsilon$ transverse
dimensions reads
\begin{align}
O^{jj^\prime} & = \frac{1}{2}\left(O^{jj^\prime} - O^{j^\prime j}\right) +
\frac{1}{2}\left(O^{jj^\prime} + O^{j^\prime j} +
\frac{1}{1-\epsilon}\, \delta^{jj^\prime} O^{kk}\right) -
\frac{1}{2(1-\epsilon)}\, \delta^{jj^\prime} O^{kk}\, .
\label{e:CollinsRenormSect6.5}
\end{align}
The three terms of this decomposition correspond to different irreducible
representations of the rotation group.  Therefore, each of them is
renormalised by a scalar renormalisation factor, and there is no mixing
between them.  To implement \eqref{op-ren} in the tensor case, we project
all tensors in $jj'$ in the same way, which in particular gives
\begin{align}
  \label{delta-decomp}
\delta^{ij}\delta^{i^\prime j^\prime} & =
\frac{1}{2}\left(\delta^{ij}\delta^{i^\prime j^\prime} -
\delta^{ij^\prime}\delta^{i^\prime j}\right) +
\frac{1}{2}\left(\delta^{ij}\delta^{i^\prime j^\prime} +
\delta^{ij^\prime}\delta^{i^\prime j} -
\frac{1}{1-\epsilon}\, \delta^{ii^\prime}\delta^{jj^\prime} \right) +
\frac{1}{2(1-\epsilon)}\, \delta^{ii^\prime} \delta^{jj^\prime}
\end{align}
for the tree-level tensors in \eqref{gluon-lo-tens}.  This explains the
appearance of the factors $1/(1-\epsilon)$ in the counterterm
\eqref{Cus-counter}.  We note that the coefficients of the diagonal,
antisymmetric and traceless symmetric tensors in that term are proportional
to the well-known DGLAP kernels for the associated gluon polarisation (apart
from the $\delta(1-x)$ terms from virtual graphs), which confirms the
consistency of the procedure.  Contracting the last term in
\eqref{delta-decomp} with the projector $\delta^{jj'}$ for unpolarised
gluons and taking the average over incoming gluon polarisations by
contracting with $\delta^{ii'} \big/ \bigl( 2 (1-\epsilon) \bigr)$, one
obtains unity, which corresponds to the correct normalisation of the
tree-level matrix element.

Notice also that the three terms in \eqref{delta-decomp} have the same
symmetry properties in the index pairs $jj'$ and $ii'$, which reflects that
there are no transitions between states corresponding to different
representations of the rotation group.

\paragraph{Quark polarisation.}

We now turn our attention to quark distributions and their renormalisation.
The graphs to be computed for the matching coefficients are as in
figure~\ref{fig:Cgg-real}, with appropriate replacements of gluons by quarks
and with eikonal lines in the fundamental representation.  The spinor
indices at the top of the graphs are to be contracted with the Dirac matrix
of the relevant operator $O_{q}$, $O_{\Delta q}$ or $O_{\delta q}$, and
corresponding contractions are done for the Dirac indices of the quark
states in the matrix element.

As is well known, the treatment of quark polarisation in this context
requires a consistent definition of $\gamma_5$ and of the $\epsilon$ tensor
in dimensional regularisation.  We use the scheme of 't Hooft, Veltman and
Breitenlohner, Maison (HVBM) \cite{tHooft:1972tcz,Breitenlohner:1977hr},
where $\gamma_5 = \frac{i}{4!}\ms \epsilon_{\lambda\mu\nu\rho} \ms
\gamma^\lambda \gamma^\mu \gamma^\nu \gamma^\rho$.  Here
$\epsilon^{}_{\lambda\mu\nu\rho}$ is the usual antisymmetric tensor if all
indices are in the four physical dimensions (with $\epsilon_{0123} = +1$)
and zero otherwise.  We also introduce its counterpart $\epsilon^{ij} =
\epsilon^{+-ij}$, where $i,j$ take values in the $D-2$ transverse
dimensions, as well as the tensor $\bar{\delta}^{ij}$, which equals
$\delta^{ij}$ if $i,j = 1,2$ and is zero for one or both of $i, j$ in the
unphysical dimensions.

For transverse quark polarisation, we avoid the use of $\gamma_5$ by taking
the Dirac matrix $\Gamma_{\delta q}^j$ in the operator
$\mathcal{O}_{\delta}$ as
\begin{align}
\label{transv-matrix}
\Gamma_{\delta q}^j &= \frac{1}{2}\ms \epsilon^{j j'} \sigma^{j' +}
     = \frac{i}{2}\ms \epsilon^{j j'} \gamma^{j'} \gamma^+ \, ,
\end{align}
which is equal to the form $\tfrac{i}{2}\ms \sigma^{j+} \gamma_5$ in
\eqref{eq:quark-proj} in the physical dimensions, $j=1,2$, whereas it
differs for values of $j$ in the unphysical sector.  The fermion trace in
the matching coefficient $C_{\delta q\ms \delta q}$ then contains two
matrices $\sigma^{j+}$ and $\sigma^{i-}$.  This gives exactly the same
result as one would obtain for the trace with $\sigma^{j+} \gamma_5$ and
$\sigma^{i-} \gamma_5$ using the ``naive dimensional regularisation''
prescription, where one assumes that $\gamma_5$ has square $1$ and
anticommutes with all $\gamma^\mu$.  That prescription is often used in the
context of transverse quark polarisation, but we prefer the formulation with
\eqref{transv-matrix}.  Note that with ``naive dimensional regularisation''
one cannot treat all polarised matching coefficients on an equal footing,
since it can only be used to compute traces with an even number of
$\gamma_5$ matrices.  The matrix $\sigma^{i+}$ has also been used to define
quark transversity in the recent work \cite{Gutierrez-Reyes:2017glx}, where
the matching coefficients for polarised TMDs were revisited in a systematic
way.

For longitudinal quark polarisation, the matrix $\gamma_5$ is
unavoidable. In the HVBM scheme, one can rewrite the Dirac matrix in the
operator $\Gamma_{\Delta q}$ as
\begin{align}
\label{long-matrix}
\Gamma_{\Delta q} &= \frac{1}{2}\ms \gamma^+ \gamma_5 = - \frac{i}{4}\ms
\epsilon^{jj'}\ms \gamma^{[j'} \gamma^{j]}\ms \gamma^{+}\, ,
\end{align}
where we define $\gamma^{[i}\gamma^{j]} = \tfrac{1}{2}\ms \bigl(\gamma^{i}
\gamma^{j} - \gamma^{j} \gamma^{i} \bigr)$.  A corresponding replacement
can be made for the matrix $\gamma^- \gamma_5$ that appears in the fermion
trace for matrix elements with incoming longitudinally polarised quarks.
Postponing the contraction with the $\epsilon$ tensor to the end of the
calculation, one can hence represent longitudinally polarised quarks by an
operator $\mathcal{O}_{\Delta q}^{jj'} = - \frac{1}{4}\ms \bar{q}\ms
\gamma^{[j'} \gamma^{j]} \gamma^+\ms q$ with two transverse indices.  This
makes it easy to see that one has mixing under renormalisation with
longitudinally polarised gluons, represented by the operator
$\mathcal{O}_{\Delta g}^{jj'} = G^{+[j'}\ms G^{+ j]}$ that has the same
transformation behaviour under rotations.  For all matching coefficients
with longitudinal polarisation, ${\Delta q \Delta q}$, ${\Delta g \Delta
  q}$, ${\Delta q \Delta g}$ and ${\Delta g \Delta g}$, one then has a
tensor matching coefficient $C_{ab}^{jj', ii'}$ that is antisymmetric in
$jj'$.  As the only terms with ultraviolet divergences in a matching
coefficient arise from the matrix elements at $\tvec{z} = \tvec{0}$, they
must be independent of $\tvec{z}$.  This leaves $\delta^{ij} \delta^{i'j'} -
\delta^{ij'}\delta^{i'j}$ as the only possible tensor structure of the UV
divergent part of the matching coefficient.  This is also the tensor
structure of the tree-level matrix element, which multiplies the one-loop
counterterm according to \eqref{matel-ren}.  For the renormalised matching
coefficient one thus has
\begin{align}
\label{tensor-match}
C^{(1)\ms jj', ii'} &= \frac{1}{2}\ms \bigl( \delta^{ij} \delta^{i'j'} -
\delta^{ij'}\delta^{i'j} \bigr)\, B^{(1)}(x,\epsilon)
 - \frac{1}{2}\ms \bigl( \delta^{ij}
\delta^{i'j'} - \delta^{ij'}\delta^{i'j} \bigr)\, Z^{(1)}(x,\epsilon)
\nonumber \\
  & \quad + \{ \text{terms finite for $\epsilon\to 0$} \}\, ,
\end{align}
where $B^{(1)}$ contains all ultraviolet divergent parts of the bare matrix
element and $Z^{(1)}$ is the relevant one-loop renormalisation factor.  The
$\overline{\text{MS}}$ prescription fixes the latter to be $S_\epsilon$
times the pole part of $B^{(1)}$.  The outcome of this discussion is that
one obtains the same counterterm (and hence the same renormalised result)
when using the tensor form \eqref{tensor-match} or when contracting it with
\begin{align}
\frac{1}{2}\, \epsilon^{jj'} \epsilon^{ii'}
  &= \frac{1}{2}\ms \bigl( \bar{\delta}^{ij} \bar{\delta}^{i'j'} -
\bar{\delta}^{ij'} \bar{\delta}^{i'j} \bigr)
\label{eps-replace}
\end{align}
and working with scalar matching coefficients, which corresponds to using
the conventional operators $\half\ms \bar{q}\ms \gamma^+ \gamma_5\ms q$ and
$i \epsilon^{jj'} G^{+j'}\ms G^{+j}$ in the HVBM scheme.  Working with
\eqref{tensor-match} has the advantage that one does not need to distinguish
between physical and unphysical dimensions during the computation.

In the computations of \cite{Echevarria:2015uaa} and
\cite{Gutierrez-Reyes:2017glx}, the tensor $\bar{\delta}$ on the r.h.s.\ of
\eqref{eps-replace} was replaced by the full transverse metric $\delta$ in
$D-2$ dimensions, as a modification of the proposal by Larin
\cite{Larin:1993tq}, where $\epsilon^{\lambda\mu\nu\rho}
\epsilon^{\lambda'\mu'\nu'\rho'}$ was replaced by products of $D$
dimensional instead of $4$ dimensional metric tensors
(cf.\ \cite{Moch:2015usa} for a recent discussion).  We see that one obtains
the same result when one contracts \eqref{tensor-match} with $\half
\bigl( {\delta}^{ij} {\delta}^{i'j'} - {\delta}^{ij'}{\delta}^{i'j} \bigr)$
instead of
\eqref{eps-replace} and imposes $\overline{\text{MS}}$ subtraction. The
essential point for this is that $B^{(1)}$ and $Z^{(1)}$ have the
\emph{same} $\epsilon$ dependent prefactor in this case.  Since poles in
$1/\epsilon$ cancel in $B^{(1)} - Z^{(1)}$ by construction, only the
$\epsilon\to 0$ limit of their prefactor enters in the renormalised matching
coefficient. The preceding argument readily generalises to order
$\alpha_s^n$, where one should replace $B^{(1)} - Z^{(1)}$ with $\sum_m
B^{(n-m)} \otimes Z^{(m)}$, where $B^{(0)} = Z^{(0)} = \delta(1-x)$.  If,
however, the divergent part of a quantity involves different tensor
structures, then the procedure just described is no longer guaranteed to
work correctly.

We note that \eqref{long-matrix} is a special case of the identity
\begin{align}
\frac{1}{2}\ms
   \bigl( \gamma_\lambda \gamma_5 - \gamma_5 \gamma_\lambda \bigr)
&= \frac{i}{3!}\, \epsilon_{\lambda\mu\nu\rho}\, \gamma^{[\mu} \gamma^\nu
   \gamma^{\rho]}
\label{axial-curr}
\end{align}
in $D$ dimensions, which has been used for a long time when discussing
the axial current in the HVBM scheme \cite{Collins:1984xc,Larin:1993tq}. The
explicit antisymmetrisation on the l.h.s.\ of \eqref{axial-curr}, necessary
to make the current Hermitian, can be omitted in \eqref{long-matrix} because
$\{\gamma^+, \gamma_5 \} = 0$. As shown in
\cite{Collins:1984xc,Larin:1993tq}, the divergence of the flavour nonsinglet
axial current is nonzero if one uses $\overline{\text{MS}}$ renormalisation
and the HVBM scheme, but a zero divergence can be achieved by an additional
finite renormalisation. The resulting nonsinglet current has zero anomalous
dimension. At order $\alpha_s$, the same finite renormalisation achieves
that the divergence of the flavour singlet axial current is given by the
Adler-Bell-Jackiw anomaly, as shown in \cite{Larin:1993tq} (at higher
orders, the required renormalisation factors differ for the flavour singlet
and nonsinglet currents). In \cite{Vogelsang:1996im,Matiounine:1998re} this
finite renormalisation was extended to the nonlocal axial current
$\mathcal{O}_{\Delta q}$ that defines longitudinally polarised quark
distributions, see also \cite{Weber:1991wd,Stratmann:1995fn}. The finite
renormalisation can be written as
\begin{align}
\label{finite-renorm}
O^{\text{NS}}_{\Delta q} &= Z_5^{\text{NS}} \otimes
O^{\text{NS}}_{\Delta q, \overline{\text{MS}}} \,,
&
O^{\text{S}}_{\Delta q} &= Z_5^{\text{S}} \otimes
O^{\text{S}}_{\Delta q, \overline{\text{MS}}} \,,
&
O^{}_{\Delta g} &= O_{\Delta g, \overline{\text{MS}}}
\end{align}
for the flavour singlet and nonsinglet operators in $x$ space, where the
currents on the r.h.s.\ are renormalised by the standard
$\overline{\text{MS}}$ prescription. At order $\alpha_s$ one has
\begin{align}
	Z_5^{\text{NS}}(x) &= Z_5^{\text{S}}(x) = 1 -
        \frac{\alpha_s}{2\pi}\, 4 C_F\ms (1-x)\, ;
\end{align}
higher orders can be found in \cite{Moch:2014sna}.\footnote{Note that $Z_5$,
  called $Z$ in \protect\cite{Moch:2014sna}, is different from $Z^5$ given
  in \protect\cite{Matiounine:1998re}. In that work, the renormalisation of
  the bare current is written in terms of the product $Z^{-1} Z^5$ of two
  renormalisation factors (or matrices in the singlet sector), where
  $Z^{-1}$ does \emph{not} correspond to the $\overline{\text{MS}}$
  prescription.}
This finite renormalisation achieves in particular that the DGLAP kernels
$P_{qq}$ and $P_{\Delta q\Delta q}$ agree up to order $\alpha_s^2$
\cite{Vogelsang:1996im,Matiounine:1998re}. It also achieves that the
hard-scattering coefficients of the Drell-Yan process in collinear
factorisation satisfy $W_{\Delta q\ms \Delta\bar{q}} = - W_{q\bar{q}}$ at
order $\alpha_s$ \cite{Weber:1991wd,Ravindran:2002na}, where $W_{\Delta q\ms
  \Delta\bar{q}}$ is for the incoming quark and antiquark both being
longitudinally polarised and $W_{q\bar{q}}$ for both being unpolarised. In
the results for the matching kernels given in the next subsection, we have
included the finite renormalisation \eqref{finite-renorm}.  At the order we
are working at, this only affects the coefficient~$C_{\Delta q \Delta q}$.


\paragraph{$\overline{\text{MS}}$ variants.}

All results given in this paper so far refer to the prescription that a
renormalisation factor $Z$ has the form
\begin{align}
Z &= \sum_{n=0}^\infty (S_\epsilon\ms \alpha_s)^n \, Z^{(n)} \,,
\end{align}
where $\alpha_s$ is the renormalised coupling and $Z^{(n)}$ for $n\ge 1$ is
a finite sum of poles in~$\epsilon$, fixed uniquely by the requirement that
the renormalised quantity be finite at $\epsilon=0$.  Equivalently, one may
express all quantities in terms of the rescaled coupling $\bar{\alpha}_s =
S_\epsilon\ms \alpha_s$ and use minimal subtraction, where all counterterms
are sums of poles.  The factor $S_\epsilon$ is given in \eqref{S-eps-def}
and was introduced in \cite{Bardeen:1978yd} to simplify the finite terms
left after renormalisation.

A variant of the scheme has been proposed in chapter 3.2.6 of
\cite{Collins:2011zzd}, where instead of $S_\epsilon$ in \eqref{S-eps-def}
one uses
\begin{align}
S_{\epsilon}^{\text{JCC}}
  & = \frac{(4\pi)^{\epsilon}}{\Gamma(1-\epsilon)}\,.
\label{e:SepsilonC}
\end{align}
The difference between the two versions is
\begin{align}
  \label{MSbar-diff}
S_{\epsilon}^{} - S_{\epsilon}^{\text{JCC}} & =
  \frac{\pi^2}{12}\, \epsilon^2 + \mathcal{O}(\epsilon^3)\,.
\end{align}
For one-loop quantities with only a single pole in $\epsilon$, this is of no
consequence, but it does matter for quantities with a double pole.  From
\eqref{double-ct} we see that the counterterm is shifted by $\pi^2/12$ times
the coefficient of the double pole if one uses $S_{\epsilon}^{\text{JCC}}$
instead of $S_\epsilon$.  For the one-loop matching coefficients, this
scheme change affects the soft factor and simply removes the term $-
\pi^2/12$ in the quantity $S_L$ defined in \eqref{double-log-def}.

Like any choice of renormalisation prescription, the choice of $S_\epsilon$
must drop out in physical quantities such as the overall cross section.  At
$\mathcal{O}(\alpha_s)$ this choice affects the matching coefficients
$\prn{R}{C}_{ab}$, the hard-scattering coefficients $C_H^2$ (with squared
logarithms in \eqref{hard-coeffs} indicating the presence of double poles
$1/\epsilon^2$ before renormalisation), as well as DPDFs in colour non-singlet
channels and DTMDs in any colour channel (via one-loop renormalisation of the
soft factors in their definitions).  Collins-Soper and DGLAP kernels, as well
as anomalous dimensions, are not affected by the choice of $S_\epsilon$ at
$\mathcal{O}(\alpha_s)$.

\subsubsection{Matching coefficients for DTMDs on DPDFs}
\label{sec:oneloop-sing}

We now give our final results for the matching coefficients for DTMDs at large
$\tvec{y}$ in the colour singlet sector, which are equal to the matching
coefficients for single parton TMDs.  We have computed them not only with the
rapidity regulator used in this paper, but also with the so-called $\delta$
regulator specified in \cite{GarciaEchevarria:2011rb}, and we find complete
agreement between the two versions.

In the gluon sector, we start from the tensor form \eqref{Cgg-final} and
project onto unpolarised or longitudinally polarised gluons, which gives
\begin{align}
C_{gg}(x,\tvec{z}) = \delta(1-x)
    & - L\,\biggl[\, P_{gg}(x) -
    \frac{1}{2}\, \delta(1-x)\, \gamma_g \,\biggr]
    + \frac{\alpha_s}{2\pi}\, C_{A}\ms S_L\, \delta(1-x) \,,
\label{e:coef_g/g_gen_pol_calc_g/g_B} \\
C_{\Delta g \Delta g}(x,\tvec{z}) = \delta(1-x)
    & - L\,\biggl[\, P_{\Delta g \Delta g}(x) -
    \frac{1}{2}\, \delta(1-x)\, \gamma_g \,\biggr]
    + \frac{\alpha_s}{2\pi}\, C_{A}\ms S_L\, \delta(1-x)
\nonumber \\
& - \frac{\alpha_s}{2\pi}\, 4C_A\, (1-x) \,,
\label{e:coef_g/g_gen_pol_calc_Dg/Dg_B}
\end{align}
where we recall that
\begin{align*}
L & = \log \frac{\mu^2 \tvec{z}^2}{b_0^2} \,,
&
S_L &= -\frac{1}{2} L^2
    + L\, \log\frac{\mu^2}{\zeta} - \sigma\ms \frac{\pi^2}{12}
\end{align*}
with $\sigma = 1$ if one takes $S_\epsilon^{}$ from \eqref{S-eps-def} for
the $\overline{\text{MS}}$ counterterms, whereas $\sigma = 0$ if one takes
$S_\epsilon^{\text{JCC}}$ from \eqref{e:SepsilonC}.  The DGLAP splitting
kernels $P_{gg}(x)$ and $P_{\Delta g \Delta g}(x)$ include a term
proportional to the anomalous dimension $\gamma_g = \beta_0\ms \alpha_s
/(2\pi) $, which comes from virtual graphs (see figure~\ref{fig:Cgg-virt}).
This term is subtracted again in the above matching coefficients, which only
receive contributions from real graphs as explained earlier.  We list all
kernels $P_{ab}$ at the end of this subsection.

Matching coefficients involving linear gluon polarisation carry tensor
indices, and we extract from \eqref{Cgg-final} the forms
\begin{align}
C_{\delta g\ms \delta g}^{jj'\!, ii'} &=
  \tau^{jj'\!, ii'}\ms C_{\delta g\ms \delta g} \,,
  \phantom{ \biggl( \frac{1}{2} \biggr) }
\\
C_{\delta g\ms g}^{jj'}
 &= \biggl(\frac{\tvec{z}^{j}\tvec{z}^{j^{\prime}}}{\tvec{z}^2} -
    \frac{1}{2}\, \delta^{jj^{\prime}} \biggr)\, C_{\delta g\ms g}^{} \,,
\label{tensor:dg_g} \\
C_{g\ms \delta g}^{ii'}
 &= \biggl(\frac{2 \tvec{z}^{i}\tvec{z}^{i^{\prime}}}{\tvec{z}^2} -
      \delta^{ii^{\prime}} \biggr)\, C_{g\ms \delta g}^{} \,,
\label{tensor:g_dg}
\end{align}
where $\tau^{jj'\!, ii'}$ is defined in \eqref{eq:taujjkk} and the scalar
coefficient functions read
\begin{align}
C_{\delta g\ms \delta g}(x,\tvec{z}) &= \delta(1-x)
    - L\,\biggl[\, P_{\delta g\ms \delta g}(x) -
    \frac{1}{2}\, \delta(1-x)\, \gamma_g \,\biggr]
    + \frac{\alpha_s}{2\pi}\, C_{A}\ms S_L\, \delta(1-x) \,,
\label{e:coef_g/g_gen_pol_calc_dg/dg_B}
\\[0.2em]
C_{\delta g\ms g}(x,\tvec{z}) & = -
\frac{\alpha_s}{2\pi}\, 2 C_A\, \frac{1-x}{x}\,,
\label{e:coef_g/g_gen_pol_calc_dg/g_B}
\\[0.5em]
C_{g\ms \delta g}(x,\tvec{z}) & = -
\frac{\alpha_s}{2\pi}\, 2 C_A\, x(1-x) \,.
\label{e:coef_g/g_gen_pol_calc_g/dg_B}
\end{align}
The prefactor in \eqref{tensor:dg_g} has been chosen such that $C_{\delta
  g\ms g}$ coincides with the corresponding matching coefficient defined
for single gluon TMDs in \cite{Echevarria:2015uaa,Gutierrez-Reyes:2017glx}.

In the quark sector, we obtain
\begin{align}
C_{qq}(x,\tvec{z}) &= \delta(1-x)
  - L\,\biggl[\, P_{qq}(x) -
       \frac{1}{2}\, \delta(1-x)\, \gamma_q \,\biggr]
  + \frac{\alpha_s}{2\pi}\, C_{F}\ms S_L\, \delta(1-x)
\nonumber \\
& \qquad \qquad \hspace{0.85em} + \frac{\alpha_s}{2\pi}\, C_{F}\, (1-x) \,,
\label{e:F_q/q} \\
C_{\Delta q\Delta q}(x,\tvec{z}) &= C_{qq}(x,\tvec{z}) \,,
\label{e:F_Dq/Dq}
\end{align}
with $\gamma_q = 3 C_F\ms \alpha_s/(2 \pi)$.  For transverse
quark polarisation, one has again a tensor valued coefficient
\begin{align}
C_{\delta q\ms \delta q}^{j, i} &= \delta^{ji}_{\phantom{\delta q}}
   C_{\delta q\ms \delta q}^{\phantom{j}}
\label{tensor:dq_dq}
\end{align}
with
\begin{align}
C_{\delta q\ms \delta q}(x,\tvec{z}) = \delta(1-x)
  & - L\,\biggl[\, P_{\delta q\ms \delta q}(x) -
       \frac{1}{2}\, \delta(1-x)\, \gamma_q \,\biggr]
  + \frac{\alpha_s}{2\pi}\, C_{F}\ms S_L\, \delta(1-x) \,.
\label{e:F_dq/dq}
\end{align}
We note that the tensor structure $\tvec{z}^j \tvec{z}^i$ would be allowed
in \eqref{tensor:dq_dq} but does not appear at $\mathcal{O}(\alpha_s)$.
For transitions between quarks and gluons, we obtain
\begin{align}
C_{qg}(x,\tvec{z}) & = -L\, P_{qg}(x)
  + \frac{\alpha_s}{2\pi}\, 2T_F \, x(1-x) \,,
\label{e:coef_q/g_calc_F} \\
C_{\Delta q\Delta g}(x,\tvec{z}) & =  -L\, P_{\Delta q\Delta g}(x)
  + \frac{\alpha_s}{2\pi}\, 2T_F (1-x) \label{e:F_Dq/Dg}
\end{align}
and
\begin{align}
C_{gq}(x,\tvec{z}) & = - L\,P_{gq}(x)
  + \frac{\alpha_s}{2\pi}\, C_{F}\, x  \,, \label{e:coef_g/q} \\
C_{\Delta g\Delta q}(x,\tvec{z}) & = - L\,P_{\Delta g\Delta q}(x)
  - \frac{\alpha_s}{2\pi}\, 2C_F (1-x) \,, \label{e:coef_Dg/Dq}
\end{align}
where we have re-introduced the normalisation factor $T_F = 1/2$.
The analogues of \eqref{e:coef_g/g_gen_pol_calc_dg/g_B} and
\eqref{e:coef_g/g_gen_pol_calc_g/dg_B} for quark-gluon transitions are
\begin{align}
C_{\delta g\ms q}(x,\tvec{z}) & = -
  \frac{\alpha_s}{2\pi}\, 2 C_F\, \frac{1-x}{x} \, ,
\label{e:F_dg/q} \\[0.2em]
C_{q\ms \delta g}(x,\tvec{z}) & =
  \frac{\alpha_s}{2\pi}\; 2T_F \, x(1-x) \,,
\label{e:F_q/dg}
\end{align}
where the scalar coefficients are defined as in \eqref{tensor:dg_g} and
\eqref{tensor:g_dg} with the subscript $g$ replaced by $q$ for the
unpolarised parton.

DPDs associated with linear gluon or transverse quark polarisation are tensor
valued and can be decomposed into basis tensors that multiply scalar (or
pseudoscalar) distributions.  We give these decompositions in
appendix~\ref{sec:decompositions} and list the resulting matching equations
between DTMDs and DPDFs in appendix~\ref{sec:match-coeffs}.

Matching coefficients $C_{ab}$ with exactly one longitudinal polarisation
($\Delta g$ or $\Delta q$) are zero due to parity invariance.  Coefficients
with exactly one transverse quark polarisation $\delta q$ vanish because the
relevant graphs involve the trace of an odd number of $\gamma$ matrices.  Due
to charge conjugation invariance, the coefficients $C_{ab}$ remain the same if
one replaces all quark indices by antiquark ones.  These symmetry
relations hold at all orders in $\alpha_s$.  At one-loop level, there are no
transitions between quarks and antiquarks or between quarks of different
flavours.  These appear starting at $\mathcal{O}(\alpha_s^2)$.

Most of the above coefficients have been calculated in the literature
before. The coefficients in \eqref{e:coef_g/g_gen_pol_calc_g/g_B},
\eqref{e:coef_g/g_gen_pol_calc_Dg/Dg_B},
\eqref{e:coef_g/g_gen_pol_calc_dg/g_B} and \eqref{e:coef_g/q} to
\eqref{e:F_dg/q} were calculated in \cite{Echevarria:2015uaa}, using SCET
and the $\delta$ regulator in the version of \cite{GarciaEchevarria:2011rb}.
We find agreement between our results and those in the arXiv version 5 of
\cite{Echevarria:2015uaa}.  Our results in
\eqref{e:F_q/q} to \eqref{e:F_Dq/Dg} agree with \cite{Aybat:2011zv} (which
contains unpolarised results only) and with \cite{Bacchetta:2013pqa}.  The
results in \eqref{e:coef_g/g_gen_pol_calc_g/g_B}, \eqref{tensor:dg_g},
\eqref{e:coef_g/q} and \eqref{e:F_dg/q} agree with \cite{Catani:2013tia} and
\cite{Becher:2012yn}.  Apart from $C_{\delta g\ms \delta g}$,
$C_{g\ms \delta g}$ and $C_{q\ms \delta g}$, all matching coefficients were
recently calculated in \cite{Gutierrez-Reyes:2017glx}, using the $\delta$
regulator briefly described in our appendix~\ref{app:delta-reg}.  We agree
with the results in the arXiv version 2 of \cite{Gutierrez-Reyes:2017glx}.

To the best of our knowledge, $C_{\delta g\ms \delta g}$, $C_{g\ms \delta
  g}$ and $C_{q\ms \delta g}$ in \eqref{e:coef_g/g_gen_pol_calc_dg/dg_B},
\eqref{e:coef_g/g_gen_pol_calc_g/dg_B} and \eqref{e:F_q/dg} have not been
given in the literature before.  They are not relevant for single parton
TMDs in a nucleon, since the PDF for a linearly polarised gluon vanishes in
that case.  This is because linear gluon polarisation corresponds to the
interference between gluons with helicity $+1$ and $-1$ in the amplitude and
its conjugate.

Let us finally list the leading-order DGLAP splitting functions, which were
first derived in~\cite{Altarelli:1977zs,Artru:1989zv}.  They are given by
\begin{align}
P_{qq}(x) & = \frac{\alpha_s}{2\pi}\, C_F\,
  \frac{1+x^2}{(1-x)_{+}} + \frac{1}{2}\,
  \delta(1-x)\, \gamma_q \,, \label{e:P_qq} \\
P_{\Delta q \Delta q}(x) & =
  P_{qq}(x), \label{e:P_DqDq} \\
P_{\delta q\ms \delta q}(x) & =
  \frac{\alpha_s}{2\pi}\, C_F\, \frac{2x}{(1-x)_{+}} + \frac{1}{2}\,
  \delta(1-x)\, \gamma_q  \label{e:P_dqdq}
\end{align}
for quark-quark transitions and by
\begin{align}
P_{gg}(x) & = \frac{\alpha_s}{2\pi}\, 2 C_A\,
  \biggl[\, \frac{x}{(1-x)_+} + \frac{(1-x)(1+x^2)}{x}
  \,\biggr] + \frac{1}{2}\, \delta(1-x)\, \gamma_g\,,
  \label{e:P_gg} \\
P_{\Delta g \Delta g}(x) & =
  \frac{\alpha_s}{2\pi}\, 2 C_A\, \biggl[\, \frac{x}{(1-x)_+} + 2 (1-x)
  \,\biggr] + \frac{1}{2}\, \delta(1-x)\, \gamma_g\,,
\label{e:P_DgDg} \\
P_{\delta g\ms \delta g}(x) &= \frac{\alpha_s}{2\pi}\, 2 C_A\,
  \frac{x}{(1-x)_+} + \frac{1}{2}\, \delta(1-x)\,
  \gamma_g  \label{e:P_dgdg}
\end{align}
for gluons.  The splitting functions mixing quarks and gluons are
\begin{align}
P_{qg}(x) & = \frac{\alpha_s}{2\pi}\, T_{F}\Big[x^2 + (1-x)^2\Big] \,,
&
P_{gq}(x) & = \frac{\alpha_s}{2\pi}\, C_F\, \frac{1+ (1-x)^2}{x} \,,
\label{e:P_gq} \\
P_{\Delta q \Delta g}(x) & = \frac{\alpha_s}{2\pi}\,
T_{F}\, \Big[x^2 - (1-x)^2\Big] \,,
&
P_{\Delta g \Delta q}(x) &
= \frac{\alpha_s}{2\pi}\, C_F\, \frac{1-(1-x)^2}{x} \,. \label{e:P_DgDq}
\end{align}
The symmetry properties of the matching coefficients $C_{ab}$ detailed below
\eqref{e:F_q/dg} also hold for the splitting functions $P_{ab}$.  In
addition, $P_{g\ms \delta g} = P_{\delta g\ms g} = 0$ because the collinear
operators $O_g(x)^{}$ and $O_{\delta g}^{jj'}(x)$ transform differently
under rotations and hence cannot mix under renormalisation.

\subsubsection{Colour non-singlet channels}
\label{sec:oneloop-nonsing}

It is easy to derive the matching coefficients in colour non-singlet channels
from the results we have just presented.  All real graphs in
figure~\ref{fig:Cgg-real} have the same colour factor, so that the colour
projections in \eqref{gen-matel} affect all graphs in the same way.  The same
holds for the graphs for quark-quark transitions and for transitions between
quarks and gluons (in the latter case, the soft factor does not contribute to
the one-loop matching coefficients and one only needs the analogues of
graphs~\ref{fig:Cgg-real}a and b).  We can therefore write
\begin{align}
	\label{LO-propto}
\prn{R}{C}^{(1)}_{ab}(x,\tvec{z};\mu,\zeta) &= c_{ab}^{}(R)\;
\pr{1}{C}^{(1)}_{ab}(x,\tvec{z};\mu,\zeta)\, .
\end{align}
The ratios $c_{ab}(R)$ of colour factors between the representation $R$ and
the singlet channel can be determined from the basic ladder graphs,
which have the topology of figure~\ref{fig:Cgg-real}a.  They are readily
obtained from the evolution kernels given in section~5.1.3 in
\cite{Diehl:2011yj} and read
\begin{align}
c_{qq}(8) &= - \frac{1}{N^2-1} \,,
&
c_{gq}(A) &= c_{qg}(A) = \sqrt{\frac{N^2}{2(N^2-1)}} \,,
\nonumber \\[0.4em]
c_{gg}(A) &= c_{gg}(S) = \frac{1}{2} \,,
&
c_{gq}(S) &= c_{qg}(S) = \sqrt{\frac{N^2-4}{2(N^2-1)}}
\end{align}
and for $N=3$
\begin{align}
c_{gg}(D) &= 0 \,, & c_{gg}(27) &= - 1/3 \, .
\end{align}
\rev{For transitions involving antiquarks one has
$c_{\bar{q}\bar{q}}(8) = c_{qq}(8)$,
$c_{g\bar{q}}(S) = c_{\bar{q}g}(S) = c_{qg}(S)$ and
$c_{g\bar{q}}(A) = c_{\bar{q}g}(A) = - c_{qg}(A)$.  The minus sign in the last
case reflects the fact that two gluons coupled to an antisymmetric octet have
negative charge parity.}

We can now derive the explicit form of the LO DGLAP kernels in a general
colour representation $R$.  For this, we use the evolution equation
\eqref{DGLAP-C} for the matching coefficients $\prn{R}{C}_{ab}$.  At order
$\alpha_s$, we can replace $\prn{R}{C}_{ab}$ on the r.h.s.\ by its
lowest-order value $\delta_{ab}\, \delta(1-x')$ and thus obtain
\begin{align}
  \label{DGLAP-from-C}
  \pr{R}{P_{ac}(x;\mu,\zeta)} &= \frac{1}{2}\,
\delta_{ac}\, \delta(1-x)\, \gamma_{F,c}(\mu,\zeta) -
\frac{1}{2}\, \frac{\partial}{\partial \log\mu}
\prn{R}{C_{ac}(x,\tvec{z};\mu,\zeta)} \,.
\end{align}
Combining this with \eqref{LO-propto}, we readily find that the leading-order
DGLAP evolution kernels of collinear DPDs in colour non-singlet channels are
given by
\begin{align}
	\label{DGLAP-R-LO}
\pr{R}{P}_{ab}(x;\mu,\zeta) &= \frac{1}{2}\, \delta_{ab}\, \delta(1-x)\,
\gamma_{F,a}(\mu,\zeta) - \frac{c_{ab}(R)}{2}\, \frac{\partial}{\partial
\log\mu}
\prn{1}{C_{ab}(x,\tvec{z};\mu,\zeta)} \nonumber
\\ &= \frac{1-c_{aa}(R)}{2}\, \delta_{ab}\, \delta(1-x)\,
\gamma_{F,a}(\mu,\zeta) + c_{ab}(R)\, \pr{1}{P}_{ab}(x;\mu)
\nonumber \\ &= - \frac{1}{2}\, \delta_{ab}\,
\delta(1-x)\, \prn{R}{\gamma}_J(\mu)\, \log\frac{\sqrt{\zeta}}{\mu} \nonumber
\\ &\quad + \frac{1-c_{aa}(R)}{2}\, \delta_{ab}\, \delta(1-x)\,
\gamma_a(\mu) + c_{ab}(R)\, \pr{1}{P}_{ab}(x;\mu) \,.
\end{align}
They are hence easily recovered from their counterparts in the colour
singlet channel.  In the last step we have brought the kernel into the form
\eqref{DGLAP-zeta-expl}, using the explicit $\zeta$ dependence
\eqref{cusp-solved} of $\gamma_{F,a}$, the general relation \eqref{AD-sum}
between anomalous dimensions, and the one-loop relation
$\prn{R}{\gamma}_{K,a} = c_{aa}(R)\, \pr{1}{\gamma}_{K,a}$, which follows
from \eqref{LO-propto}.  We recall that the colour singlet kernels
$\pr{1}{P}_{aa}(x;\mu)$ contain a contribution $\delta(1-x)\, \gamma_a /2$
from virtual graphs, see \eqref{e:P_qq} to \eqref{e:P_dgdg}.  The last line
of \eqref{DGLAP-R-LO} simply reflects that these graphs have the same colour
factor for all channels $R$.

Using \eqref{DGLAP-R-LO}, one can rewrite the $\zeta$ independent
distributions $\widehat{F}$ introduced in \eqref{coll-zeta-expl} as
\begin{align}
\pr{R}{\widehat{F}_{a_1 a_2,\, \mu_0,\zeta_0}(x_i,\tvec{y};\mu_i)}
& = \exp\,\biggl[ \bigl(1-c_{a_1 a_1}(R)\bigr) \int_{\mu_{0}}^{\mu_1}
\frac{d\mu}{\mu}\, \gamma_{a_1}(\mu)
\nonumber \\[0.6em]
& \qquad\,
  + \bigl(1-c_{a_2 a_2}(R)\bigr) \int_{\mu_{0}}^{\mu_2} \frac{d\mu}{\mu}\,
  \gamma_{a_2}(\mu) \ms\biggr]\;
  \pr{R}{\widetilde{F}_{a_1 a_2,\,
    \mu_0,\zeta_0}(x_i,\tvec{y};\mu_i)}\, ,
\end{align}
where $\widetilde{F}$ satisfies the evolution equation \eqref{DGLAP-red} with
kernels $c_{a_1 b_1}(R)\, \pr{1}{P_{a_1 b_1}(x_1';\mu_1^{},\mu_1^2)}$ instead
of $\pr{R}{P_{a_1 b_1}(x_1';\mu_1^{},\mu_1^2)}$.  One can thus use
numerical code for the one-loop evolution of colour singlet DPDs by rescaling
the evolution kernels.

\subsection{Splitting kernels for DPDs}
\label{sec:kernels_for_splitting}

Both DTMDs and DPDFs can be matched on single-parton distributions at small
$\tvec{y}$ as specified in \eqref{split-TMD}, \eqref{split-TMD-coll} and
\eqref{coll-DPD-exp}.  At $\mathcal{O}(\alpha_s)$, all three matching
equations involve the same kernels $\prn{R}{T}_{a_0\to a_1 a_2}^{ll'}$, which
were computed in section~5.2.2 of \cite{Diehl:2011yj}.  We list them here
in the notation of \cite{Diehl:2017kgu} (arXiv version 2), where compared to
\protect\cite{Diehl:2011yj} the kernels $T$ have the opposite order of
indices and include a colour factor.  Thus one has for instance
\begin{align}
\prn{1}{T}^{ll'}_{g\to q\bar{q}} \,\big|_{\text{here}}
& = T_F^{}\ms T^{l' \bs l}_{g\to q\bar{q}}
  \,\big|_{\text{Ref.~\smash{\protect\cite{Diehl:2011yj}}}} \,,
&
\prn{1}{T}^{ll'}_{q\to g\ms\Delta {q}} \,\big|_{\text{here}}
& = C_F^{}\ms T^{l' \bs l}_{q\to g\ms\Delta {q}}
\,\big|_{\text{Ref.~\smash{\protect\cite{Diehl:2011yj}}}} \,.
\end{align}
The new assignment of indices is such that $l$ refers to the amplitude and
$l'$ to its complex conjugate.  We must note a mistake in equation (5.62) of
\cite{Diehl:2011yj}, where the correct order of indices in the kernels $T$
and $U$ is $l'\bs l$ and not $ll'$.

We only give the kernels for the splitting of an unpolarised parton $a_0$
here.  In the matching equation \eqref{split-TMD} for DTMDs on TMDs, one can
also have transverse quark or linear gluon polarisation for $a_0$, but this
possibility is absent if one matches on collinear PDFs in an unpolarised
proton.

We start with the kernels for DPDs in the colour singlet channel.  For a
gluon splitting into a quark and an antiquark, one has
\begin{alignat}{4}
\prn{1}{T}_{g\to q\bar{q}}^{ll'} ( u ) & = {}-
\prn{1}{T}_{g\to \Delta q \Delta \bar{q}}^{ll'} (u) &&= T_F \ms
(u^2+\bar{u}^2)\, \delta^{ll'} \,,
\nn\\[0.4em]
\prn{1}{T}_{g\to \Delta q\ms \bar{q}}^{ll'} (u) & = {}-
\prn{1}{T}_{g\to q \Delta \bar{q}}^{ll'}(u) &&= - i T_F\ms
(u-\bar{u})\, \epsilon^{ll'} \,,
\nn\\
& \hspace{1em}
  \Bigl[\prn{1}{T}_{g\to \delta q\ms \delta \bar{q}}^{ll'}(u)\Bigr]^{jj'}
&& = - 2 T_F\ms u \bar{u}\; \delta^{ll'} \delta^{jj'} \,,
\end{alignat}
where $\bar{u} = 1-u$.  For a quark splitting into a gluon and a quark, the
kernels are given by
\begin{align}
\label{split-q-gq}
\prn{1}{T}_{q\to gq}^{ll'} (u) &= C_F\ms
\frac{1+\bar{u}^2}{u}\, \delta^{ll'} \,,
&
\prn{1}{T}_{q\to \Delta g\ms q}^{ll'} (u) & = i C_F\ms
\frac{1+\bar{u}^2}{u}\, \epsilon^{ll'} \,,
\nn\\
\prn{1}{T}_{q\to \Delta g \Delta q}^{ll'} (u) & = C_F\ms
(1+\bar{u})\, \delta^{ll'} \,,
&
\prn{1}{T}_{q\to g\Delta q}^{ll'} (u) & = i C_F\ms
(1+\bar{u})\, \epsilon^{ll'} \,,  \phantom{\frac{1^2}{1}}
\nn\\
\left[ \prn{1}{T}_{q\to \delta g\ms q}^{ll'} (u) \right]^{jj'}
  & = 2 C_F\ms \frac{\bar{u}}{u}\, \tau^{ll',jj'}
\end{align}
with $\tau^{ll',jj'}$ defined in \eqref{eq:taujjkk}.  The kernels for an
antiquark splitting into a gluon and antiquark have the same form due to
charge conjugation invariance.  Finally, for a gluon splitting into two
gluons one has
\begin{align}
\prn{1}{T}_{g\to gg}^{ll'}(u) & = 2 C_A
  \left( \frac{u}{\bar{u}} +
  \frac{\bar{u}}{u} + u \bar{u}\right)\, \delta^{ll'} \,,
\nn\\[0.2em]
\prn{1}{T}_{g\to \Delta g \Delta g}^{ll'} (u) & = 2 C_A
  (2 - u\bar{u})\, \delta^{ll'} \,, \phantom{\frac{1}{2}}
&
\prn{1}{T}_{g\to g\Delta g}^{ll'} (u) & = 2i C_A
  \left(2\bar{u} + \frac{u}{\bar{u}}\right)\, \epsilon^{ll'} \,,
\nn\\[0.2em]
\left[\prn{1}{T}_{g\to \delta g\ms \delta g}^{ll'}(u)\right]^{jj',kk'} & =
  C_A\ms u\bar{u}\; \delta^{ll'} \tau^{jj',kk'} \,,
&
\left[\prn{1}{T}_{g\to g\ms \delta g}^{ll'} (u)\right]^{kk'}
 & = 2 C_A\ms \frac{u}{\bar{u}}\, \tau^{ll',kk'} \,.
\end{align}
All other kernels that cannot be obtained from those above by interchanging
$a_1$ and $a_2$ or by changing quarks into antiquarks in \eqref{split-q-gq}
are zero.
In appendix~\ref{sec:split-coeffs} we give the matching equations that
follow from these results for the scalar and pseudoscalar DTMDs or DPDFs
defined in appendix~\ref{sec:decompositions}.

Each of the above kernels corresponds to exactly one Feynman graph, so that
the results for other colour channels can be obtained by changing the
overall factor as
\begin{align}
  \label{split-col-ratio}
\prn{R}{T}_{a_0\to a_1 a_2}
&= c_{a_0\to a_1 a_2}(R)\; \prn{1}{T}_{a_0\to a_1 a_2}
\end{align}
with
\begin{align}
c_{q \to gq}(A) & = -\frac{N}{\sqrt{2}}\,,
&
c_{q \to gq}(S) & = \sqrt{\frac{N^2-4}{2}} \,,
&
c_{g \to q\bar{q}}(8) & = -\frac{1}{\sqrt{N^2-1}}\,,
\nn\\[0.3em]
c_{g \to gg}(A) & = - \frac{\sqrt{N^2-1}}{2} \,,
&
c_{g \to gg}(S) &= - c_{g \to gg}(A) \,,
\nn\\[0.3em]
c_{g \to gg}(D) & = 0 \,, & c_{g \to gg}(27) & = -\sqrt{3} \,.
  \phantom{ \sqrt{\frac{N^2-4}{2}} }
\end{align}
For the colour representations $D$ and $27$, we have given the results for
$N=3$.  Notice that in all cases except $c_{g \to q\bar{q}}(8)$ and $c_{g
  \to gg}(D)$, the colour non-singlet channel is enhanced over the singlet
one for SU(3).

\section{Summary}
\label{sec:conclusions}

We have performed a systematic analysis of double parton scattering for
measured transverse momenta, extending the framework for factorisation and
resummation formulated in \cite{Collins:2011zzd} for single parton
scattering with colourless particles in the final state.  A major challenge
is the description of soft gluon exchange and the associated rapidity
evolution, which has a much richer colour structure in DPS than in SPS.  We
handle this structure by projecting suitable pairs of partons onto definite
colour representations.

Based on perturbative results up to two loops \cite{Vladimirov:2016qkd}, we
have proposed equation \eqref{CS-for-S} for the rapidity dependence of the
soft factor in DPS, which is a Collins-Soper equation for matrices in colour
space.  Assuming the validity of that equation, we have given a concise
definition of DPDs in \eqref{F-sub-def} and derived their general
properties.  Applied to SPS, our construction provides an alternative form
for the square root factor in the definition of single-parton TMDs by
Collins \cite{Collins:2011zzd}.
Transverse momentum dependent DPDs (DTMDs) follow the evolution equations
\eqref{RG-TMD-again} and \eqref{CS-TMD}, which can be solved in closed
analytic form as given in \eqref{DTMD-evolved}.  The rapidity evolution of
DTMDs mixes different colour channels, and its solution involves a matrix
exponential in the space of colour representations.  One-loop expressions of
the evolution kernels and anomalous dimensions are given in
section~\ref{sec:CS_kernels} for all parton combinations and colour channels.

In collinear factorisation, i.e.\ when the transverse momenta of the
final-state particles are integrated over, the soft factor for DPS simplifies
considerably: it becomes diagonal in colour representation space and
independent of the parton type.  This can be shown using only colour algebra
and the fact that pairs of Wilson lines associated with initial and final
state partons cancel as $W W^\dagger = \one$.  Important consequences of this
result are the simple structure \eqref{col-Xsimp2} of the cross section for
DPS processes producing coloured particles, as well as the fact that evolution
in rapidity does not mix different colour
channels for collinear DPDs (called DPDFs here).  The corresponding evolution
equations can be found in \eqref{CS-coll-RG} and \eqref{DGLAP-zeta}, and an
analytic expression that makes the rapidity dependence explicit is provided in
\eqref{coll-zeta-expl}.  At one-loop accuracy, the evolution kernels in colour
non-singlet channels are related in a simple way to the usual DGLAP kernels,
as specified in \eqref{DGLAP-R-LO}.

If the size $q_T$ of the measured transverse momenta in a DPS process is
large compared with nonperturbative scales $\Lambda$ (but still small
compared with the scale $Q$ of the hard-scattering processes), one can use
TMD factorisation with DTMDs expressed in terms of perturbative kernels and
simpler hadronic matrix elements.  There are two regimes for this
short-distance matching, depending on the relative size of the transverse
distance $\tvec{y}$ between the two partons and the distances $\tvec{z}_1$
and $\tvec{z}_2$ that are conjugate to the measured transverse momenta in
the cross section formula.

In the first regime, when $\tvec{y}$ is much larger than $\tvec{z}_1$ and
$\tvec{z}_2$, the matching is very similar to the familiar matching of
single parton TMDs onto single parton PDFs.  The resulting expressions are
given in \eqref{small-z-evolved} and \eqref{W-large-y}.  Apart from a
suppression factor for colour non-singlet DPDFs, which is controlled by
their rapidity evolution kernel $\pr{R}{J}(\tvec{y})$, we obtain a product
of Sudakov exponentials and of matching kernels (one for each parton) with
DPDFs.  At one-loop order, the matching kernels $\pr{R}{C}(x,\tvec{z})$ are
due to ladder-type graphs and coincide with the ones for single-parton TMDs
in the colour singlet channel $R=1$.  Most of these kernels have been given
in the literature before, but we have recomputed them in
section~\ref{sec:Coefficient_functions} and extended the calculation to
provide results for all combinations of parton type, polarisation and colour
needed for DPS.  We use the HVBM scheme for defining $\gamma_5$ and the
$\epsilon$ tensor in dimensional regularisation, but point out a way to
avoid these quantities before taking the number of dimensions to $D=4$.

In the second regime, when $\tvec{y}$ is of the same size as $\tvec{z}_1$
and $\tvec{z}_2$, one can match the DTMDs onto collinear distributions of
twist two (i.e.\ usual PDFs), twist three or twist four.  We neglect
collinear twist-three distributions since in an unpolarised proton they are
chiral-odd and hence expected to play a minor role in the regime of small
parton momentum fractions typical of DPS processes.  The matching onto
twist-four distributions is presently only known at lowest order, where it
is trivial and given in \eqref{intr-TMD-tw4}.  The contribution that matches
onto ordinary PDFs corresponds to the splitting of one parton into two.  At
leading order it has the form \eqref{split-TMD-coll} with splitting kernels
given in section~\ref{sec:kernels_for_splitting}.  An important point is
that in the DPS cross section, terms matching onto collinear distributions
of different twist come with different powers of $\Lambda/q_T$ and of
$\alpha_s$ as specified in \eqref{power-small-y}.

To combine the expressions for matching at large or small $\tvec{y}$, we adapt
the subtraction formalism of \cite{Collins:2011zzd}.  In this formalism,
double counting is avoided by a subtraction term in the cross section.  This
term can be easily obtained from the small-$\tvec{y}$ or the large-$\tvec{y}$
expression without further computation, but it requires some care when
choosing scales as discussed in section~\ref{sec:combine-y}.  The same
subtraction formalism is used to combine the cross sections for DPS, SPS and
their interference as specified in sections~\ref{sec:TMD-UV-region} and
\ref{sec:reg_sub_simp}.  In the regime where $\Lambda \ll q_T \ll Q$ one needs
a limited set of nonperturbative quantities for computing the cross section:
the DPDFs $\pr{R}{F}(x_i,\tvec{y})$ and their rapidity evolution kernel
$\pr{R}{J}(\tvec{y})$, collinear twist-four distributions (which arise from
DPDs in the short-distance limit) and ordinary single-parton PDFs.  The
colour octet kernel $\pr{8}{J}(\tvec{y})$ turns out to be equal to the
Collins-Soper kernel for single-gluon TMDs, providing a surprising connection
between two rather different areas of factorisation.

In the multi-scale regime $\Lambda \ll q_T \ll Q$ we thus have a
significantly increased predictive power of the theory.  Rather than a huge set
of transverse-momentum dependent distributions, the TMD cross section involves
the same nonperturbative functions as the cross section integrated over $q_T$.
All other ingredients in the factorisation formula are of perturbative
nature.  In most (although not all) cases they are known at least to the
first nontrivial order in $\alpha_s$.  The results presented in this work
provide a starting point for phenomenological investigations, for instance of
electroweak diboson production with measured transverse boson momenta.  An
important task will be to assess which types of correlations and which regions
of $\tvec{y}$ are important in a given process and kinematical setting.

\rev{Before concluding, we wish to discuss the question of scheme dependence
in the treatment of large gluon rapidities.  Matrix elements of Wilson lines
along lightlike directions typically have rapidity divergences, which
originate from regions of gluon momenta with infinite rapidity.  In the
present paper, we follow the approach of Collins \cite{Collins:2011zzd} and
use Wilson line directions away from the light cone to regulate these
divergences.  The regulated Wilson lines then depend on rapidity variables.
In the DPD construction laid out in section~\ref{sec:comb-soft-coll}, the
parameter $\zeta$ is defined in terms of a central rapidity $Y_C$ that
specifies the range of gluon rapidities effectively included in the DPD.
Several other schemes for regulating rapidity divergences have been proposed
and used in the literature, see
\cite{GarciaEchevarria:2011rb,Echevarria:2015byo,Echevarria:2016scs,%
Chiu:2011qc,Chiu:2012ir,Li:2016axz} and \cite{Becher:2010tm}.  Quantities that
depend on a regulator variable will in general differ between schemes.  By
contrast, quantities that enter a cross section formula in the same way in two
schemes must be equal.  This holds in particular for TMDs, and hence also for
the associated Collins-Soper kernels and matching coefficients onto PDFs.
Discussions of equivalence between specific schemes can be found in
\cite{Collins:2012uy,Echevarria:2012js,Collins:2017oxh}.}

\rev{It follows that our results in sections~\ref{sec:colour} to
\ref{sec:one-loop} are independent of the rapidity regulator
scheme.\footnote{A few equations in these sections are obviously specific to
the regulator we use, such as the definition \eqref{zeta-def} of $\zeta$ in
terms of $Y_C$.}
Of course, this only holds if one defines the variables $\zeta$ and
$\bar{\zeta}$ in a consistent way across different schemes, see for instance
appendix~\ref{app:delta-reg} here and section~6.1 in
\cite{Vladimirov:2017ksc}.  We have checked by explicit calculation that the
one-loop matching coefficients presented in
section~\ref{sec:Coefficient_functions} remain the same when one uses the
$\delta$ regulator specified in \cite{GarciaEchevarria:2011rb}.  By contrast,
our arguments in section~\ref{sec:collinear_DPDs} are specifically formulated
for Wilson lines with finite rapidities and would need to be adapted to other
schemes, using an appropriate translation between the regulator variables as
given in appendix~\ref{app:delta-reg} for a particular case.}

\rev{As already noted, we absorb the soft factor into the DPDs, corresponding
to what is done for SPS in several TMD factorisation approaches
\cite{Collins:2011zzd,GarciaEchevarria:2011rb}. A different route was taken in
\cite{Manohar:2012jr}, where the soft factor explicitly appears in the final
DPS cross section formula.  Such an approach is often taken in the SCET
literature in situations when the soft factor can be computed in perturbation
theory (e.g.\ for single Drell-Yan production at large $q_T$).  In DPS,
however, one needs the soft factor at large distances $\tvec{y}$.  Including
it into the DPDs therefore limits the number of nonperturbative functions
required to compute the cross section.}

Given the results of the present paper and the arguments regarding Glauber
gluon exchange in \cite{Diehl:2015bca}, we consider that TMD factorisation for
DPS is now established at the same level of rigour as for single hard
scattering, apart from the following issues.  \rev{$(i)$:~As already
mentioned, there is no complete all-order proof of the properties we assumed
for the soft factor in section~\ref{sec:soft-fact}, but significant
process in this direction has been made in \cite{Vladimirov:2017ksc}.}
$(ii)$:~There is no all-order proof of the nonabelian Ward identities required
for factorising soft gluon exchange between left and right moving partons into
the soft matrix.  $(iii)$:~As noted below \eqref{match-gg-combine}, we have
omitted Wilson line self interactions in our calculation, following the logic
discussed for SPS in chapter~13.7 of \cite{Collins:2011zzd}.  A more thorough
understanding of this issue, in particular for DPS, has not yet been achieved.
Progress on any of these three points will constitute a further step towards a
rigorous treatment of double parton scattering in QCD.

\acknowledgments

We gratefully acknowledge discussions with Alessandro Bacchetta, John Collins,
Tom van Daal, Miguel G. Echevarr\'{i}a, Jonathan Gaunt, Zhongbo Kang, Piet
Mulders, Riccardo Nagar, Alexei Prokudin, Ted Rogers, Maximilian Stahlhofen,
Iain Stewart, Frank Tackmann, Alexey Vladimirov, Werner Vogelsang and Wouter
Waalewijn.  TK is supported by the European Community under the ``Ideas''
program QWORK (contract 320389).  He acknowledges the hospitality of the
\mbox{Munich} Institute for Astro- and Particle Physics (MIAPP) of the DFG
cluster of excellence ``Origin and Structure of the Universe''.

All figures were made using JaxoDraw~\cite{Binosi:2003yf,Binosi:2008ig}.
For calculations we have used the FeynCalc
package~\cite{Mertig:1990an,Shtabovenko:2016sxi}, the ColorMath
package~\cite{Sjodahl:2012nk} and
FORM~\cite{Vermaseren:2000nd,Kuipers:2012rf}.

\appendix
\section{Matrix manipulations for the soft factor}
\label{app:matrix-algebra}

Referring to section~\ref{sec:soft-fact}, we now show that the set of
properties 1a -- 1c for the soft factor is equivalent with properties 2a
-- 2b.  As a corollary, we find that the soft matrix $S(Y)$ is positive
definite for arbitrary large $Y$ if these properties hold.

\paragraph{Properties 2a -- 2b follow from properties 1a -- 1c.}

The matrix $s(0)$ is nonsingular by assumption (property 1b), so that one
can define a matrix $\widehat{K}$ by
\begin{align}
	\label{K-hat-def}
\widehat{K} s(0) &= s(0) K \,.
\end{align}
The differential equation \eqref{CS-for-s} (property 1a) is thus solved by
\begin{align}
	\label{s-CS-solved}
s(Y) &= s(0)\ms e^{Y K} = e^{Y \widehat{K}}\ms s(0)
\end{align}
for all $Y$.  We now define
\begin{align}
	\label{S-hat-def}
\widehat{S}(Y) &= s(Y/2)\, s^\dagger(Y/2)\, .
\end{align}
Using \eqref{S-decomp} (property 1c) for $Y_0 = Y/2$, we see that
$\widehat{S}(Y)$ approximates $S(Y)$ for $Y\gg 1$.

For any complex vector $a$ one has
\begin{align}
	\label{quad-form}
a^\dagger\, \widehat{S}(Y)\ms a &= b^\dagger\ms b \, ,
\end{align}
with $b = s^\dagger(Y/2)\, a$.  The expression in \eqref{quad-form} is
obviously positive or zero, and since $s^\dagger(Y/2)$ is nonsingular by
assumption (property 1b), it is zero only for $a=0$.  Therefore
$\widehat{S}(Y)$ is positive definite for all $Y$.  This implies property
2b, and it also implies that $S(Y)$ is positive for all $Y$ where it can
be approximated by $\widehat{S}(Y)$.

Multiplying \eqref{K-hat-def} with $s^\dagger(0)$ on the right, and using
that $K$ is Hermitian, we obtain
\begin{align}
  \label{Khat-dag}
\widehat{K} \widehat{S}(0) = \widehat{S}(0) \widehat{K}^\dagger \, .
\end{align}
We thus finally have
\begin{align}
\widehat{S}(Y) &= e^{Y \widehat{K} /2}\ms \widehat{S}(0)
					e^{Y \widehat{K}^\dagger /2}
	= e^{Y \widehat{K}} \widehat{S}(0)\, ,
\end{align}
where we used \eqref{s-CS-solved} and \eqref{S-hat-def} in the first step
and \eqref{Khat-dag} in the second one.  This implies property 2a for $Y
\gg 1$, where $\widehat{S}(Y)$ approximates $S(Y)$.


\paragraph{Properties 1a -- 1c follow from properties 2a -- 2b.}

We start by defining the matrix $\widehat{S}(Y)$ for all $Y$ by
\begin{align}
\label{S-hat-def-2}
\widehat{S}(Y) &= \lim_{Y_0 \to \infty}e^{(Y - Y_0) \widehat{K}} S(Y_0)\, .
\end{align}
This obviously satisfies
\begin{align}
	\label{CS-for-S-hat}
\frac{\partial}{\partial Y}\, \widehat{S}(Y)
  &= \widehat{K}\ms \widehat{S}(Y)
\end{align}
exactly. For $Y, Y_0 \gg 1$ one has $S(Y) = e^{(Y - Y_0) \widehat{K}}
S(Y_0)$ by virtue of \eqref{CS-for-S} (property 2a), which implies
$\widehat{S}(Y) = S(Y)$ for $Y\gg 1$.

Because $S(Y)$ is Hermitian according to \eqref{soft-symm}, it follows
from \eqref{CS-for-S} that $\widehat{K} S(Y) = S(Y)\ms
\widehat{K}^\dagger$ for $Y\gg 1$.  Using this, one deduces from
\eqref{S-hat-def-2} that $\widehat{S}(Y)$ is Hermitian as well.  With
\eqref{CS-for-S-hat} we then have
\begin{align}
\widehat{K} \widehat{S}(Y) = \widehat{S}(Y) \widehat{K}^\dagger
\end{align}
for all $Y$. The solution of \eqref{CS-for-S-hat} can thus be written as
\begin{align}
	\label{S-hat-decomp}
\widehat{S}(Y) &= e^{(Y-Y_1) \widehat{K}}\ms \widehat{S}(Y_1)
	= e^{(Y-Y_1-Y_0) \widehat{K}}\ms 
		\widehat{S}(Y_1)\ms e^{Y_0 \widehat{K}^\dagger}
\end{align}
for arbitrary $Y_0$ and $Y_1$.

We now recall the definition of the square root ${M}^{1/2}$ of a positive
definite matrix $M$.  One can write $M = V^\dagger D V$, where $V$ is
unitary and $D$ is diagonal with positive entries $d_{ii}$.  One defines
${D}^{1/2}$ as the diagonal matrix with entries $\sqrt{d_{ii}}$, and
furthermore $\sqrt{M} = V^\dagger {D}^{1/2} V$.  If $M$ is real then $V$
can be taken as an orthogonal matrix, so that ${M}^{1/2}$ is real as well.
Since $\widehat{S}(Y_1)$ is positive definite for some $Y_1$ (property
2b), one can use \eqref{S-hat-decomp} with $Y_0 = (Y-Y_1)/2$ and write
$\widehat{S}(Y) = t(Y)\, t^\dagger (Y)$, where $t(Y) = e^{(Y-Y_1)
  \widehat{K}/2}\ms \bigl[ \widehat{S}(Y_1)\bigr]{}^{1/2}$ is a
nonsingular matrix.  With the same argument as given below
\eqref{S-hat-def}, it then follows that $\widehat{S}(Y)$ is positive for
all $Y$.

We can now define
\begin{align}
	\label{s-def}
s(Y) &= e^{Y \widehat{K}}\ms \bigl[ \widehat{S}(0) \bigr]{}^{1/2}\ms U \,,
\end{align}
where $U$ is a unitary matrix.  $s(Y)$ is nonsingular (property 1b) and
satisfies the decomposition \eqref{S-decomp} (property 1c) by virtue of
\eqref{S-hat-decomp} with $Y_1 = 0$.  Defining
\begin{align}
	\label{K-def}
K &= U^\dagger\ms \bigl[ \widehat{S}(0) \bigr]{}^{-1/2}
	\widehat{K}\, \bigl[ \widehat{S}(0) \bigr]{}^{1/2}\ms U\, ,
\end{align}
one readily finds that $K^\dagger = K$ and
\begin{align}
	\label{eq:s_expre}
s(Y) &= \bigl[ \widehat{S}(0) \bigr]{}^{1/2}\ms U\ms e^{Y K}\,,
\end{align}
which implies \eqref{CS-for-s} (property 1a).  The definitions
\eqref{s-def} and \eqref{K-def} are consistent with \eqref{K-hat-def}
above.  We note that a more general definition of $s(Y)$ and $K$ is
obtained by replacing $\widehat{S}(0) \to \widehat{S}(Y_1)$ in
\eqref{s-def} to \eqref{eq:s_expre}, as well as $Y \to (Y - Y_1/2)$ in the
exponentials there.


\paragraph{Restrictions on $U$.}

In principle, the matrix $U$ can be chosen freely as a smooth function of
$\tvec{z}_{1}$, $\tvec{z}_{2}$ and $\tvec{y}$, but independent of $Y$ and of
the renormalisation scale $\mu$.  We restrict ourselves to choices such that
$U=1$ for $\tvec{y}=\tvec{0}$.  Since $S(Y)$ is diagonal in that case (see
section~\ref{sec:soft-simple}), the same then holds for $\widehat{S}(Y)$,
$s(Y)$ and for $K = \widehat{K}$.

We furthermore impose a number of restrictions on $U$ in different partonic
channels.  We thus demand that
\begin{align}
  \label{U-star}
U_{qq}^{} = & U_{\bar{q}\bar{q}}^* \,,
&
U_{q\bar{q}}^{} &= U_{\bar{q}q}^* \,,
&
U_{qg}^{} &= U_{\bar{q}g}^* \,,
&
U_{gg}^{} &= U_{gg}^* \,,
\end{align}
which is necessary to obtain the corresponding relations \eqref{s-star} for
$s_{a_1 a_2}$ from those for $S_{a_1 a_2}$.  As discussed in
section~\ref{sec:soft-symm}, we also require that $\pr{RR'}{U}$, defined in
analogy to \eqref{soft-proj}, is a real valued matrix in representation space
and satisfies
\begin{align}
  \label{U-qbar}
\pr{RR'}{U}_{qq}(\tvec{z}_1, \tvec{z}_2, \tvec{y})
&= \pr{RR'}{U}_{q\bar{q}}(\tvec{z}_1, -\tvec{z}_2, \tvec{y})
\nonumber \\
&= \pr{RR'}{U}_{\bar{q}q}(- \tvec{z}_1, \tvec{z}_2, \tvec{y})
 = \pr{RR'}{U}_{\bar{q}\bar{q}}(- \tvec{z}_1, -\tvec{z}_2, \tvec{y}) \,.
\end{align}
We note that the trivial choice $U_{a_1 a_2} = \one$ satisfies all
requirements just listed and gives $\pr{RR'}{U}_{a_1 a_2} = \delta_{RR'}$.

\section{Comparison with the \texorpdfstring{$\delta$}{delta} regulator
  scheme} 
\label{app:delta-reg}

\rev{As emphasised in section~\ref{sec:conclusions}, our treatment of soft and
collinear factors uses Wilson line directions away from the light cone to
regulate rapidity divergences, as was done by Collins in
\cite{Collins:2011zzd}.  The regulated Wilson lines then depend on large but
finite rapidities.}

\rev{The two-loop calculation of the DPS soft factor in
\cite{Vladimirov:2016qkd} was performed using the $\delta$~regulator scheme
specified in \cite{Echevarria:2015byo,Echevarria:2016scs}.  Given the
particular relevance of this calculation for our assumptions on the soft
factor in section~\ref{sec:soft-fact}, we now compare the variables in that
scheme with the ones in the scheme of Collins.  For simplicity we perform the
comparison for the SPS soft factor.  It is easy to extend the following
arguments to DPS, including the appropriate rescaling of $\zeta$, see
\eqref{zeta-def-TMD} and \eqref{zeta-def}.  We note that there are two earlier
variants of the $\delta$ regulator, described in
\cite{GarciaEchevarria:2011rb} and \cite{Echevarria:2012js}, and compared with
the Collins scheme in \cite{Collins:2012uy} and \cite{Echevarria:2012js},
respectively.}

In the $\delta$ regulator scheme of
\cite{Echevarria:2015byo,Echevarria:2016scs}, Wilson lines are taken along
lightlike paths but modified by an exponential damping at large distances
$z^-$ ($z^+$).  This damping is controlled by a parameter $\delta^+$
($\delta^-$), which transforms like the plus (minus) component of a vector
under longitudinal boosts.  One has a correspondence of variables
\begin{align}
  \label{Y-corr}
Y_R & \,\leftrightarrow\, \log \frac{\mu}{\delta^-} \,,
&
Y_L & \,\leftrightarrow\, \log \frac{\delta^+}{\mu} \,.
\end{align}
Due to boost invariance, the soft factor $S$ only depends on $Y_R-Y_L
\leftrightarrow - \log(\delta^+ \delta^- /\mu^2)$.  The correspondence
indicated by $\leftrightarrow$ is valid when taking derivatives, so that
$\mu$ in \eqref{Y-corr} could be replaced by another quantity that has
dimension of mass and is boost invariant.  In our formalism, the soft
factor can be split into
\begin{align}
S(Y_R-Y_L) &= \sqrt{S(2 Y_R - 2 Y_C)}\, \sqrt{S(2 Y_C - 2 Y_L)} \,,
\end{align}
and the same splitting is performed with the $\delta$ regulator, with
the following correspondence of variables:
\begin{align}
  \label{zeta-corr}
2 (Y_R - Y_C) &\,\leftrightarrow\, \log\frac{\mu^2}{(\delta^-)^2}\,
      \frac{2 (\bar{k}^-)^2}{\bar{\zeta}\rule{0pt}{2.2ex}} \,,
&
2 (Y_C - Y_L) &\,\leftrightarrow\, \log\frac{\mu^2}{(\delta^+)^2}\,
      \frac{2 (k^+)^2}{\zeta} \,.
\end{align}
Here the momentum components $k^+$ and $\bar{k}^-$ correspond to the two
partons initiating a hard scattering, and we have normalised the rapidity
parameters as $\zeta \bar{\zeta} = (2 k^+ \bar{k}^-)^2$, in accordance to
the convention \eqref{zeta-def-TMD} used in the present paper.\footnote{The
  normalisation of $\zeta$ and $\bar{\zeta}$ is not relevant for the Collins
  Soper equation, but it does matter for the argument of the anomalous
  dimensions $\gamma_{F,a}$.}
On both sides of \eqref{zeta-corr}, the two expressions add up to twice
the argument of the original soft factor in the relevant formalism.
After the square roots of the soft factor are combined with unsubtracted
collinear matrix elements (see \eqref{eq:unsub_single}), one removes the
rapidity regulator in either scheme, taking $\delta^- \to 0$ and $\delta^+
\to 0$ or $Y_R\to \infty$ and $Y_L\to -\infty$.  \rev{The resulting
distributions depend on the rapidity variable $\zeta$ or $\bar{\zeta}$.}

\section{Combining SPS and DPS for different hard scales}
\label{app:diff_scales}

In this appendix, we discuss the generalisation of the scale setting discussed
in section~\ref{sec:reg_sub_simp} to the case where the scales of the two DPS
processes ($\mu_1$ and $\mu_2$) and of the SPS process~($\mu_h$) are
different.  In the four contributions \eqref{eq:x-secs_onescale} to the
cross section, we now take
\begin{align}
  \label{eq:x-secs_scales}
\sigma_{\text{SPS}}: && f(x_1+x_2,\tvec{Z})
        &\sim E_2(\mu_h;\mu_Z)\, f(\mu_Z) \,,
\nonumber \\
\sigma_{\text{DPS/SPS}}: && D(x_i,\tvec{y}_+,\tvec{Z})
	&\sim E_3(\mu_1,\mu_2|\mu_h;\mu_Z)\, U(\mu_Z)\, f(\mu_Z) \,,
\nonumber \\
\sigma_{\text{SPS/DPS}}: && D(x_i,\tvec{y}_-,\tvec{Z})
	&\sim E_3(\mu_h|\mu_1,\mu_2;\mu_Z)\, U^*(\mu_Z)\, f(\mu_Z) \,,
\nonumber \\
\sigma_{\text{DPS}}: && F(x_i,\tvec{z}_i,\tvec{y})
	&\sim E_4(\mu_1,\mu_2| \mu_1, \mu_2;\mu_Z)\,
         U(\mu_Z)\, U^*(\mu_Z)\, f(\mu_Z)
 + F_{\text{int}}(\mu_1,\mu_2) \,.
\end{align}
In $E_3$ and $E_4$, we now allow different scales for all three or four
parton legs (see the remarks at the ends of sections~\ref{sec:ren-DTMDs}
and~\ref{sec:TMD-UV-region}).  Scale arguments in the amplitude and in its
complex conjugate are separated by a vertical bar ``$|$''.  If only a single
scale is given, such as $\mu_Z$ in \eqref{eq:x-secs_scales}, then all scales
are taken equal.  In the subtraction terms for the SPS/DPS interference, we
now take
\begin{align}
\sigma_{\text{DPS/SPS}, y_+\to 0}: D & \sim 
	E_3(\bar{\mu}_1,\bar{\mu}_2|\mu_h;\hat{\mu}_h)\,
        U(\hat{\mu}_h)\, E_2(\hat{\mu}_h;\mu_Z)\, f(\mu_Z)
\nonumber \\[0.1em]
& \qquad \mbox{with~ } \hat{\mu}_h = p(\nu\ms |\tvec{y}_+|; \mu_Z, \mu_h)
         \mbox{ ~and~ } \bar{\mu}_i = p(\nu\ms |\tvec{y}_+|; \mu_i, \mu_h) \,,
\nonumber \\[0.3em]
\sigma_{\text{SPS/DPS}, y_-\to 0}: D &\sim 
	E_3(\mu_h|\bar{\mu}_1,\bar{\mu}_2;\hat{\mu}_h)\, 
        U^*(\hat{\mu}_h)\, E_2(\hat{\mu}_h;\mu_Z)\, f(\mu_Z)
\nonumber \\[0.1em]
& \qquad\mbox{with~ } \hat{\mu}_h = p(\nu\ms |\tvec{y}_-|; \mu_Z, \mu_h)
        \mbox{ ~and~ } \bar{\mu}_i = p(\nu\ms |\tvec{y}_-|; \mu_i, \mu_h)
\end{align}
instead of \eqref{interf-subtr-terms}.
The profile scales $\hat{\mu}$ are the same as in the single-scale case of
section~\ref{sec:reg_sub_simp}, whilst $\bar{\mu}$ interpolates between the
different hard scales.  The relation \eqref{amp-matching} between the
hard-scattering amplitudes for SPS and DPS holds for equal scales of
$\alpha_s$ in all its terms.  In all subtraction terms, the scale of
$\alpha_s$ in the DPS amplitudes $H_{\alpha_1 \beta_1}$ and $H_{\alpha_2
  \beta_2}$ should be taken as $\bar{\mu}_i$.  This is for instance relevant
for $gg\to H$, which involves the strong coupling already at leading order.
With the choices just specified, one finds that the limiting relation
\eqref{interf-limits} for the subtraction term still holds.

Generalising \eqref{1v1-subtr-terms}, the subtraction terms for DPS
can be taken as follows:
\begin{align}
  \label{1v1-subtr-terms-diff}
\sigma_{\text{DPS}, y_-\to 0}: && F & \sim
E_4(\bar{\mu}_1,\bar{\mu}_2 | \mu_1,\mu_2;
    \hat{\mu}_1,\hat{\mu}_2 | \hat{\mu}_h,\hat{\mu}_h)\,
U^*(\hat{\mu}_h)\, E_3(\hat{\mu}_1,\hat{\mu}_2 | \hat{\mu}_h; \mu_Z)\,
U(\mu_Z)\, f(\mu_Z) \,,
\nonumber \\
\sigma_{\text{DPS}, y_+\to 0}: && F & \sim
E_4(\mu_1,\mu_2|\bar{\mu}_1,\bar{\mu}_2;
\hat{\mu}_h,\hat{\mu}_h | \hat{\mu}_1, \hat{\mu}_2)\,
U(\hat{\mu}_h)\, E_3(\hat{\mu}_h | \hat{\mu}_1, \hat{\mu}_2; \mu_Z)\,
U^*(\mu_Z)\, f(\mu_Z) \,,
\nonumber \\
\sigma_{\text{DPS}, y_\pm\to 0}: && F & \sim
E_4(\bar{\mu}_1,\bar{\mu}_2 | \bar{\mu}_1,\bar{\mu}_2; \hat{\mu}_h)\,
U(\hat{\mu}_h)\ U^*(\hat{\mu}_h)\,
E_2(\hat{\mu}_h;\mu_Z)\, f(\mu_Z)
\end{align}
with
\begin{align}
\hat{\mu} &= p\bigl( \nu \min\{ |\tvec{y}_+|, |\tvec{y}_-| \};
\mu_Z, \mu_h \bigr)
&
\mbox{ and }
&&
\bar{\mu}_i &= p\bigl( \nu \min\{ |\tvec{y}_+|, |\tvec{y}_-| \};
\mu_i, \mu_h \bigr) \,.
\end{align}
For $|\tvec{y}_+| \gg 1/\nu$ and $|\tvec{y}_-| \gg 1/\nu$ one finds the
limiting behaviour in \eqref{DPS-limits}, whereas all three terms in
\eqref{1v1-subtr-terms-diff} tend to the 1v1 contribution to DPS if one or
both of the transverse distances is of order $1/\nu$.  Overall, we thus find
that the subtraction formalism can be adapted to work in the multi-scale case.

\section{Feynman rules}
\label{sec:Feynman_rules}

In this appendix we give the Feynman rules used in our calculation of the
matching coefficients in section~\ref{sec:Coefficient_functions}.  We
compute cut graphs and thus need in particular the rules for propagators and
vertices to the right of the cut (corresponding to the complex conjugate
amplitude).  This requires special care for the three-gluon vertex.  A
derivation of these rules can for instance be found in
\cite{Belitsky:1997ay} for the QCD Lagrangian (but not for eikonal lines).
The rules given here correspond to the conventions in the chapters 3 and 7
of \cite{Collins:2011zzd}.  In particular, the coupling constant is defined
such that the covariant derivative reads $D_{\mu} = \partial_{\mu} + i g
A^{a}{}_{\!\!\!\mu}\, t^{a}$.

\begin{figure}[!tb]
\centering
\includegraphics[width=0.81\textwidth]{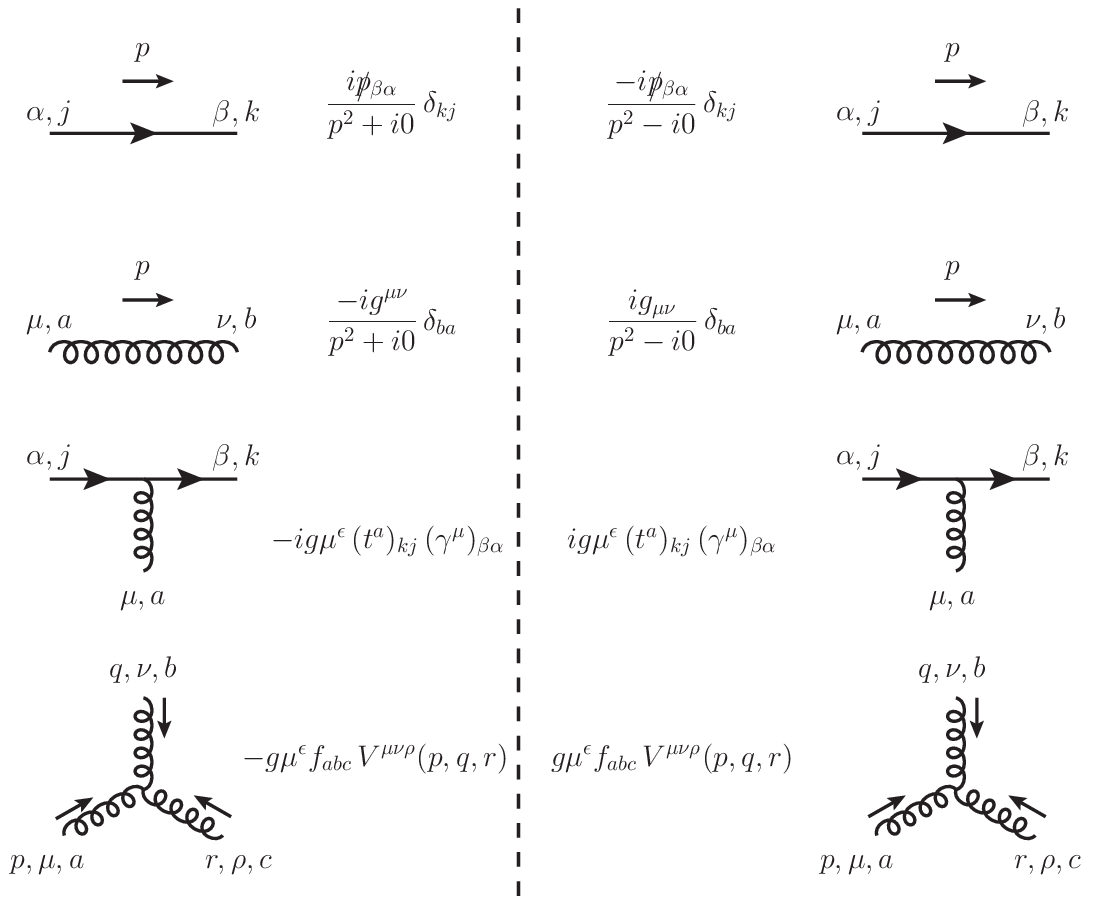} \\
\hspace{7.2em}
\includegraphics[width=0.43\textwidth]{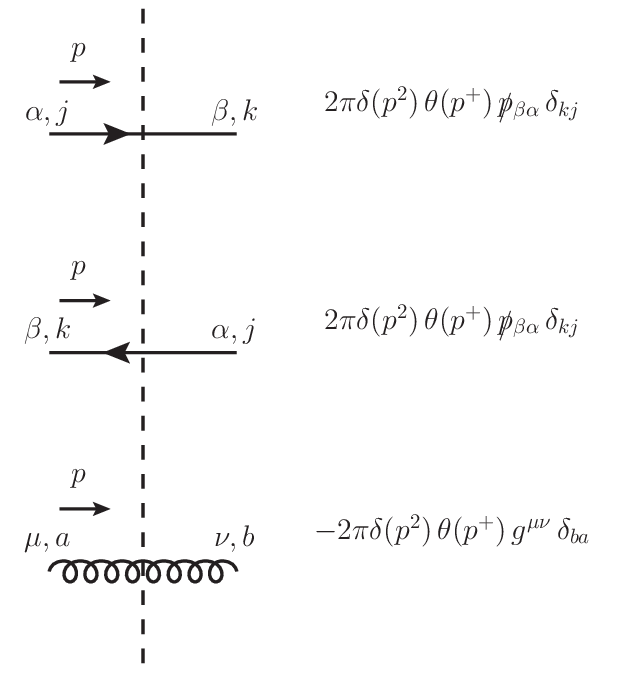}
\caption{\label{fig:Feynman_rules_QCD} Feynman rules for propagators and
vertices in cut graphs.  We use Feynman gauge and set the quark mass to zero.
$V^{\mu\nu\rho}(p,q,r)$ is defined in \protect\eqref{vert}.  With these
rules for cut quark and antiquark lines, there is no minus sign for a
closed fermion loop going across the cut.}
\end{figure}

The Feynman rules for the propagators and vertices arising from the QCD
Lagrangian in Feynman gauge can be found in
figure~\ref{fig:Feynman_rules_QCD}, where we use the notation
\begin{align}
  \label{vert}
V^{\mu\nu\rho}(p,q,r) & = 
 (p-q)^{\rho}g^{\mu\nu} + (q-r)^{\mu}g^{\nu\rho} + (r-p)^{\nu}g^{\rho\mu}\,. 
\end{align}
The four-gluon vertex is not given here since it does not appear in our
calculations.

\begin{figure}[!tb]
\begin{center}\centering
\includegraphics[width=0.8\textwidth]{%
  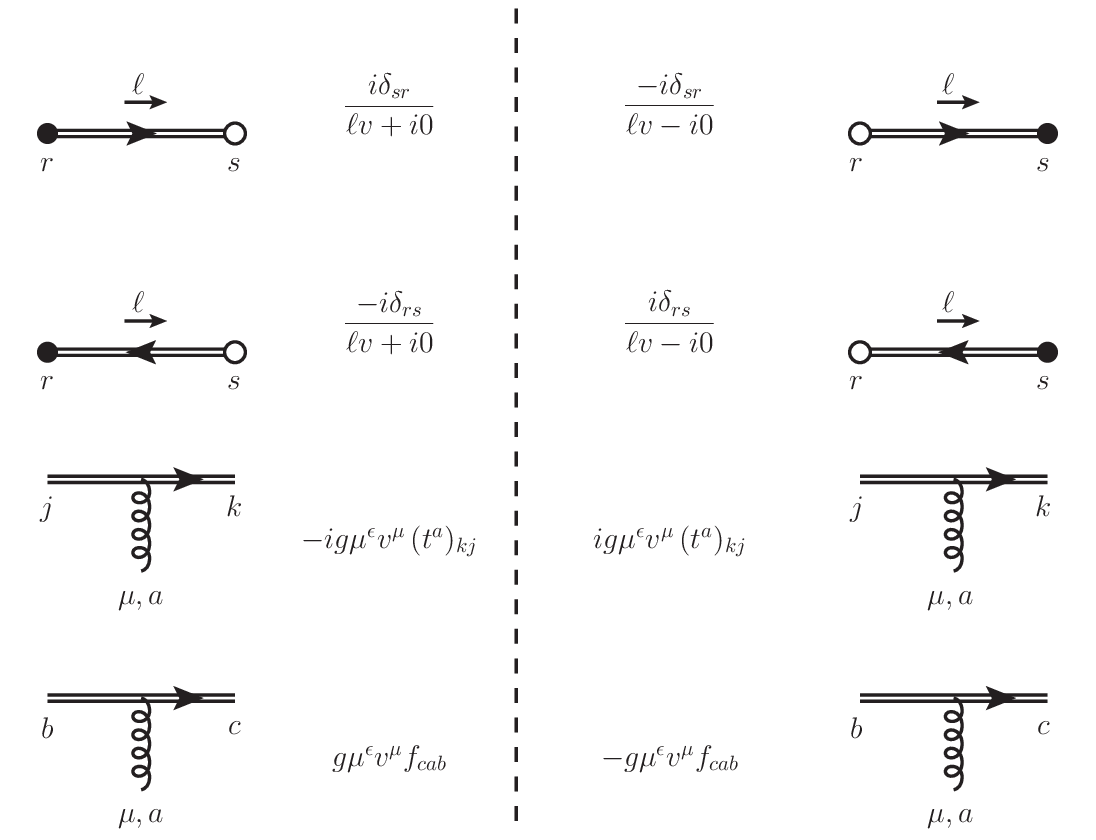} \\
\caption{\label{fig:Feynman_rules_Eik} Feynman rules involving eikonal lines
  along the direction $v$.  The colour indices $r$ and $s$ refer to either
  the fundamental or the adjoint representation, whereas $j$ and $k$ are
  colour triplet and $a$, $b$ and $c$ are colour octet indices.}
\end{center}

\vspace{0.7em}

\begin{center}
\includegraphics[width=0.71\textwidth]{%
  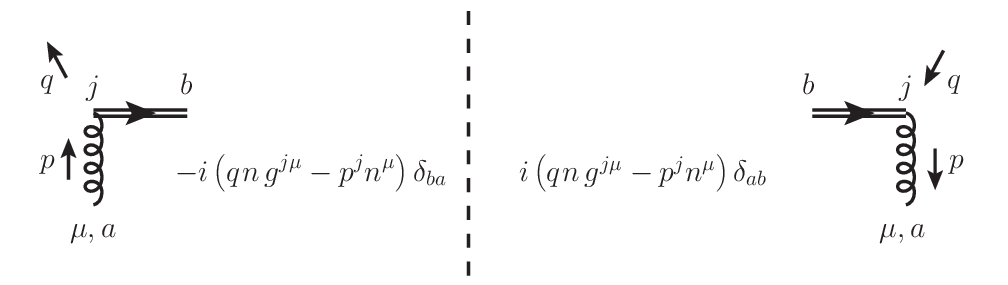} \\
\caption{\label{fig:Feynman_rules_gluon} Feynman rule corresponding to the
  operator in a gluon distribution when an eikonal line along $n$ is
  attached to the gluon.  The Lorentz index $j$ is transverse w.r.t.\ the
  lightlike direction~$n$.  The rule for the graph without the eikonal line
  is obtained by setting $p=q$.}
\end{center}
\end{figure}

The Feynman rules for the propagators and couplings involving eikonal lines
are given in figure~\ref{fig:Feynman_rules_Eik}.  They arise from the
expansion of Wilson line operators, given by~\eqref{WL-def} and its analogue
for the adjoint representation.  We use there the notation of
\cite{Diehl:2011yj}, where open and closed circles at the ends of eikonal
lines (in addition to arrows on and above them) were introduced as a way to
make the correspondence between graphs and mathematical expressions unique.
Let us briefly explain this.

First of all, the full circle indicates the (relative) past and the open
circle the (relative) future time direction when considering the path of the
Wilson line in space-time.  This determines the sign of $i 0$ in the eikonal
propagators.  Secondly, the arrow \emph{above} the eikonal line propagator
corresponds to the momentum flow, just as for quark and gluon propagators,
and is necessary because the eikonal propagator is linear in the momentum
carried by the line.  Finally, the arrow \emph{on} the eikonal line fixes
the overall sign of the propagator and determines the order of contraction
of colour indices, with matrix multiplication going against the direction of
the arrow.  This is the same convention as for Dirac indices on fermion
lines.  For eikonal lines in the fundamental colour representation, this
arrow indicates the canonical order of multiplication, with expressions
involving Gell-Mann matrices $t_a$ rather than their transpose
$(t_a)^T$. For adjoint eikonal lines, the order in which the indices are
arranged can be chosen freely (writing e.g.\ $f_{abc}$ or $-f_{acb}$).

The vertices for a gluon coupling to a fundamental or an adjoint eikonal line
are given in the third and fourth line of
figure~\ref{fig:Feynman_rules_Eik}, respectively.  In the last row, the
relative sign between the vertices to the left and the right of the cut can
be understood from the relation $f_{cab} = -i (T^{a})_{cb}$, where $T^a$ acts
as a generator of the colour group.  The arrow on the line is required to
distinguish the role of the indices $c$ and $b$, given that $f_{cab}$ is
antisymmetric in its indices.\footnote{We note that in figure~7.10 of
  \cite{Collins:2011zzd} a minus sign 
  is missing in the coupling of a gluon to an adjoint eikonal line on the
  right of the cut.  Otherwise, we agree with the Feynman rules given
  there.}

In figure~\ref{fig:Feynman_rules_gluon} we give the Feynman rule for the
operator in a gluon distribution when an eikonal line is attached to the
gluon.  This rule was derived in section~7.6 of \cite{Collins:2011zzd}, see
also \cite{Collins:1981uw,Collins:2008sg}.  It can be obtained by expanding
the operator $n_\mu\ms F^{\mu j}(z)\, W(z,n)$ in $g$.  The Lorentz index $j$
in the gluon field strength is restricted to be transverse to the lightlike
vector $n$.  Note that this rule corresponds to the gluon operator
\eqref{eq:gluon-ops} only if the Wilson line is along $v_L = n$.

\section{Useful integrals}
\label{sec:Useful_relations}

In this appendix, we give a number of Fourier integrals that are useful for
the calculations in section~\ref{sec:Coefficient_functions}.  We start with
\begin{align}
\int \frac{d^{2-2\epsilon}\tvec{k}}{\left(\tvec{k}^{2}\right)^{\alpha}}\,
e^{i\tvec{k} \tvec{z}}
& = \left(\frac{\tvec{z}^2}{4\pi}\right)^{\epsilon+\alpha-1}
  \frac{\pi^\alpha\ms
  \Gamma(1-\epsilon-\alpha)}{\Gamma(\alpha)}, 
\label{e:FT_power}
\\
\int\frac{d^{2-2\epsilon}\tvec{k}} {\left(\tvec{k}^{2}\right)^{\alpha}}\,
  e^{i\tvec{k} \tvec{z}}\, \log(\tvec{k}^2)
& = \left(\frac{\tvec{z}^2}{4\pi}\right)^{\epsilon+\alpha-1}
  \frac{\pi^\alpha\ms
  \Gamma(1-\epsilon-\alpha)}{\Gamma(\alpha)} 
\nonumber \\
& \qquad \times \biggl[\ms \log \frac{4}{\tvec{z}^2}
  +\psi(\alpha)+\psi(1-\alpha-\epsilon) \ms\biggr] \,, 
\label{e:FT_power_and_log}
\end{align}
where $ \psi(x) = \frac{d}{dz} \log\Gamma(z)$ is the digamma function.  One
can obtain \eqref{e:FT_power_and_log} by differentiation of
\eqref{e:FT_power} with respect to $\alpha$.  A useful relation for
integrals involving tensors is given by
\begin{align}
\int\,\frac{d^{2-2\epsilon}{\tvec{k}}}{\left(\tvec{k}^{2}\right)^\alpha}\,
e^{i{\tvec{k}} \tvec{z}}\, \tvec{k}^{j}\tvec{k}^{j'}
& = -\frac{\pi^{\alpha -  1}\,\Gamma(3-\alpha-\epsilon)}{\Gamma(\alpha)}
\left(\frac{\tvec{z}^2}{4\pi}\right)^{\epsilon
  + \alpha -2} \biggl[ \frac{\tvec{z}^{j}\tvec{z}^{j'}}{\tvec{z}^2} +
\frac{\delta^{jj'}}{2(\epsilon + \alpha - 2)} \biggr] \,.
\label{e:FT_two_open_indices_C}
\end{align}
We obtain \eqref{e:FT_two_open_indices_C} by rewriting $\tvec{k}^{j}
\tvec{k}^{j'}$ as $- \partial^2 /(\partial \tvec{z}^j \partial
\tvec{z}^{j'})$ on the l.h.s.\ of \eqref{e:FT_two_open_indices_C} and then
performing these derivatives on the r.h.s.\ of \eqref{e:FT_power}.  From the
above equations, it follows that
\begin{align}
\int
\frac{d^{2-2\epsilon}\tvec{k}}{\tvec{k}^{2}}\, e^{i\tvec{k} \tvec{z}}\,
\biggl[\frac{k^{jj'}}{\tvec{k}^2} + f(\epsilon)\, \delta^{jj'}\biggr]
 & = - \pi\Gamma(1-\epsilon)\left(\frac{\tvec{z}^2}{4\pi}\right)^{\epsilon}
\biggl[\frac{z^{jj'}}{\tvec{z}^2}
  + \frac{f(\epsilon)}{\epsilon}\, \delta^{jj'} \biggr] \,, 
\label{e:FT_tensor}
\end{align}
where the traceless symmetric tensors $k^{jj'}$ and $z^{jj'}$ are defined by
\eqref{tens-def} and $f(\epsilon)$ is an arbitrary function.  The integral
over $k^{jj'}/\tvec{k}^4$ is thus finite, whereas the one over $f(\epsilon)
/\tvec{k}^2$ has an ultraviolet pole if $f(\epsilon)$ is finite at $\epsilon
= 0$.

\section{Tensor decomposition of DPDs}
\label{sec:decompositions}

In this appendix, we give the tensor decompositions of DTMDs and DPDFs in
terms of scalar and pseudoscalar functions.  We emphasise that these
decompositions are only meant to be complete in the $2$ physical transverse
dimensions.  In calculations that have to be done in $2-2\epsilon$
transverse dimensions, one must either avoid the decompositions below (as we
do) or extend them appropriately.

DTMDs depend on the three transverse vectors $\tvec{y}$, $\tvec{z}_1$ and
$\tvec{z}_2$, but one can construct the tensors necessary for their
decomposition from  $\y$ and
\begin{align}
  \label{ytilde-def}
\ytilde^j = \epsilon^{jj'} \y^{j'}
\end{align}
together with the invariants $\delta^{jj'}$ and $\epsilon^{jj'}$.  This is
because in
2 dimensions there are only two linearly independent vectors.  The choice of
the above two vectors results in the most straightforward correspondence
between the DTMDs and the DPDFs, where only the vectors $\y$ and $\ytilde$
can be used.  The decomposition of the double quark distributions
\cite{Diehl:2011yj} reads
\begin{align}\label{eq:qq_decomp}
F_{qq}(x_i,\tvec{z}_i,\tvec{y}) &= f_{qq}(x_i,\tvec{z}_i,\tvec{y}) \,,\nn\\
F_{\D q \D q} & = f_{\D q \D q} \,,\nn\\
F_{\D q\ms q} & = g_{\D q\ms q} \,,\nn\\
F_{q \D q} & = g_{q \D q} \,,\nn\\
F_{\d q\ms q}^j & = \ytilde^j M f_{\d q\ms q} + \y^j M g_{\d q\ms q} \,,\nn\\
F_{q\ms \d q}^j & = \ytilde^j M f_{q\ms \d q} + \y^j M g_{q\ms \d q} \,,\nn\\
F_{\d q\ms \d q}^{jj'} & =
 \delta^{jj'} f_{\d q\ms \d q} + 2\tau^{jj'\!,\y\y} M^2 f_{\d q\ms \d q}^t
\nn\\
	& \quad + 2\tau^{jj'\!,\y\ytilde} M^2 g_{\d q\ms \d q}^s
 + (\y^j \ytilde^{j'}-\ytilde^j\y^{j'})\, M^2 g_{\d q\ms \d q}^a\,,
\end{align}
where $M$ is the proton mass and the tensor $\tau^{jj'\!,kk'}$ is defined in
\eqref{eq:taujjkk}.  Note that one has
$\tau^{jj'\!,kk'} \tau^{kk'\!,ll'} = \tau^{jj'\!,ll'}$.  We employ a shorthand
notation where vectors $\y$ or $\ytilde$ appearing as an index of $\tau$
denote contraction, i.e.\
$\tau^{jj'\!,\y\y} = \tau^{jj'\!,kk'}\, \y^k \y^{k'}$ etc. $f$ and $g$ denote
scalar and pseudo-scalar functions respectively.  The pseudoscalar functions
are absent in the decomposition of DPDFs because one cannot construct a
pseudoscalar from $\y$ alone.  Decompositions analogous to
\eqref{eq:qq_decomp} hold for quark-antiquark distributions and for
distributions of two antiquarks.

For quark-gluon distributions we write
\begin{align}
	\label{eq:gq_decomp}
F_{qg}(x_i,\tvec{z}_i,\tvec{y}) &= f_{qg}(x_i,\tvec{z}_i,\tvec{y}) \,,\nn\\
F_{\D q \D g} & = f_{\D q \D g} \,,\nn\\
F_{\D q\ms g} & = g_{\D q\ms g} \,,\nn\\
F_{q \D g} & = g_{q \D g} \,,\nn\\
F_{\d q\ms g}^j & = \ytilde^j M f_{\d q\ms g} + \y^j M g_{\d q\ms g} \,,\nn\\
F_{q\ms \d g}^{jj'} & = \tau^{jj'\!,\y\y} M^2 f_{q\ms \d g}
 + \tau^{jj'\!,\y\ytilde} M^2 g_{q\ms \d g}\,, \nn\\
F_{\d q\ms \d g}^{j,kk'} & = -\tau^{\ytilde j,kk'} M f_{\d q\ms \d g}
 - (\ytilde^j \tau^{kk'\!,\y\y} + \y^j \tau^{kk'\!,\y\ytilde})\,
  M^3 f_{\d q\ms \d g}^t \nn\\
	& \quad - \tau^{\y j,kk'} M g_{\d q\ms \d g} - (\y^j \tau^{kk'\!,\y\y}
 - \ytilde^j \tau^{kk'\!,\y\ytilde})\, M^3g_{\d q\ms \d g}^t\, .
\end{align}
Analogous decompositions hold when the quark is replaced by an antiquark.
Two-gluon distributions are decomposed as
\begin{align}
	\label{eq:gg_decomp}
F_{gg} (x_i,\tvec{z}_i,\tvec{y})&= f_{gg} (x_i,\tvec{z}_i,\tvec{y})\,,\nn\\
F_{\D g \D g} & = f_{\D g \D g}\,, \nn\\
F_{\D g\ms g} & = g_{\D g\ms g} \,,\nn\\
F_{g \D g} & = g_{g \D g} \,,\nn\\
F_{\d g\ms g}^{jj'} & = \tau^{jj'\!,\y\y} M^2 f_{\d g\ms g}
 + \tau^{jj'\!,\y\ytilde} M^2 g_{\d g\ms g} \,,\nn\\
F_{g\ms \d g}^{jj'} & = \tau^{jj'\!,\y\y} M^2 f_{g\ms \d g}
 + \tau^{jj'\!,\y\ytilde} M^2 g_{g\ms \d g} \,,\nn\\
F_{\d g\ms \d g}^{jj',kk'} & = \tau^{jj'\!,kk'} f_{\d g\ms \d g} /2
 + (\tau^{jj'\!,\y\ytilde} \tau^{kk'\!,\y\ytilde}
 - \tau^{jj'\!,\y\y} \tau^{kk'\!,\y\y})\, M^4 f_{\d g\ms \d g}^t \nn\\
	& \quad + (\tau^{jj'\!,\y\ytilde} \tau^{ kk'\!,\y\y}
 + \tau^{jj'\!,\y\y}\tau^{kk'\!,\y\ytilde} )\, M^4 g_{\d g\ms \d g}^s \nn\\
 & \quad+ (\tau^{jj'\!,\y\ytilde} \tau^{kk'\!,\y\y}
 - \tau^{jj'\!,\y\y}\tau^{kk'\!,\y\ytilde} )\, M^4 g_{\d g\ms \d g}^a \,.
\end{align}
We note that in the decomposition of each DTMD, the tensors multiplying
different terms are orthogonal to each other.

The decompositions for the nonzero DPDFs in terms of real-valued scalar
functions have already been given in \cite{Diehl:2011yj,Diehl:2013mla}. They
can be directly obtained from the DTMD decompositions above by setting all
pseudo-scalar functions (i.e.\ all $g$'s) equal to zero (and removing the
arguments $\tvec{z}_i$ in $F$ and $f$).

\section{Matching coefficients for (pseudo)scalar DPDs}
\label{sec:match-coeffs}

Using the decompositions in the previous subsection and the results for the
one-loop matching coefficients given in section~\ref{sec:oneloop-sing}, we can
perform the necessary tensor contractions and obtain a full set of matching
equations at the level of scalar and pseudoscalar distributions.  We emphasise
that this is done in the two physical space-time dimensions, since all
intermediate calculations requiring dimensional regularisation have been
completed at this point.
For definiteness, let us recall that the tensor-valued coefficients in
section~\ref{sec:oneloop-sing} appear in the matching relations as
\begin{align}
\label{match-tensor}
\pr{R}{F^{jj'\!,kk'}_{\delta g\ms \delta g}(x_i, \tvec{z}_i, \tvec{y}; \zeta)}
&= \pr{R\,}{C^{jj'\!, ll'}_{\delta g\ms \delta g}(x_1',
      \tvec{z}_1; x_1\zeta/x_2)}
   \underset{x_1}{\otimes}
   \pr{R\,}{C^{kk'\!,mm'}_{\delta g\ms \delta g}(x_2',
     \tvec{z}_2; x_2\ms\zeta/x_1)}
\nonumber \\[0.1em]
&\quad
\underset{x_2}{\otimes}
\pr{R}{F^{ll'\!,mm'}_{\delta g\ms \delta g}(x_i', \tvec{y}; \zeta)}
 + \{ \text{quark-gluon mixing terms} \}
\end{align}
for the distribution $F_{\delta g\ms \delta g}$ and in an analogous manner in
other relations involving DPDs that carry polarisation indices.  In an
expansion up to $\mathcal{O}(\alpha_s)$, one keeps the one-loop terms only in
one of the two matching coefficients.

In the following we give the terms with matching coefficients for the parton
with momentum fraction $x_1$; the terms for the parton with momentum fraction
$x_2$ are fully analogous.  We omit colour labels $R$ for brevity.  With the
tensor structure of the kernels in section~\ref{sec:oneloop-sing} we then
obtain for two-quark distributions
\begin{align}
f_{qq} & = C_{qq} \oone f_{qq} + C_{qg} \oone f_{gq} - \y^2M^2 \omega_1
C_{q\ms \d g} \oone f_{\d g\ms q}\,,
\nn\\
f_{\D q \D q} & = C_{\D q \D q} \oone f_{\D q \D q} + C_{\D q \D g} \oone f_{\D g \D q} \,,
\nn\\
g_{\D q\ms q} & = g_{q \D q} = 0\,, \phantom{\oone}
\nn\\
|\y|M f_{\d q\ms q} & = |\y|M C_{\d q\ms \d q} \oone f_{\d q\ms q} \,,
\nn\\
|\y|M g_{\d q\ms q} & = 0\,, \phantom{\oone}
\nn\\
|\y|M f_{q\ms \d q} & = |\y|M C_{qq} \oone f_{q\ms \d q} + |\y|MC_{q g} \oone f_{g\ms \d q}
	+ |\y|M \omega_1 C_{q\ms \d g} \oone ( f_{\d g\ms \d q} + \y^2 M^2 f_{\d g\ms \d q}^t) \,,
\nn\\
|\y| M g_{q\ms \d q} & = - |\y|M \tilde{\omega}_1 C_{q\ms \d g} \oone (f_{\d g\ms \d q} - \y^2 M^2 f_{\d g\ms \d q}^t) \,,
\nn\\
f_{\d q\ms \d q} & = C_{\d q\ms \d q} \oone f_{\d q\ms \d q} \,,
\nn\\
\y^2 M^2 f_{\d q\ms \d q}^t & = \y^2 M^2 C_{\d q\ms \d q} \oone f_{\d q\ms \d q}^t \,,
\nn\\
g_{\d q\ms \d q}^s & = g_{\d q\ms \d q}^a = 0 \,,
\end{align}
where we have defined
\begin{align}
\omega_1 & = - \frac{2\tau^{\tvec{y}\tvec{y}, \tvec{z}_1\! \tvec{z}_1}}{%
  \tvec{y}^2 \ms \tvec{z}_1^2} = - \cos(2 \varphi_1) \,,
\nonumber\\
\tilde{\omega}_1 & = -
\frac{2\tau^{\tvec{y}\tilde{\tvec{y}}, \tvec{z}_1\! \tvec{z}_1}}{%
  \tvec{y}^2 \ms \tvec{z}_1^2} = \sin(2 \varphi_1)
\end{align}
with $\ytilde$ given in \eqref{ytilde-def}.  Here $\varphi_1$ is the angle
between $\tvec{y}$ and $\tvec{z}_1$ in the transverse plane, oriented such
that
$\epsilon^{jj'}_{\phantom{1}} \tvec{y}^j_{\phantom{1}} \tvec{z}^{j'}_1 =
|\tvec{y}|\ms |\tvec{z}_1| \sin\varphi_1$.
For mixed quark and gluon distributions the matching reads
\begin{align}
f_{gq} & = C_{gg} \oone f_{gq} - \y^2M^2 \omega_1 C_{g\ms \d g} \oone f_{\d g\ms q} + C_{gq} \oone f_{qq} \,,
\nn\\
f_{\D g \D q} & = C_{\D g \D g} \oone f_{\D g \D q} + C_{\D g \D q} \oone f_{\D q \D q} \,,
\nn\\
g_{\D g\ms q} & = g_{g \D q} = 0 \,, \phantom{\oone}
\nn\\
\y^2M^2 f_{\d g\ms q} & = \y^2M^2C_{\d g\ms \d g} \oone f_{\d g\ms q} - \omega_1 C_{\d g\ms g} \oone f_{gq} - \omega_1 C_{\d g\ms q} \oone f_{qq} \,,
\nn\\
\y^2 M^2 g_{\d g\ms q} & = - \tilde{\omega}_1 C_{\d g\ms g} \oone f_{gq} - \tilde{\omega}_1 C_{\d g\ms q} \oone f_{qq} \,,
\nn\\
|\y|Mf_{g\ms \d q} & = |\y|M C_{g g} \oone f_{g\ms \d q} + |\y|M \omega_1
C_{g\ms \d g} \oone (f_{\d g\ms \d q}
	+ \y^2 M^2 f_{\d g\ms \d q}^t) + |\y|MC_{gq} \oone f_{q\ms \d q} \,,
\nn\\
|\y|M g_{g\ms \d q} & = - |\y|M \tilde{\omega}_1 C_{g\ms \d g} \oone ( f_{\d g\ms \d q} - \y^2M^2 f_{\d g\ms \d q}^t) \,,
\nn\\
2 |\y| M f_{\d g\ms \d q} & = 2 |\y| M C_{\d g\ms \d g} \oone f_{\d g\ms \d q}
	+ |\y|M \omega_1 C_{\d g\ms g} \oone f_{g\ms \d q} + |\y|M \omega_1
        C_{\d g\ms q} \oone f_{q\ms \d q} \,,
\nn\\
2 |\y|^3 M^3 f_{\d g\ms \d q}^t & = 2 |\y|^3 M^3 C_{\d g\ms \d g} \oone f_{\d g\ms \d q}^t
	+ |\y|M \omega_1 C_{\d g\ms g} \oone f_{g\ms \d q}
	+ |\y|M \omega_1 C_{\d g\ms q} \oone f_{q\ms \d q} \,,
\nn\\
2|\y|M g_{\d g\ms \d q} & = - |\y|M \tilde{\omega}_1 (C_{\d g\ms g} \oone f_{g\ms \d q}
	+ C_{\d g\ms q} \oone f_{q\ms \d q}) \,,
\nn\\
2|\y|^3M^3 g_{\d g\ms \d q}^t & = - |\y|M \tilde{\omega}_1 (C_{\d g\ms g}
\oone f_{g\ms \d q} + C_{\d g\ms q} \oone f_{q\ms \d q})
\end{align}
and
\begin{align}
f_{qg} & = C_{qq} \oone f_{qg} + C_{qg} \oone f_{gg} - \y^2M^2 \omega_1
C_{q\ms \d g} \oone f_{\d g\ms g} \,,
\nn\\
f_{\D q \D g} & = C_{\D q \D q} \oone f_{\D q \D g} + C_{\D q \D g} \oone f_{\D g \D g} \,,
\nn\\
g_{\D q\ms g} & = g_{q \D g} = 0 \,, \phantom{\oone}
\nn\\
|\y|M f_{\d q\ms g} & = |\y|M C_{\d q\ms \d q } \oone f_{\d q\ms g} \,,
\nn\\
|\y| M g_{\d q\ms g} & = 0 \,, \phantom{\oone}
\nn\\
\y^2 M^2 f_{q\ms \d g} & = \y^2 M^2 C_{qq} \oone f_{q\ms \d g} + \y^2 M^2 C_{qg} \oone f_{g\ms \d g}
	- \omega_1 C_{q\ms \d g} \oone ( f_{\d g\ms \d g} - \y^4 M^4 f_{\d g\ms \d g}^t) \,,
\nn\\
\y^2 M^2 g_{q\ms \d g} & = - \tilde{\omega}_1 C_{q\ms \d g} \oone ( f_{\d g\ms \d g} + \y^4M^4 f_{\d g\ms \d g}^t ) \,,
\nn\\
|\y| M f_{\d q\ms \d g} & = |\y| M C_{\d q\ms \d q} \oone f_{\d q\ms \d g} \,,
\nn\\
|\y^3| M^3 f_{\d q\ms \d g}^t & = |\y|^3 M^3 C_{\d q\ms \d q} \oone f_{\d q\ms \d g}^t \,,
\nn\\
g_{\d q\ms \d g} & = g_{\d q\ms \d g}^t = 0 \,.
\end{align}
For two-gluon distributions we have
\begin{align}
	f_{gg} & = C_{gg} \underset{x_1}{\otimes} f_{gg}
		- \tvec{y}^2 M^2 \omega_1 C_{g\ms \delta g} \underset{x_1}{\otimes} f_{\delta g\ms g}
		+ C_{gq} \underset{x_1}{\otimes} f_{qg} \,,
\nonumber\\
	f_{\Delta g \Delta g} & = C_{\Delta g \Delta g} \underset{x_1}{\otimes} f_{\Delta g \D g}
		+ C_{\D g \D q} \oone f_{\D q \D g} \,,
	\nn\\
	g_{\D g\ms g} & = g_{g \D g} = 0 \,, \phantom{\oone}
		\nn\\
	\tvec{y}^2M^2 f_{\d g\ms g} & = \y^2 M^2 C_{\d g\ms \d g} \oone f_{\d g\ms g}
			- \omega_1 C_{\d g\ms g} \oone f_{gg} - \omega_1 C_{\d g\ms q} \oone f_{qg}
		\,,
		\nn\\
	\y^2M^2 g_{\d g\ms g} & = - \tilde{\omega}_1 C_{\d g\ms g} \oone f_{gg} - \tilde{\omega}_1 C_{\d g\ms q} \oone f_{qg} \,,
		\nn\\
	\y^2 M^2 f_{g\ms \d g} & = \y^2 M^2 C_{gg} \oone f_{g\ms \d g} - \omega_1 C_{g\ms \d g} \oone (f_{\d g\ms \d g} - \y^4 M^4 f_{\d g\ms \d g}^t) \,,
		\nn\\
	\y^2M^2 g_{g\ms \d g} & = - \tilde{\omega}_1 C_{g\ms \d g} \oone (f_{\d g\ms \d g} + \y^4M^4 f_{\d g\ms \d g}^t) \,,
		\nn\\
	2 f_{\d g\ms \d g} & = 2 C_{\d g\ms \d g} \oone f_{\d g\ms \d g}
		- \y^2M^2 \omega_1 C_{\d g\ms g} \oone f_{g\ms \d g}
		- \y^2 M^2 \omega_1 C_{\d g\ms q} \oone f_{q\ms \d g} \,,
		\nn\\
	2\y^4M^4 f_{\d g\ms \d g}^t & = 2 \y^4 M^4 C_{\d g\ms \d g} \oone f_{\d g\ms \d g}^t
		+ \y^2 M^2 \omega_1 C_{\d g\ms g} \oone f_{g\ms \d g}
		+ \y^2 M^2 \omega_1 C_{\d g\ms q} \oone f_{q\ms \d g} \,,
		\nn\\
	2 \y^4 M^4 g_{\d g\ms \d g}^s & = - \y^2 M^2 \tilde{\omega}_1 C_{\d g\ms g} \oone f_{g\ms \d g}
		- \y^2 M^2 \tilde{\omega}_1 C_{\d g\ms q} \oone f_{q\ms \d g} \,,
		\nn\\
	g_{\d g\ms \d g}^a & = g_{\d g\ms \d g}^s \,.
\end{align}
Further relations are obtained by replacing quark by antiquark labels.

Let us comment on the $\tvec{y}$ dependent factors in the above relations.
The factors of $|\tvec{y} M|^n$ are such that they can be completely absorbed
into the distributions (such rescaled distributions were denoted by $h$ in
\cite{Diehl:2013mla}).  The factors $\omega_1$ and $\tilde{\omega}_1$ appear
in the matching between DTMDs and DPDFs carrying different polarisations; the
azimuthal dependence they provide is required by the conservation of angular
momentum along the $z$ axis.  The factor $\tilde{\omega}_1$ appears in the
matching between pseudoscalar and scalar functions and ensures that the
r.h.s.\ of the matching equations is odd under parity.

\section{Splitting kernels for (pseudo)scalar DPDs}
\label{sec:split-coeffs}

In this appendix, we give the matching of scalar and pseudoscalar DTMDs onto
PDFs in the region of small $\tvec{y}$ and $\tvec{z}_i$.  The following
results readily follow from the splitting kernels given in
section~\ref{sec:kernels_for_splitting} and the tensor decompositions in
appendix~\ref{sec:decompositions}.  The scalar matching kernels
$\prn{R}{T}_{a_0\rightarrow a_1a_2}$ used in the following are obtained from
those in section~\ref{sec:kernels_for_splitting} by removing all tensors
$\delta$, $\epsilon$ and $\tau$, so that one has
$\prn{1}{T}_{g\to q\bar{q}} = T_F\ms (u^2+\bar{u}^2)$,
$\prn{1}{T}_{g\to \Delta q\ms \bar{q}} = - i T_F\ms (u-\bar{u})$,
$\prn{1}{T}_{g\to \delta q\ms \delta\bar{q}} = -2 T_F\ms u\bar{u}$,
\mbox{$\prn{1}{T}_{q\to \delta g\ms q} = 2 C_F\ms \bar{u}/u$}, etc.
The only exception to this rule is that we define
$\prn{1}{T}_{g\to \delta g\ms \delta g} = 2 C_A\ms u \bar{u}$.  Then the
coefficients $\prn{1}{T}_{a_0\to a_1 a_2}$ used here agree with the
coefficients $T_{a_0\to a_1 a_2}$ in appendix~B of \cite{Diehl:2014vaa} for
all channels.  The kernels for other colour representations are obtained
from \eqref{split-col-ratio}.

The matching between (pseudo)scalar DTMDs and PDFs then reads
\begin{align}
\pr{R}{f}_{\text{spl},\, q \bar{q}}(x_i,\tvec{z}_i,\tvec{y})
	&= \frac{\tvec{y}_{+} \tvec{y}_{-}}{%
			\tvec{y}_{+}^{2}\ms \tvec{y}_{-}^{2}}\,
	\frac{\alpha_s }{2\pi^2}\,
	\pr{R\,}{T}_{g\to q \bar{q}} \,
	\frac{f_{g}(x_1+x_2)}{x_1+x_2}\,,
	\nn\\
\pr{R}{f}_{\text{spl},\, \Delta q \Delta \bar{q}}(x_i,\tvec{z}_i,\tvec{y})
	&= \frac{\tvec{y}_{+} \tvec{y}_{-}}{%
			\tvec{y}_{+}^{2}\ms \tvec{y}_{-}^{2}}\,
	\frac{\alpha_s }{2\pi^2}\,
	\pr{R\,}{T}_{g\to \Delta q\ms \Delta \bar{q}} \,
	\frac{f_{g}(x_1+x_2)}{x_1+x_2}\,,
	\nn\\
\pr{R}{g}_{\text{spl},\, \Delta q\ms \bar{q}}(x_i,\tvec{z}_i,\tvec{y})
	&= \frac{\tilde{\tvec{y}} (\tvec{z}_1-\tvec{z}_2)}{%
			\tvec{y}_{+}^{2}\ms \tvec{y}_{-}^{2}}\,
	\frac{\alpha_s }{2\pi^2}\,
	\pr{R\,}{T}_{g\to \Delta q\ms \bar{q}} \,
	\frac{f_{g}(x_1+x_2)}{x_1+x_2}\,,
	\nn\\
\pr{R}{g}_{\text{spl},\, q\ms \Delta \bar{q}}(x_i,\tvec{z}_i,\tvec{y})
	 &= \frac{\tilde{\tvec{y}} (\tvec{z}_1-\tvec{z}_2)}{%
		\tvec{y}_{+}^{2}\ms \tvec{y}_{-}^{2}}\,
	\frac{\alpha_s }{2\pi^2}\,
	\pr{R\,}{T}_{g\to q\ms \Delta \bar{q}} \,
	\frac{f_{g}(x_1+x_2)}{x_1+x_2}\,,
	\nn\\
\pr{R}{f}_{\text{spl},\, \delta q\ms \delta \bar{q}}(x_i,\tvec{z}_i,\tvec{y})
 	&= \frac{ \tvec{y}_+ \tvec{y}_-}{%
		\tvec{y}_{+}^{2}\ms \tvec{y}_{-}^{2}}\,
	\frac{\alpha_s }{2\pi^2}\,
	\pr{R\,}{T}_{g\to \delta q\ms \delta \bar{q}} \,
	\frac{f_{g}(x_1+x_2)}{x_1+x_2}
\end{align}
for quark-antiquark DTMDs. For splitting into gluon-quark distributions we
have
\begin{align}
\pr{R}{f}_{\text{spl},\, g q}(x_i,\tvec{z}_i,\tvec{y})
	&= \frac{\tvec{y}_{+} \tvec{y}_{-}}{%
		\tvec{y}_{+}^{2}\ms \tvec{y}_{-}^{2}}\,
	\frac{\alpha_s }{2\pi^2}\,
	\pr{R\,}{T}_{q\to g q } \,
	\frac{f_{q}(x_1+x_2)}{x_1+x_2}\,,
	\nn\\
\pr{R}{f}_{\text{spl},\, \Delta g \Delta q}(x_i,\tvec{z}_i,\tvec{y})
	&= \frac{\tvec{y}_{+} \tvec{y}_{-}}{%
		\tvec{y}_{+}^{2}\ms \tvec{y}_{-}^{2}}\,
	\frac{\alpha_s }{2\pi^2}\,
	\pr{R\,}{T}_{q\to \Delta g \Delta q } \,
	\frac{f_{q}(x_1+x_2)}{x_1+x_2}\,,
	\nn\\
\pr{R}{g}_{\text{spl},\, \Delta g\ms q}(x_i,\tvec{z}_i,\tvec{y})
	&= \frac{\tilde{\tvec{y}} (\tvec{z}_1-\tvec{z}_2)}{%
		\tvec{y}_{+}^{2}\ms \tvec{y}_{-}^{2}}\,
	\frac{\alpha_s }{2\pi^2}\,
	\pr{R\,}{T}_{q\to \Delta g\ms q } \,
	\frac{f_{q}(x_1+x_2)}{x_1+x_2}\,,
	\nn\\
\pr{R}{g}_{\text{spl},\, g\ms \Delta q}(x_i,\tvec{z}_i,\tvec{y})
	&= \frac{\tilde{\tvec{y}} (\tvec{z}_1-\tvec{z}_2)}{%
		\tvec{y}_{+}^{2}\ms \tvec{y}_{-}^{2}}\,
	\frac{\alpha_s }{2\pi^2}\,
	\pr{R\,}{T}_{q\to g\ms \Delta q } \,
	\frac{f_{q}(x_1+x_2)}{x_1+x_2}\,,
	\nn\\
(\tvec{y}^2 M^2)\, \pr{R}{f}_{\text{spl},\,
          \delta g\ms q}(x_i,\tvec{z}_i,\tvec{y})
	&= \frac{2 (\tvec{y}_{+}\tvec{y})(\tvec{y}_{-}\tvec{y}) - \tvec{y}^2 (\tvec{y}_+\tvec{y}_-)}{%
		\tvec{y}_{+}^{2}\ms \tvec{y}_{-}^{2}}\, \frac{1}{\tvec{y}^2}\;
	\frac{\alpha_s }{2\pi^2}\,
	\pr{R\,}{T}_{q\to \delta g\ms q } \,
	\frac{f_{q}(x_1+x_2)}{x_1+x_2}\,,
	\nn\\
(\tvec{y}^2 M^2)\, \pr{R}{g}_{\text{spl},\,
          \delta g\ms q}(x_i,\tvec{z}_i,\tvec{y})
	&= \frac{(\tvec{y}_{+}\tvec{y})(\tvec{y}_{-}\tilde{\tvec{y}}) + (\tvec{y}_{-}\tvec{y})(\tvec{y}_{+}\tilde{\tvec{y}})}{%
		\tvec{y}_{+}^{2}\ms \tvec{y}_{-}^{2}}\, \frac{1}{\tvec{y}^2}\;
	\frac{\alpha_s }{2\pi^2}\,
	\pr{R\,}{T}_{q\to \delta g\ms q } \,
	\frac{f_{q}(x_1+x_2)}{x_1+x_2}
\end{align}
and for two-gluon DTMDs
\begin{align}
\pr{R}{f}_{\text{spl},\, g g}(x_i,\tvec{z}_i,\tvec{y})
	&= \frac{\tvec{y}_{+} \tvec{y}_{-}}{%
		\tvec{y}_{+}^{2}\ms \tvec{y}_{-}^{2}}\,
	\frac{\alpha_s }{2\pi^2}\,
	\pr{R\,}{T}_{g\to g g } \,
	\frac{f_{g}(x_1+x_2)}{x_1+x_2}\,,
	\nn\\
\pr{R}{f}_{\text{spl},\, \Delta g \Delta g}(x_i,\tvec{z}_i,\tvec{y})
	&= \frac{\tvec{y}_{+} \tvec{y}_{-}}{%
		\tvec{y}_{+}^{2}\ms \tvec{y}_{-}^{2}}\,
	\frac{\alpha_s }{2\pi^2}\,
	\pr{R\,}{T}_{g\to \Delta g \Delta g } \,
	\frac{f_{g}(x_1+x_2)}{x_1+x_2}\,,
	\nn\\
\pr{R}{g}_{\text{spl},\, g\ms \Delta g}(x_i,\tvec{z}_i,\tvec{y})
	&= \frac{\tilde{\tvec{y}}(\tvec{z}_1-\tvec{z}_2)}{%
		\tvec{y}_{+}^{2}\ms \tvec{y}_{-}^{2}}\,
	\frac{\alpha_s }{2\pi^2}\,
	\pr{R\,}{T}_{g\to g\ms \Delta g } \,
	\frac{f_{g}(x_1+x_2)}{x_1+x_2}\,,
	\nn\\
(\tvec{y}^2 M^2)\, \pr{R}{f}_{\text{spl},\,
          g\ms \delta g}(x_i,\tvec{z}_i,\tvec{y})
	&= \frac{2(\tvec{y}_{+}\tvec{y})(\tvec{y}_{-}\tvec{y}) - \tvec{y}^2 (\tvec{y}_+\tvec{y}_-)}{%
		\tvec{y}_{+}^{2}\ms \tvec{y}_{-}^{2}}\, \frac{1}{\tvec{y}^2}\;
	\frac{\alpha_s }{2\pi^2}\,
	\pr{R\,}{T}_{g\to g\ms \delta g } \,
	\frac{f_{g}(x_1+x_2)}{x_1+x_2}\,,
	\nn\\
(\tvec{y}^2 M^2)\, \pr{R}{g}_{\text{spl},\,
          g\ms \delta g}(x_i,\tvec{z}_i,\tvec{y})
	&= \frac{(\tvec{y}_{+}\tvec{y})(\tvec{y}_{-}\tilde{\tvec{y}}) + (\tvec{y}_{-}\tvec{y})(\tvec{y}_{+}\tilde{\tvec{y}})}{%
		\tvec{y}_{+}^{2}\ms \tvec{y}_{-}^{2}}\, \frac{1}{\tvec{y}^2}\;
	\frac{\alpha_s }{2\pi^2}\,
	\pr{R\,}{T}_{g\to g\ms \delta g } \,
	\frac{f_{g}(x_1+x_2)}{x_1+x_2}\,,
	\nn\\
\pr{R}{f}_{\text{spl},\, \delta g\ms \delta g}(x_i,\tvec{z}_i,\tvec{y})
 	&= \frac{\tvec{y}_+\tvec{y}_-}{%
		\tvec{y}_{+}^{2}\ms \tvec{y}_{-}^{2}}\,
	\frac{\alpha_s }{2\pi^2}\,
	\pr{R\,}{T}_{g\to \delta g\ms \delta g } \,
	\frac{f_{g}(x_1+x_2)}{x_1+x_2}\,.
\end{align}



\phantomsection
\addcontentsline{toc}{section}{References}

\bibliographystyle{JHEP}
\bibliography{match}


\newcommand{\tenbar}{\scalebox{0.7}{$\overline{10}$}}
\newcommand{\ten}{\scalebox{0.7}{$10\phantom{\overline{1}}\hspace{-1ex}$}}
\newcommand{\tenbarten}{\tenbar\, \ten}
\newcommand{\tentenbar}{\ten\ms \tenbar}

\newcommand{\SA}{S\bs A}

\newcommand{\Rp}{R^{\ms \prime}}
\newcommand{\Rpp}{R^{\ms \prime\prime}}

\newcommand{\Rbar}{\overline{R}}
\newcommand{\Rpbar}{\overline{R}{}^{\ms \prime}}
\newcommand{\Rppbar}{\overline{R}{}^{\ms \prime\prime}}

\newcommand{\Pro}[2]{P_{\rule{0pt}{1.6ex}#1}^{#2}}

\newcommand{\epstwo}[2]{\varepsilon(#1)\, \varepsilon(#2)}
\newcommand{\etatwo}[2]{\eta(#1)\, \eta(#2)}

\newcommand{\conv}[1]{\underset{#1}{\otimes}}

\newpage

\section{Erratum}
\subsection{Colour structure of DPDs}

The analysis of the colour structure of two-gluon DPDs is incomplete in the original manuscript and needs to be extended.  Consider a quantity with four adjoint indices that transforms like an overall colour singlet.  One can first couple the indices pairwise to one of the combinations $R=1,A,S, 10, \overline{10}, 27$.  An overall singlet is then obtained if the colour representations two index pairs are in conjugate colour representations.  This includes the cases where one pair is in the antisymmetric octet and the other in the symmetric octet.  These cases were omitted in \cite{Mekhfi:1985dv}, see equations (6.a), (6.b) and (7) in that work.  This mistake was repeated in equation (2.121) of \cite{Diehl:2011yj}.

In appendix A of \cite{Kasemets:2014yna} it was shown that the two-gluon distributions corresponding to $R=10$ and $\overline{10}$ in the previous construction are equal.  The proof given there makes use of the colour decompositions given in \cite{Mekhfi:1985dv,Diehl:2011yj} and breaks down once the missing combinations with one symmetric and one antisymmetric octet are included.  We hence must extend the colour decompositions in the present work by adding the mixed octet combinations and restoring separate quantities for $R=10$ and $\overline{10}$.

In the list of projection operators $P_R$ in \eqref{gluon-proj}, we hence replace $P_D^{} = P_{\ten} + P_{\ms\tenbar}$ with the separate projectors
\begin{align}
\label{decupl-def}
P_{\ten}^{a a'\, b b'}
   &= \frac{1}{4}\ms \bigl( \delta^{a b}
      \delta^{a' b'} - \delta^{a b'} \delta^{a' b} \bigr)
      - \frac{1}{2}\, P_{A}^{a a'\, b b'}
      - \frac{i}{4}\, \bigl( d^{a b c} f^{a' b' c}
         + f^{a b c} d^{a' b' c} \bigr) \,,
\nonumber \\
P_{\tenbar}^{a a'\, b b'}
   &= \frac{1}{4}\ms \bigl( \delta^{a b}
      \delta^{a' b'} - \delta^{a b'} \delta^{a' b} \bigr)
      - \frac{1}{2}\, P_{A}^{a a'\, b b'}
      + \frac{i}{4}\, \bigl( d^{a b c} f^{a' b' c}
         + f^{a b c} d^{a' b' c} \bigr)
\end{align}
for the decuplet and antidecuplet.  These projectors agree with $P^{\ten}_{a a' b b'}$ and $P^{\tenbar}_{a a' b b'}$ in equation (1.19) of \cite{Keppeler:2012ih}.  They also agree with the corresponding projectors in equation (12) of \cite{Mekhfi:1985dv}, provided that we identify $\langle a' b' | P_{R} | a b \rangle = P_{R}^{a b\, a' b'}$ in analogy to equation (2) of that paper.  Note that the form of these projectors is correct for SU($N$) and not restricted to $N=3$~\cite{Keppeler:2012ih}.
The symmetry relation \eqref{proj-sym} is not valid for decuplet projectors and must be corrected to read
\begin{align}
  \label{proj-sym-corr}
\Pro{R}{\ul{r}\, \ul{s}} &= \Pro{\Rbar}{\ul{s}\, \ul{r}} \,,
\end{align}
where $\Rbar$ is the conjugate representation of $R$.  It is understood that $\overline{R} = 10$ for $R = \overline{10}$ and that $\Rbar = R$ for the singlet, all octets, and for $R=27$.  The projector relation \eqref{proj-prop} remains valid.

For the mixed gluon octet channels, we introduce the tensors
\begin{align}
\Pro{AS}{a a'\, b b'} &= \frac{1}{\sqrt{N^2-4}}\ms f^{a a' c} d^{b b' c} \,,
&
\Pro{\SA}{a a'\, b b'} &= \frac{1}{\sqrt{N^2-4}}\ms d^{a a' c} f^{b b' c} \,,
\end{align}
which satisfy the relations
\begin{align}
\label{Ptilde-basics}
\Pro{AS}{\ul{a}\, \ul{b}} &= \Pro{\SA}{\ul{b}\, \ul{a}} \,,
&
\Pro{AS}{\ul{a}\, \ul{a}} &= \Pro{\SA}{\ul{a}\, \ul{a}} = 0 \,.
\end{align}
They are not projectors, since they do not satisfy \eqref{proj-prop}.

We also note that \eqref{proj-prop} is incorrect for the octet tensors with mixed fundamental and adjoint indices: the contraction $P_{A}^{a a'\, i i'} P_{S}^{i' i\, b b'}$ is not zero but equal to $P_{AS}^{a a'\, b b'}$.  The octet tensors in \eqref{mixed-proj} are therefore not projectors either.  Fortunately, the incorrect instances of \eqref{proj-prop} were not used in our calculations.

To have a uniform notation for calculations, we write the projectors as
\begin{align}
\Pro{R}{\ul{r}\, \ul{s}} &= \Pro{R \ms \Rbar}{\ul{r}\, \ul{s}} \,,
\end{align}
where the double indices $\ul{r}$ and $\ul{s}$ are both in the fundamental or both in the adjoint representation.  We furthermore write $P_{R_1 R_2\rule{0pt}{1.3ex}}^{\ul{i}\, \ul{a}} = P_{R_2\ms R_1\rule{0pt}{1.3ex}}^{\ul{a}\, \ul{i}}$ for the tensors in \eqref{mixed-proj}, with $R_1 = 1, 8$ for the fundamental indices and $R_2 = 1, A, S$ for the adjoint ones.
We then have the simple rules
\begin{align}
\label{P-rules}
\Pro{R_1 R_2}{\ul{r}\, \ul{s}} &= \Pro{R_2 R_1}{\ul{s}\, \ul{r}} \,,
&
\Pro{R_1 R_2}{\ul{r}\, \ul{s}} \, \Pro{R_3 R_4}{\ul{s}\, \ul{t}}
   &= \delta_{R_2\ms \Rbar_3}^{} \, \Pro{R_1 R_4}{\ul{r}\, \ul{t}} \,.
\end{align}
The corrected form of \eqref{PR-norm} hence reads
\begin{align}
\label{PR-norm-corr}
\Pro{R_1 R_2}{\ul{r}\, \ul{s}}\, \Pro{R_3 R_4}{\ul{r}\, \ul{s}}
   &= \delta_{R_2\ms \Rbar_4}^{} \, \Pro{R_1 R_3}{\ul{r}\, \ul{r}}
    = \delta_{R_1 \Rbar_3}^{} \, \delta_{R_2\ms \Rbar_4}^{} \; m(R_1) \,,
\end{align}
where the definition \eqref{mR-def} of the multiplicity $m(R)$ remains valid as it stands.\footnote{For general $N$, one obtains multiplicities $m(8) = N^2 -1$ for the octet and $m(10) = (N^2 - 1) (N^2 - 4) /4$ for the decuplet representations.}

The tensor $P_{R_1 R_2}$ couples two double-index objects in the representations $R_1$ and $R_2$ to an overall singlet.  Hence, the two representations must have the same multiplicity.  Formally one can impose this by defining $P_{R \Rp} = 0$ for $m(R) \neq m(\Rp)$.  One should then also define $P_{\ten \ten} = P_{\tenbar\, \tenbar} = 0$.

An object with four colour indices that transforms as an overall colour singlet can now be represented in terms of the tensors $P_{R \Rp}$.  The corrected form of the decomposition \eqref{eq:mat-dec} reads
\begin{align}
\label{mat-dec-corr}
M^{\ul{r}\, \ul{s}}
   &= \sum_{R \Rp} \frac{1}{m(R)}\,
   \Pro{R\ms \Rp}{\ul{r}\, \ul{s}}\;
   \Bigl( \Pro{\Rbar\, \Rpbar}{\ul{t}\, \ul{u}} \ms
      M^{\ul{t}\, \ul{u}}_{\phantom{1}} \Bigl) \,.
\end{align}
A simple example illustrating the necessity to include $R \Rp = AS$ in the sum over representation pairs is $M^{a a'\, b b'} = f^{a a' c} \ms d^{b b' c}$.  The relation \eqref{eq:contr_proj} correctly reads
\begin{align}
	\label{eq:contr_proj-corr}
M_1^{\ul{r}\, \ul{s}}\, M_2^{\ul{r}\, \ul{s}}
&= \sum_{R \Rp} \frac{1}{m(R)}\,
\bigl( \Pro{R\ms \Rp}{\ul{r}\, \ul{s}} \, M_1^{\ul{r}\, \ul{s}} \bigr)\,
\bigr( \Pro{\Rbar\, \Rpbar}{\ul{t}\, \ul{u}} \, M_2^{\ul{t}\, \ul{u}} \bigr)
\,.
\end{align}
Note that, for each contracted double index, the corresponding representations in $M_1$ and $M_2$ are conjugate to each other as a consequence of \eqref{PR-norm-corr}.

Various derivations and results in the paper must be adjusted to reflect the change from one to two representation labels and the conjugation of representations (which is trivial for all representations apart from the decuplet and antidecuplet).  We present the necessary changes in the order in which they appear in the manuscript.

The corrected form for the colour decomposition \eqref{dpd-colour-decomp} of a DPD is
\begin{align}
\label{dpd-colour-decomp-corr}
F_{a_1 a_2}^{\ul{r}_1 \ul{r}_2} &=
\sum_{R_1 R_2}
  \frac{1}{\mathcal{N}_{a_1} \mathcal{N}_{a_2}}\,
  \frac{1}{\epstwo{R_1}{R_2}}\,
  \frac{1}{\sqrt{m(R_1)}}\;
  \pr{R_1 R_2}{F}_{a_1 a_2}\, \Pro{R_1 R_2}{\ul{r}_1 \ul{r}_2}
\end{align}
with
\begin{align}
\label{proj-dpd-def-corr}
\pr{R_1 R_2}{F}_{a_1 a_2} &=
  \mathcal{N}_{a_1} \mathcal{N}_{a_2}\, \epstwo{R_1}{R_2}\,
  \frac{1}{\sqrt{m(R_1)}}\;
  \Pro{\Rbar_1\ms \Rbar_2}{\ul{s}_1 \ul{s}_2}
   F_{a_1 a_2 \phantom{R}}^{\ul{s}_1 \ul{s}_2} \,,
\end{align}
where instead of $\varepsilon_a(R)$ from \eqref{eps-def} we now use
\begin{align}
  \label{eps-def-corr}
\varepsilon(A) &= i \,,
&
\varepsilon(R) &= 1 \text{~~for } R \neq A .
\end{align}
With our new notation, the definition \eqref{col-proj-op} of colour projected twist-two operators reads
\begin{align}
  \label{col-proj-op-corr}
\prn{R}{O}^{\ms\ul{r}}_{a\rule{0pt}{1.15ex}}
 &= \mathcal{N}_a\, \varepsilon(R)\, \Pro{R\ms \Rbar}{\ul{r} \ul{s}}\,
    O^{\ms\ul{s}}_{a\rule{0pt}{1.15ex}} \,,
\end{align}
and \eqref{col-proj-matel} must be modified to
\begin{align}
  \label{col-proj-matel-corr}
2\pi \delta(p^+ - p'^+)\, 2p^+\, \pr{R_1 R_2}{F}_{\us, a_1 a_2}
&=  \frac{1}{\sqrt{m(R_1)}}\, \Pro{\Rbar_1\ms \Rbar_2}{\ul{r}\, \ul{s}} \;
    \big\langle p' \big|\, \prn{R_1}{O}^{\ms\ul{r}}_{a_1}\,
       \prn{R_2}{O}^{\ms\ul{s}}_{a_2}\, \big| p \big\rangle \,.
\end{align}
For $R_2 = \Rbar_1$, one can eliminate the projector on the r.h.s.\ by using that $P_{R \Rbar}^{\ul{r}\, \ul{s}} \, \prn{R}{O}^{\ms\ul{s}}_{a\rule{0pt}{1.15ex}} = \prn{R}{O}^{\ms\ul{r}}_{a\rule{0pt}{1.15ex}}$.

We modify the colour projection \eqref{soft-proj} of the soft matrix as
\begin{align}
\label{soft-proj-corr}
\prn{R_1^{} R_2^{}, \Rp_1 \Rp_2}{S_{a_1 a_2}} &=
  \frac{\epstwo{R_1}{R_2}}{\epstwo{\Rp_1}{\Rp_2}}
  \frac{1}{\sqrt{m(R_1)\, m(\Rp_1)}}\, \Pro{\Rbar_1 \Rbar_2}{\ul{r}_1
  \ul{r}_2}\, S_{a_1 a_2 \phantom{R}}^{\ul{r}_1 \ul{r}_2, \ul{s}_1
  \ul{s}_2}\, \Pro{\Rpbar_1 \Rpbar_2}{\ul{s}_1 \ul{s}_2} \,.
\end{align}
The same holds for the colour space matrices $S^{-1}$, $s$, $s^{-1}$, and $K$.  Together with \eqref{eq:contr_proj-corr}, this leads to a modification of
the multiplication rule for matrices in representation space, which we define in general by
\begin{align}
\label{mat-mult-def}
\prn{R_1^{} R_2^{}, \Rpp_{1} \Rpp_2}{(M_1 \cdot M_2)}
&= \sum_{\Rp_1 \Rp_2}
   \prn{R_1^{} R_2^{}, \Rpbar_1 \Rpbar_2}{M_1}\;
      \prn{\Rp_1 \Rp_2, \Rpp_1 \Rpp_2}{M_2} \,,
\end{align}
where the projections of $M_1$ and $M_2$ on representations are defined as in \eqref{soft-proj-corr}.  The correct form of \eqref{S-inverse-proj} then reads
\begin{align}
   \label{S-inverse-proj-corr}
\prn{R_1^{} R_2^{}, \Rpp_1 \Rpp_2}{(S_{a_1 a_2}^{-1} \cdot S_{a_1 a_2}^{})}\;
& =
\prn{R_1^{} R_2^{}, \Rpp_1 \Rpp_2}{(s_{a_1 a_2}^{-1} \cdot s_{a_1 a_2}^{})}\;
  = \delta_{R_1^{} \Rppbar_1}\, \delta_{R_2^{} \Rppbar_2} \,,
\end{align}
and corresponding changes are to be made in \eqref{S-decomp-proj} and \eqref{sub-unsub}.
Again as a consequence of \eqref{eq:contr_proj-corr}, the expression \eqref{CS-Xsect-final} for the cross section is modified to
\begin{align}
	\label{CS-Xsect-final-corr}
X &= \frac{H_{a_1
    b_1}}{\mathcal{N}_{a_1}\ms \mathcal{N}_{b_1}}\, \frac{H_{a_2
    b_2}}{\mathcal{N}_{a_2}\ms \mathcal{N}_{b_2}}\,
    \sum_{R_1 R_2} \etatwo{R_1}{R_2}\;
    \pr{\Rbar_1 \Rbar_2}{F_{b_1 b_2}} \pr{R_1^{} R_2^{}}{F_{a_1 a_2}}
\end{align}
with
\begin{align}
\eta(R) &= \varepsilon^2(R) \,,
\end{align}
so that $\eta(A) = -1$ and $\eta(R) = 1$ for $R \neq A$.

We now discuss the necessary adjustments to the symmetry properties derived in section \ref{sec:soft-symm}.  Using that $\Pro{R_1 R_2}{a a'\, b b'} = \bigl[ \Pro{\Rbar_1 \Rbar_2}{a a'\, b b'} \ms\bigr]^*$, one finds that \eqref{S-hermit} and \eqref{S-anti} must be corrected to
\begin{align}
  \label{S-hermit-corr}
\prn{R_1^{} R_2^{}, \Rpbar_1 \Rpbar_2}{S}_{a_1 a_2}
   &= \left( \prn{\Rp_1 \Rp_2, \Rbar_1^{} \Rbar_2^{}}{S}_{a_1 a_2} \right)^* \,,
\\
  \label{S-anti-corr}
\prn{R_1^{} R_2^{}, \Rpbar_1 \Rpbar_2}{S}_{a_1 a_2}
 &= \frac{\etatwo{\Rp_1}{\Rp_2}}{\etatwo{R_1}{R_2}} \,
  \left( \prn{\Rbar_1^{} \Rbar_2^{}, \Rp_1 \Rp_2}{S}_{\bar{a}_1 \bar{a}_2}
  \right)^*
\nonumber \\
 &= \frac{\etatwo{\Rp_1}{\Rp_2}}{\etatwo{R_1}{R_2}} \;
  \prn{\Rpbar_1 \Rpbar_2, R_1^{} R_2^{}}{S}_{\bar{a}_1 \bar{a}_2} \,.
\end{align}
An analogous correction is to be made for \eqref{K-anti}.  The result that the soft matrix in representation space is real valued remains true, so that the complex conjugation in \eqref{S-hermit-corr} and \eqref{S-anti-corr} can be omitted.  Doing this in the first line of \eqref{S-anti-corr}, one finds that
\begin{align}
\label{soft-tens}
\prn{\tentenbar, \tenbarten}{S}_{g g}
   &= \prn{\tenbarten, \tentenbar}{S}_{g g} \,,
&
\prn{\tentenbar, \tentenbar}{S}_{g g}
   &= \prn{\tenbarten, \tenbarten}{S}_{g g} \,,
\end{align}
which is expected because charge conjugation interchanges the representations $10$ and $\overline{10}$.  Furthermore, one finds that $\prn{R_1^{} R_2^{}, \Rp_1 \Rp_2}{S}_{g g} = 0$ if there is an odd number of representations $A$ and no representation is equal to $10$ or $\overline{10}$.  This reflects the fact that the antisymmetric and the symmetric gluon octets transform under charge conjugation with a relative minus sign.  Corresponding results hold for the Collins-Soper Kernel $K_{g g}$.

Using that $\epstwo{R_1}{R_2} \ms P_{\rule{0pt}{1.3ex}\Rbar_1 \Rbar_2}^{a a'\, b b'} = \bigl[\ms \epstwo{R_1}{R_2} \ms P_{\rule{0pt}{1.3ex}R_1 R_2}^{a' \! a\; b'\bs b} \,\bigr]^*$, one finds that $\pr{R_1 R_2}{F}_{a_1 a_2}(x_i, \tvec{y})$ is real valued in all colour channels except for the decuplets, for which one has $\pr{\tentenbar}{F}_{g g}(x_i, \tvec{y}) = \bigl[\ms \pr{\tenbarten}{F}_{g g}(x_i, \tvec{y}) \bigr]^*$.


\paragraph{The soft factor for collinear DPDs.}

The relations \eqref{comm-glu} and \eqref{comm-mixed} become
\begin{align}
   \label{comm-glu-mixed-corr}
W^{\phantom{\dagger}}_{a b}\, W^\dagger_{b' a'}\, \Pro{R \Rp}{b b'\,c c'}
  &= \Pro{R \Rp}{a a'\,b b'}\,
      W^{\phantom{\dagger}}_{b c}\, W^\dagger_{c' b'} \,,
\nonumber \\
W^{\phantom{\dagger}}_{a b}\, W^\dagger_{b' a'}\, \Pro{R \Rp}{b b'\,k' k}
   &= \Pro{R \Rp}{a a'\,j' j}\,
      W^{\phantom{\dagger}}_{j k}\, W^\dagger_{k' j'} \,,
&
W^{\phantom{\dagger}}_{i j}\, W^\dagger_{j' i'}\, \Pro{R \Rp}{j j'\, a a'}
   &= \Pro{R \Rp}{i i'\, b b'}\,
      W^{\phantom{\dagger}}_{ab}\, W^\dagger_{b' a'} \,,
\end{align}
so that \eqref{comm-soft} generalises to
\begin{align}
  \label{comm-soft-corr}
(S_{a_1 a_2})^{\cdots\, \ul{r} \,\cdots}_{\cdots\, \ul{s} \,\cdots} \;
   \Pro{R \Rp}{\ul{s} \,\ul{t}}
&= \Pro{R \Rp}{\ul{r} \,\ul{s}}\;
   (S_{a_1 a_2})^{\cdots\, \ul{s} \,\cdots}_{\cdots\, \ul{t} \,\cdots} \,.
\end{align}
Note that these relations do \emph{not} involve conjugation of representation labels.  As a corollary, one obtains
\begin{align}
\Pro{R \Rp}{\ul{r}_1 \ul{s}}\;
   S_{a_1 a_2 \phantom{R}}^{\ul{s}\, \ul{r}_2, \ul{t}_1 \ul{t}_2}
&= S_{a_1 a_2 \phantom{R}}^{\ul{r}_1 \ul{r}_2, \ul{s}\, \ul{t}_2}\;
   \Pro{R \Rp}{\ul{s}\, \ul{t}_1}
\end{align}
and its analogue for the second parton.
Using these results and the rules in \eqref{P-rules}, one can generalise equations \eqref{comm-soft} to \eqref{coll-soft-final} to all valid combinations of representations.  We find that for the collinear soft factor the first two representations must be conjugate to the second two,
\begin{align}
\label{soft-factor-diag-corr}
\prn{R_1^{} R_2^{}, \Rp_{1} \Rp_2}{S}_{a_1 a_2}
   &= \delta_{R_1^{} \Rpbar_1} \, \delta_{R_2^{} \Rpbar_2} \,
      \prn{R_1^{} R_2^{}, \Rbar_1^{} \Rbar_2^{}}{S}_{a_1 a_2} \,,
\end{align}
and that for singlet and octet channels, the soft factors are equal for the quark and gluon representations.  We thus get unity in the singlet sector,
\begin{align}
  \label{coll-soft-sing-corr}
\pr{1}{S} &= \pr{11, 11}{S_{q q}} = \pr{11, 11}{S_{q g}} = \pr{11, 11}{S_{g q}}
           = \pr{11, 11}{S_{g g}} = 1 \,,
\intertext{and three non-trivial factors}
  \label{coll-soft-final-corr}
\pr{8}{S} &= \pr{88, 88}{S_{q q}}
   = \pr{8A, 8A}{S_{q g}} = \pr{8S, 8S}{S_{q g}}
   = \pr{A8, A8}{S_{g q}} = \pr{S8, S8}{S_{g q}}
\nonumber \\
  &= \pr{AA, AA}{S_{g g}} = \pr{SS, SS}{S_{g g}}
   = \pr{AS, AS}{S_{g g}} = \pr{\SA, \SA}{S_{g g}}\,,
\nonumber \\
\pr{10}{S} &= \pr{\tentenbar, \tenbar\ms \ten}{S_{g g}}
            = \pr{\tenbarten, \ten\, \tenbar}{S_{g g}} \,,
\nonumber \\
\pr{27}{S} &= \pr{27\, 27, 27\, 27}{S_{g g}} \,,
\end{align}
where for the decuplet sector we used \eqref{soft-tens}.  Analogous results hold for the Collins-Soper Kernel $J$ for collinear DPDs.  On the left-hand side of \eqref{coll-soft-final-corr}, we have labelled the soft factors by one representative of the different labels on the right-hand sides.  Taking a different representative, one has $\pr{8}{S} = \pr{A}{S} = \pr{S}{S}$ and $\pr{10}{S} = \pr{\overline{10}}{S}$.

The extended soft factor can be expressed in terms of the one for colour singlet production, and in generalisation of \eqref{soft-factor-comm}, we have
\begin{align}
	\label{soft-factor-comm-corr}
& \Pro{\Rbar_1^{} \Rbar_2^{}}{\ul{r}_1 \ul{r}_2}\,
  \Pro{\Rpbar_1 \Rpbar_2}{\ul{u}_1 \ul{u}_2}\,
  \Pro{\Rbar_3^{} \Rpbar_3}{\ul{s}_1 \ul{t}_1}\,
  \Pro{\Rbar_4^{} \Rpbar_4}{\ul{s}_2 \ul{t}_2}\,
  (S_{a_1 a_2})^{\ul{r}_1 \ul{r}_2, \ul{u}_1 \ul{u}_2}_{\ul{s}_1
      \ul{s}_2, \ul{t}_1 \ul{t}_2}
\nonumber \\
&\qquad\qquad =
   \delta_{R_{1} \Rbar_3}^{}\, \delta_{R_2 \Rbar_4}^{}\,
   \delta_{\Rp_{1} \Rpbar_3}\, \delta_{\Rp_2 \Rpbar_4}\;
   m(R_1)\, \pr{R_1}{S} \,,
\end{align}
where the combination of tensors on the left and Kronecker deltas on the right implies that all representations must have the same multiplicity.  This can be used to correct the argument in section~\ref{sec:interlude} for the production of final states with net colour.  In the definition \eqref{col-Xsect-start}, the order of colour indices in $H_{a b}$ must be interchanged (see figure~\ref{fig:jets-fact}), so that
\begin{align}
	\label{col-Xsect-start-corr}
X = H_{a_1 b_1}^{\ul{t}_1 \ul{s}_1}\, H_{a_2 b_2}^{\ul{t}_2 \ul{s}_2}\, &
\bigl[ F^T_{\us, b_1 b_2}(Y_R) \, S^{-1}(Y_R-Y_L) \bigr]^{ \ul{r}_1
  \ul{r}_2} \,
\nonumber \\
 & \times \bigl[ S(Y_R-Y_L) \bigr]^{\ul{r}_1
  \ul{r}_2, \ul{u}_1 \ul{u}_2}_{ \ul{s}_1 \ul{s}_2, \ul{t}_1 \ul{t}_2} \,
\bigr[ S^{-1}(Y_R-Y_L)\, F^{}_{\us, a_1 a_2}(Y_L) \bigr]^{ \ul{u}_1
  \ul{u}_2} \,.
\end{align}
Correcting \eqref{col-Xsect-int} to include all relevant colour channels, we obtain as a replacement of \eqref{col-Xsimp2} the final expression
\begin{align}
\label{col-Xsimp2-corr}
X &= \sum_{R_1^{} R_2^{} \Rp_1 \Rp_2}
   \prn{\Rbar_1^{} \Rp_{1}}{H_{a_1 b_1}}\,
   \prn{\Rbar_2^{} \Rp_2}{H_{a_2 b_2}}\,
   \prn{\Rpbar_1 \Rpbar_2}{F_{b_1 b_2}}\,
   \prn{R_1^{} R_2^{}}{F_{a_1 a_2}}
\end{align}
with hard-scattering factors
\begin{align}
  \label{H-col-def-corr}
\prb{R \Rp}{H}_{a b} &=
   \frac{1}{\mathcal{N}_{a}\, \mathcal{N}_{b}}\,
   \frac{1}{\epstwo{R}{\Rp}}\, \frac{1}{m(R)}\,
   P_{\Rbar^{}\ms \Rpbar}^{\ul{s}\, \ul{t}}\,
   H_{a b}^{\ul{s}\, \ul{t}} \,.
\end{align}
For colour singlet final states, one has $\prb{R \Rp}{H}_{a b} = \delta_{R \Rpbar}\, \eta(R)\, H_{a b} \big/ (\mathcal{N}_{a}\, \mathcal{N}_{b})$ with $H_{a b}$ defined in \eqref{hard-scatt-sing}.  Then \eqref{col-Xsimp2-corr} reduces to \eqref{CS-Xsect-final-corr}.


\paragraph{Evolution of DPDs.}
The colour structure of DPDF and DTMD evolution equations and their solutions must be adjusted for the number of representation labels and the difference between a representation and its conjugate.

The Collins-Soper kernel for DTMDs must have four representation labels, as does the soft factor from which it is derived.  The Collins-Soper equation \eqref{CS-TMD} then becomes
\begin{align}
	\label{CS-TMD-corr}
& \frac{\partial}{\partial\log \zeta} \prn{R_1 R_2}{F_{a_1
    a_2}}(x_i,\tvec{z}_i,\tvec{y}; \mu_i,\zeta)
\nonumber \\
&\qquad = \frac{1}{2} \sum_{\Rp_1 \Rp_2}
\prn{R_1^{} R_2^{}, \Rpbar_1 \Rpbar_2}{K_{a_1 a_2}(\tvec{z}_i,\tvec{y}; \mu_i)}\, \prn{\Rp_1 \Rp_2}{F_{a_1 a_2}(x_i,\tvec{z}_i,\tvec{y}; \mu_i,\zeta)}
\end{align}
with a kernel satisfying
\begin{align}
	\label{CS-TMD-RG-corr}
\frac{\partial}{\partial \log\mu_1}\, \prn{R_1^{} R_2^{}, \Rp_1 \Rp_2}{K_{a_1
    a_2}(\tvec{z}_i,\tvec{y}; \mu_i)}
    &= {}- \gamma_{K, a_1}^{}(\mu_1)\,
      \delta_{R_1^{} \Rpbar_1}\, \delta_{R_2^{} \Rpbar_2}
\end{align}
and a corresponding equation for $\mu_2$.  The matrix exponential in \eqref{CS-TMD-sol} is now defined as the exponential series with the matrix multiplication in \eqref{mat-mult-def}.  Using the analogue of \eqref{S-anti-corr} for $K_{a_1 a_2}$, the relation \eqref{rewrite-M-exp} becomes
\begin{align}
\label{rewrite-M-exp-corr}
& \etatwo{R_1}{R_2}\;
\pr{\Rbar_1^{} \Rbar_2^{}, \Rp_1 \Rp_2}{\exp}\,\bigl[ M_{b_1 b_2} L
    \bigr]\, \prn{\Rpbar_1 \Rpbar_2}{F}_{b_1 b_2}
\nonumber \\
& \qquad =
\etatwo{\Rp_1}{\Rp_2}\; \prn{\Rpbar_1 \Rpbar_2}{F}_{b_1 b_2}\;
    \pr{\Rp_1 \Rp_2, \Rbar_1^{} \Rbar_2^{}}{\exp}\,\bigl[ M_{a_1 a_2} L
    \bigr] \,,
\end{align}
where we have abbreviated the logarithm by $L$.  The matrix $M_{a_1 a_2}$ is defined by the updated version of \eqref{split-K-M},
\begin{align}
	\label{split-K-M-corr}
& \prn{R_1^{} R_2^{}, \Rp_1 \Rp_2}{K_{a_1 a_2}(\tvec{z}_i,\tvec{y}; \mu_i)}
\nonumber \\
&\qquad = \delta_{R_1^{} \Rpbar_1}\, \delta_{R_2^{} \Rpbar_2}\,
\bigl[ \pr{11}{K}_{a_1}(\tvec{z}_1;\mu_1)
     + \pr{11}{K}_{a_2}(\tvec{z}_2;\mu_2) \bigr]
 + \prn{R_1^{} R_2^{}, \Rp_1 \Rp_2}{M_{a_1 a_2}(\tvec{z}_i,\tvec{y})}\,,
\end{align}
where $\pr{11}{K}_{a}$ is the Collins-Soper kernel for a single-parton TMD.  The corrected form of \eqref{W-generic} will be given below.

The kernel for the rapidity evolution of DPDFs derives from the collinear soft factor and hence is labelled by a single representation.  We thus have
\begin{align}
	\label{CS-coll-corr}
\frac{\partial}{\partial\log \zeta}\,
\prn{R_1 R_2}{F}(x_i,\tvec{y}; \mu_i,\zeta)
&= \frac{1}{2}\, \prb{R_1}{J(\tvec{y}; \mu_i)}\,
   \prn{R_1 R_2}{F(x_i,\tvec{y}; \mu_i,\zeta)}
\end{align}
as update of \eqref{CS-coll}, where instead of $\prb{R_1}{J}$ we could also write $\prb{R_2}{J}$.
By contrast, the DGLAP kernels for DPDFs carry two colour indices, and the evolution equation for the first parton scale has the structure
\begin{align}
\frac{\partial}{\partial \log\mu_1}\,
\prn{R_1 R_2}{F_{a_1 a_2}}
&= 2 \sum_{b_1, \Rp_1} \prn{R_1^{} \Rpbar_1}{P_{a_1 b_1}} \conv{x_1}
   \prn{\Rp_1 R_2^{}}{F_{b_1 a_2}} \,.
\end{align}
The Collins-Soper equation \eqref{DGLAP-CS} for the evolution kernel should be modified to
\begin{align}
	\label{DGLAP-CS-corr}
\frac{\partial}{\partial \log\zeta}\, \prb{R \Rp}{P_{a b}(x;\mu,\zeta)} &= -
\frac{1}{4}\, \delta_{R \Rpbar}\, \delta_{ab}^{}\, \delta(1-x)\,
   \prn{R}{\gamma_J}(\mu) \,,
\end{align}
and corresponding changes are to be made in \eqref{DGLAP-zeta-expl}.
As a consequence of charge conjugation invariance, one has
\begin{align}
\pr{\tentenbar}{P}_{g g} &= \pr{\tenbarten}{P}_{g g} \,,
&
\pr{AS}{P}_{g g} &= \pr{\SA}{P}_{g g} = 0 \,,
\end{align}
see the discussion after equation \eqref{S-anti-corr}.


\paragraph{Short-distance matching.}
We now specify how the definitions and main results for the short-distance matching of of DTMDs in section \ref{sec:short} are to be adjusted.  The necessary adjustments of intermediate steps and of corollaries are easy to deduce.  Starting in section~\ref{sec:small_zi}, we define colour projections
\begin{align}
\label{CS-proj-def}
\prn{R}{C}_{S, a}(\tvec{z}) &= \Pro{\Rbar\ms R}{\ul{s}\, \ul{t}} \;
   C_{S, a}^{\ul{s}\, \ul{t}}(\tvec{z}) \big/ m(R)
\end{align}
for the matching kernel of the TMD soft factor for small $\tvec{z}$.  The two representation labels of $P_{}^{\ul{s}\, \ul{t}}$ in \eqref{CS-proj-def} must be conjugate to each other, because $P_{\rule{0pt}{1.25ex}AS}^{\ul{s}\, \ul{t}} \; C_{S, g}^{\ul{s}\, \ul{t}} = P_{\rule{0pt}{1.25ex}\SA}^{\ul{s}\, \ul{t}} \; C_{S, g}^{\ul{s}\, \ul{t}} = 0$ due to charge conjugation invariance, and because the double indices $\ul{s}$ and $\ul{t}$ are either both in the fundamental or both in the adjoint representation.  From \eqref{soft-op-start} we then obtain
\begin{align}
\label{soft-fact-match-corr}
\prn{R_1^{} R_2^{}, \Rp_{1} \Rp_2}{S_{a_1 a_2}(\tvec{z}_i, \tvec{y})}
&= \delta_{R_1^{} \Rpbar_1}\, \delta_{R_2^{} \Rpbar_2}
\prn{R_1}{C_{S, a_1}(\tvec{z}_1)}\, \prn{R_2}{C_{S, a_2}(\tvec{z}_2)}\;
\pr{R_1}{S(\tvec{y})}
\end{align}
instead of \eqref{soft-fact-match}.  The matching of the twist-two operators at small $\tvec{z}$ reads
\begin{align}
  \label{x-op-match-indices-corr}
{O}^{\ul{r}}_a(x,\tvec{y},\tvec{z})
  &= \sum_{b} C_{\us, a b}^{\ul{r}\, \ul{s}}(x',\tvec{z})\,
     \underset{x}{\otimes} {O}^{\ul{s}}_b(x',\tvec{y}) \,.
\end{align}
In terms of the colour projected matching coefficients
\begin{align}
\pr{R \Rp}{C}_{\us, a b} &=
   \frac{\mathcal{N}_a}{\mathcal{N}_b}\,
   \frac{\varepsilon(R)}{\varepsilon(\Rp)}\,
   \frac{1}{m(R)}\, \Pro{\Rbar\, \Rpbar}{\ul{r}\, \ul{s}}\,
   C_{\us, a b}^{\ul{r}\, \ul{s}} \,,
\end{align}
one obtains the corrected form of \eqref{x-op-match} as
\begin{align}
  \label{x-op-match-indices}
\prn{R}{O}^{\ul{r}}_a(x,\tvec{y},\tvec{z})
&= \sum_{b, \Rp}
  \pr{R \Rpbar}{C}_{\us, a b}(x',\tvec{z})\,
  \underset{x}{\otimes} \Pro{R \Rpbar}{\ul{r}\, \ul{s}} \,
  \pr{R'}{O}^{\ul{s}}_b(x',\tvec{y})
\end{align}
for the colour projected operators defined in \eqref{col-proj-op-corr}.  The possibility of quark-gluon mixing requires two colour labels $R$ and $\Rp$ for $C_{\us, a b}$, in contrast to $C_{S, a}$.  Adjusting the following steps in the manuscript and defining
\begin{align}
	\label{full-C-def-corr}
\pr{R \Rp}{C_{a b}(x, \tvec{z};\mu,Y_C)} &= \lim_{Y_L \to -\infty}
   \frac{\pr{R \Rp}{C_{\text{us}, a b}(x, \tvec{z}; \mu,Y_L)}}{
   \sqrt{\prn{R}{C}_{S, a}(\tvec{z}; \mu,2Y_C - 2Y_L)\rule{0pt}{1.9ex}}} \,,
\end{align}
one finds that the correct colour structure for the matching of a DTMD onto a DPDF is
\begin{align}
	\label{full-match-color}
\prn{R_1 R_2}{F_{a_1 a_2}(\tvec{z}_i, \tvec{y})}
 &= \sum_{b_1 b_2, \Rp_1 \Rp_2}
    \prn{R_1^{} \Rpbar_1}{C_{a_1 b_1}(\tvec{z}_1)}
    \underset{x_1}{\otimes}
    \prn{R_2^{} \Rpbar_2}{C_{a_2 b_2}(\tvec{z}_2)}
    \underset{x_2}{\otimes}
    \prn{\Rp_1 \Rp_2}{F_{b_1 b_2}(\tvec{y})} \,,
\end{align}
where for brevity, we only give the transverse position arguments.  The full version is given in \eqref{full-match-corr} below.  Due to charge conjugation invariance, one has
\begin{align}
\prn{\tentenbar}{C}_{g g} &= \prn{\tenbarten}{C}_{g g} \,,
&
\prn{AS}{C}_{g g} &= \prn{\SA}{C}_{g g} = 0 \,,
\end{align}
and analogous relations for $C_{\us, g g}$.

The definitions of the kernel $\pr{R}{K}_a$ for the rapidity dependence of $\prn{R}{C}_{S, a}$ and of the associated anomalous dimension $\pr{R}{\gamma}_a$  remain as given in \eqref{C-soft-CS} and \eqref{gamma-K-def}.  Equation \eqref{CS-coeff} hence becomes
\begin{align}
	\label{CS-coeff-corr}
\frac{\partial}{\partial \log\zeta}\,
   \pr{R \Rp}{C_{ab}(x, \tvec{z}; \mu,\zeta)}
 &= \frac{1}{2}\, \pr{R}{K_a}(\tvec{z}; \mu)\,
    \pr{R \Rp}{C_{a b}(x, \tvec{z}; \mu,\zeta)} \,,
\end{align}
whereas the limiting expression \eqref{CS-gen-match} for small $\tvec{z}_i$ should be replaced with
\begin{align}
	\label{CS-gen-match-corr}
& \pr{R_1^{} R_2^{}, \Rp_1 \Rp_2}{K_{a_1 a_2}(\tvec{z}_i,\tvec{y}; \mu_i)}
\nonumber \\
&\qquad = \delta_{R_1^{} \Rpbar_1}\, \delta_{R_2^{} \Rpbar_2}\,
   \bigl[\ms \pr{R_1}{K_{a_1}(\tvec{z}_1;\mu_1)} +
             \pr{R_2}{K_{a_2}(\tvec{z}_2;\mu_2)} +
             \prb{R_1}{J(\tvec{y}; \mu_i)} \bigr] \,.
\end{align}
Charge conjugation invariance implies $\prn{\ten}{C}_{S,g} = \prn{\tenbar}{C}_{S, g}$ and hence $\prn{\ten}{K}_g = \prn{\tenbar}{K}_g$.
The representation labels in \eqref{small-z-start} and \eqref{CS-evolve-C} should be adapted as in \eqref{CS-gen-match-corr} and \eqref{CS-coeff-corr}, respectively.  The corrected forms of \eqref{DGLAP-C}, \eqref{small-z-evolved} and \eqref{W-large-y} will be given below.  The functions $g_F$ and $g_K$ introduced in section~\ref{sec:large-z} must have separate colour labels for the two partons, corresponding to the functions on the r.h.s.\ of their definitions in \eqref{g-function-defs}.

The result \eqref{soft-fact-small-y} for the expansion of the soft factor at small $\tvec{z}_1$, $\tvec{z}_2$ and $\tvec{y}$ correctly reads
\begin{align}
  \label{soft-fact-small-y-corr}
\prn{R_1^{} R_2^{}, \Rp_1 \Rp_2}{S_{a_1 a_2}}(\tvec{z}_i, \tvec{y}; \mu_i,Y)
&= \prn{R_1^{} R_2^{},
   \Rp_1 \Rp_2}{C_{S, a_1 a_2}}(\tvec{z}_i, \tvec{y}; \mu_i,Y)\,.
\end{align}
The kernel $T_{a_0 \to a_1 a_2}$ for parton-level splitting in \eqref{split-TMD-coll} and \eqref{coll-DPD-exp} should have separate labels $R_1, R_2$ for $a_1, a_2$, as do the DPDs.  The same holds for the twist-four function $G_{a_1 a_2}$ in \eqref{intr-TMD-tw4} and \eqref{coll-DPD-exp}.

In section~\ref{sec:gen-proc}, one should replace \eqref{gen-matel} with
\begin{align}
  \label{gen-matel-corr}
2\pi \delta(p^+ - p'^+)\, 2p^+ \,
\prb{R \Rp}{\mathcal{M}}_{ab}(x,\tvec{z})
& = \frac{1}{\mathcal{N}_b}\, \frac{1}{m(R)}\,
   \Pro{\Rpbar \Rbar}{\ul{r}\, \ul{s}}
   \bigl\langle b,p',r' \big|\ms \prn{R}{O}^{\ms\ul{s}}_a(x,\tvec{y},\tvec{z})
   \ms\big| b,p,r \bigr\rangle
\nonumber \\[0.2em]
&= \varepsilon(R)\; \frac{\mathcal{N}_a}{\mathcal{N}_b}\,
   \frac{1}{m(R)}\, \Pro{\Rpbar \Rbar}{\ul{r}\, \ul{s}}\,
   \bigl\langle b,p',r' \big|\ms O^{\ms\ul{s}}_a(x,\tvec{y},\tvec{z})
   \ms\big| b,p,r \bigr\rangle \,,
   \hspace{4em}
\nonumber \\[0.3em]
\pr{R}{\mathcal{M}}_{S,a}(\tvec{z})
&= \frac{1}{m(R)}\, \Pro{\Rbar R}{\ul{r}\, \ul{s}}\,
   \bigl\langle 0 \big|\ms
     O_{S,a}^{\ms \ul{r}, \ul{s}}(\tvec{y},\tvec{z})
     \ms\big|\ms 0 \bigr\rangle
\end{align}
and the first equation in \eqref{match-matel} with
\begin{align}
  \label{match-matel-corr}
\prb{R \Rpp}{\mathcal{M}}^{}_{a c}(x,\tvec{z})
&= \sum_{b, \Rp} \pr{R \Rpbar}{C}^{}_{\us, a b}(x',\tvec{z})\,
   \underset{x}{\otimes} \prb{\Rp \Rpp}{\mathcal{M}}^{}_{b c}(x') \,,
\end{align}
whilst the second equation in \eqref{match-matel} remains valid as it stands.
The adjustment of the remaining equations in that section is straightforward.


\paragraph{One-loop kernels.}
The kernel matrix \eqref{M-gg-LO} for Collins-Soper evolution of two-gluon DTMDs should be extended to include all colour channels.  We refrain from giving the corresponding expressions here.
In the limit $|\tvec{z}_1|, |\tvec{z}_2| \ll |\tvec{y}|$, the elements of this matrix can be expressed in terms of the kernels $\pr{R}{K}_g$ and $\pr{R}{J}_g$ according to \eqref{CS-gen-match-corr}.  We find that $\pr{10}{K}_g = 0$ and $\pr{10}{J} = \prb{D}{J}$ with $\prb{D}{J}$ given in \eqref{J-LO-higher}.

The colour factors $c_{a b}$ in \eqref{LO-propto} and $c_{a_0 \to a_1 a_2}$ in \eqref{split-col-ratio} now depend on two representation labels $(R_1 R_2)$ referring to the parton pairs $(a b)$ or $(a_1 a_2)$.  In addition to the results given in the manuscript, we find that both factors are zero for $R_1 R_2 = 10\, \overline{10}$ and $\overline{10}\, 10$.  This holds for general number of colours $N$.  Both factors are also zero for $R_1 R_2 = AS$ and $\SA$ due to charge conjugation invariance.


\subsection{Definition and rescaling of the rapidity parameter \texorpdfstring{$\zeta$}{zeta}}

The definition \eqref{zeta-def} of the rapidity parameter in DPDs refers to the plus-momenta $x_1 p^+$ and $x_2\ms p^+$ of the two extracted partons, in generalisation of the definition \eqref{zeta-def-TMD} for single-parton TMDs.  This leads to a problem in convolutions of DPDFs, which was overlooked and led to a number of mistakes in the original manuscript.  The problem concerns the evolution equation \eqref{DGLAP-zeta} of DPDFs and the matching equation \eqref{full-match} of DTMDs onto DPDFs.  The DPDFs on the right-hand-sides of these equations are to be taken at a fixed central rapidity $Y_C$.  This is explicit in the steps leading to \eqref{full-match}, and a corresponding derivation can be given for \eqref{DGLAP-zeta}.  This implies that in integrals over a momentum fraction of a DPD, the variable $\zeta$ does not remain constant but must change such that $Y_C$ remains constant.  Such a rescaling in all convolution integrals is possible, but we find it rather awkward.  To avoid it, we express the rapidity dependence of DPDFs by the parameter
\begin{align}
\label{zeta-p-def}
\zeta_p &= 2 (p^+)^2\, e^{-2 Y_C} \,,
\end{align}
which refers to the proton momentum.  We use this parameter also for DTMDs, given their close connection with DPDFs.  The handling of the rapidity dependence in  DTMDs is then different from the one for single-parton TMDs in the modern literature, notably in \cite{Collins:2011zzd}.  We note, however, that \eqref{zeta-p-def} corresponds to the definition of the $\zeta$ parameter in the original work of Collins and Soper \cite{Collins:1981uk}.
The parameter $\zeta_p$ is then to be used instead of $\zeta$ from \eqref{zeta-def} as argument for DTMDs and DPDFs throughout the paper.  In the following, we point out equations that change in a non-trivial way.

Defining the analogue of \eqref{zeta-p-def} for the left-moving proton as
\begin{align}
\label{zeta-bar-def}
\zeta_{\bar{p}} &= 2 (\bar{p}^-)^2\, e^{2 Y_C}\,,
\end{align}
the product of rapidity parameters is fixed to
\begin{align}
\zeta_p \ms \zeta_{\bar{p}} &= s^2 \,,
\end{align}
where $s = 2 p^+ \bar{p}^{-}$ is the squared c.m.\ energy of the proton-proton collision (neglecting proton mass corrections).  In terms of the invariant masses produced by the two hard scatters, one has $s = Q_1^2 \ms\big/ (x_1^{} \bar{x}_1^{})  = Q_2^2 \ms\big/ (x_2^{}\ms \bar{x}_2^{})$.

The definitions \eqref{zeta-p-def} and \eqref{zeta-bar-def} refer to the Collins regulator for rapidity divergences.  If one works in a different scheme, such as the $\delta$ regulator discussed in appendix \ref{app:delta-reg}, the replacement of $\zeta$ by $\zeta_p = \zeta /(x_1 x_2)$ as argument of DPDs must be implemented accordingly.


\paragraph{DTMD evolution.}  Equations that involve DTMDs but no DPDFs are correct in the original manuscript and just need to be rewritten to reflect the change of argument in the distributions.  In terms of the new rapidity parameter, the renormalisation group equation \eqref{RG-TMD-again} for DTMDs reads
\begin{align}
\label{RG-TMD-corr}
\frac{\partial}{\partial \log\mu_1}\,
  \prn{R_1 R_2}{F_{a_1 a_2}(x_i,\tvec{z}_i,\tvec{y};\mu_i,\zeta_p)}
&= \gamma_{F, a_1}(\mu_1, x_1^2 \ms \zeta_p)\,
   \prn{R_1 R_2}{F_{a_1 a_2}(x_i,\tvec{z}_i,\tvec{y};\mu_i,\zeta_p)}
\end{align}
and in analogy for the derivative w.r.t.\ $\log \mu_2$.  The rescaled rapidity parameters $x_1 \zeta /x_2$ and $x_2\ms \zeta/ x_1$ in \eqref{RG-TMD-sol} should hence be replaced with $x_1^2 \ms \zeta_p$ and $x_2^2 \ms \zeta_p$, respectively.

Some attention is needed when choosing initial conditions for Collins-Soper evolution.  In the matching relations for small $\tvec{y}$ discussed in section~\ref{sec:small-y}, the rapidity parameter dependence of the DTMD arises from a short-distance matching kernel, which does not know about the proton momentum and hence can depend on $x_1 x_2 \ms \zeta_p$ but not on $\zeta_p$.  Taking that kernel at a fixed scale $\zeta_0$ so as to minimise corrections from higher orders thus gives the DTMD at $\zeta_p = \zeta_0 / (x_1 x_2)$.  Assuming that a corresponding initial condition is also useful at large $\tvec{y}$, we therefore rewrite \eqref{DTMD-evolved} in the form
\begin{align}
  \label{DTMD-evolved-corr}
  & \prn{R_1 R_2}{F}_{a_1 a_2}(x_i,\tvec{z}_i,\tvec{y};\mu_i,\zeta_p)
\nonumber \\
  & \quad = \exp\,\biggl\{ \int_{\mu_{01}}^{\mu_1} \frac{d\mu}{\mu}\,
        \biggl[ \gamma_{a_1}(\mu) - \gamma_{K,a_1}(\mu)\,
          \log\frac{x_1 \sqrt{\zeta_p}}{\mu} \ms\biggr] +
        \pr{11}{K}_{a_1}(\tvec{z}_1;\mu_{01})
    \log\frac{\sqrt{x_1 x_2\ms \zeta_p}}{\sqrt{\zeta_0}}
\nonumber \\
  & \qquad \hspace{1.5em} + \int_{\mu_{02}}^{\mu_2} \frac{d\mu}{\mu}\,
        \biggl[ \gamma_{a_2}(\mu) - \gamma_{K,a_2}(\mu)\,
          \log\frac{x_2 \sqrt{\zeta_p}}{\mu} \ms\biggr] +
        \pr{11}{K}_{a_2}(\tvec{z}_2;\mu_{02})
    \log\frac{\sqrt{x_1 x_2\ms \zeta_p}}{\sqrt{\zeta_0}} \,\biggr\}
\nonumber \\
  & \qquad \times \sum_{\Rp_1 \Rp_2}
      \pr{R_1^{} R_2^{}\, \Rpbar_1 \Rpbar_2}{\exp}\,\biggl[ M_{a_1
            a_2}(\tvec{z}_i,\tvec{y})
          \log\frac{\sqrt{x_1 x_2\ms \zeta_p}}{\sqrt{\zeta_0}} \ms\biggr]
\nonumber \\[0.1em]
  & \qquad\quad \times
        \prn{\Rp_1 \Rp_2}{F}_{a_1 a_2}\bigl( x_i,\tvec{z}_i,\tvec{y};
         \mu_{01},\mu_{02},\zeta_0 / (x_1 x_2) \bigr)\,.
\end{align}
Correspondingly, the expression \eqref{W-generic} for the cross section level becomes
\begin{align}
  \label{W-generic-corr}
& W_{a_1 a_2 b_1 b_2} =
  \exp\,\biggl\{ \int_{\mu_{01}}^{\mu_1} \frac{d\mu}{\mu}\,
        \biggl[ \gamma_{a_1}(\mu) - \gamma_{K,a_1}(\mu)\,
          \log\frac{Q_1^2}{\mu^2} \biggr] +
        \pr{11}{K}_{a_1}(\tvec{z}_1;\mu_{01})
    \log\frac{Q_1 Q_2}{\zeta_0}
\nonumber \\
  & \quad \hspace{5.5em} + \int_{\mu_{02}}^{\mu_2} \frac{d\mu}{\mu}\,
        \biggl[ \gamma_{a_2}(\mu) - \gamma_{K,a_2}(\mu)\,
          \log\frac{Q_2^2}{\mu^2} \biggr] +
        \pr{11}{K}_{a_2}(\tvec{z}_2;\mu_{02})
    \log\frac{Q_1 Q_2}{\zeta_0} \,\biggr\}
\nonumber \\[0.5em]
  & \quad \times \Phi(\nu \tvec{y}_+)\, \Phi(\nu \tvec{y}_-)
    \sum_{R_1^{} R_2^{} \Rp_1 \Rp_2}
\etatwo{R_1^{}}{R_2^{}}\; \prn{\Rbar_1 \Rbar_2}{F}_{b_1 b_2}\bigl(\bar{x}_i,
      \tvec{z}_i,\tvec{y}; \mu_{01},\mu_{02},
      \zeta_0 / (\bar{x}_1 \bar{x}_2) \bigr)\,
\nonumber \\
  & \qquad \times
     \pr{R_1^{} R_2^{}, \Rpbar_1 \Rpbar_2}{\exp}\,
     \biggl[ M_{a_1 a_2}(\tvec{z}_i,\tvec{y})
          \log\frac{Q_1 Q_2}{\zeta_0} \,\biggr]\,
        \prn{\Rp_1 \Rp_2}{F}_{a_1 a_2}\bigl(x_i,\tvec{z}_i,\tvec{y};
            \mu_{01},\mu_{02}, \zeta_0 / (x_1 x_2) \bigr)\,.
\end{align}
Analogous variable substitutions should be made in equations \eqref{CS-TMD-sol}, \eqref{small-z-start}, \eqref{b-star-DPDs}, \eqref{small-yz-evolved} and \eqref{W-small-y}.
Appropriate modifications should also be made for the rapidity parameters in the last sentence of section~\ref{sec:TMD-UV-region} and throughout section~\ref{sec:reg_sub_simp}, including equations \eqref{TMD-evol}, \eqref{TMD-tw3-evol}, and \eqref{zeta-hat}.


\paragraph{DPDF evolution.}
The corrected DGLAP equation \eqref{DGLAP-zeta} for DPDFs is
\begin{align}
	\label{DGLAP-zeta-corr}
& \frac{\partial}{\partial \log\mu_1}\,
\prn{R_1 R_2}{F_{a_1 a_2}(x_1,x_2, \tvec{y};\mu_i,\zeta_p)}
\nonumber \\
&\qquad = 2 \sum_{b_1, \Rp_1}
   \pr{R_1^{} \Rpbar_1}{P_{a_1 b_1}(x_1';\mu_1^{},x_1^2\ms \zeta_p)}
   \conv{x_1}
   \prn{\Rp_1 R_2^{}}{F_{b_1 a_2}(x_1',x_2^{},\tvec{y};\mu_i,\zeta_p)} \,,
\end{align}
where the rapidity argument of the evolution kernel ${P}$ is rescaled in the same way as the anomalous dimensions in \eqref{RG-TMD-corr}, thus referring to the plus-momentum of the parton associated with the renormalization scale $\mu_1$.  An equation analogous to \eqref{DGLAP-zeta-corr} holds of course for the derivative w.r.t.\ $\log \mu_2$.

With the corrected DGLAP equations for DPDFs, equations \eqref{coll-zeta-expl} to \eqref{DGLAP-red-initial} must be modified as well.  A correct way to separate the $\zeta$ dependence is
\begin{align}
  \label{coll-zeta-expl-corr}
& \prn{R_1 R_2}{F_{a_1 a_2}(x_i,\tvec{y};\mu_i,\zeta_p)}
\nonumber \\[0.5em]
&\quad = \exp\,\biggl[\ms \prb{R_1}{J}(\tvec{y};\mu_i) \,
      \log\frac{\sqrt{\zeta_p}}{\sqrt{\zeta_0 \rule{0pt}{1.7ex}}}
   \ms\biggr]\;
   \prn{R_1 R_2}{F_{a_1 a_2}(x_i,\tvec{y};\mu_i,\zeta_0)}
\nonumber \\
&\quad
= \exp\,\biggl[\ms \prb{R_1}{J}(\tvec{y};\mu_i) \,
      \log\frac{\sqrt{\zeta_p}}{\sqrt{\zeta_0 \rule{0pt}{1.7ex}}}
   - \int_{\mu_{0}}^{\mu_1} \frac{d\mu}{\mu}\, \prn{R_1}{\gamma_J}(\mu)\,
     \log\frac{\sqrt{\zeta_0}}{\mu}
   - \int_{\mu_{0}}^{\mu_2} \frac{d\mu}{\mu}\, \prn{R_1}{\gamma_J}(\mu)\,
     \log\frac{\sqrt{\zeta_0}}{\mu} \ms\biggr]
\nonumber \\
&\qquad \times
   \prn{R_1 R_2}{\widehat{F}_{a_1 a_2,\, \mu_0,\zeta_0}(x_i,\tvec{y};\mu_i)} \,,
   \phantom{\frac{1}{1}}
\end{align}
where ${\widehat{F}}$ obeys the evolution equation
\begin{align}
	\label{DGLAP-red-corr}
& \frac{\partial}{\partial \log\mu_1}\, \prn{R_1 R_2}{\widehat{F}_{a_1 a_2,\,
            \mu_0,\zeta_0}(x_1,x_2,\tvec{y};\mu_1,\mu_2)}
\nonumber \\[0.2em]
& \qquad = - \;\prn{R_1}{\gamma_J}(\mu_1) \log x_1 \,
  \prn{R_1 R_2}{\widehat{F}_{a_1 a_2,\,
      \mu_0,\zeta_0}(x_1,x_2,\tvec{y};\mu_1,\mu_2)}
\nonumber \\[0.3em]
& \qquad\quad
  + 2 \sum_{b_1, \Rp_1}
  \pr{R_1^{} \Rpbar_1}{P_{a_1 b_1}(x_1';\mu_1^{},\mu_1^2)} \conv{x_1}
  \prn{\Rp_1 R_2^{}}{\widehat{F}_{b_1 a_2,\,
      \mu_0,\zeta_0}(x_1',x_2^{},\tvec{y};\mu_1,\mu_2)}
\end{align}
and its analogue for $\mu_2$, with the initial condition
\begin{align}
   \label{DGLAP-red-initial-corr}
& \prn{R_1 R_2}{\widehat{F}_{a_1 a_2,\,
      \mu_0,\zeta_0}(x_i,\tvec{y};\mu_0,\mu_0)}
 = \prn{R_1 R_2}{F_{a_1 a_2}(x_i,\tvec{y};\mu_0,\mu_0,\zeta_0)}
\nonumber \\
&\quad
= \exp\,\Bigl[\ms \prb{R_1}{J}(\tvec{y};\mu_0, \mu_0) \,
      \log \sqrt{x_1 x_2 \rule{0pt}{1.6ex}} \ms\Bigr] \,
  \prn{R_1 R_2}{F_{a_1 a_2}\bigl( x_i,\tvec{y};
      \mu_0,\mu_0,\zeta_0 /(x_1 x_2) \bigr)} \,.
\end{align}
In \eqref{coll-zeta-expl-corr} it is essential that $\zeta_0$ does \emph{not} depend on $x_1$ or $x_2$, so that the initial condition \eqref{DGLAP-red-initial-corr} is taken at constant $\zeta_p = \zeta_0$.  In the second line of that equation, we have indicated that this can of course be related to an initial condition at $\zeta_p = \zeta_0 /(x_1 x_2)$ by another step of Collins-Soper evolution.

Reflecting these changes, the last sentence in section \ref{sec:oneloop-nonsing} should be modified as follows:
``One can thus adapt numerical code for the one-loop evolution of colour singlet DPDs by rescaling the evolution kernels and adding the term with $\prn{R_1}{\gamma_J}$ from~\eqref{DGLAP-red-corr}.''


\paragraph{Matching DTMDs to DPDFs.}
The corrected master formula \eqref{full-match} for the matching of a DTMD at large $\tvec{y}$ reads
\begin{align}
	\label{full-match-corr}
& \prn{R_1 R_2}{F_{a_1 a_2}(x_i, \tvec{z}_i, \tvec{y};\mu_i,\zeta_p)}
  = \sum_{b_1 b_2, \Rp_1 \Rp_2}
    \prn{R_1^{} \Rpbar_1}{C_{a_1 b_1}(x_1\!\bs\smash{'},
      \tvec{z}_1;\mu_1, x_1^2 \ms \zeta_p)}
\nonumber \\[0.2em]
 & \qquad\qquad \underset{x_1}{\otimes}
    \prn{R_2^{} \Rpbar_2}{C_{a_2 b_2}(x_2\!\bs\smash{'},
      \tvec{z}_2;\mu_2, x_2^2 \ms \zeta_p)} \underset{x_2}{\otimes}
    \prn{\Rp_1 \Rp_2}{F_{b_1 b_2}(x_i\!\bs\smash{'},
         \tvec{y}; \mu_i,\zeta_p)} \,.
\end{align}
Corresponding substitutions for the rescaled rapidity parameters must be made in \eqref{CS-evolve-C} and \eqref{match-tensor}.  Moreover, equation \eqref{DGLAP-C} is incorrect and should be replaced with
\begin{align}
  \label{DGLAP-C-corr}
& \frac{\partial}{\partial \log\mu}
  \pr{R \Rpp}{C_{ac}(x,\tvec{z};\mu,\zeta)}
 = \gamma_{F,a}(\mu,\zeta) \, \pr{R \Rpp}{C_{ac}(x,\tvec{z};\mu,\zeta)}
\nonumber \\
& \qquad\qquad
   - 2 \sum_{b, \Rp} \; \int_x^1 \frac{d x'}{x'}\;
  \pr{R \Rpbar}{C_{ab}(x',\tvec{z};\mu,\zeta)} \;
  \prb{\Rp\bs \Rpp}{P_{b c}\biggl( \frac{x}{x'};
                  \mu,\frac{\zeta}{x'^{\ms 2}} \ms\biggr)}
  \,,
\end{align}
which involves a special rescaling of the rapidity parameter in the evolution kernel ${P}$.
The corrected form of \eqref{small-z-evolved} is
\begin{align}
	\label{small-z-evolved-corr}
 & \prn{R_1 R_2}{F_{a_1 a_2}(x_i,\tvec{z}_i,\tvec{y};\mu_i,\zeta_p)}
\nonumber \\[0.2em]
 &\quad = \exp\, \Biggl\{ \int_{\mu_{01}}^{\mu_1} \frac{d\mu}{\mu}\, \biggl[
\gamma_{a_1}(\mu) - \gamma_{K,a_1}(\mu)
\log\frac{x_1 \sqrt{\zeta_p}}{\mu} \ms\biggr] +
\pr{R_1}{K}_{a_1}(\tvec{z}_1;\mu_{01}) \log\frac{x_1 \sqrt{\zeta_p}}{\mu_{01}}
\nonumber \\
 & \qquad\quad\;\; + \int_{\mu_{02}}^{\mu_2} \frac{d\mu}{\mu}\,
\biggl[ \gamma_{a_2}(\mu) - \gamma_{K,a_2}(\mu)
\log\frac{x_2 \sqrt{\zeta_p}}{\mu} \ms\biggr] +
\pr{R_2}{K}_{a_2}(\tvec{z}_2;\mu_{02}) \log\frac{x_2 \sqrt{\zeta_p}}{\mu_{02}}
\,\Biggr\}
\nonumber \\[0.5em]
 & \qquad \times \sum_{b_1 b_2, \Rp_1 \Rp_2}
   \prn{R_1^{} \Rpbar_1}{C}_{a_1 b_1}(x_1',\tvec{z}_1^{};
      \mu_{01}^{},\mu_{01}^2)
   \underset{x_1}{\otimes}
   \prn{R_2^{} \Rpbar_2}{C}_{a_2 b_2}(x_2',\tvec{z}_2^{};
      \mu_{02}^{},\mu_{02}^2)
\nonumber \\[-0.3em]
 & \qquad\qquad \underset{x_2}{\otimes}
 \exp\, \biggl[ \prb{R_1}{J}(\tvec{y};\mu_{0i})
 \log\frac{\sqrt{x_1' x_2'\ms \zeta_p}}{\sqrt{\zeta_0}} \ms\biggr] \,
 \prn{\Rp_1 \Rp_2}{F_{b_1 b_2}\bigl( x_i',\tvec{y};
   \mu_{0i}^{},\zeta_0^{} / (x_1' x_2') \bigr)} \,.
\end{align}
This is quite similar to \eqref{DTMD-evolved-corr}, but now the initial condition for the rapidity parameter depends on the momentum fractions $x_1'$ and $x_2'$ of the DPD under the convolution integrals with the matching kernels ${C}$.  In \eqref{eq:colzeta} one should replace $\zeta$ with $\zeta_p$ and $\zeta_{0}$ with $\zeta_0 / (x_1' x_2')$ in the function arguments and in the explicit logarithm.  In the cross-section level result \eqref{W-large-y}, one should then replace the last line by
\begin{align}
 & \underset{x_2}{\otimes}
   \bigl[ \Phi(\nu \tvec{y}) \bigr]^2\, \exp\,\biggl[
   \prb{R_1}{J}(\tvec{y};\mu_{0i})\ms
      \log\frac{s \ms \sqrt{x_1' x_2' \ms \bar{x}_1' \bar{x}_2'}}{\zeta_0}
   \,\biggr]\,
\nonumber \\[0.2em]
 & \qquad \times
   \prn{\Rp_3 \Rp_4}{F_{d_1 d_2}\bigl( \bar{x}_i',\tvec{y};
      \mu_{0i}^{},\zeta_0^{} / (\bar{x}_1' \bar{x}_2') \bigr)}\,
   \prn{\Rp_1 \Rp_2}{F_{c_1 c_2}\bigl( x_i',\tvec{y};
      \mu_{0i}^{},\zeta_0^{}  / (x_1' x_2') \bigr)} \,,
\end{align}
where as in \eqref{small-z-evolved-corr} the constant $\zeta_0$ is rescaled by the momentum fractions of the DPDs under the convolution integrals with the matching kernels.

Given the above modifications regarding ${\widehat{F}_{a_1 a_2,\, \mu_0,\zeta_0}}$, we do not see a corrected form of \eqref{eq:coll_rap_sep} that would be particularly useful, so this equation and its discussion should be discarded.


\paragraph{Acknowledgements.}
We thank Florian Fabry, Jo Gaunt, Peter Pl\"o{\ss}l, and Lorenzo Zoppi for valuable feedback.


\end{document}